%% file: main.tex
\documentclass[phd,icsa,twoside,logo]{infthesis}

\usepackage{color, colortbl}
\usepackage{comment}
\usepackage{algorithm,algpseudocode}
\usepackage{amssymb}
\usepackage{rotating}
\usepackage{beramono} 
\usepackage{amsmath}
\usepackage{enumitem}
\usepackage{url}
\usepackage{cite}
\usepackage{tabularx} 
\usepackage{multirow}
\usepackage{stackengine} 
\usepackage{CJKutf8}
\usepackage{tikz}
\usepackage{subfigure}
\usepackage{smartdiagram}
\usepackage{longtable}
\usepackage[scr=boondoxo]{mathalfa}
\usepackage{hyperref}
\usepackage{pifont}
\usepackage{afterpage}
\newcommand{\Deoxys}{\includegraphics[scale=0.20, trim={0cm 0.85cm 0cm 1cm}]{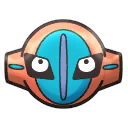}}%
\newcommand{\Sign}{\includegraphics[scale=0.12, trim={0cm 0.85cm 0cm 1cm}]{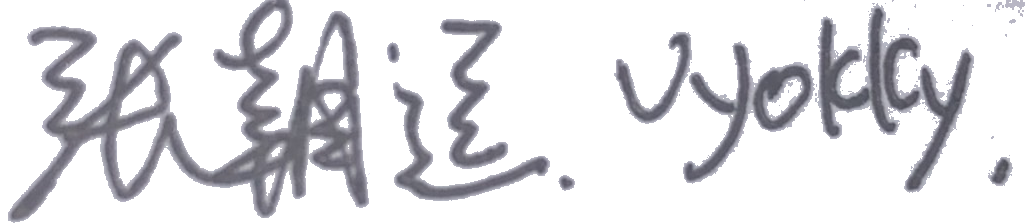}}
\definecolor{Gray}{gray}{0.8}

\algnewcommand{\Inputs}[1]{%
  \State \textbf{Inputs:}
  \Statex \hspace*{\algorithmicindent}\parbox[t]{.8\linewidth}{\raggedright #1}
}
\algnewcommand{\Initialize}[1]{%
  \State \textbf{Initialize:}
  \Statex \hspace*{\algorithmicindent}\parbox[t]{.8\linewidth}{\raggedright #1}
}

\definecolor{mediumaquamarine}{rgb}{0.4, 0.8, 0.67}
\definecolor{bittersweet}{rgb}{1.0, 0.44, 0.37}
\definecolor{sandybrown}{rgb}{0.96, 0.64, 0.38}
\definecolor{saffron}{rgb}{0.96, 0.77, 0.19}
\definecolor{salmonpink}{rgb}{1.0, 0.57, 0.64}
\definecolor{limegreen}{rgb}{0.2, 0.8, 0.2}
\definecolor{persianorange}{rgb}{0.85, 0.56, 0.35}
\definecolor{darkgray}{rgb}{0.66, 0.66, 0.66}

\newcommand{\nes}{\textsc{NES}\xspace}
\newcommand{\bd}{\textsc{Boundary}\xspace}
\newcommand{\pw}{\textsc{Pointwise}\xspace}
\newcommand{\hj}{\textsc{HopSkipJumpAttack}\xspace}
\newcommand{\opt}{\textsc{Opt-Attack}\xspace}

\newcommand{\cz}[1]{\textcolor{black}{#1}}
\newcommand{\rev}[1]{\textcolor{black}{#1}}
\newcommand{\edit}[1]{\textcolor{black}{#1}}
\newcommand{\rv}[1]{\textcolor{black}{#1}}

\newcommand{\eg}{{e.g.},\xspace}
\newcommand{\ie}{{i.e.},\xspace}
\newcommand{\etal}{{et al.},\xspace}
\newcommand{\etc}{{etc.}}
\newcommand{\dc}{$\mathcal{D}$-Conv\xspace} 
\DeclareMathOperator*{\argmax}{argmax} 
\newcommand{\oast}{\stackMath\mathbin{\stackinset{c}{0ex}{c}{0ex}{\ast}{\bigcirc}}}
\newcommand{\ostar}{\stackMath\mathbin{\stackinset{c}{0ex}{c}{0ex}{\star}{\bigcirc}}}

\newcommand{\name}{\textsc{Microscope}\xspace}
\newcommand{\tita}{\textsc{Tiki-Taka}\xspace}
\newcommand{\PreserveBackslash}[1]{\let\temp=\\#1\let\\=\temp}
\newcolumntype{C}[1]{>{\PreserveBackslash\centering}p{#1}}
\newcolumntype{R}[1]{>{\PreserveBackslash\raggedleft}p{#1}}
\newcolumntype{L}[1]{>{\PreserveBackslash\raggedright}p{#1}}

\title{Deep Neural Mobile Networking}
\author{Chaoyun Zhang}

\abstract{%
    The next generation of mobile networks is set to become increasingly complex, as these struggle to accommodate tremendous data traffic demands generated by ever-more connected devices that have diverse performance requirements in terms of throughput, latency, and reliability. This makes monitoring and managing the multitude of network elements intractable with existing tools and impractical for traditional machine learning algorithms that rely on hand-crafted feature engineering. In this context, embedding machine intelligence into mobile networks becomes necessary, as this enables systematic mining of valuable information from mobile big data and automatically uncovering correlations that would otherwise have been too difficult to extract by human experts. In particular, deep learning based solutions can automatically extract features from raw data, without human expertise. The performance of artificial intelligence (AI) has achieved in other domains draws unprecedented interest from both academia and industry in employing deep learning approaches to address technical challenges in mobile networks.
   
    This thesis attacks important problems in the mobile networking area from various perspectives by harnessing recent advances in deep neural networks. As a preamble, we bridge the gap between deep learning and mobile networking by presenting a survey on the crossovers between the two areas.  Secondly, we design dedicated deep learning architectures to forecast mobile traffic consumption at city scale. In particular, we tailor our deep neural network models to different mobile traffic data structures (i.e. data originating from urban grids and geospatial point-cloud antenna deployments) to deliver precise prediction. Next, we propose a mobile traffic super resolution (MTSR) technique to achieve coarse-to-fine grain transformations on mobile traffic measurements using generative adversarial network architectures. This can provide  insightful knowledge to mobile operators about mobile traffic distribution,  while effectively reducing the data post-processing overhead. Subsequently, the mobile traffic decomposition (MTD) technique is proposed to break the aggregated mobile traffic measurements into service-level time series, by using a deep learning based framework. With MTD, mobile operators can perform more efficient resource allocation for network slicing (i.e, the logical partitioning of physical infrastructure) and alleviate the privacy concerns that come with the extensive use of deep packet inspection. Finally, we study the robustness of network specific deep anomaly detectors with a realistic black-box threat model  and  propose reliable solutions for defending against attacks that seek to subvert existing network deep learning based intrusion detection systems (NIDS).
    
    Lastly, based on the results obtained, we identify important research directions that are worth pursuing in the future, including \emph{(i)} serving deep learning with massive high-quality data \emph{(ii)} deep learning for spatio-temporal mobile data mining \emph{(iii)} deep learning for geometric mobile data mining \emph{(iv)} deep unsupervised learning in mobile networks, and \emph{(v)} deep reinforcement learning for mobile network control. Overall, this thesis demonstrates that deep learning can underpin powerful tools that address data-driven problems in the mobile networking domain. With such intelligence, future mobile networks can be monitored and managed more effectively and thus higher user quality of experience can be guaranteed.
}

\begin{document}

\begin{preliminary}
   
\maketitle
\begin{laysummary}
Due to the ever-growing mobile traffic demands and increasing types of mobile services, the architectures of next generation mobile network inevitably become more complex. Therefore, managing the multitude of network elements becomes intractable, if merely relying on human-driven management mechanisms. Using machine learning to automate analysis and management of mobile network functions becomes promising. Triggered by recent advanced techniques of parallel computing, deep learning demonstrated its superior ability in automatic feature engineering, which traditionally requires human expertise. 

In this thesis, we attack a range of important problems in the mobile networking area from different perspectives, using deep learning approaches. As a preamble, we bridge the gap between deep learning and mobile networking by presenting a survey on their crossovers. 

We then examine the problem of mobile traffic forecasting, which is increasingly important for resource management, network slicing and public transportation support. We design dedicated deep learning architectures to forecast mobile traffic at the city scale tailored to different types of network deployments, to effectively extract spatio-temporal correlations and deliver precise prediction for both aggregated and service-level measurements. With our approaches, mobile operators can manage the mobile networks more effectively and thus offering better services to mobile users.

Mobile traffic engineering requires fine-grained knowledge of traffic distribution over space and time. Obtaining fine-grained traffic measurements is however costly, as it relies on dedicated probes and expensive data post-processing. Drawing inspiration from image super resolution, we propose a mobile traffic super resolution (MTSR) technique to complete coarse-to-fine-grain transformation on mobile traffic measurement using generative adversarial network (GAN) architectures. After training, the GAN can take the raw coarse measurements as input and infer precisely their fine-grained counterparts, to serve various applications. This significantly reduces the measurements overhead and delivers deeper insights to the mobile operators. 

Knowledge of traffic consumption at per mobile service level is important for network slicing purpose, \ie logical partitioning of the physical infrastructure among services with different requirements. However, current approaches heavily rely on deep packet inspection (DPI), which is difficult to deploy at large scale and has substantial privacy problems. To mitigate this issue, we propose the mobile traffic decomposition (MTD) technique to break the aggregated mobile traffic measurements into the service level measurement using a deep learning based frameworks called \name. \name can capture the spatio-temporal correlations in mobile traffic and deliver precise per-service traffic estimates, which can be an important complement of (DPI).

Deep neural networks are becoming increasingly popular for network intrusion detection systems, as they can achieve high detection accuracy with limited feature engineering. However, such sophisticated models are vulnerable to adversarial attacks, which can compromise the detectors by introducing dedicated subtle perturbation to time-based traffic features. We introduce \tita, to study the robustness of deep anomaly detectors against a realistic black-box adversarial attack threat model, and propose defense solutions which help minimize to risk of those attacks.

We lastly conclude this thesis and pinpoint future research directions which are promising to pursue. Overall, this thesis explores a new territory that empowers the future mobile networks with deep neural network based machine intelligence. 
\end{laysummary}

\begin{acknowledgements}
Time flies, as my Ph.D. journey passed in the blink of an eye. Tremendous thanks to my principal supervisor, Paul Patras, who constantly provided generous guidance, support and encouragement throughout the years, making my Ph.D. journey so meaningful, rewarding and unforgettable. It was a great pleasure to work with Paul, as he treated me more as a friend and brother, instead of a student. I feel so fortunate to had Paul as my supervisor. Without him, I could never complete this thesis.

Secondly, I would also like to thank my second supervisor Rik Sarkar, and to Subramanian Ramamoorthy who provided valuable feedback on my annual reviews. Their comments and suggestions helped me shape research ideas. Many thanks to collaborators through the years, including Xi Ouyang, Marco Fiore, Hamed Haddadi, Iain Muarry and Zongzuo Wang. Their insightful views, valuable advice and inspiration really improve my career. Thanks to Eiko Yoneki and Mirella Lapata, for examining my thesis and provide valuable feedback. Special thanks to Xavier Costa-Perez, who provided valuable guidance, supervision and help during my internship at NEC Laboratories Europe, Heidelberg.

Great gratitude towards our group members in ICSA, including Rui Li, Haoyu Liu, Yini Fang, Aboubacar Diallo, Rajkarn Singh, Valentin Radu, Abhirup Ghosh and Praveen Tammana. I really enjoy working and discussing with them, as they inspire my research and we improve each other. I also miss and thank other friends in the school: Jianpeng Cheng, Xingxing Zhang, Li Dong, Bowen Li, Yang Liu, Jiangming Liu, Xinnuo Xu and Mingcan Zhu. We have spent a lot of good time together, which I will never forget.

Special thanks to my parents, for their support, consideration and unconditional love. Many thanks to my girlfriend Xinru Cao, for her love and encouragement. I dedicate this thesis to my family. Lastly, great appreciation to the University of Edinburgh and Electronics and Telecommunications Research Institute (ETRI), South Korea, for providing scholarship and funding to support my study.
\end{acknowledgements}

 \begin{declaration}
   I declare that this thesis was composed by myself, that the work contained herein is my own except where explicitly stated otherwise in the text, and that this work has not been submitted for any other degree or professional qualification except as specified.
   Parts of this work have been published in or submitted to academic conferences and journals, including:
   
   \begin{enumerate}
   \item \textbf{C. Zhang}, M. Fiore, I. Murray and P. Patras. ``CloudLSTM: A Recurrent Neural Model for Spatiotemporal Point-cloud Stream Forecasting", (Technical report, Under review) \cite{zhang2019cloudlstm}.
    \item \textbf{C. Zhang} M. Fiore, C. Ziemlicki and P. Patras. ``Microscope: Mobile Service Traffic Decomposition for Network Slicing as a Service", (Technical report, Under review) \cite{zhang2019service}.
    \item \textbf{C. Zhang}, M. Fiore and P. Patras. ``Multi-Service Mobile Traffic Forecasting via Convolutional Long Short-Term Memories", IEEE International Symposium on Measurements and Networking, Catania, Italy, 2019 \cite{Patr1907:Multi}.
   \item \textbf{C. Zhang}, P. Patras, H. Haddadi.``Deep Learning in Mobile and Wireless Networking: A Survey", IEEE Communications Surveys \& Tutorials, 2019 \cite{zhang2018deep}.
    \item\textbf{C. Zhang}, P. Patras. ``Long-Term Mobile Traffic Forecasting Using Deep Spatio-Temporal Neural Networks",  in Proceeding of the Nineteenth International Symposium on Mobile Ad Hoc Networking and Computing (ACM MobiHoc) 2018, Los Angeles, USA \cite{zhang2018long}.
    \item\textbf{C. Zhang}, M. Zhong, Z. Wang, N. Goddard, C. Sutton. ``Sequence-to-Point Learning with Neural Networks for Nonintrusive Load Monitoring",  in Proceeding of the 32\textsuperscript{nd} AAAI Conference on Artificial Intelligence, New Orleans, USA, 2018 \cite{zhang2018sequence}.
    \item \textbf{C. Zhang}, X. Ouyang, P. Patras, ``ZipNet-GAN: Inferring Fine-grained Mobile Traffic Patterns via a Generative Adversarial Neural Network'', in Proceeding of the 13th International Conference on Emerging Networking Experiments and Technologies (ACM CoNEXT), Seoul/Incheon, South Korea, Dec. 2017 \cite{zhang2017zipnet}.
    \item \textbf{C. Zhang}, X. Costa-Perez, P. Patras, ``\tita: Attacking and Defending Deep Learning-based Intrusion Detection Systems'', (Technical report, Under review) \cite{zhang2019tikitaka}.
   \end{enumerate}
   
   \par 
  \vspace{0.2in}
   \begin{flushright}
   \Sign
   \end{flushright}
    \raggedleft({\em Chaoyun Zhang\/})
\end{declaration}

\begin{CJK*}{UTF8}{gbsn}
\dedication{$\mathscr{To~}$ $\mathscr{My~}$ $\mathscr{Family}$. \\
~~~~献给我的家人。\\
--~~$\mathscr{VYOKKY}$~\Deoxys~--
}
\end{CJK*}

\tableofcontents

\newpage
\begin{abbreviation}
{\large \centering 
\begin{longtable}[h!]{L{3.5cm}L{12cm}}
5G                 & 5th Generation Mobile System                       \\
A+                 & Adjusted Anchored Neighborhood Regression          \\
ACK                & Acknowledgment                                    \\
AE                 & Autoencoder                                        \\
AI                 & Artificial Intelligence                            \\
ARIMA              & Autoregressive Integrated Moving Average           \\
ASR                & Attack Success Rate                                \\
BN                 & Batch Normalization                                \\
BSC                & Base Station Controller                            \\
CNN/ConvNet        & Convolutional Neural Network                       \\
CDR                & Call Detail Record                                 \\
CE                 & Cross Entropy                                      \\
CloudGRU           & Convolutional Point Cloud Gated Recurrent Unit     \\
CloudLSTM          & Convolutional Point Cloud Long Short Term Memory   \\
CloudRNN           & Convolutional Point Cloud Recurrent neural Network \\
ConvLSTM           & Convolutional Long Short Term Memory               \\
C-RAN              & Centralized Radio Access Network                   \\
D2D                & Device-to-Device                                   \\
DAGGER             & Dataset Aggregation                                \\
DBN                & Deep Belief Network                                \\
DNN                & Deep Neural Network                                \\
DNS                & Domain Name System                                 \\
$\mathcal{D}$-Conv & Dynamic Convolution                                \\
DDoS               & Distributed Denial of Service                      \\
DefCNN             & Deformable Convoutional Neural Network             \\
DoS                & Denial of Service                                  \\
DPI                & Deep Packet Inspection                             \\
DRL                & Deep Reinforcement Learning                        \\
DSE                & Deep Similarity Encoder                            \\
D-STN              & Double Spatio-Temporal neural Network              \\
EAT                & Ensemble Adversarial Training                      \\
ExpS               & Exponential Smoothing                              \\
FGSM               & Fast Gradient Sign Method                          \\
FLOPs              & Floating Point Operations                          \\
GAN                & Generative Adversarial Network                     \\
GP                 & Gaussian Process                                   \\
GPU                & Graphic Processing Unit                            \\
GPRS               & General Packet Radio Services                      \\
GTP                & GPRS Tunneling Protocol user plane                 \\
IoT                & Internet of Things                                 \\
LSTM               & Long Short Term Memory                             \\
LReLU              & Leaky Rectified Linear Unit                        \\
HW-ExpS            & Holt-Winters Exponential Smoothing                 \\
IDS                & Intrusion Detection System                         \\
I-FGSM             & Iterative Fast Gradient Sign Method                \\
ISP                & Internet Service Provider                          \\
KaFFPa             & Karlsruhe Fast Flow Partitioning                   \\
KL                 & Kullback-Leibler                                   \\
MAC                & Media Access Control                               \\
MAE                & Mean Absolute Error                                \\
MANO               & Management and Orchestration                       \\
MAPE               & Mean Absolute Percentage Error                     \\
MEC                & Mobile Edge Computing                              \\
ML                 & Machine Learning                                   \\
MI-FGSM            & Momentum Iterative Fast Gradient Sign Method       \\
MLP                & Multilayer Perceptron                              \\
MIMO               & Multi-Input Multi-Output                           \\
MSE                & Mean Squared Error                                 \\
MTD                & Mobile Traffic Decomposition                       \\
MTSR               & Mobile Traffic Super Resolution                    \\
NFV                & Network Function Virtualization                    \\
NLP                & Natural Language Processing                        \\
NID                & Network Intrusion Detection                        \\
NIDS               & Network Intrusion Detection System                 \\
NMAE               & Normalized Mean Absolute Error                     \\
NN                 & Neural Network                                     \\
NRMSE              & Normalized Root Mean Square Error                  \\
NSP                & Network Service Provider                           \\
NVF                & Network Virtualization Function                    \\
ONOS               & Open Network Operating System                      \\
OSM                & Open Source Mano                                   \\
OTS                & Ouroboros Training Scheme                          \\
PGW                & Packet Gateway                                     \\
PR                 & Public Relations                                   \\
PSNR               & Peak Signal-to-Noise Ratio                         \\
QoE                & Quality of Experience                              \\
QoS                & Quality of Service                                 \\
RAN                & Radio Access Network                               \\
RAT                & Radio Access Technology                            \\
RBM                & Restricted Boltzmann Machine                       \\
ReLU               & Rectified Linear Unit                              \\
ResNet             & Residual Neural Network                            \\
RGF                & Random Gradient-Free                               \\
RMSE               & Root Mean Square Error                             \\
RNC                & Radio Network Controller                           \\
RNN                & Recurrent Neural Network                           \\
SC                 & Sparse Coding                                      \\
S-GW               & Service Gateway                                    \\
Seq2Seq            & Sequence-to-Sequence                               \\
SGD                & Stochastic Gradient Descent                        \\
SLA                & Service-Level Agreement                            \\
SNI                & Server Name Indication                             \\
SON                & Self-Organizing Network                            \\
SR                 & Super Resolution                                   \\
SRCNN              & Super Resolution Convolutional Neural Network      \\
SSIM               & Structural Similarity                              \\
STN                & Spatio-Temporal neural Network                     \\
SVM                & Support Vector Machine                             \\
TCP                & Transmission Control Protocol                      \\
TLS                & Transport Layer Security                           \\
t-SNE              & t-distributed Stochastic Neighbor Embedding        \\
VAE                & Variation Autoencoder                              \\
VIM                & Virtual Infrastructure Manager                     \\
VNF                & Virtual Network Function                           \\
ZipNet             & Zipper Network                                     \\
ZOO                & Zeroth Order Optimization     
\end{longtable}
}

\end{abbreviation}

\end{preliminary}

\include{chap1}

\include{chap2}
\include{chap3}
\include{chap4}
\include{chap5}
\include{chap6}
\include{chap7}
\include{chap8}


\bibliographystyle{unsrt}


\bibliography{main.bib}

\end{document}

%% file: chap1.tex
\chapter{Introduction}

Internet connected mobile devices are penetrating every aspect of individuals' life, work, and entertainment. The increasing number of smartphones and the emergence of evermore diverse applications trigger a surge in mobile data traffic. Indeed, the latest industry forecasts indicate that the annual worldwide IP traffic consumption will reach 4.8 zettabytes (10\textsuperscript{15} MB) by 2022, with smartphone traffic exceeding PC traffic one year earlier~\cite{cisco2017}. Given the shift in user preference towards wireless connectivity,
Internet Service Providers (ISPs) must develop \emph{intelligent} heterogeneous architectures and tools that can serve the 5\textsuperscript{th} generation of mobile systems (5G) and gradually meet more stringent end-user application requirements~\cite{agiwal2016next, gupta2015survey}. 

The growing diversity and complexity of mobile network architectures has made monitoring and managing the multitude of network elements intractable. Therefore, embedding versatile machine intelligence into future mobile networks is drawing unparalleled research interest~\cite{zheng2016big, jiang2017machine}. This trend is reflected in machine learning (ML) based solutions to problems ranging from radio access technology (RAT) selection~\cite{nguyen2017reinforcement} to malware detection~\cite{narudin2016evaluation}. 
ML enables systematic mining of valuable information from traffic data and automatically uncover correlations that would otherwise have been too complex to extract by human experts~\cite{anareport}. As the flagship of machine learning, deep learning has achieved remarkable performance in areas such as computer vision \cite{zhang2015convolutional} and natural language processing (NLP)~\cite{socher2012deep}. Networking researchers are also beginning to recognize the power and importance of deep learning, and are exploring its potential to solve problems specific to the mobile networking domain \cite{specialissue, wang2017machine}. 

\section{Research Challenges\label{sec:challenge}} 
5G mobile network infrastructures and services have already been deployed in many densely populated areas to support high-speed, low-latency and reliable wireless communication. Though recent progress appears promising, there remain several research challenges to be addressed. Specifically:

\begin{enumerate}
    \item The rapid uptake of mobile devices and the rising popularity of mobile applications and services pose unprecedented demands on mobile and wireless networking infrastructure. Upcoming 5G systems are evolving to support exploding mobile traffic volumes, real-time extraction of fine-grained analytics, and agile management of network resources, so as to maximize user experience. Fulfilling these tasks is challenging, as mobile environments are increasingly complex, heterogeneous, and evolving.
    \item Large-scale mobile traffic analytics is becoming essential to digital infrastructure provisioning, public transportation, events planning, and other domains. Monitoring city-wide mobile traffic is however a complex and costly process that relies on dedicated probes. Some of these probes have limited precision or coverage, others gather tens of gigabytes of logs daily, which independently offer limited insights. Extracting fine-grained patterns involves expensive spatial aggregation of measurements, storage, and post-processing.
    \item Network slicing aligns mobile network operation to this emerging context, as it allows operators to isolate and customize network resources on a \emph{per-service} basis. A key input for provisioning resources to slices is the real-time information about the traffic demands generated by individual services. Acquiring such knowledge is challenging: legacy approaches based on in-depth inspection of traffic streams have high computational costs, which inflate with the widening adoption of encryption over data and control traffic.
    \item Neural networks are becoming increasingly important in the development of network intrusion detection system (NIDS), as they have the potential to achieve high detection accuracy while requiring limited feature engineering. Deep learning-based detectors are however vulnerable to adversarial examples, via which attackers that are oblivious to the precise mechanics of the targeted NIDS aim to evade identification, by adding subtle perturbations to time-based traffic features. Defending against such black-box adversarial attacks remains a challenging task.
\end{enumerate}{}

Embedding deep learning into the 5G mobile and wireless networks is well justified. In particular, data generated by mobile environments are increasingly heterogeneous, as these are usually collected from various sources, have different formats, and exhibit complex correlations \cite{alsheikh2016mobile}. As a consequence, a range of specific problems become too difficult or impractical for traditional machine learning tools (e.g., shallow neural networks). This is because \emph{(i)} their performance does not improve if provided with more data~\cite{Goodfellow-et-al-2016} and \emph{(ii)} they cannot handle highly dimensional state/action spaces in control problems~\cite{mnih2015human}. In contrast, big data fuels the performance of deep learning, as it eliminates domain expertise and instead employs hierarchical feature extraction. In essence this means information can be distilled efficiently and increasingly abstract correlations can be obtained from the data, while reducing the pre-processing effort. Graphics Processing Unit (GPU)-based parallel computing further enables deep learning to make inferences within milliseconds. This facilitates network analysis and management with high accuracy and in a timely manner, overcoming the run-time limitations of traditional mathematical techniques (\eg convex optimization, game theory, meta heuristics).

\section{Advantages of Deep Learning in Mobile Networking\label{sec:adv}}

With the growth of complexity of future mobile network, we recognize several benefits of employing deep learning to handle and analyze the huge volume of associated mobile data generated per day, thereby improving the services of mobile networks. 
\begin{enumerate}
\item It is widely acknowledged that, while vital to the performance of traditional ML algorithms, feature engineering is costly \cite{Domingos:2012}. A key advantage of deep learning is that it can automatically extract high-level features from data that has complex structure and inner correlations. The learning process does not need to be designed by a human, which tremendously simplifies prior feature handcrafting~\cite{lecun2015deep}. The importance of this is amplified in the context of mobile networks, as mobile data is usually generated by heterogeneous sources, is often noisy, and exhibits non-trivial spatial/temporal patterns~\cite{alsheikh2016mobile}, whose labeling would otherwise require outstanding human effort.
\item Secondly, deep learning is capable of handling large amounts of data. Mobile networks generate high volumes of different types of data at fast pace. Training traditional ML algorithms (e.g., Support Vector Machine (SVM) \cite{tsang2005core} and Gaussian Process (GP)~\cite{rasmussen2006gaussian}) sometimes requires to store all the data in memory, which is computationally infeasible under big data scenarios. Furthermore, the performance of ML does not grow significantly with large volumes of data and plateaus relatively fast~\cite{Goodfellow-et-al-2016}. In contrast, Stochastic Gradient Descent (SGD) employed to train NNs only requires sub-sets of data at each training step, which guarantees deep learning's scalability with big data. Deep neural networks further benefit as training with big data prevents model over-fitting.

\item Traditional supervised learning is only effective when sufficient labeled data is available. However, most current mobile systems generate unlabeled or semi-labeled data~\cite{alsheikh2016mobile}. Deep learning provides a variety of methods that allow exploiting unlabeled data to learn useful patterns in an unsupervised manner, e.g., Variational Autoencoder (VAE)~\cite{pu2016variational}, Generative Adversarial Network (GAN)~\cite{goodfellow2014generative}. Applications include clustering~\cite{schroff2015facenet}, data distributions approximation~\cite{goodfellow2014generative}, un/semi-supervised learning~\cite{kingma2014semi, stewart2017label}, and one/zero shot learning~\cite{rezende2016one, socher2013zero}, among others. 

\item Compressive representations learned by deep neural networks can be shared across different tasks, while this is limited or difficult to achieve in other ML paradigms (e.g., linear regression, random forest, etc.). Therefore, a single model can be trained to fulfill multiple objectives, without requiring complete model retraining for different tasks. We argue that this is essential for mobile network engineering, as it reduces computational and memory requirements of mobile systems when performing multi-task learning applications~\cite{georgiev2017low}. 

\item \rev{Deep learning is effective in handing geometric mobile data \cite{monti2017geometric}, while this is a conundrum for other ML approaches. Geometric data refers to multivariate data represented by coordinates, topology, metrics and order \cite{le2004geometric}. Mobile data, such as mobile user location and network connectivity can be naturally represented by point clouds and graphs, which have important geometric properties. These data can be effectively modeled by dedicated deep learning architectures, such as PointNet++ \cite{qi2017pointnet} and Graph CNN \cite{kipf2016semi}. Employing these architectures has great potential to revolutionize the geometric mobile data analysis \cite{wang2018spatio}.}
\end{enumerate}
\rv{These advantages make deep learning a powerful tool for addressing problems specific to mobile networking, which have been either considered intractable previously or require significant effort if tackled with existing theoretical methods or heuristics.}

\section{\rev{Limitations of Deep Learning in Mobile Networking} \label{sec:limit}}
However, deep learning is not a silver bullet to every application in the mobile networking area, as it also has several shortcomings, which partially restricts its applicability in this domain. Specifically,
\begin{enumerate}
    \item \rev{In general, deep learning (including deep reinforcement learning) is vulnerable to adversarial examples \cite{nguyen2015deep, behzadan2017vulnerability}. These refer to artifact inputs that are intentionally designed by an attacker to fool machine learning models into making mistakes \cite{nguyen2015deep}. While it is difficult to distinguish such samples from genuine ones, they can trigger mis-adjustments of a model with high likelihood. Deep learning, especially CNNs are vulnerable to these types of attacks. This may also affect the applicability of deep learning in mobile systems. For instance, hackers may exploit this vulnerability and construct cyber attacks that subvert deep learning based detectors \cite{madani2018robustness}. Constructing deep models that are robust to adversarial examples is imperative, but remains challenging. }
    \item \rev{Deep learning algorithms are largely black boxes and have low interpretability. Their major breakthroughs are in terms of accuracy, as they significantly improve performance of many tasks in different areas. However, although deep learning enables creating ``machines'' that have high accuracy in specific tasks, we still have limited knowledge as of why NNs make certain decisions. This limits the applicability of deep learning, e.g. in network economics. Therefore, businesses would rather continue to employ statistical methods that have high interpretability, whilst sacrificing on accuracy. Researchers have recognized this problem and investing continuous efforts to address this limitation of deep learning (e.g. \cite{bau2017network, wu2017beyond, chakraborty2017interpretability}).}
    \item \rev{Deep learning is heavily reliant on data, which sometimes can be more important  than the model itself. Deep models can further benefit from training data augmentation \cite{perez2017effectiveness}. This is indeed an opportunity for mobile networking, as networks generates tremendous amounts of data. However, data collection may be costly, and face privacy concern, therefore it may be difficult to obtain sufficient information for model training. In such scenarios, the benefits of employing deep learning may be outweigh by the costs.}
    \item Deep learning can be computationally demanding. Advanced parallel computing (e.g. GPUs, high-performance chips) fostered the development and popularity of deep learning, yet deep learning also heavily relies on these. Deep NNs usually require complex structures to obtain satisfactory accuracy performance. However, when deploying NNs on embedded and mobile devices, energy and capability constraints have to be considered. Very deep NNs may not be suitable for such scenario and this would inevitably compromise accuracy. 
    \item \rev{Deep neural networks usually have many hyper-parameters and finding their optimal configuration can be difficult. For a single convolutional layer, we need to configure at least hyper-parameters for the number, shape, stride, and dilation of filters, as well as for the residual connections. The number of such hyper-parameters grows exponentially with the depth of the model and can highly influence its performance. Finding a good set of hyper-parameters can be similar to looking for a needle in a haystack. The AutoML platform\footnote{\rev{AutoML -- training high-quality custom machine learning models with minimum effort and machine learning expertise. \url{https://cloud.google.com/automl/}}} provides a first solution to this problem, by employing progressive neural architecture search \cite{liu2017progressive}. This task, however, remains costly.}
\end{enumerate}
These pitfalls need to be avoided when deploying deep learning in real mobile network systems.

\section{Thesis Contribution}
This thesis attacks important problems in the mobile networking domain using deep learning approaches from different perspectives. In particular, we harness the unique advantages of deep learning and tailor those to individual mobile networking problem, which leads the following contributions:

\subsection{Survey on Deep Learning in Mobile Networking}

First, we bridge the gap between deep learning and mobile and wireless networking research, by a survey of the crossovers between the two areas. We provide a review of mobile and wireless networking research based on deep learning, categorized by different applications domains. Part of this work has been published in:
\begin{enumerate}
    \item \textbf{C. Zhang}, P. Patras and H. Haddadi.``Deep Learning in Mobile and Wireless Networking: A Survey", IEEE Communication Survey \& Tutorial, 2019.
\end{enumerate}

\subsection{Long-term Mobile Traffic Forecasting on City Grids}
Forecasting with high accuracy the volume of data traffic that mobile users will consume is becoming increasingly important for precision traffic engineering, demand-aware network resource allocation, as well as public transportation. Measurements collection in dense urban deployments is however complex and expensive, and the post-processing required to make predictions is highly non-trivial, given the intricate spatio-temporal variability of mobile traffic due to user mobility. 

Second, we harness the exceptional feature extraction abilities of deep learning and propose a \underline{S}patio-\underline{T}emporal neural \underline{N}etwork (STN) architecture purposely designed for \emph{precise network-wide mobile traffic forecasting} over city grids. We present a mechanism that fine tunes the STN and enables its operation with only limited ground truth observations. We then introduce a Double STN technique (D-STN), which uniquely combines the STN predictions with historical statistics, thereby making faithful \emph{long-term} mobile traffic projections. 

Experiments we conduct with real-world mobile traffic datasets, collected over 60 days in both urban and rural areas, demonstrate that the proposed (D-)STN schemes perform up to 10-hour long predictions with remarkable accuracy, irrespective of the time of day when they are triggered. Specifically, \rv{our solutions achieve up to 61\% smaller prediction errors as compared to widely used forecasting approaches, while operating with up to 600 times shorter measurement intervals.} Part of this work has been published in:
\begin{enumerate}
    \item\textbf{C. Zhang} and P. Patras. ``Long-Term Mobile Traffic Forecasting Using Deep Spatio-Temporal Neural Networks",  in Proceeding of the Nineteenth International Symposium on Mobile Ad Hoc Networking and Computing (ACM MobiHoc) 2018, Los Angeles, USA.
\end{enumerate}

\subsection{Multi-service Mobile Traffic Forecasting on Point-cloud Antennas}
The third contribution of the thesis attacks the multi-service mobile traffic forecasting problem on geospatial point-cloud structural antennas, where traditional CNN-based neural networks cannot be directly employed. This is distinct from the forecasting over grids in the previous contribution. To handle the point cloud data structure, we introduces CloudLSTM, a new branch of recurrent neural network models tailored to forecasting over data streams generated by geospatial point-cloud sources. We design a Dynamic Convolution (\dc) operator as the core component of CloudLSTMs, which allows performing convolution operations directly over point-clouds and extracts local spatial features from sets of neighboring points that surround different elements of the input. This maintains the permutation invariance of sequence-to-sequence learning frameworks, while enabling learnable neighboring correlations at each time step -- an important aspect in spatio-temporal predictive learning. The \dc operator resolves the grid-structural data requirements of existing spatio-temporal forecasting models (\eg ConvLSTM) and can be easily plugged into traditional LSTM architectures with sequence-to-sequence learning and attention mechanisms. We perform antenna-level forecasting of the data traffic generated by mobile services, demonstrating that the proposed CloudLSTM achieves state-of-the-art performance with measurement datasets collected in operational metropolitan-scale mobile network deployments on multi-service mobile traffic forecasting, \rv{by achieving up to 45.9\% lower prediction errors than other deep learning baselines}. Part of this work has been published or is under review, as follows:
\begin{enumerate}
    \item\textbf{C. Zhang}, M. Fiore, I. Murray and P. Patras. ``CloudLSTM: A Recurrent Neural Model for Spatiotemporal Point-cloud Stream Forecasting", (Under review).
   \item \textbf{C. Zhang}, M. Fiore and P. Patras. ``Multi-Service Mobile Traffic Forecasting via Convolutional Long Short-Term Memories", IEEE International Symposium on Measurements and Networking, Catania, Italy, 2019.
\end{enumerate}

\subsection{Mobile Traffic Super Resolution}
Next, we propose \cz{an original mobile traffic super-resolution technique} that overcomes these problems by inferring narrowly localized traffic consumption from coarse measurements. We draw inspiration from image processing and design a deep-learning architecture tailored to mobile networking, which combines \underline{Zip}per \underline{Net}work (ZipNet) and Generative Adversarial neural Network (GAN) models. This enables to uniquely capture spatio-temporal relations between traffic volume snapshots routinely monitored over broad coverage areas (`low-resolution') and the corresponding consumption at 0.05 km\textsuperscript{2} level (`high-resolution') usually obtained after intensive computation. 

\rv{Experiments we conduct with a real-world dataset demonstrate that the proposed ZipNet(-GAN) infers traffic consumption with up to 100$\times$ higher granularity as compared to standard probing, irrespective of the coverage and the position of the probes. Importantly, our solutions outperform existing traditional and deep-learning based interpolation methods, as we achieve up to 78\% lower reconstruction errors, 40\% higher fidelity of reconstructed traffic patterns, and improve the structural similarity by 36.4$\times$. To our knowledge, this is the first time super-resolution concepts are applied to large-scale mobile traffic analysis and our solution is the first to infer fine-grained urban traffic patterns from coarse aggregates.} Part of this work has been published in:
\begin{enumerate}
   \item \textbf{C. Zhang}, X. Ouyang, P. Patras, ``ZipNet-GAN: Inferring Fine-grained Mobile Traffic Patterns via a Generative Adversarial Neural Network", in Proceeding of the 13th International Conference on Emerging Networking Experiments and Technologies (ACM CoNEXT), Seoul/Incheon, South Korea, Dec. 2017.
\end{enumerate}

\subsection{Mobile Traffic Decomposition}
We present an \cz{original approach to service-level demand estimation for network slicing}, which hinges on \textit{decomposition}, \ie the inference of per-service demands from traffic aggregates. By operating on total traffic volumes only, our approach overcomes the complexity and limitations of legacy traffic classification techniques, and provides an output suitable for recent `Network Slice as a Service' models.
We implement decomposition through \name, a dedicated framework based on novel 3D Deformable Convolutional Neural Networks (3D-DefCNNs) designed to handle spatial distortion in the input data due to irregular radio access deployment and coverage, and to exploit hidden spatio-temporal features in traffic aggregates. Experiments with metropolitan-scale measurements collected in an operational network demonstrate that \rv{\name accurately infers per-service traffic demands, with estimation errors below 1.2\%.} Further, we offer a practical perspective on the performance of \name, showing that resource allocations based on decomposition bear affordable additional costs for the operator, compared to the ideal case where  perfect knowledge of per-service traffic is available. Part of this work has been published or is under review, as follows:

\begin{enumerate}
    \item \textbf{C. Zhang} M. Fiore, C. Ziemlicki and P. Patras. ``Microscope: Mobile Service Traffic Decomposition for Network Slicing as a Service", (Under review).
    \item\textbf{C. Zhang}, M. Zhong, Z. Wang, N. Goddard, C. Sutton. ``Sequence-to-Point Learning with Neural Networks for Nonintrusive Load Monitoring",  in Proceeding of the 32\textsuperscript{nd} AAAI Conference on Artificial Intelligence, New Orleans, USA, 2018.
\end{enumerate}

\subsection{Attacking and Defending Deep Learning-based Intrusion Detection Systems}
We then introduce \tita, a framework for \emph{(i)} assessing the robustness of state-of-the-art deep learning-based Network Intrusion Detection Systems (NIDS) against black-box adversarial manipulations, and which \emph{(ii)} incorporates defense mechanisms that we propose to increase the resistance to attacks employing such evasion techniques. Specifically, we select five different cutting-edge black-box adversarial attack mechanisms to subvert three popular malicious traffic detectors that employ neural networks. We experiment with two publicly available datasets and consider both one-to-all and one-to-one classification scenarios, \ie discriminating illicit vs benign traffic, and respectively identifying a specific type of anomalous traffic among many considered. \rv{The results obtained suggest that attackers can fool NIDS with up to 35.7\% success rates}, by only altering the time-based features of the traffic generated. To counteract these weaknesses, we propose three defense mechanisms, namely model voting ensembling, ensembling adversarial training, and query detection. We demonstrate that when employing our proposed methods, \rv{the reinforced NIDS can defend against adversarial attacks with nearly 100\% success rates, with most types of malicious traffic}. This confirms their effectiveness and makes the case for adoption in more robust and reliable deep anomaly detectors. \rv{To the best of our knowledge, our work is the first to propose defenses against adversarial attacks targeting NIDS.} The outcome of the research has been summarized into a paper as follows:

\begin{enumerate}
    \item \textbf{C. Zhang}, X. Costa-Perez and  P. Patras . ``\tita: Attacking and Defending Deep Learning-based Intrusion Detection Systems", (Under review).
\end{enumerate}

\section{Thesis Organization}
We show the high-level overview of the thesis in Fig.\ref{fig:skeleton}, and organize the rest of the thesis as follows:

\begin{figure}[htb]
\begin{center}
\includegraphics[width=1.05\textwidth]{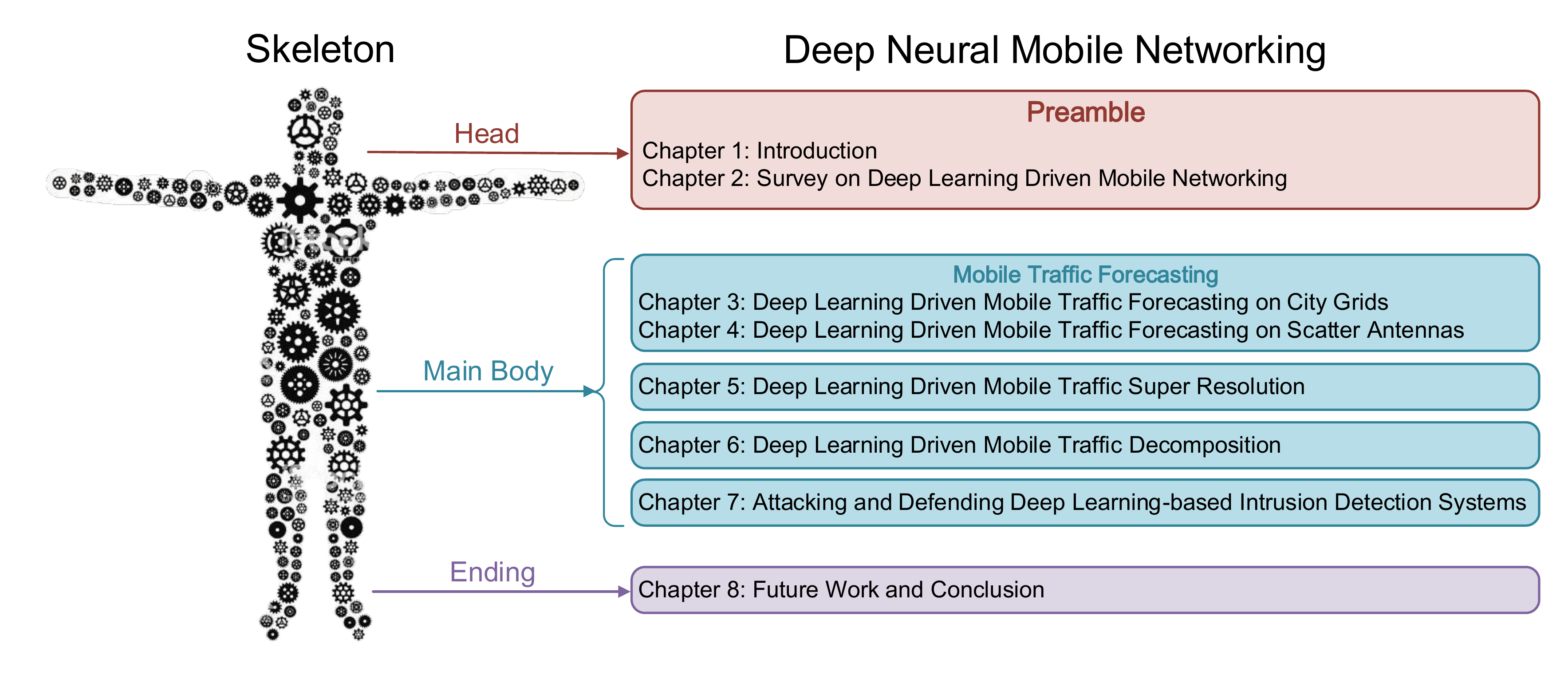}
\end{center}
\caption{\label{fig:skeleton} The skeleton of the thesis.}
\end{figure}

\textbf{Chapter \ref{chap:sur}} surveys deep learning applications in mobile networking, which can be regarded as a preamble of the thesis.

\textbf{Chapter \ref{chap:fore1}} focuses on using a dedicated deep neural network architecture to perform long-term aggregated mobile traffic forecasting at city scale on grid structural data. We start with one-step forecasting, then present two important approaches to extend the reliable predicting duration. With our approaches, we further demonstrate that the employed architecture can be trained on dataset collected in one city, while generalizing well on a different area which has different traffic scale and spatial topology.

\textbf{Chapter \ref{chap:fore2}} investigates the mobile traffic forecasting in multi-service scenarios in the network slicing context. The employed dataset is collected at the antenna level, which are non-uniformly distributed over the city. This form sequences of geospatial point clouds, which differ from the grid structural format handled in Chapter \ref{chap:fore1}. We develop a novel neural model named CloudLSTM tailored to the point cloud data to perform multi-service mobile traffic forecasting. We show that our CloudLSTM can deliver precise prediction and outperform state-of-the-art approaches.

\textbf{Chapter \ref{chap:mtsr}} introduces a mobile traffic super resolution (MTSR) approach to reduce the cost of fine-grained mobile traffic measurements by performing coarse-to-fine grained transformation using a generative adversarial network (GAN). By training the generator and discriminator iteratively, our method can improve the granularity of traffic measurement by up to 100$\times$ with high accuracy, while remaining robust to anomalous traffic.

\textbf{Chapter \ref{chap:decompose}} presents an original mobile traffic decomposition (MTD) technique to perform service-level demand estimation given aggregated traffic using deep neural network architectures. This overcomes the complexity and limitations of traditional traffic classification techniques, \ie deep packet inspection. We test our proposal at different network levels at metropolitan-scale measurements with 10 popular mobile services, and demonstrate its superior effectiveness and efficiency. 

\textbf{Chapter \ref{chap:tita}} studies the mobile network security with a realistic black-box threat model. We introduce \tita, to first \emph{(i)} evaluate the robustness of deep learning NIDS against black-box adversarial attacks, and \emph{(ii)} propose defense mechanisms to protect them from those adversarial intrusion. By employing our proposed defense methods, the successful rates of each adversarial attack drop to nearly 0\% on most types of malicious traffic. This enables to construct more robust and reliable deep anomaly detectors against adversarial attacks.

\textbf{Chapter \ref{chap:conclude}} concludes this thesis, and pinpoints several open research issues and promising directions, which may lead to valuable future research results.

%% file: chap2.tex
\chapter{Survey of Deep Learning in Mobile Networking\label{chap:sur}}

As a preamble, this chapter fills the gap between deep learning and mobile and wireless networking, by presenting an up-to-date survey of research that lies at the intersection between these two fields.  Specifically, 
\begin{enumerate}
\item \textbf{High-level Articles of Deep Learning in Mobile Networking} reviews survey, tutorial and magazine papers that discuss deep learning and mobile networking in a bigger pictures, without touching a specific problem.
\item \textbf{Deep Learning Driven Network-Level Mobile Data Analysis} focuses on deep learning applications built on mobile big data collected within the network, including network prediction, traffic classification, and Call Detail Record (CDR) mining.
\item \textbf{Deep Learning Driven Network Security} presents work that leverages deep learning to improve  network security, which we cluster by focus as infrastructure, software, and privacy related.

\end{enumerate} 

\section{Related High-level Articles}\label{sec:related}
Mobile networking and deep learning problems have been researched mostly independently. Only recently crossovers between the two areas have emerged. Several notable works paint a comprehensives picture of the deep learning and/or mobile networking research landscape. We categorize these works into \emph{(i)} pure overviews of deep learning techniques, \emph{(ii)} reviews of analyses and management techniques in modern mobile networks, and \emph{(iii)} reviews of works at the intersection between deep learning and computer networking. We discuss the most representative publications in each class. \\ 

\noindent\textbf{Overviews of Deep Learning and its Applications}
The era of big data is triggering wide interest in deep learning across different research disciplines~\cite{chen2014big, najafabadi2015deep, gheisari2017survey, hordri2017systematic} and a growing number of surveys and tutorials are emerging (e.g. \cite{deng2014deep, deng2014tutorial}). LeCun \emph{et al.} give a milestone overview of deep learning, introduce several popular models, and look ahead at the potential of deep neural networks \cite{lecun2015deep}. 
Schmidhuber undertakes an encyclopedic survey of deep learning, likely the most comprehensive thus far, covering the evolution, methods, applications, and open research issues~\cite{schmidhuber2015deep}. Liu \emph{et al.} summarize the underlying principles of several deep learning models, and review deep learning developments in selected applications, such as speech processing, pattern recognition, and computer vision \cite{liu2017survey}. 

Arulkumaran \emph{et al.} present several architectures and core algorithms for deep reinforcement learning, including deep Q-networks, trust region policy optimization, and asynchronous advantage actor-critic~\cite{kai2017brief}. Their survey highlights the remarkable performance of deep neural networks in different control problem (e.g., video gaming, Go board game play, etc.). Similarly, deep reinforcement learning has also been surveyed in~\cite{li2017deeprl}, where the authors shed more light on applications. Zhang \emph{et al.} survey developments in deep learning for recommender systems~\cite{zhang2017deep}, which have potential to play an important role in mobile advertising. As deep learning becomes increasingly popular, Goodfellow \emph{et al.} provide a comprehensive tutorial of deep learning in a book that covers prerequisite knowledge, underlying principles, and popular applications~\cite{Goodfellow-et-al-2016}. \\

\noindent\textbf{Surveys on Future Mobile Networks}
The emerging 5G mobile networks incorporate a host of new techniques to overcome the performance limitations of current deployments and meet new application requirements. Progress to date in this space has been summarized through surveys, tutorials, and magazine papers (e.g. \cite{andrews2014will, agiwal2016next, gupta2015survey, panwar2016survey, mao2017survey}).  
Andrews \emph{et al.} highlight the differences between 5G and prior mobile network architectures, conduct a comprehensive review of 5G techniques, and discuss research challenges facing future developments\edit{\cite{andrews2014will}}. Agiwal \emph{et al.} review new architectures for 5G networks, survey emerging wireless technologies, and point out research problems that remain unsolved~\cite{agiwal2016next}. Gupta \emph{et al.} also review existing work on 5G cellular network architectures, subsequently proposing a framework that incorporates networking ingredients such as Device-to-Device (D2D) communication, small cells, cloud computing, and the IoT \cite{gupta2015survey}.

Intelligent mobile networking is becoming a popular research area and related work has been reviewed in the literature (e.g. \cite{jiang2017machine, buda2016can, keshavamurthy2016conceptual, alsheikh2014machine, li2017intelligent, bkassiny2013survey, bui2017survey, valente2017survey}). Jiang \emph{et al.} discuss the potential of applying machine learning to 5G network applications including massive MIMO and smart grids~\cite{jiang2017machine}. This work further identifies several research gaps between ML and 5G that remain unexplored. Li \emph{et al.} discuss opportunities and challenges of incorporating artificial intelligence (AI) into future network architectures and highlight the significance of AI in the 5G era~\cite{li2017intelligent}. Klaine \emph{et al.} present several successful ML practices in Self-Organizing Networks (SONs), discuss the pros and cons of different algorithms, and identify future research directions in this area~\cite{valente2017survey}. \edit{Potential exists to apply AI and exploit big data for energy efficiency purposes~\cite{wu2016big}. Chen \emph{et al.} survey traffic offloading approaches in wireless networks, and propose a novel reinforcement learning based solution \cite{chen2015energy}. This opens a new research direction toward embedding machine learning towards greening cellular networks.}

\noindent\textbf{Deep Learning Driven Networking Applications}
A growing number of papers survey recent works that bring deep learning into the computer networking domain. 
Alsheikh \emph{et al.} identify benefits and challenges of using big data for mobile analytics and propose a Spark based deep learning framework for this purpose~\cite{alsheikh2016mobile}. Wang and Jones discuss evaluation criteria, data streaming and deep learning practices for network intrusion detection, pointing out research challenges inherent to such  applications~\cite{wang2017big}. Zheng \emph{et al.} put forward a big data-driven mobile network optimization framework in 5G networks, to enhance QoE performance~\cite{zheng2016big}. More recently, Fadlullah \emph{et al.} deliver a survey on the progress of deep learning in a board range of areas, highlighting its potential application to network traffic control systems \cite{fadlullah2017state}. Their work also highlights several unsolved research issues worthy of future study. 

Ahad \emph{et al.} introduce techniques, applications, and guidelines on applying neural networks to wireless networking problems~\cite{ahad2016neural}. Despite several limitations of neural networks identified, this chapter focuses largely on old neural networks models, ignoring recent progress in deep learning and successful applications in current mobile networks. 
Lane \emph{et al.} investigate the suitability and benefits of employing deep learning in mobile sensing, and emphasize on the potential for accurate inference on mobile devices~\cite{lane2015can}.
Ota \emph{et al.} report novel deep learning applications in mobile multimedia. Their survey covers state-of-the-art deep learning practices in mobile health and well-being, mobile security, mobile ambient intelligence, language translation, and speech recognition. Mohammadi \emph{et al.} survey recent deep learning techniques for Internet of Things (IoT) data analytics \cite{mohammadi2018deep}. They overview comprehensively existing efforts that incorporate deep learning into the IoT domain and shed light on current research challenges and future directions. \edit{Mao \emph{et al.} focus on deep learning in wireless networking~\cite{mao2018deep}. Their work surveys state-of-the-art deep learning applications in wireless networks, and discusses research challenges to be solved in the future.}

\section{Deep Learning Driven Network-level Mobile Data Analysis}\label{sec:netdata}
Notably, The development of mobile technology (e.g. smartphones, augmented reality, etc.) are forcing mobile operators to evolve mobile network infrastructures. As a consequence, both the cloud and edge side of mobile networks are becoming increasingly sophisticated to cater for users who produce and consume huge amounts of mobile data daily. These data can be either generated by the sensors of mobile devices that record individual user behaviors (app-level), or from the mobile network infrastructure (network-level), which reflects dynamics in urban environments. Appropriately mining these data can benefit multidisciplinary research fields and the industry in areas such mobile network management, social analysis, public transportation, personal services provision, and so on~\cite{cheng2017exploiting}. Network operators, however, could become overwhelmed when managing and analyzing massive amounts of heterogeneous mobile data \cite{ahmed2018recent}. Deep learning is probably the most powerful methodology that can overcoming this burden.

Network-level mobile data refers broadly to logs recorded by Internet service providers, including infrastructure metadata, network performance indicators and call detail records (CDRs). The recent remarkable success of deep learning ignites global interests in exploiting this methodology for mobile network-level data analysis, so as to optimize mobile networks configurations, thereby improving end-uses' QoE. These work can be categorized into four types: network state prediction, network traffic classification, CDR mining and radio analysis. In what follows, we review work in these directions.\\

\noindent \textbf{Network State Prediction} refers to inferring mobile network traffic or performance indicators, given historical measurements or related data. Pierucci and Micheli investigate the relationship between key objective metrics and QoE \cite{pierucci2016neural}. They employ MLPs to predict users' QoE in mobile communications, based on average user throughput, number of active users in a cells, average data volume per user, and channel quality indicators, demonstrating high prediction accuracy. Network traffic forecasting is another field where deep learning is gaining importance. By leveraging sparse coding and max-pooling, Gwon and Kung develop a semi-supervised deep learning model to classify received frame/packet patterns and infer the original properties of flows in a WiFi network \cite{gwon2014inferring}. Their proposal demonstrates superior performance over traditional ML techniques. Nie~\emph{et al.} investigate the traffic demand patterns in wireless mesh network \cite{nie2017network}. They design a DBN along with Gaussian models to precisely estimate traffic distributions. 

\begin{figure}[htb]
\begin{center}
\includegraphics[width=\columnwidth]{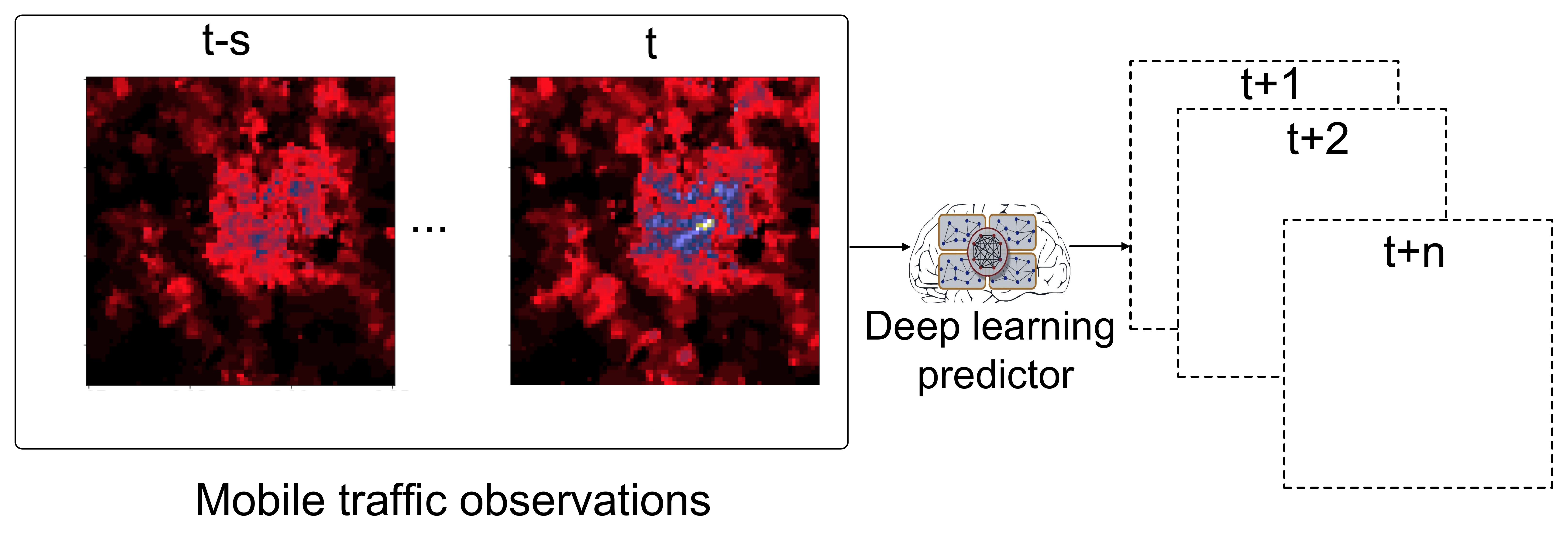}
\end{center}
\caption{\label{fig:forecasting} \rev{The underlying principle of city-scale mobile traffic forecasting. The deep learning predictor takes as input a sequence of mobile traffic measurements in a region (snapshots $t-s$ to $t$), and forecasts how much mobile traffic will be consumed in the same areas in the future $t+1$ to $t+n$ instances.} }
\end{figure}

\rev{In addition to the above, several researchers employ deep learning to forecast mobile traffic at city scale, by considering spatio-temporal correlations of geographic mobile traffic measurements. We illustrate the underlying principle in Fig.~\ref{fig:forecasting}.} In \cite{wangspatiotemporal}, Wang \emph{et al.} propose to use an AE-based architecture and LSTMs to model spatial and temporal correlations of mobile traffic distribution, respectively. In particular, the authors use a global and multiple local stacked AEs for spatial feature extraction, dimension reduction and training parallelism. Compressed representations extracted are subsequently processed by LSTMs, to perform final forecasting. Experiments with a real-world dataset demonstrate superior performance over SVM and the Autoregressive Integrated Moving Average (ARIMA) model.  \edit{Deep learning was also employed in \cite{huang2017study, alawe2018improving, feng2018deeptp} and \cite{chen2018deep0}, where the authors employ CNNs and LSTMs to perform mobile traffic forecasting. By effectively extracting spatio-temporal features, their proposals gain significantly higher accuracy than traditional approaches, such as ARIMA.} \rev{Wang \emph{et al.} represent spatio-temporal dependencies in mobile traffic using graphs, and learn such dependencies using Graph Neural Networks \cite{wang2018spatio}. Beyond the accurate inference achieved in their study, this work also demonstrates potential for precise social events inference.}


\noindent \textbf{Traffic Classification}
is aimed at identifying specific applications or protocols among the traffic in networks. Wang recognizes the powerful feature learning ability of deep neural networks and uses \rev{a deep AE} to identify protocols in a TCP flow dataset, achieving excellent precision and recall rates~\cite{wang2015applications}. Work in~\cite{wang2017end} proposes to use a 1D CNN for encrypted traffic classification. The authors suggest that this structure works well for modeling sequential data and has lower complexity, thus being promising in addressing the traffic classification problem. Similarly, Lotfollahi \emph{et al.} present Deep Packet, which is based on a CNN, for encrypted traffic classification \cite{lotfollahi2017deep}. Their framework reduces the amount of hand-crafted feature engineering and achieves great accuracy. \rev{An improved stacked AE is employed in~\cite{8553650}, where Li \emph{et al.} incorporate Bayesian methods into AEs to enhance the inference accuracy in network traffic classification.} \edit{More recently, Aceto \emph{et al.} employ MLPs, CNNs, and LSTMs to perform encrypted mobile traffic classification \cite{aceto2018mobile}, arguing that deep NNs can automatically extract complex features present in mobile traffic. As reflected by their results, deep learning based solutions obtain superior accuracy over RFs in classifying Android, IOS and Facebook traffic.}
CNNs have also been used to identify malware traffic, where work in \cite{wang2017malware} regards traffic data as images and unusual patterns that malware traffic exhibit are classified by representation learning. Similar work on mobile malware detection will be further discussed in subsection \ref{sec:security}. 
\\

\noindent\textbf{CDR Mining} involves extracting knowledge from specific instances of telecommunication transactions such as phone number, cell ID, session start/end time, traffic consumption, etc. Using deep learning to mine useful information from CDR data can serve a variety of functions. For example, Liang \emph{et al.} propose Mercury to estimate metro density from streaming CDR data, using RNNs \cite{liang2016mercury}. They take the trajectory of a mobile phone user as a sequence of locations; RNN-based models work well in handling such sequential data. Likewise, Felbo \emph{et al.} use CDR data to study demographics \cite{felbo2016using}. They employ a CNN to predict the age and gender of mobile users, demonstrating the superior accuracy of these structures over other ML tools. More recently, Chen \emph{et al.} compare different ML models to predict tourists' next locations of visit by analyzing CDR data \cite{chen2017comprehensive}. Their experiments suggest that RNN-based predictors significantly outperform traditional ML methods, including Naive Bayes, SVM, RF, and MLP.\\

\textbf{Lessons learned}: Network-level mobile data, such as mobile traffic, usually involves essential spatio-temporal correlations. These correlations can be effectively learned by CNNs and RNNs, as they are specialized in modeling spatial and temporal data (e.g., images, traffic series). An important observation is that large-scale mobile network traffic can be processed as sequential snapshots, as suggested in \cite{zhang2017zipnet, zhang2018long}, which resemble images and videos \cite{ouyang20193d, zhang2018driver}. Therefore, potential exists to exploit image processing techniques for network-level analysis. Techniques previously used for imaging usually, however, cannot be directly employed with mobile data. Efforts must be made to adapt them to the particularities of the mobile networking domain. We expand on this future research direction in Sec.~\ref{sec:st-traffic}. 

\rev{On the other hand, although deep learning brings precision in network-level mobile data analysis, making causal inference remains challenging, due to limited model interpretability. For example, a NN may predict there will be a traffic surge in a certain region in the near future, but it is hard to explain why this will happen and what triggers such a surge. Additional efforts are required to enable explanation and confident decision making. At this stage, the community should rather use deep learning algorithms as intelligent assistants that can make accurate inferences and reduce human effort, instead of relying exclusively on these. }

\section{Deep Learning Driven Network Security}\label{sec:security}
With the increasing popularity of wireless connectivity, protecting users, network equipment and data from malicious attacks, unauthorized access and information leakage becomes crucial. Cyber security systems guard mobile devices and users through firewalls, anti-virus software, and Intrusion Detection Systems (IDS) \cite{buczak2016survey}. The firewall is an access security gateway that allows or blocks the uplink and downlink network traffic, based on pre-defined rules. Anti-virus software detects and removes computer viruses, worms and Trojans and malware. IDSs identify unauthorized and malicious activities, or rule violations in information systems. Each performs its own functions to protect network communication, central servers and edge devices.

Modern cyber security systems benefit increasingly from deep learning \cite{kwon2017survey}, since it can enable the system to \emph{(i)}~automatically learn signatures and patterns from experience and generalize to future intrusions (supervised learning); or \emph{(ii)} identify patterns that are clearly differed from regular behavior (unsupervised learning). This dramatically reduces the effort of pre-defined rules for discriminating intrusions. Beyond protecting networks from attacks,  deep learning can also be used for attack purposes, bringing huge potential to steal or crack user passwords or information. \edit{In this subsection, we review deep learning driven network security from three perspectives, namely infrastructure, software, and user privacy. Specifically, infrastructure level security work focuses on detecting anomalies that occur in the physical network and software level work is centered on identifying malware and botnets in mobile networks. From the user privacy perspective, we discuss methods to protect from how to protect against private information leakage, using deep learning. To our knowledge, no other reviews summarize these efforts.}.\\

\noindent\edit{\textbf{Infrastructure level security:} We mostly focus on anomaly detection at the infrastructure level, i.e. identifying network events (e.g., attacks, unexpected access and use of data) that do not conform to expected behaviors.} Many researchers exploit the outstanding unsupervised learning ability of AEs \cite{yousefi2017autoencoder}. For example, Thing investigates features of attacks and threats that exist in IEEE 802.11 networks \cite{thing2017ieee}. The author employs a stacked AE to categorize network traffic into 5 types (i.e. legitimate, flooding, injection and impersonation traffic), achieving 98.67\% overall accuracy. The AE is also exploited in \cite{aminanto2016detecting}, where Aminanto and Kim use an MLP and stacked AE for feature selection and extraction, demonstrating remarkable performance. Similarly, Feng \emph{et al.} use AEs to detect abnormal spectrum usage in wireless communications~\cite{feng2016anomaly}. Their experiments suggest that the detection accuracy can significantly benefit from the depth of AEs.

Distributed attack detection is also an important issue in mobile network security. Khan \emph{et al.} focus on detecting flooding attacks in wireless mesh networks \cite{khan2016distributed}. They simulate a wireless environment with 100 nodes, and artificially inject moderate and severe distributed flooding attacks, to generate a synthetic dataset. Their deep learning based methods achieve excellent false positive and false negative rates. Distributed attacks are also studied in \cite{diro2017distributed}, where the authors focus on an IoT scenario. Another work in \cite{saied2016detection} employs MLPs to detect distributed denial of service attacks. By characterizing typical patterns of attack incidents, the proposed model works well in detecting both known and unknown distributed denial of service attacks. \rev{More recently, Nguyen \emph{et al.} employ RBMs to classify cyberattacks in the mobile cloud in an online manner \cite{nguyen2018cyberattack}. Through unsupervised layer-wise pre-training and fine-tuning, their methods obtain over 90\% classification accuracy on three different datasets, significantly outperforming other machine learning approaches.}

Martin \emph{et al.} propose a conditional VAE to identify intrusion incidents in IoT \cite{lopez2017conditional}. In order to improve detection performance, their VAE infers missing features associated with incomplete measurements, which are common in IoT environments. The true data labels are embedded into the decoder layers to assist final classification. Evaluations on the well-known NSL-KDD dataset \cite{tavallaee2009detailed} demonstrate that their model achieves remarkable accuracy in identifying \edit{denial of service}, probing, remote to user and user to root attacks, outperforming traditional ML methods by 0.18 in terms of F1 score. \edit{Hamedani \emph{et al.} employ MLPs to detect malicious attacks in delayed feedback networks \cite{hamedani2018reservoir}. The proposal achieves more than 99\% accuracy over 10,000 simulations.}\\

\noindent\textbf{Software level security:} Nowadays, mobile devices are carrying considerable amount of private information. This information can be stolen and exploited by malicious apps installed on smartphones for ill-conceived purposes \cite{tam2017evolution}. Deep learning is being exploited for analyzing and detecting such threats.

Yuan \emph{et al.} use both labeled and unlabeled mobile apps to train an RBM \cite{yuan2014droid}. By learning from 300 samples, their model can classify Android malware with remarkable accuracy, outperforming traditional ML tools by up to 19\%. Their follow-up research in \cite{yuan2016droiddetector} named Droiddetector further improves the detection accuracy by 2\%. Similarly, Su \emph{et al.} analyze essential features of Android apps, namely requested permission, used permission, sensitive application programming interface calls, action and app components \cite{su2016deep}. They employ DBNs to extract features of malware and an SVM for classification, achieving high accuracy and only requiring 6 seconds per inference instance.

Hou \emph{et al.} attack the malware detection problem from a different perspective. Their research points out that signature-based detection is insufficient to deal with sophisticated Android malware \cite{hou2016deep4maldroid}. To address this problem, they propose the Component Traversal, which can automatically execute code routines to construct weighted directed graphs. By employing a Stacked AE for graph analysis, their framework Deep4MalDroid can accurately detect Android malware that intentionally repackages and obfuscates to bypass signatures and hinder analysis attempts to their inner operations. This work is followed by that of Martinelli \emph{et al.}, who exploit CNNs to discover the relationship between app types and extracted syscall traces from real mobile devices \cite{martinelli2017evaluating}. The CNN has also been used in \cite{mclaughlin2017deep}, where the authors draw inspiration from NLP and take the disassembled byte-code of an app as a text for analysis. Their experiments demonstrate that CNNs can effectively learn to detect sequences of opcodes that are indicative of malware. Chen \emph{et al.} incorporate location information into the detection framework and exploit an RBM for feature extraction and classification \cite{chen2017deep123}. Their proposal improves the performance of other ML methods.

 \edit{Botnets are another important threat to mobile networks.} A botnet is effectively a network that consists of machines compromised by bots. These machine are usually under the control of a botmaster who takes advantages of the bots to harm public services and systems \cite{rodriguez2013survey}. Detecting botnets is challenging and now becoming a pressing task in cyber security. 
Deep learning is playing an important role in this area. For example, Oulehla \emph{et al.} propose to employ neural networks to extract features from mobile botnet behaviors \cite{oulehla2016detection}. They design a parallel detection framework for identifying both client-server and hybrid botnets, and demonstrate encouraging performance. Torres \emph{et al.} investigate the common behavior patterns that botnets exhibit across their life cycle, using LSTMs \cite{torres2016analysis}. They employ both under-sampling  and  over-sampling to address the class imbalance between botnet and normal traffic in the dataset, which is common in anomaly detection problems. Similar issues are also studies in \cite{eslahi2016mobile} and \cite{alauthaman2016p2p}, where the authors use standard MLPs to perform mobile and peer-to-peer botnet detection respectively, achieving high overall accuracy.\\

\noindent\textbf{User privacy level:} Preserving user privacy during training and evaluating a deep neural network is another important research issue \cite{liu2016collaborative}. Initial research is conducted in \cite{shokri2015privacy}, where the authors enable user participation in the training and evaluation of a neural network, without sharing their input data. This allows to preserve individual's privacy while benefiting all users, as they collaboratively improve the model performance. Their framework is revisited and improved in \cite{aono2017privacy}, where another group of researchers employ additively homomorphic encryption, to address the information leakage problem ignored in \cite{shokri2015privacy}, without compromising model accuracy. This significantly boosts the security of the system.  \rev{More recently, Wang \emph{et al.} \cite{wang2018not} propose a framework called ARDEN to preserve users' privacy while reducing communication overhead in mobile-cloud deep learning applications. ARDEN partitions a NN across cloud and mobile devices, with heavy computation being conducted on the cloud and mobile devices performing only simple data transformation and perturbation, using a differentially private mechanism. This simultaneously guarantees user privacy, improves inference accuracy, and reduces resource consumption.}

Osia \emph{et al.} focus on privacy-preserving mobile analytics using deep learning. They design a client-server framework based on the Siamese architecture \cite{chopra2005learning}, which accommodates a feature extractor in mobile devices and correspondingly a classifier in the cloud \cite{ossia2017hybrid}. By offloading feature extraction from the cloud, their system offers strong privacy guarantees. An innovative work in \cite{abadi2016deep} implies that deep neural networks can be trained with differential privacy. The authors introduce a differentially private SGD to avoid disclosure of private information of training data. Experiments on two publicly-available image recognition datasets demonstrate that their algorithm is able to maintain users privacy, with a manageable cost in terms of complexity, efficiency, and performance. This approach is also useful for edge-based privacy filtering techniques such as Distributed One-class Learning~\cite{shamsabadi2018}. 

Servia \emph{et al.} consider training deep neural networks on distributed devices without violating privacy constraints \cite{servia2017personal}. Specifically, the authors retrain an initial model locally, tailored to individual users. This avoids transferring personal data to untrusted entities, hence user privacy is guaranteed. Osia \emph{et al.} focus on protecting user's personal data from the inferences' perspective. In particular, they break the entire deep neural network into a feature extractor (on the client side) and an analyzer (on the cloud side) to minimize the exposure of sensitive information. Through local processing of raw input data, sensitive personal information is transferred into abstract features, which avoids direct disclosure to the cloud.
Experiments on gender classification and emotion detection suggest that this framework can effectively preserve user privacy, while maintaining remarkable inference accuracy.

\edit{Deep learning has also been exploited for cyber attacks, including attempts to compromise private user information and guess passwords.}
In \cite{hitaj2017deep}, Hitaj \emph{et al.} suggest that learning a deep model collaboratively is not reliable. By training a GAN, their attacker is able to affect such learning process and lure the victims to disclose private information, by injecting fake training samples. Their GAN even successfully breaks the differentially private collaborative learning in \cite{abadi2016deep}. The authors further investigate the use of GANs for password guessing. In \cite{hitaj2017passgan}, they design PassGAN, which learns the distribution of a set of leaked passwords. Once trained on a dataset, PassGAN is able to match over 46\% of passwords in a different testing set, without user intervention or cryptography knowledge. This novel technique has potential to revolutionize current password guessing algorithms. 

Greydanus breaks a decryption rule using an LSTM network \cite{greydanus2017learning}. They treat decryption as a sequence-to-sequence translation task, and train a framework with large enigma pairs. The proposed LSTM demonstrates remarkable performance in learning polyalphabetic ciphers. Maghrebi \emph{et al.} exploit various deep learning models (i.e. MLP, AE, CNN, LSTM) to construct a precise profiling system and perform side channel key recovery attacks \cite{maghrebi2016breaking}. Surprisingly, deep learning based methods demonstrate overwhelming performance over other template machine learning attacks in terms of efficiency in  breaking both unprotected and protected Advanced Encryption Standard implementations. \rev{In \cite{ning2018deepmag}, Ning \emph{et al.} demonstrate that an attacker can use a CNN to infer with over 84\% accuracy what apps run on a smartphone and their usage, based on magnetometer or orientation data. The accuracy can increase to 98\% if motion sensors information is also taken into account, which jeopardizes user privacy. To mitigate this issue, the authors propose to inject Gaussian noise into the magnetometer and orientation data, which leads to a reduction in inference accuracy down to 15\%, thereby effectively mitigating the risk of privacy leakage.}\\

\edit{\textbf{Lessons learned:} Most deep learning based solutions focus on existing network attacks, yet new attacks emerge every day. As these new attacks may have different features and appear to behave `normally', old NN models may not easily detect them. Therefore, an effective deep learning technique should be able to \emph{(i)} rapidly transfer the knowledge of old attacks to detect newer ones; and \emph{(ii)} constantly absorb the features of newcomers and update the underlying model. Transfer learning and lifelong learning are strong candidates to address this problems. Research in this directions remains shallow, hence we expect more efforts in the future.}

Another issue to which attention should be paid is the fact that NNs are vulnerable to adversarial attacks. This has been briefly discussed in Sec.~\ref{sec:limit}. Although formal reports on this matter are lacking, hackers may exploit weaknesses in NN models and training procedures to perform attacks that subvert deep learning based cyber-defense systems. We will introduce solutions for defending those attacks in Chapter~\ref{chap:tita}.

\section{Tailoring Deep Learning to Mobile Networks}\label{sec:tailor}

Although deep learning performs remarkably in many mobile networking areas, the No Free Lunch (NFL) theorem indicates that there is no single model that can work universally well in all problems \cite{wolpert1997no}. This implies that for any specific mobile and wireless networking problem, \edit{we may need to adapt different deep learning architectures to achieve the best performance.} In this section, we look at how to tailor deep learning to mobile networking applications from three perspectives, namely, mobile devices and systems, distributed data centers, and changing mobile network environments.
\subsection{Tailoring Deep Learning to Mobile Devices and Systems}
The ultra-low latency requirements of future 5G networks demand runtime efficiency from all operations performed by mobile systems. This also applies to deep learning driven applications. However, current mobile devices have limited hardware capabilities, which means that implementing complex deep learning architectures on such equipment may be computationally unfeasible, unless appropriate model tuning is performed. To address this issue, ongoing research improves existing deep learning architectures \cite{cheng2017survey}, such that the inference process does not violate latency or energy constraints \cite{lane2017squeezing, tang2017enabling}, nor raise any privacy concern \cite{wang2018deep}. We outline discuss their key contributions next.

\edit{Iandola \emph{et al.} design a compact architecture for embedded systems named SqueezeNet, which has similar accuracy to that of AlexNet \cite{krizhevsky2012imagenet}, a classical CNN, yet 50 times fewer parameters \cite{iandola2017squeezenet}. SqueezeNet is also based on CNNs, but its significantly smaller model size \emph{(i)} allows more efficiently training on distributed systems; \emph{(ii)} reduces the transmission overhead when updating the model at the client side; and \emph{(iii)} facilitates deployment on resource-limited embedded devices.} Howard \emph{et al.} extend this work and introduce an efficient family of streamlined CNNs called MobileNet, which uses depth-wise separable convolution operations to drastically reduce the number of computations required and the model size~\cite{howard2017mobilenets}. This new design can run with low latency and can satisfy the requirements of mobile and embedded vision applications. The authors further introduce two hyper-parameters to control the  width and resolution of multipliers, \edit{which can help strike an appropriate trade-off between accuracy and efficiency.} The ShuffleNet proposed by Zhang \emph{et al.} improves the accuracy of MobileNet by employing point-wise group convolution and channel shuffle, while retaining similar model complexity~\cite{zhang2017shufflenet}. In particular, the authors discover that more groups of convolution operations can reduce the computation requirements. 

Zhang \emph{et al.} focus on reducing the number of parameters of structures with fully-connected layers for mobile multimedia features learning \cite{zhang2017tucker}. This is achieved by applying Trucker decomposition to weight sub-tensors in the model, while maintaining decent reconstruction capability. The Trucker decomposition has also been employed in \cite{huynh2017deepmon}, where the authors seek to approximate a model with fewer parameters, in order to save memory. Mobile optimizations are further studied for RNN models. In \cite{cao2017mobirnn}, Cao \emph{et al.} use a mobile toolbox called RenderScript\footnote{Android Renderscript \url{https://developer.android.com/guide/topics/renderscript/compute.html.}} to parallelize specific data structures and enable mobile GPUs to perform computational accelerations. Their proposal reduces the latency when running RNN models on Android smartphones. Chen \emph{et al.} shed light on implementing CNNs on iOS mobile devices \cite{chen2016deep}. In particular, they reduce the model executions latency, through space exploration for data re-usability and kernel redundancy removal. The former alleviates the high bandwidth requirements of convolutional layers, while the latter reduces the memory and computational requirements, with negligible performance degradation. 

Rallapalli \emph{et al.} investigate offloading \emph{very deep} CNNs from clouds to edge devices, by employing memory optimization on both mobile CPUs and GPUs \cite{rallapalli2016very}. Their framework enables running at high speed deep CNNs with large memory requirements in mobile object detection applications. Lane \emph{et al.} develop a software accelerator, DeepX, to assist deep learning implementations on mobile devices. The proposed approach exploits two inference-time resource control algorithms, i.e., runtime layer compression and deep architecture decomposition \cite{lane2016deepx}. The runtime layer compression technique controls the memory and computation runtime during the inference phase, by extending model compression principles. This is important in mobile devices, since offloading the inference process to edge devices is more practical with current hardware platforms. Further, the deep architecture designs ``decomposition plans'' that seek to optimally allocate data and model operations to local and remote processors. By combining these two, DeepX enables maximizing energy and runtime efficiency, under given computation and memory constraints. \rev{Yao \emph{et al.} \cite{yao2018fastdeepiot} design a framework called FastDeepIoT, which fisrt learns the execution time of NN models on target devices, and subsequently conducts model compression to reduce the runtime without compromising the inference accuracy. Through this process, up to
78\% of execution time and 69\% of energy consumption is reduced, compared to state-of-the-art compression algorithms.}

\rev{More recently, Fang \emph{et al.} design a framework called NestDNN, to provide flexible resource-accuracy trade-offs on mobile devices \cite{fang2018nestdnn}. To this end, the NestDNN first adopts a model pruning and recovery scheme, which  translates deep NNs to single compact multi-capacity models. With this approach up to 4.22\% inference accuracy can be achieved with six mobile vision applications, at a 2.0$\times$ faster video frame processing rate and reducing energy consumption by 1.7$\times$. In \cite{xu2018deepcache}, Xu \emph{et al.} accelerate deep learning inference for mobile vision from the caching perspective. In particular, the proposed framework called DeepCache stores recent input frames as cache keys and recent feature maps for individual CNN layers as cache values. The authors further employ reusable region lookup and reusable region propagation, to enable a region matcher to only run once per input video frame and load cached feature maps at all layers inside the CNN. This reduces the inference time by 18\% and energy consumption by 20\% on average. Liu \emph{et al.} develop a usage-driven framework named AdaDeep, to select a combination of compression techniques for a specific deep NN on mobile platforms \cite{liu2018ondemand}. By using a deep Q learning optimizer, their proposal can achieve appropriate trade-offs between accuracy, latency, storage and energy consumption.} 

Beyond these works, researchers also successfully adapt deep learning architectures through other designs and sophisticated optimizations, such as parameters quantization \cite{wu2016quantized, zen2016fast}, model slimming \cite{li2018deeprebirth}, sparsification and separation \cite{bhattacharya2016sparsification}, representation and memory sharing \cite{georgiev2017low, falcao2017evaluation}, convolution operation optimization \cite{cho2017mec}, pruning \cite{guo2017pruning}, cloud assistance \cite{li2017fitcnn} and compiler optimization \cite{chen2018tvm}. These techniques will be of great significance when embedding deep neural networks into mobile systems.


\subsection{Tailoring Deep Learning to Distributed Data Containers}

\edit{Mobile systems generate and consume massive volumes of mobile data every day. This may involve similar content, but which is distributed around the world. Moving such data to centralized servers to perform model training and evaluation inevitably introduces communication and storage overheads, which does not scale. However, neglecting characteristics embedded in mobile data, which are associated with local culture, human mobility, geographical topology, etc., during model training can compromise the robustness of the model and implicitly the performance of the mobile network applications that build on such models. The solution is to offload model execution to distributed data centers or edge devices, to guarantee good performance, whilst alleviating the burden on the cloud.}

\edit{As such, one of the challenges facing parallelism, in the context of mobile networking, is that of training neural networks on a large number of mobile devices that are battery powered, have limited computational capabilities and in particular lack GPUs. The key goal of this paradigm is that of training with a large number of mobile CPUs at least as effective as with GPUs. The speed of training remains important, but becomes a secondary goal.}

\begin{figure*}[h!]
\begin{center}
\includegraphics[width=1\textwidth]{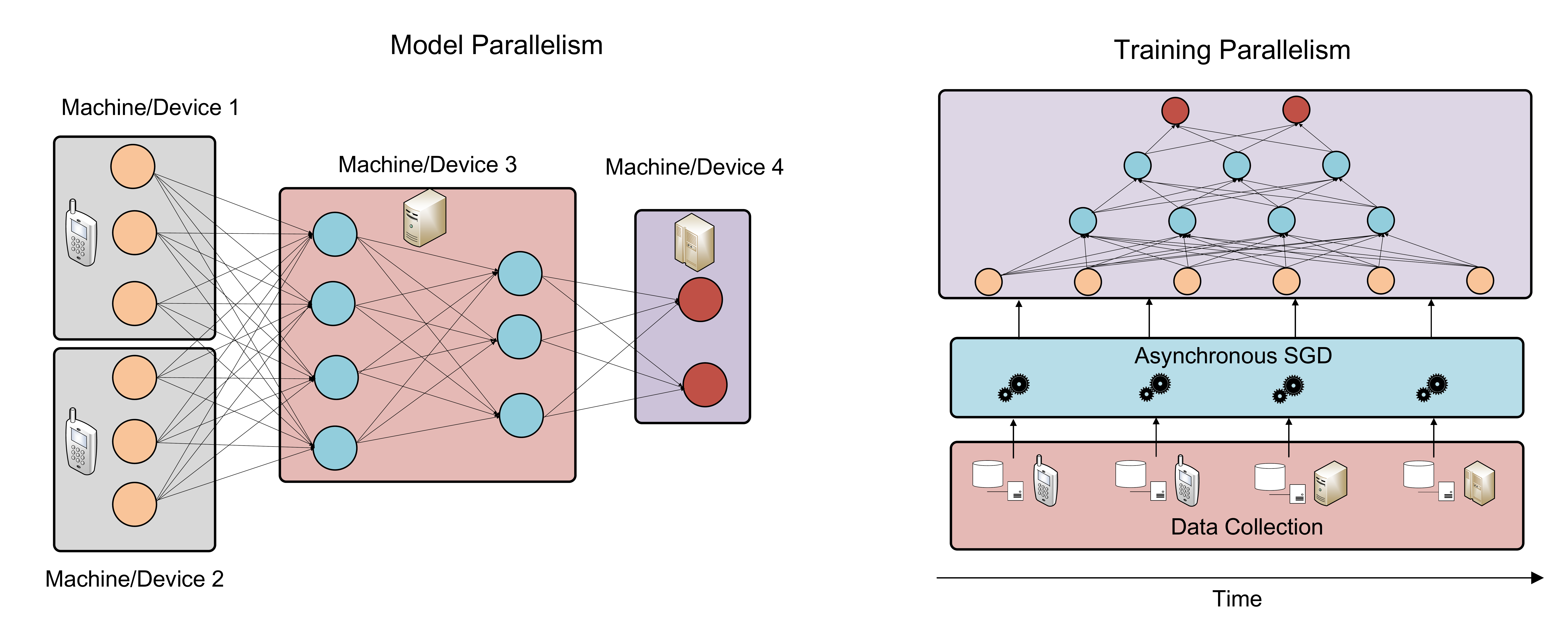}
\end{center}
\vspace*{-0,5em}
\caption{\label{fig:distributed} The underlying principles of model parallelism (left) and training parallelism (right).}
\end{figure*}

Generally, there are two routes to addressing this problem, namely, \emph{(i)} decomposing the model itself, to train (or make inference with) its components individually; or \emph{(ii)} scaling the training process to perform model update at different locations associated with data containers. Both schemes allow one to train a single model without requiring to centralize all data. We illustrate the principles of these two approaches in Fig. \ref{fig:distributed}.\\

\noindent \textbf{Model Parallelism.} Large-scale distributed deep learning is first studied in \cite{dean2012large}, where the authors develop a framework named \emph{DistBelief}, which enables training complex neural networks on thousands of machines. In their framework, the full model is partitioned into smaller components and distributed over various machines. Only nodes with edges (e.g. connections between layers) that cross boundaries between machines are required to communicate for parameters update and inference. This system further involves a parameter server, which enables each model replica to obtain the latest parameters during training. Experiments demonstrate that the proposed framework can be training significantly faster on a CPU cluster, compared to training on a single GPU, while achieving state-of-the-art classification performance on ImageNet~\cite{ILSVRC15}.

Teerapittayanon \emph{et al.} propose deep neural networks tailored to distributed systems, which include cloud servers, fog layers and geographically distributed devices \cite{teerapittayanon2017distributed}. The authors scale the overall neural network architecture and distribute its components hierarchically from cloud to end devices. The model exploits local aggregators and binary weights, to reduce computational storage, and communication overheads, while maintaining decent accuracy. Experiments on a multi-view multi-camera  dataset demonstrate that this proposal can perform efficient cloud-based training and local inference. Importantly, without violating latency constraints, the deep neural network obtains essential benefits associated with distributed systems, such as fault tolerance and privacy.

Coninck \emph{et al.} consider distributing deep learning over IoT for classification applications \cite{de2016distributed}. Specifically, they deploy a small neural network to local devices, to perform coarse classification, which enables fast response filtered data to be sent to central servers. If the local model fails to classify, the larger neural network in the cloud is activated to perform fine-grained classification. The overall architecture maintains good accuracy, while significantly reducing the latency typically introduced by large model inference.

Decentralized methods can also be applied to deep reinforcement learning. In \cite{omidshafiei2017deep}, Omidshafiei \emph{et al.} consider a multi-agent system with partial observability and limited communication, which is common in mobile systems. They combine a set of sophisticated methods and algorithms, including hysteresis learners, a deep recurrent Q network, concurrent experience replay trajectories and distillation, to enable multi-agent coordination using a single joint policy under a set of decentralized partially observable MDPs. Their framework can potentially play an important role in addressing control problems in distributed mobile systems. 
\\

\noindent \textbf{Training Parallelism} is also essential for mobile system, as mobile data usually come asynchronously from different sources. Training models effectively while maintaining consistency, fast convergence, and accuracy remains however challenging~\cite{gupta2016model}. 

A practical method to address this problem is to perform asynchronous SGD. The basic idea is to enable the server that maintains a model to accept delayed information (e.g. data, gradient updates) from workers. At each update iteration, the server only requires to wait for a smaller number of workers. This is essential for training a deep neural network over distributed machines in mobile systems. The asynchronous SGD is first studied in \cite{recht2011hogwild}, where the authors propose a lock-free parallel SGD named HOGWILD, which demonstrates significant faster convergence over locking counterparts. The Downpour SGD in \cite{dean2012large} improves the robustness of the training process when work nodes breakdown, as each model replica requests the latest version of the parameters. Hence a small number of machine failures will not have a significant impact on the training process. A similar idea has been employed in \cite{goyal2017accurate}, where Goyal \emph{et al.} investigate the usage of a set of techniques (i.e. learning rate adjustment, warm-up, batch normalization), which offer important insights into training large-scale deep neural networks on distributed systems. Eventually, their framework can train an network on ImageNet within 1 hour, which is impressive in comparison with traditional algorithms.

Zhang \emph{et al.} argue that most of asynchronous SGD algorithms suffer from slow convergence, due to the inherent variance of stochastic gradients \cite{zhang2016asynchronous}. They propose an improved SGD with variance reduction to speed up the convergence. Their algorithm outperforms other asynchronous SGD approaches in terms of convergence, when training deep neural networks on the Google Cloud Computing Platform. The asynchronous method has also been applied to deep reinforcement learning. In \cite{mnih2016asynchronous}, the authors create multiple environments, which allows agents to perform asynchronous updates to the main structure. The new A3C algorithm breaks the sequential dependency and speeds up the training of the traditional Actor-Critic algorithm significantly. In \cite{hardy2017distributed}, Hardy \emph{et al.} further study distributed deep learning over cloud and edge devices. In particular, they propose a training algorithm, \emph{AdaComp}, which allows to compress worker updates of the target model. This significantly reduce the communication overhead between cloud and edge, while retaining good fault tolerance.

Federated learning is an emerging parallelism approach that enables mobile devices to collaboratively learn a shared model, while retaining all training data on individual devices \cite{pmlr-v54-mcmahan17a, mcmahan2017federated}. Beyond offloading the training data from central servers, this approach performs model updates with a Secure Aggregation protocol \cite{cryptoeprint:2017:281}, which decrypts the average updates only if enough users have participated, without inspecting individual updates. Based on this idea, Google recently build a prototype system using federated Learning in the domain of mobile devices \cite{bonawitz2019towards}. This fulfills the objective that ``bringing the code to the data, instead of the data to the code", which protects individuals' privacy. 

\subsection{Tailoring Deep Learning to Changing Mobile Network Environments}\label{sec:changing}
\begin{figure*}[htb]
\begin{center}
\includegraphics[width=1\textwidth]{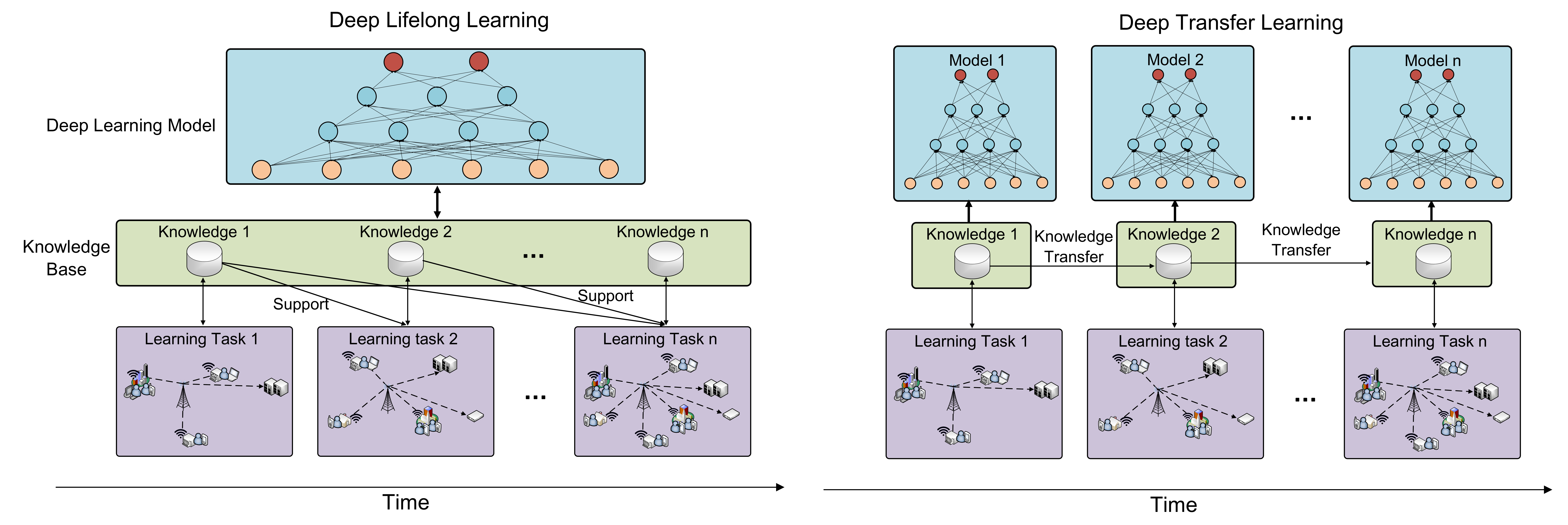}
\end{center}
\caption{\label{fig:changeable} The underlying principles of deep lifelong learning (left) and deep transfer learning (right). Lifelong learning retains the knowledge learned while transfer learning exploits labeled data of one domain to learn in a new target domain.}
\end{figure*}

Mobile network environments often exhibit changing patterns over time. For instance, the spatial distributions of mobile data traffic over a region may vary significantly between different times of the day \cite{furno2017joint}. Applying a deep learning model in changing mobile environments requires lifelong learning ability to continuously absorb new features, without forgetting old but essential patterns. Moreover, new smartphone-targeted viruses are spreading fast via mobile networks and may severely jeopardize users' privacy and business profits. These pose unprecedented challenges to current anomaly detection systems and anti-virus software, as such tools must react to new threats in a timely manner, using limited information. To this end, the model should have transfer learning ability, which can enable the fast transfer of knowledge from pre-trained models to different jobs or datasets. This will allow models to work well with limited threat samples (one-shot learning) or limited metadata descriptions of new threats (zero-shot learning). Therefore, both lifelong learning and transfer learning are essential for applications in ever changing mobile network environments. We illustrated these two learning paradigms in Fig. \ref{fig:changeable} and review essential research in this subsection.\\

\noindent \textbf{Deep Lifelong Learning} mimics human behaviors and seeks to build a machine that can continuously adapt to new environments, retain as much knowledge as possible from previous learning experience~\cite{chen2016lifelong}.
There exist several research efforts that adapt traditional deep learning to lifelong learning. For example, Lee \emph{et al.} propose a dual-memory deep learning architecture for lifelong learning of everyday human behaviors over non-stationary data streams \cite{lee2016dual}. To enable the pre-trained model to retain old knowledge while training with new data, their architecture includes two memory buffers, namely a deep memory and a fast memory. The deep memory is composed of several deep networks, which are built when the amount of data from an unseen distribution is accumulated and reaches a threshold. The fast memory component is a small neural network, which is updated immediately when coming across a new data sample. These two memory modules allow to perform continuous learning without forgetting old knowledge. Experiments on a non-stationary image data stream prove the effectiveness of this model, as it significantly outperforms other online deep learning algorithms. The memory mechanism has also been applied in \cite{graves2016hybrid}. In particular, the authors introduce a differentiable neural computer, which allows neural networks to dynamically read from and write to an external memory module. This enables lifelong lookup and forgetting of knowledge from external sources, as humans do.

Parisi \emph{et al.} consider a different lifelong learning scenario in \cite{parisi2017lifelong}. They abandon the memory modules in \cite{lee2016dual} and design a self-organizing architecture with recurrent neurons for processing time-varying patterns. A variant of the Growing When Required network is employed in each layer, to to predict neural activation sequences from the previous network layer. This allows learning time-vary correlations between inputs and labels, without requiring a predefined number of classes. Importantly, the framework is robust, as it has tolerance to missing and corrupted sample labels, which is common in mobile data.

Another interesting deep lifelong learning architecture is presented in \cite{tessler2017deep}, where Tessler \emph{et al.} build a DQN agent that can retain learned skills in playing the famous computer game Minecraft. The overall framework includes a pre-trained model, Deep Skill Network, which is trained a-priori on various sub-tasks of the game. When learning a new task, the old knowledge is maintained by incorporating reusable skills through a Deep Skill module, which consists of a Deep Skill Network array and a multi-skill distillation network. These allow the agent to selectively transfer knowledge to solve a new task. Experiments demonstrate that their proposal significantly outperforms traditional double DQNs in terms of accuracy and convergence. This technique has potential to be employed in solving mobile networking problems, as it can continuously acquire new knowledge.\\

\noindent \textbf{Deep Transfer Learning:} Unlike lifelong learning, transfer learning only seeks to use knowledge from a specific domain to aid learning in a target domain. Applying transfer learning can accelerate the new learning process, as the new task does not require to learn from scratch. This is essential to mobile network environments, as they require to agilely respond to new network patterns and threats. A number of important applications emerge in the computer network domain \cite{valente2017survey}, such as Web mining \cite{lopez2018deep}, caching \cite{bacstuug2015transfer} and base station sleep strategies \cite{li2014tact}. 

There exist two extreme transfer learning paradigms, namely one-shot learning and zero-shot learning. One-shot learning refers to a learning method that gains as much information as possible about a category from only one or a handful of samples, given a pre-trained model \cite{fei2006one}. On the other hand, zero-shot learning does not require any sample from a category \cite{palatucci2009zero}. It aims at learning a new distribution given meta description of the new category and correlations with existing training data. Though research towards deep one-shot learning \cite{rezende2016one, vinyals2016matching} and deep zero-shot learning \cite{changpinyo2016synthesized, oh2017zero} is in its infancy, both paradigms are very promising in detecting new threats or traffic patterns in mobile networks.
\vspace{-1em}
\section{Summary}\label{sec:conclusion}
\vspace{-1em}
Deep learning is playing an increasingly important role in the mobile and wireless networking domain. In this chapter, we provided a survey of recent work that lies at the intersection between deep learning and mobile networking. 

%% file: chap3.tex
\chapter{Deep Learning Driven Mobile Traffic Forecasting on City Grids\label{chap:fore1}}
 Precision traffic engineering and demand-aware allocation of cellular network resources is becoming essential to support emerging applications, including augmented/virtual reality, autonomous vehicles, and digital healthcare. These tasks require real-time traffic analysis and accurate prediction capabilities \cite{xu2016big}, which are challenging to implement with existing tools~\cite{naboulsi:2016}. In particular, mobile network monitoring currently relies on specialized equipment, e.g. probes \cite{keysight}. Deploying these at each base station is expensive and involves storing locally massive amounts of logs that later have to be transferred for analysis. If monitoring is instead exclusively employed at selected locations in the core network, it requires substantial processing power. Timely and exact mobile traffic forecasting is further complicated by the complex spatio-temporal patterns of user demand~\cite{wang2015understanding, furno:2017}.

\begin{figure}[!t]
\centering\includegraphics[width=\columnwidth]{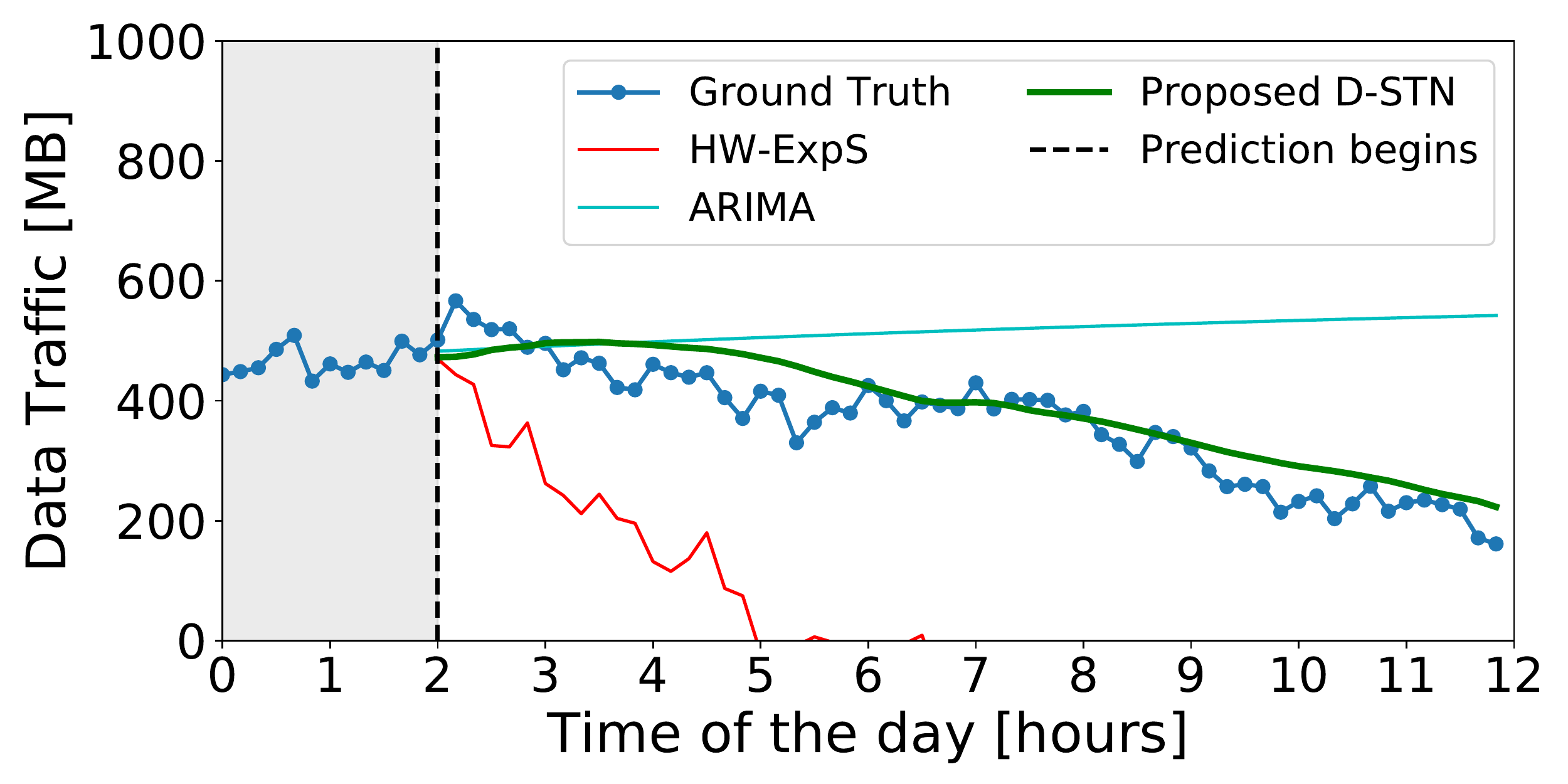}
\caption{Traffic volume in one cell forecast over 10 hours (after 2-hour observations) with Holt-Winters Exponential Smoothing (HW-ExpS), ARIMA, and the original deep learning based approach we introduce in this chapter (D-STN), which captures spatio-temporal correlations. Experiments with the Telecom Italia dataset for Milan~\cite{dataset}.} \label{fig:example}
\end{figure}

Alarm based monitoring systems enable engineers to react to abrupt changes in traffic volume only \emph{a posteriori}, which impacts negatively on the latency perceived by interactive applications. While long-term network traffic forecasting techniques have been proposed to overcome this problem in wired broadband networks (e.g. \cite{Krithikaivasan:2007,papagiannaki2003long, fiandrino2019ml}), mobile networks have received less attention \cite{li2014prediction}. Existing mobile traffic prediction mechanisms (e.g. \cite{tikunov2007traffic,Kim2011}) work well in estimating \emph{trends}, though largely undertake na\"ive predictive modeling across time series observed at \emph{individual} base stations; therefore, they are ill-suited to network-wide forecasting. Further, they ignore important spatial correlations associated with user movement, hence  are predominantly limited to short-term predictions. Such is the case of common practice exponential smoothing (ExpS)~\cite{tikunov2007traffic} and Auto-Regressive Integrated Moving Average (ARIMA)~\cite{Kim2011} techniques, whose accuracy degrades considerably over time when the data series exhibit frequent fluctuations, as we exemplify in Fig.~\ref{fig:example}. 
In the figure we also illustrate the behavior of an original deep learning based approach that we introduce in this chapter. The deep learning based solution exploits cross-spatial and cross-temporal features of the city-wide traffic consumption, and operates substantially longer in the absence of timely measurements, significantly outperforming prior schemes in terms of prediction accuracy.


\textbf{Contributions:} We propose an original deep \underline{S}patio-\underline{T}emporal neural \underline{N}etwork (STN) architecture that exploits important correlations between user traffic patterns at different locations and times, to achieve \emph{precise network-wide mobile traffic \mbox{forecasting}} and overcome the limitations of commonly used prediction techniques. Specifically, we harness the ability of ConvLSTM and 3D-ConvNet structures to model long-term trends and short-term variations of the mobile traffic volume, respectively. We build an ensembling system that exploits the benefits of both models by using two layers that fuse the features extracted by each model from previous traffic measurements. 
We then employ a multilayer perceptron (MLP)~\cite{Goodfellow-et-al-2016} to map the output of the second fusion layer onto final predictions of future mobile traffic volumes. To the best of our knowledge, this is the first time such neural network structures are fused and employed for the purpose of mobile traffic forecasting. The STN is demonstrably effective in spatio-temporal features extraction through convolution operations, while being less complex than traditional neural networks such Restricted Boltzmann Machines (RBM), which require a considerably larger number of parameters to be stored in memory~\cite{salakhutdinov2007restricted}.

Secondly, to enable \emph{long-term traffic forecasting with only limited observations}, we propose an Ouroboros Training Scheme (OTS). This fine tunes the deep neural network such that earlier predictions can be used as input, while the difference between its output and the ground truth is minimized. Our intuition is that re-training the neural network on actual traffic measurements combined with one-step predictions made over these will \cz{enhance} long-term predictions' quality when ground truth becomes (partly) unavailable. 

Thirdly, we specify a Double STN (D-STN) solution that combines the proposed STN with a decay mechanism, which mitigates accumulating prediction errors by mixing the predictions with an empirical mean of the locally observed traffic. This approach ensures mobile traffic volume forecasts remain within reasonable bounds and their duration is substantially extended. Unlike recent work that combines LSTMs and autoencoders for traffic prediction purposes~\cite{wang2017spatiotemporal}, the proposed (D-)STN are not limited to 1-step inferences. Instead, we achieve practical and reliable multi-step forecasting without requiring to train separate neural networks for each base station.

Finally, we implement the proposed (D-)STN prediction techniques on a GPU cluster and conduct experiments on publicly available real-world mobile traffic datasets collected over 60 days and released through the Telecom Italia's Big Data Challenge~\cite{dataset}. The results obtained demonstrate that, once trained, our solutions provide high-accuracy long-term (10-hour long) traffic predictions, while operating with short observation intervals (2 hours) and irrespective of the time of day when they are triggered. Importantly, our models outperform commonly used prediction methods (HW-ExpS and ARIMA), as well as the traditional multilayer perceptron (MLP) and Support Vector Machine (SVM), reducing the normalized root mean square error (NRMSE) by up to 61\% and requiring up to 600 times shorter measurement intervals.

\section{The Mobile Traffic Forecasting Problem}
\label{sec:problem}
Our objective is to make accurate long-term forecasts of the volume of mobile data traffic users consume at different locations in a city, following measurement-based observations. We formally express network-wide mobile traffic consumption observed over a time interval $T$ as a spatio-temporal sequence of data points $\mathcal{D} = \{D_1,D_2,...,D_T\}$, where $D_t$ is a snapshot at time $t$ of the mobile traffic volume in a geographical region represented as an $X\times Y$ grid, i.e. 
\begin{equation}
\begin{aligned}
D_t = \begin{bmatrix}
d^{(1,1)}_t &\cdots  & d^{(1,Y	)}_t\\
 \vdots&\vdots  &\vdots \\
 d^{(X,1)}_t& \cdots & d^{(X,Y)}_t\
\end{bmatrix},
\end{aligned}
\end{equation}
where $d_t^{(x,y)}$ measures the data traffic volume in a square cell with coordinates $(x,y)$ and the sequence can be regarded as a tensor $\mathcal{D}\in \mathbb{R}^{T\times X\times Y}$. From a machine learning perspective, the spatio-temporal traffic forecasting problem is to predict the most likely $K$-step sequence of data points, given previous $S$ observations. That means solving
\begin{align}
&\begin{aligned}
&\widetilde{D}_{t+1},\ldots,\widetilde{D}_{t+K} = \\
& \quad \mathop{\arg\max}\limits_{D_{t+1},\ldots,D_{t+K}} p\left(D_{t+1},\ldots,D_{t+K}|D_{t-S+1},\ldots,D_{t}\right).
\end{aligned}
\end{align}
There is growing evidence that important spatio-temporal correlations exist between traffic patterns \cite{wang2015understanding,furno:2017}, though the value at any location, $d_{t+1}^{(x,y)}$, largely depends only on the traffic in neighboring cells and information associated with distant cells could be neglected \cite{li2014prediction}. That is, statistical dependence exists between proximate cells, while the traffic patterns at distant cells provide little insight into how traffic consumption will evolve at a `target' cell. Therefore, confining consideration to the traffic in $(r+1)\times (r+1)$ adjacent cells allows us to simplify the problem and express one-step predictions as
\begin{align}
\label{eq.indepedance}
&\begin{aligned}
&p\left(D_{t+1}|D_{t-S+1},\ldots,D_{t}\right) \approx \\
& \qquad \qquad \prod_{x=1}^X\prod_{y=1}^Y  p\left(d_{t+1}^{(x,y)}|F^{(x,y)}_{t-S+1},\ldots,F^{(x,y)}_{t}\right),
\end{aligned}
\end{align}
where
\begin{equation}
\begin{aligned}
F^{(x,y)}_{t} = \begin{bmatrix}
d^{(x-\frac{r}{2},y-\frac{r}{2})}_t &\cdots  & d^{(x+\frac{r}{2},y-\frac{r}{2})}_t\\
 \vdots& d_t^{(x,y)}  &\vdots \\
 d^{(x-\frac{r}{2},y+\frac{r}{2})}_t& \cdots & d^{(x+\frac{r}{2},y+\frac{r}{2})}_t\
\end{bmatrix}
\end{aligned}
\label{eq:matrix}
\end{equation}
is the data traffic matrix at time $t$ in an $(r+1)\times (r+1)$ region adjacent to location $(x,y)$. Then the prediction of $D_{t+1}$ can be expressed as the set 
\begin{equation}
\widetilde{D}_{t+1} = \left\{ \widetilde{d}_{t+1}^{(x,y)} ~|~ x =1, \ldots, X; y =1, \ldots, Y \right\},
\end{equation}
where $\widetilde{d}_{t+1}^{(x,y)}$ is the prediction for $d_{t+1}^{(x,y)}$ and is obtained by solving
\begin{equation}
\begin{aligned}
\widetilde{d}_{t+1}^{(x,y)} &= \mathop{\arg\max}\limits_{d_{t+1}^{(x,y)}} p\left(d_{t+1}^{(x,y)}|F^{(x,y)}_{t-S+1},\ldots,F^{(x,y)}_{t}\right).
\end{aligned}
\label{eq:prediction}
\end{equation}
\rv{The above predictions depend only on a marginal distribution, which is precisely the distribution we will model with the Spatio-Temporal Network (STN) we propose next to forecast mobile traffic across different locations and times. Forecasting points at all locations will be concatenated to reconstruct the entire traffic snapshot.}

\section{The Spatio-Temporal Network}
\begin{figure}[t]
\begin{center}
\includegraphics[width=1\textwidth]{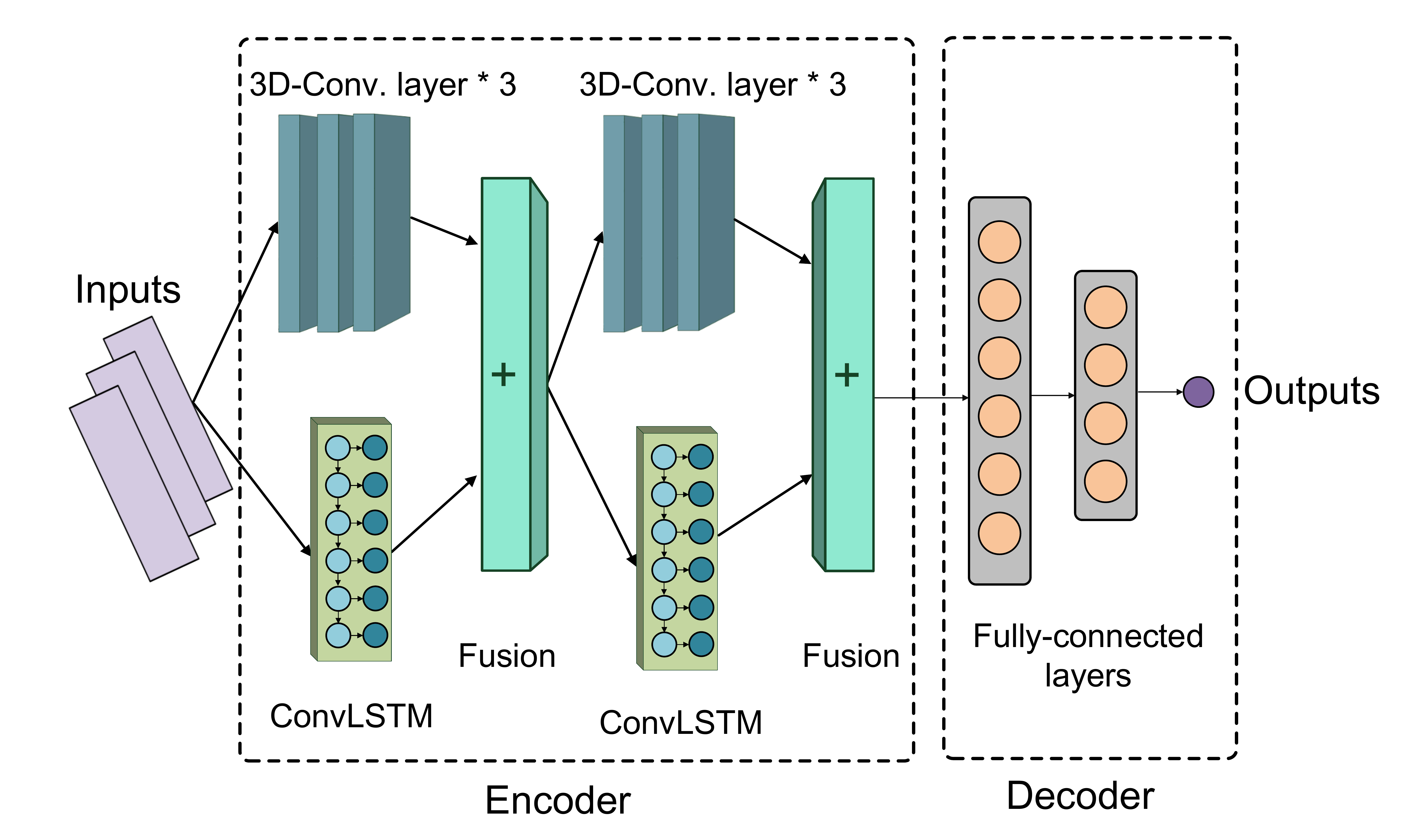}
\end{center}
\caption{\label{Fig:stn} The encoder-decoder architecture of the proposed STN. Measured traffic matrices are processed by ConvLSTMs and 3D-ConvNets to encode spatio-temporal features, then fused. Subsequently, the intermediary output is decoded into predictions by a stack of fully-connected layers (MLP).}
\end{figure}

We design a deep neural network architecture, which we name Spatio-Temporal neural Network (STN), to solve the traffic forecasting problem posed in Sec.~\ref{sec:problem}. The proposed STN follows an encoder-decoder paradigm, where we combine a stack of Convolutional Long Short-Term Memory (ConvLSTM) and three-dimensional Convolutional Network (3D-ConvNet) elements, as illustrated in Fig.~\ref{Fig:stn} and detailed next. Our intuition is that the ability of these structures to handle time series data with spatial dependencies as already demonstrated in e.g. video applications, could be exploited for accurate mobile traffic forecasting. In our case each of these elements are fed with traffic matrices, formally expressed as in (\ref{eq:matrix}), and embed this input through hidden layers into several feature maps. We then fuse and return these features at the output of the encoder. The decoder is a multi-layer perceptron (MLP), which is a supervised learning technique that takes as input the output of the encoder and makes the final predictions through fully-connected layers, as specified in~(\ref{eq:prediction}). The key benefit of employing the MLP lies within the model's ability to solve complex regression problems. To our knowledge, \emph{the problem of precise mobile traffic forecasting has not been tackled previously by fusing ConvLSTM and \mbox{3D-ConvNet} neural networks}, as we propose. In what follows we outline the operation of these structures and explain how they are fused.

\textbf{ConvLSTMs:} 
The Long Short-Term Memory (LSTM) is a special recurrent neural network (RNN) that remedies the vanishing gradient problem characteristic to RNNs, by introducing a set of ``gates" \cite{gers2000learning}. The gates in standard LSTMs are usually fully connected and as a consequence the number of parameters is large, requiring substantial memory and computation time for training purposes. Hence, although proven powerful when working with sequential data, this model is highly complex and frequently turns overfitted. 

ConvLSTMs solve this problem by replacing the inner dense connections with convolution operations \cite{xingjian2015convolutional}. This approach reduces significantly the number of parameters in the model and enhances its ability to handle spatio-temporal data, which is particularly important to our problem. Further, ConvLSTMs can capture long-term trends present in sequences of data points, which makes them particularly well-suited to making inferences about mobile data traffic, as it is known this exhibits important spatio-temporal correlations~\cite{xu2016big,wang2015understanding}.

Given a sequence of 3-D inputs (i.e. multiple ``frames'' of data traffic measurements across 2-D grids/feature maps) denoted $\mathbf{X} = \{ X_1, X_2,...,X_T \}$, we specify the operations of a single ConvLSTM in~(\ref{eq:convlstm}). Here `$\odot$' denotes the Hadamard product, `$*$' the 2-D convolution operator, and $\sigma(\cdot)$ is a sigmoid function. Since each hidden element of this neural network is represented as a 2-D map, we effectively capture cross-spatial traffic correlations through the convolution operations.
\begin{equation}
\begin{aligned}
&i_t = \sigma(W_{xi}*X_t + W_{hi}*H_{t-1}+W_{ci}\odot C_{t-1} + b_i),\\
&f_t = \sigma(W_{xf}*X_t + W_{hf}*H_{t-1}+W_{cf}\odot C_{t-1}+b_f),\\
&C_t = f_t\odot C_{t-1} + i_t \odot \tanh(W_{xc}*X_t + W_{hc}*H_{t-1}+b_c),\\
&o_t = \sigma(W_{xo}*X_t + W_{ho}*H_{t-1}+W_{co} \odot C_t +b_o),\\
&H_t = o_t\odot \tanh(C_t).
\end{aligned}
\label{eq:convlstm}
\end{equation}
In the above, $W_{(\cdot \cdot)}$ and $b_{(\cdot)}$ denote weights and biases we obtain through model training, which we perform on the complete architecture using a stochastic optimization algorithm, as we detail towards the end of this section.  
Note that the inputs $X_t$, cell outputs $C_t$, hidden states $H_t$, input gates $i_t$, forget gates $f_t$, and output gates $o_t$ in the \mbox{ConvLSTM's} inner structure are all 3-D tensors. The first two dimensions of the tensors form the spatial dimension, while the third is the number of feature maps. The input-to-state, cell-to-state, and cell-to-cell transitions are element-wise controlled by each gate $i_t, o_t$, and $f_t$, which allows the model to ``learn to forget" in the spatio-temporal dimension. This property dramatically improves the model's ability to capture spatio-temporal trends \cite{xingjian2015convolutional}.
\vspace*{0.5em}

\textbf{3D-ConvNets:}
Our STN architecture further includes 3D-Conv-Net elements (see Fig.~\ref{Fig:stn}), which extend standard ConvNet models with a temporal dimension \cite{ji20133d}. This choice is motivated by recent results showing 3D-ConvNet perform remarkably in terms of spatio-temporal feature learning~\cite{ji20133d, guo20153d}. Further, they also capture well local dependencies, as seen in minor fluctuations of traffic sequences that are triggered by stochastic human mobility. Given a sequence of spatio-temporal data with $N$ feature maps  $\mathbf{X} = \{ X_1, X_2,...,X_N\}$, the output of a \mbox{3-D} convolutional layer will consist of $H_1, \ldots H_M$ convoluted feature maps, given by
\begin{equation}
\begin{aligned}
H_m = \text{act} \left(\sum_{n=1}^N X_{n} * W_{mn} + b_{m}\right),
\end{aligned}
\end{equation}
where `$*$' corresponds now to a 3-D convolution operator and $\text{act}(\cdot)$ denotes an activation function, whose objective is to increase the model's non-linearity. Functions such as the rectified linear unit (ReLU) and the sigmoid are commonly used for this purpose. 
We note that 3D-ConvNets differ from ConvLSTMs primarily because they do not involve back-propagation through time (BPTT), which in the latter happens between cells. Instead 3D-ConvNets require more layers to attain similar results. On the other hand, the 3-D convolutions enable the model to also capture cross-temporal traffic correlations essential in our problem. The 3D-ConvNet shares weights across different locations in the input, allowing to maintain the relation between neighboring input points and spatio-temporal locality in feature representations. This property enables 3D-ConvNets to capture better short-term traffic fluctuations and overall improves the model's generalization abilities.
\vspace*{0.5em}

\textbf{STN -- Fusing ConvLSTMs and 3D-ConvNets:}
To leverage the ability of both ConvLSTM and 3D-ConvNet to learn spatio-temporal features and forecast mobile data traffic with high accuracy, the STN architecture we propose blends the output of both models through two ``fusion'' layers (see Fig.~\ref{Fig:stn}). The goal of the fusion operation is to build an ensembling system which includes two dedicated deep spatio-temporal models, which has been proven to enhance the model's performance~\cite{veit2016residual}. By averaging intermediate outputs of both ConvLSTM and 3D-ConvNet twice, the proposed STN reinforces the ensembling system. 
Thus through the fusion operations we jointly exploit the advantages of both ConvLSTM (capturing long-term trends) and 3D-ConvNet (capturing local fluctuations), which leads to superior prediction performance, as compared to simply employing any of the two individually. This is demonstrated by the results we present in Sec.~\ref{sec:experiments} and is a key novelty of our purpose-built neural network architecture. Importantly, unlike other deep learning approaches proposed recently \cite{wang2017spatiotemporal}, where different structures extract spatial and temporal patterns separately, our solution can jointly distil spatio-temporal traffic features. This enables us to train our architecture in an end-to-end manner, instead of individually training each of its \cz{components}, as required in~\cite{wang2017spatiotemporal}. Further, the proposed STN has excellent generalization abilities, as we will demonstrate it is sufficient to train the STN for a single geographic area, before applying it for inference to others.

In our design the input data (measurements) is first processed in parallel by one ConvLSTM and one 3D-ConvNet, then their outputs are aggregated by a fusion layer that performs the following element-wise addition:
\begin{equation}
\begin{aligned}
\mathrm{H}(\Theta_H;X) = h_{C}(\Theta_1;X) + h_{L}(\Theta_2;X),
\end{aligned}
\end{equation}
where $h_{C}$ and $h_{L}$ are the outputs of a single 3D-ConvNet and respectively a ConvLSTM, and $\Theta_H = \{\Theta_1, \Theta_2\}$ denotes the set of their parameters (weights and biases). \rv{We choose the addition operation instead of concatenation, to maintain the dimension in the hidden layer.} We employ this procedure twice to encode the spatio-temporal embedding of the data sequence given as input. Subsequently, we decode the obtained features and perform prediction via an MLP. The MLP connects the outputs from the second fusion layer over every time step, to achieve an ``attention''-like mechanism \cite{sutskever2014sequence}. This allows the model to make use of all temporal features learned in order to produce final predictions, rather than merely relying on the last state. It is important to note that, unlike traditional \cz{r}estricted Boltzmann Machines (RBMs) employed for time series forecasting \cite{Kuremoto201447}, the proposed STN shares weights between inputs, therefore it requires to configure considerably fewer parameters. 

The STN models the expected value of the marginal distribution $p(d_{t+1}^{(x,y)}| F^{(x,y)}_{t-S+1},$ $\ldots,F^{(x,y)}_{t})$, i.e. it takes a ``local'' spatio-temporal traffic matrix to predict the volume of traffic at the following time instance $t+1$, and its output consists only of one regression value corresponding to an $(x,y)$ location. The prediction is repeated multiple times to encompass all the locations covered by the cellular network grid.

The encoder contains six 3D-convolutional layers with (3, 3, 3) and respectively (6, 6, 6) feature maps, and two ConvLSTM layers with 3 and 6 feature maps. In our design, we set the length of the input $S=12$ (corresponding to 2 hours, with traffic volume snapshots every 10 mins) and $r=10$ (i.e. considering spatial correlation among 10$\times$10 adjacent cells). \rv{The length of the input is sufficient to capture the evolution trend of traffic behavior, while retaining a reasonable model complexity.} Thus each input is a $11\times11\times12$ tensor and the STN model will predict the traffic volume $\widetilde{d}_{t+1}$ at the center of a $11\times11$ map in the next time step. 

To obtain these predictions, we effectively train a neural network model $\mathcal{M}$ parameterized by $\Theta$, which takes a set of inputs $\mathbf{F}^{(x,y)}_{t} := \left\{F^{(x,y)}_{t-S+1},F^{(x,y)}_{t-S+2},...,F^{(x,y)}_{t}\right\}$ and outputs traffic volume forecasts $\widetilde{d}_{t+1}^{(x,y)}$, i.e.
\begin{equation}
\widetilde{d}_{t+1}^{(x,y)} = \mathcal{M}\left(\Theta;\mathbf{F}^{(x,y)}_{t}\right).
\end{equation}
We adopt the maximum likelihood estimation (MLE) method to train this model by minimising the following $\mathrm{L}_2$ loss function:
\begin{equation}
\begin{aligned}
\mathrm{L}(\Theta) = \frac{1}{T \cdot X \cdot Y} \sum_{t=1}^T\sum_{x=1}^X\sum_{y=1}^Y||\mathcal{M}(\Theta;\mathbf{F}^{(x,y)}_{t}) - d^{(x,y)}_{t}||^2.
\end{aligned}
\label{eq:loss1}
\end{equation}
Stochastic Gradient Descent (SGD) based methods are widely used to optimize this loss function. In our work, we choose the Adam optimizer \cite{kingma2014adam}, which commonly yields faster convergence compared to traditional SGD. We employ standard configuration ($\beta_1 = 0.9, \beta_2 = 0.999, \epsilon = 10^{-8}$), setting the initial learning rate to 0.005.

Note that STN only performs \emph{one-step} predictions based on complete ground truth information available at every step. This is however problematic when multi-step prediction is desired, as soon as ground truth information becomes unavailable. In the next section we extend our design to achieve multi-step predictions and \emph{forecast mobile traffic long-term with only limited ground truth observations}. 

\section{Long-Term Mobile Traffic Forecasting}
\begin{figure}[t]
\centering\includegraphics[width=\columnwidth]{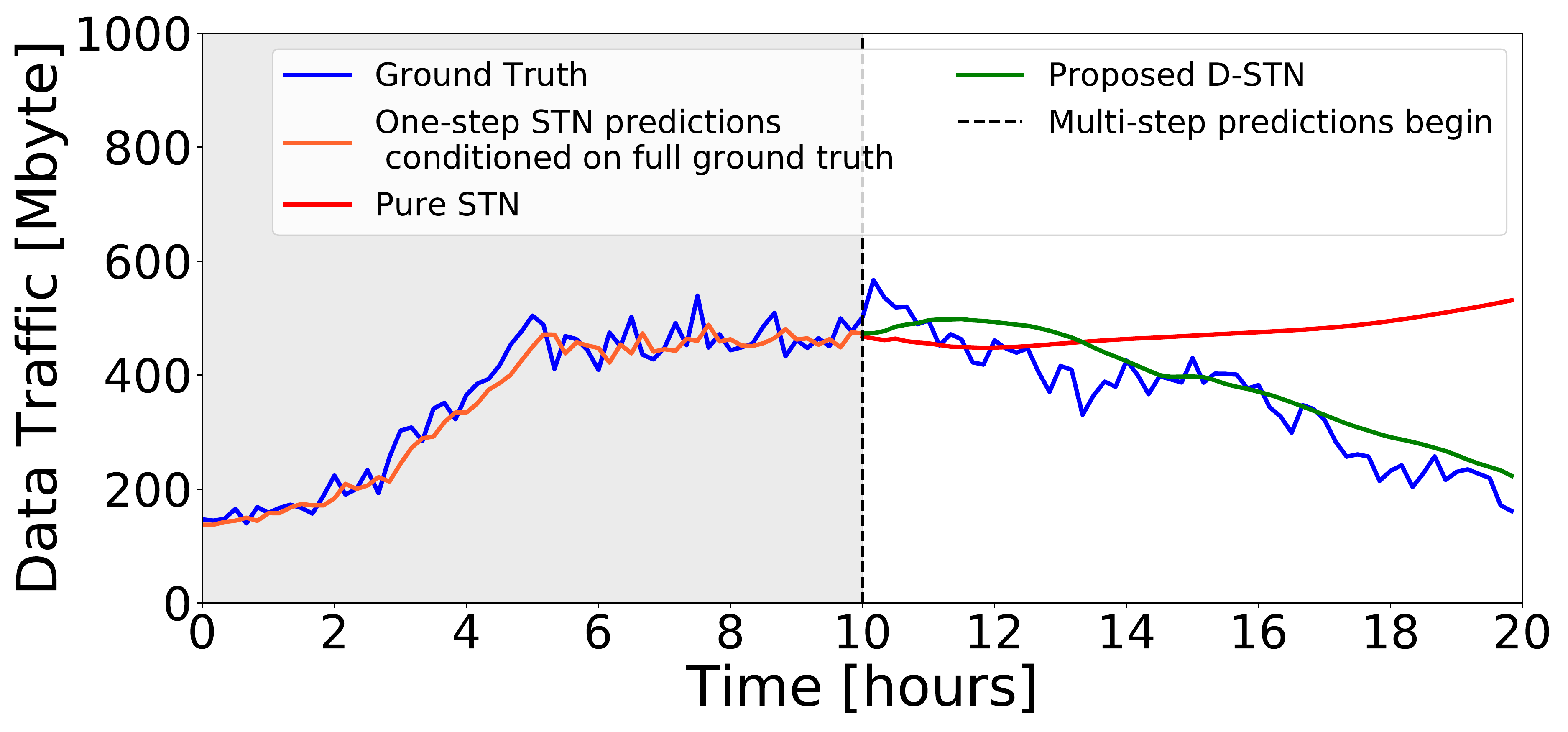}
\caption{Traffic volume predicted by pure STN and D-STN. First (shaded), the STN makes one-step predictions based on fully observable ground truth, then both methods perform multi-step predictions as ground truth becomes unavailable.} \label{fig:extend}

\end{figure}

In the case of multi-step predictions, ground truth becomes rapidly unavailable, while still necessary for conditioning in the prediction problem posed in (\ref{eq:prediction}). An intuitive way to address this issue is to recursively reuse recent predictions as input for the following prediction step. Inevitably, this leads to prediction errors that are likely to accumulate over time and the results obtained may be modest. For instance, the predictions the proposed STN makes will be very accurate when ground truth measurements are always available, as we illustrate in the shaded region of Fig.~\ref{fig:extend}. However, when network measurements are suspended, the pure STN's output is rapidly unable to follow the actual evolution of the data traffic (observe the red curve in the light region shown in Fig.~\ref{fig:extend}).  

To address this problem and achieve reliable long-term forecasting, we propose a Double STN scheme (D-STN), which comprises two key enhancements. Specifically, (1) we introduce an Ouroboros Training Scheme (OTS) that fine tunes the neural network, allowing earlier predictions to be fed as inputs when ground truth becomes unavailable, while keeping prediction errors small; (2) D-STN blends the newly trained STN with historical statistics that essentially summarise prior knowledge of long-term trends, and thus enables accurate forecasting for extended periods of time (up to 10h). We exemplify the D-STN's ability to track the real traffic in Fig.~\ref{fig:extend} and detail its operation next.

\subsection{An Ouroboros Training Scheme}
Turning attention again to (\ref{eq:prediction}), we can see that ground truth observations become unavailable as early as at the second prediction step. If working with $S=12$, by the twelfth step, no measurements will be available for prediction. To deal with this issue, one possibility is to substitute the missing information with earlier model predictions. However, simply conditioning on inferred data will unavoidably introduce errors that accumulate over time. Intuitively, the reason is that the model only works well when the training data and test data follow the same distributions. \rv{When the model operates based on its own predictions, any prediction error, however small, will lead to diverging distributions of input and respectively real data.} Similar behavior was previously observed in imitation learning problems~\cite{ross2010efficient}.
Examining again Fig.~\ref{fig:extend}, observe the traffic volume experiences an abrupt rise when the multi-step prediction is activated. As a result, the pure STN, which completely relies on the model's predictions, fails to capture this initial trend and ultimately the error grows considerably with the prediction step. 

\begin{algorithm}[h]
  \caption{The Ouroboros Training Scheme}
  \label{Alg.OTS}
  \begin{algorithmic}[1]
    \Inputs{Time series training data $\mathcal{D} = \{D_1,D_2,...,D_T\}$}
    \Initialize{A pre-trained model $\mathcal{M}$ with parameters~$\Theta$.}
    \For{$e = 1$ to $E$}
        \For{$t=1$ to $T-S$}
            \If{$t=1$}
                \State{$Q \leftarrow \{D_1,D_2,...,D_S\}$ \quad
               $\blacktriangleright$ Generate input queue using first $S$ ground truth measurements.}
            \Else
               \State{Pop the first element out of $Q$,}
               \State{Predict $D'_{S+t-1}$ by $\mathcal{M}$ with input $Q$,}
               \State{Push $D'_{S+t-1}$ to the end of $Q$.}
            \EndIf
            \State{Generate the target input $T \leftarrow D_{S+t}$.}
            \State{Train $\mathcal{M}$ with input $Q$ and target $T$ by SGD.}
        \EndFor
    \EndFor
    \end{algorithmic}
\end{algorithm}

We solve this problem by introducing an Ouroboros training scheme (OTS), which draws inspiration from the DAGGER technique that mixes ground truth and self-generated sequences in the model training phase~\cite{ross2011reduction}. Our goal is to enable the neural network to forecast precisely, irrespective of whether conditioning on ground truth or on predictions it made earlier. Therefore, apart from the original training data, the OTS recursively stores the model predictions and uses these for a second round of training, to \emph{mimic the prediction behavior}. We summarize this procedure in Algorithm ~\ref{Alg.OTS}, where at every epoch $e$, we regard the inputs fed to the model as a queue $Q$. First ($t=1$) we initialize this with $S$ ground truth observations (line 6). Subsequently, at every sub-step~$t$, we pop the oldest frame in $Q$ (line 8) and push the prediction made at the last step (lines 9--10), thereby rebuilding the model's input. When used, $Q$ is partitioned into $X\times Y$ data points, each of which is an $(r+1) \times (r+1) \times S$ tensor (in our design these are 10,000 tensors with dimension $11 \times 11 \times 12$). We feed the model with $Q$ and corresponding ground truth~$T$, then preform Stochastic Gradient Descent (SGD) based training \cite{kingma2014adam} (line 13). An epoch will stop when it exhausts all training data, i.e. $t$ reaches $T-S$. This procedure resembles the behavior of a mythological snake called Ouroboros, which perpetually eats its own tail (hence the name of the proposed scheme). We work with $E=1$ epoch, which is sufficient for our problem. 

The OTS algorithm works particularly well, as it broadens the horizon of the model by extending the training data with its predictions and enlarging the support set of the input distribution. This can be regarded as a data augmentation technique, which was proven powerful when training complex models~\cite{krizhevsky2012imagenet}. In our case OTS suppresses the overestimation tendency, as the retraining forces the predictions to be substantially closer to the ground truth.

\subsection{Blending Predictions \& Historical Statistics}
Employing the OTS will improve the accuracy of multi-step predictions, though note that uncertainty may grow over time and thus limit accuracy when long-term (e.g. >5h long) forecasting is desired. In what follows we propose a further improved Double STN (D-STN) forecasting system that addresses this issue. 

\begin{figure}[!t]
\begin{center}
\includegraphics[width=\columnwidth]{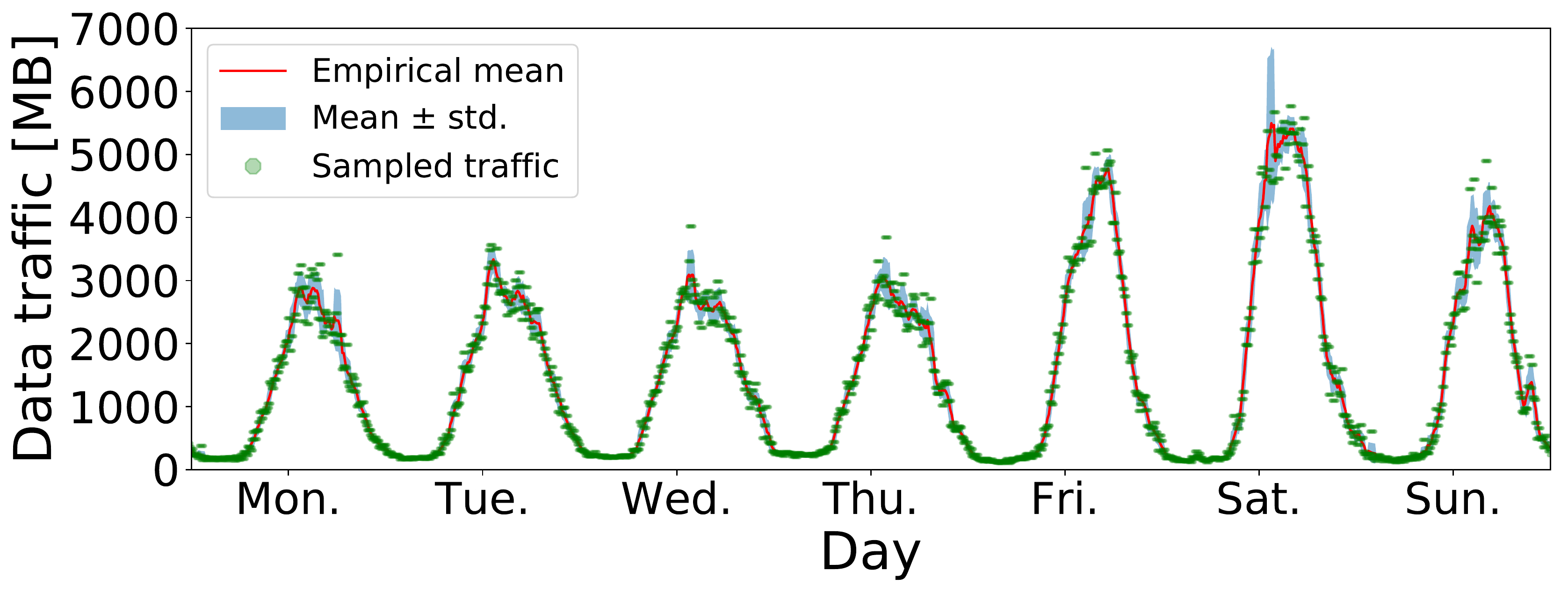}
\end{center}
\caption{\label{Fig:mean} Traffic volume sampled over 7 weeks (1\textsuperscript{st} Nov -- 12\textsuperscript{th} Dec 2013), weekly empirical mean, and standard deviation in cell (48, 60) of the Milan grid.}

\end{figure}

The D-STN design stems from two important properties of mobile data traffic that we observed. Namely, data traffic exhibits certain periodicity (in both daily and weekly patterns) and relatively flat averages, if observed over long intervals \cite{li2014prediction}. This is indeed the case also for the city of Milan, as we illustrate in Fig.~\ref{Fig:mean}, where we plot the weekly empirical mean of mobile traffic volume in a selected cell between 1 Nov and 12 Dec 2013 (7 weeks). The figure also shows the actual sampled traffic and the standard deviation in the same period. Observe that despite a few outliers, traffic volume samples are close to the empirical mean. We conjecture that incorporating prior knowledge of averages into the model can reduce uncertainty and improve the prediction performance, if utilized appropriately. On the other hand, there is little correlation between traffic volumes measured at intervals far (e.g. hours) apart, which has also been observed for other similar time series \cite{che2017recurrent}.

Therefore, we propose to combine information of the weekly empirical mean with the STN predictions, using a light-weight decay mechanism that reduces the weight of the predictions over time. In our experiments we use the weekly empirical mean of mobile traffic volume computed over 7 weeks (1 Nov -- 12 Dec 2013). Assume we trigger forecasting at time $t$ and denote $D_{t+h}^{M}$ and $D_{t+h}^{O}$ the predictions at time $t+h$ of the original STN without fine tuning and respectively with the OTS enabled. Denoting $\overline{D}_{t+h}$ the weekly empirical mean of actual measurements, we define the $h$-step ahead forecasting of our D-STN as:
\begin{equation}
\begin{aligned}
D'_{t+h} := \gamma(h)\left[\alpha(h) D_{t+h}^{M} + (1-\alpha(h))D_{t+h}^{O}\right] + (1-\gamma(h))\overline{D}_{t+h},
\end{aligned}
\end{equation}
where
\begin{equation}
\begin{aligned}
\gamma(h) = 1-\frac{1}{1+e^{-(wh+b)}}
\end{aligned}
\end{equation}
is a sigmoid decay function which is rectified between $(0,1)$. $\gamma(h)$ controls the weighting of model predictions and empirical means. It will non-linearly degenerate to zero over $h$, meaning the empirical mean will dominate the prediction's result if $h$ is large. The hyper-parameters $w$ and $b$ control the changing rate and the initial state of $\gamma(h)$, and are  set to $w=0.01$ and $b=-5$ based on \textbf{cross validation over the training set}. Note that $b$ must be a large negative value to guarantee that the model's predictions will mostly contribute at the beginning of the process.
In practice, the empirical mean $\overline{D}_{t+h}$ can also be updated in an on-line manner, as new measurements are conducted. 

In addition, predictions made by the two models $D_{t+h}^{M}$ and $D_{t+h}^{O}$ are weighted by $\alpha(h)$, which is dynamically updated as follows:
\begin{equation}
\begin{aligned}
\alpha(h) = \max\left(1-\frac{h(1-\delta)}{S}  ,\delta\right).
\end{aligned}
\end{equation}
Recall that $S$ denotes the temporal length of the input sequence and $\delta$ is a fixed lower bound for $\alpha$. Since the fine tuned STN is trained using OTS with the model predictions, the original STN deserves a heavier weight in the initial steps, where the model's input still consists partially of ground truth observations. To ensure $D_{t+h}^{M}$ and $D_{t+h}^{O}$ contribute equally as time advances, we set $\delta = 0.5$. The weights of $D_{t+h}^{M}, D_{t+h}^{O}$ and $\overline{D}_{t+h}$, sum up to 1, i.e.
\begin{equation}\label{normalise}
\begin{aligned}
 \gamma(h)\times\alpha(h) + \gamma(h)\times(1-\alpha(h)) + 1- \gamma(h) = 1,
\end{aligned}
\end{equation}
which normalises the output of the D-STN.
Note that the original and the OTS trained models share the same input when making predictions, i.e.
\begin{equation}\label{inputs}
\begin{aligned}
\mathcal{D} = \left\{\begin{matrix}
\left \{ D_{t+h-S},...,D_{t} \right \}\cup \left \{D'_{t+1},...,D'_{t+h-1} \right \}, \hfill h\leqslant S;\\ 
\left \{ D'_{t+h-S},...,D'_{t+h-1} \right \}, \hfill h>S,
\end{matrix}\right.
\end{aligned}
\end{equation}
where $D_{t}$ is the observable ground truth at time $t$. 

Next we compare the performance of STN, D-STN, commonly used network traffic prediction schemes, and other deep learning based predictors, demonstrating substantial accuracy gains over state-of-the-art techniques.

\section{Performance Evaluation}
\label{sec:experiments}
In this section we first briefly describe our implementation of the proposed neural network models. Then we evaluate the performance of STN and D-STN by conducting multi-step predictions and comparing their accuracy with that of Holt-Winters Exponential Smoothing (HW-ExpS), Auto-Regressive Integrated Moving Average (ARIMA), the Multi-Layer Perceptron (MLP), standard ConvLSTM and 3D-ConvNet based predictors, and recent deep learning-based short-term predictors.  
\rv{We note that we do not compare with the seasonal ARIMA \cite{williams2003modeling}, as the HW-ExpS already embraces a seasonal component and the seasonal ARIMA does not capture the spatial correlations.}

\subsection{Mobile Traffic Datasets\label{sec:grid_dataset}}
We experiment with publicly available real-world mobile traffic datasets released through Telecom Italia's Big Data Challenge~\cite{dataset}. These contain network activity measurements in terms of total cellular traffic volume observed over 10-minute intervals, for the city of Milan and the Trentino region, collected between 1 Nov 2013 and 1 Jan 2014 (2 months). The two areas are of different sizes, have different population, and therefore exhibit dissimilar traffic patterns. Milan's coverage area is partitioned into $100\times 100$ squares of $0.055$km\textsuperscript{2} (i.e. 235m $\times$ 235m). Trentino's coverage is composed of $117\times98$ cells of $1$km\textsuperscript{2} each. 

\begin{figure*}[t]
\centering
\includegraphics[width=2.83in]{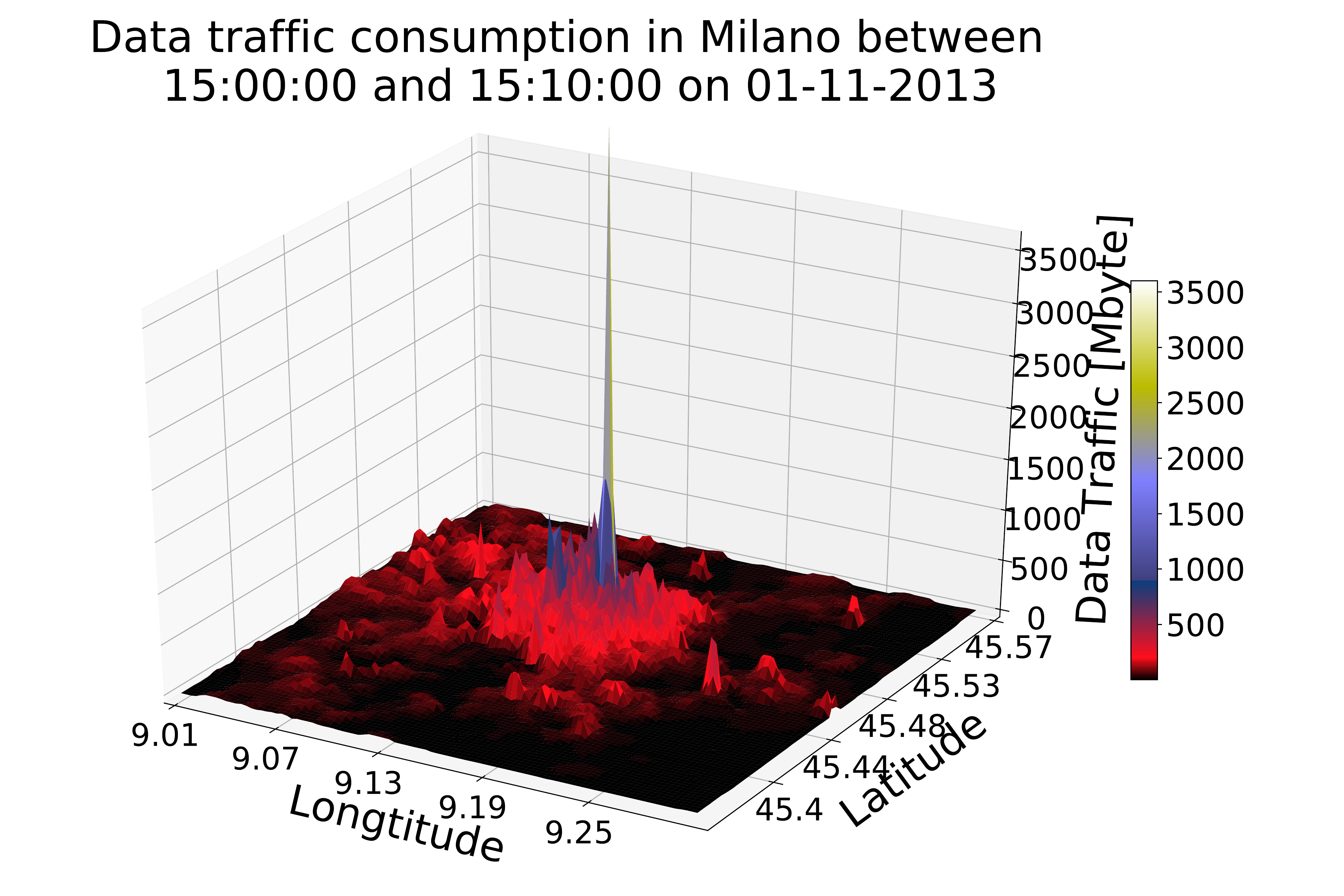}
\includegraphics[width=2.83in]{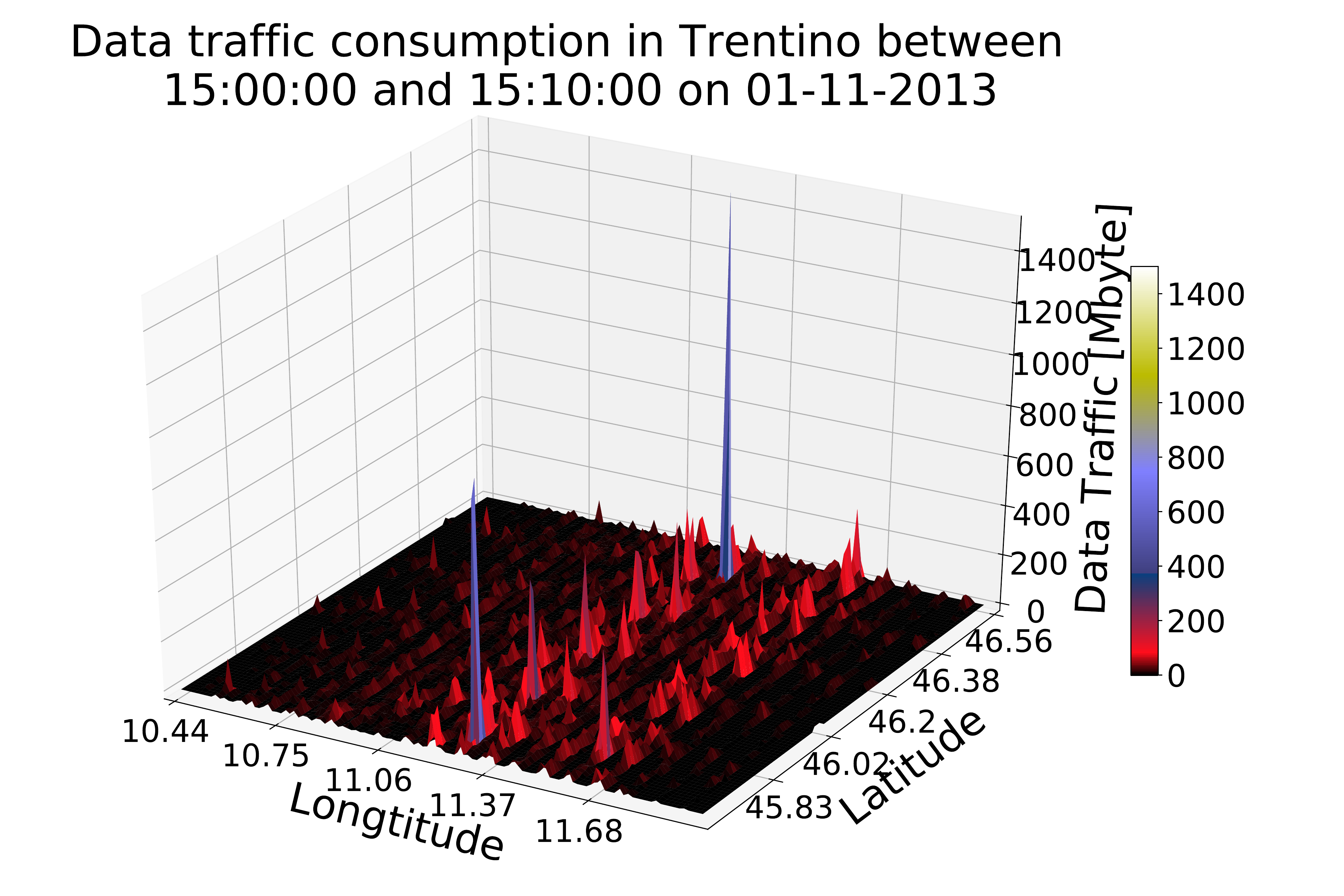}
\caption{Snapshots of the spatial distribution of mobile data traffic in Milan (left) and Trentino (right). Full dataset corresponds to measurements collected over 2 months.}
\label{Fig:topology0}
\end{figure*}

A snapshot of the traffic's spatial distribution in each of the two areas is shown in Fig.~\ref{Fig:topology0}. Observe that the traffic intensity is higher in the center of Milan and overall can be regarded as almost following a Gaussian distribution in the spatial dimension. \rv{This represents traffic patterns in a metropolitan area.} In contrast, \rv{the traffic in Trentino is representative for areas where smaller towns are clustered, yet laid out at some distance from each other. Thus the traffic} exhibits multiple clusters of smaller volume, and as expected, the overall total consumption is less than that in Milan. 

\subsection{Implementation}
We implement the proposed (D-)STN forecasting schemes as well as the conventional ConvLSTM, 3D-ConvNet and MLP using the open-source Python libraries TensorFlow~\cite{tensorflow2015-whitepaper} and TensorLayer \cite{tensorlayer}. We train the models using a GPU cluster comprising 15 nodes, each equipped with 1-2 NVIDIA TITAN X and Tesla K40M computing accelerators (3584 and respectively 2280 cores). To evaluate their prediction performance, we use one machine with one NVIDIA GeForce GTX 970M GPU for acceleration. We implement, train, and evaluate the ARIMA model using the {\ttfamily statsmodels} Python package without GPU acceleration. Training the HW-ExpS technique is fairly straightforward, as the procedure involves a simple loop to compute the sequence of smoothed values, best estimates of the linear trend, and seasonal correction factors \cite{winters60}.

\subsection{Prediction Accuracy Evaluation and Comparison with Existing Techniques}

Next we evaluate the performance of our proposals against widely used time series prediction techniques, namely HW-ExpS configured with $\alpha = 0.9, \beta = 0.1,$ and $\gamma = 0.001$, ARIMA using $p=3, d=1,$ and $q=2$, and respectively MLP with 2 layers. The hyper-parameters of HW-ExpS and ARIMA are selected based on \textbf{cross validation on training sets}. To achieve a fair comparison, we stop feeding HW-ExpS and ARIMA with ground truth for further tuning when prediction is triggered. We also evaluate the performances of two components of the STN, namely the traditional ConvLSTM and 3D-ConvNet. Arguably other traditional machine learning tools such as Deep Belief Networks (DBNs -- stacks of Restricted Boltzmann Machines) and Support Vector Machine (SVMs) could also be employed for comparison. However, we give limited consideration to these models, since DBNs are essentially MLPs that perform layer-wise pre-training to initialize the weights, which requires substantially more time for the benefit of reducing over-fitting (this is not viewed as essential, especially when the data used for training is sufficient). The standard SVM training algorithm has $O(n^3)$ and $O(n^2)$ time and space complexities, where $n$ is the size of the training set (in our case, $n = 57,480,000$ points), and is thus computationally infeasible on very large datasets. We do however compare the prediction performance of our schemes against that of an SVM trained on a sub-set of the original data. For completeness, we also compare our proposal against a recent deep learning approach \cite{wang2017spatiotemporal}, which comprises auto-encoders (AEs) and an LSTM. Unfortunately, this solution cannot operate when ground truth data is (partially) unavailable and is only able to perform one-step predictions. Therefore in this case we limit the comparison to such short-term forecasts.

We train all deep learning models, i.e. (D-)STN, \mbox{ConvLSTM}, 3D-ConvNet,  MLP, and AE+LSTM with data collected in Milan between 1 Nov 2013 and 10 Dec 2013 (40 days, \rv{66.7\% of the data}), validate on the data in the same city for the following 10 days (\rv{16.7\% of the data}), and we evaluate their performances on both Milan and Trentino datasets collected between 20--30 Dec 2013 (\rv{16.7\% of the data}). \rv{This is close to the widely-used 60-20-20 data partition for training, validation and testing. In addition, the training and validation data covers rich traffic activities caused by different social events (\eg black Friday, football matches) and therefore will allow the model to capture unusual traffic patterns. Note that all of the neural network models are reinforced with OTS and empirical mean.} We train HW-ExpS and ARIMA on both datasets with measurements collected during the first 50 days and test on the same sets used for evaluating the deep learning models.

We quantify the accuracy of the proposed (D-)STN and existing prediction methods through the Normalized Root Mean Square Error (NRMSE) as given below:
\begin{equation}
\begin{aligned}
\text{NRMSE}=\frac{1}{\overline{d}}\sqrt{\sum_{k=1}^N \frac{(\widetilde{d}_k-d_k)^2}{N}}, \label{eq:nrmse}
\end{aligned}
\end{equation}
where $\widetilde{d}_k$ are the predicted values, $d_k$ are the corresponding ground truth values, $N$ denotes the total number of measurement points over space and time, and $\overline{d}$ is their mean. NRMSE is frequently used for the comparison between datasets or models with different scales. The smaller the NRMSE, the more accurate the predictions of the model are.

\begin{figure*}[!t]
\begin{center}
\includegraphics[width=1.03\columnwidth]{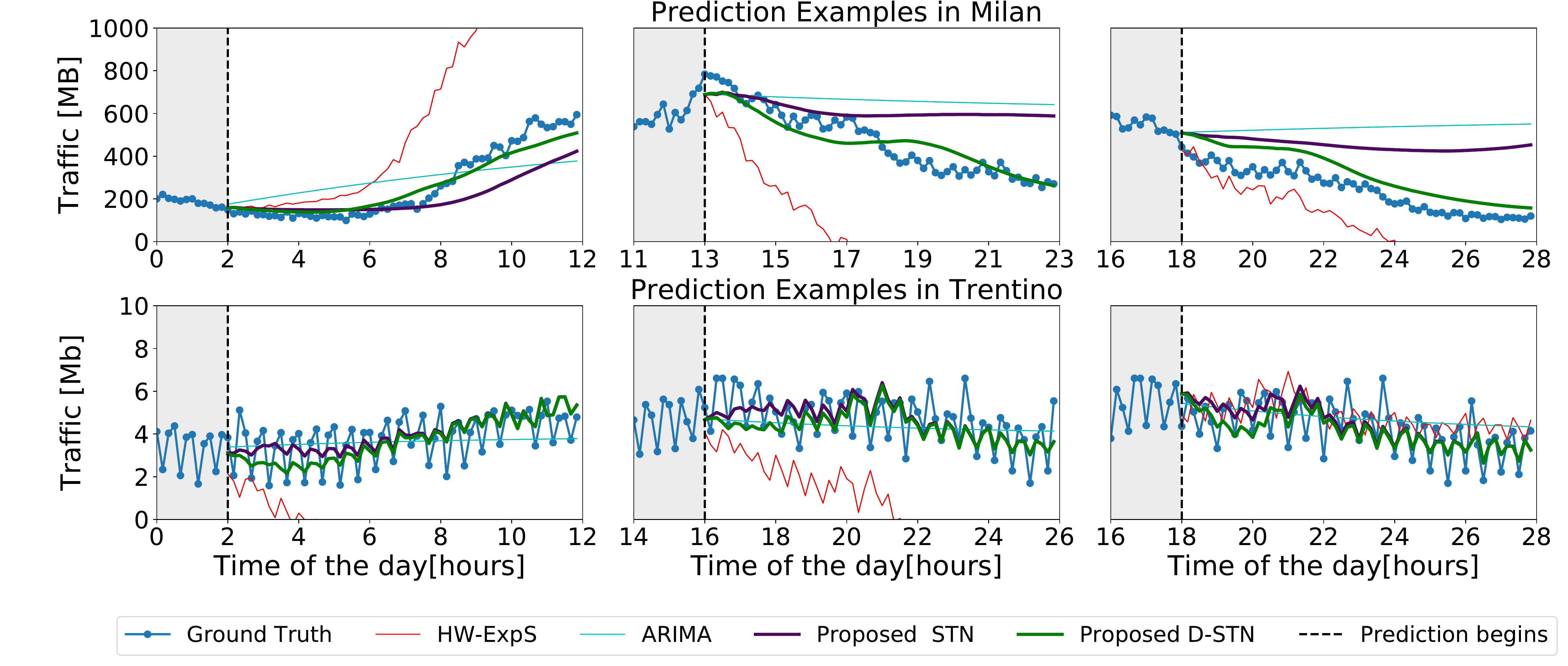}
\end{center}
\caption{\label{Fig:exa_1} Traffic predicted by the proposed (D-)STN and conventional approaches, as well as ground truth measurements in Milan (above) and Trentino (below). Curves in each plot correspond to predictions started at different time of day. Experiments performed at location index (46,57) in Milan during 23\textsuperscript{rd} Dec 2013 and location index (50,62) in Trentino on 21\textsuperscript{st} Dec 2013.}
\end{figure*}

For a fair comparison, (except for the AE+LSTM approach that is limited to 1-step predictions), \emph{the output of all deep learning based approaches evaluated, including the STN, and standard MLP, ConvLSTM, 3D-ConvNet, and SVM, are mixed with the empirical mean $\overline{D}_{t+h}$ of the available ground truth}. That is
\begin{equation}
\begin{aligned}
D'_{t+h} := \gamma(h) D_{t+h}^{M} + (1-\gamma(h))\overline{D}_{t+h}.
\end{aligned}
\end{equation}

We illustrate the mobile data traffic volume predicted by (D-)STN and conventional forecasting techniques, alongside with ground truth observations for a weekday in Milan (above) and a weekend day in Trentino (below) in Fig.~\ref{Fig:exa_1}.  Each sub-figure shows a different scenario, in terms of the time of day when the predictors are triggered. For the deep learning methods, observations are carried out for 2 hours (data shown before the vertical dotted lines) and predictions are performed for a total duration of 10 hours. Note that the 2h-long observations are used as \emph{input} for prediction and not for actual training. HW-ExpS and ARIMA are continuously fed with ground truth data, as required. 

Observe that in predictions in the city of Milan, 
\textbf{STN and D-STN yield the best performance among other approaches, especially as the prediction duration grows.}
In particular, traditional HW-ExpS largely over-/under-estimates future traffic and ARIMA gives almost linear and slowly increasing estimates. 
HW-ExpS employs a constant weight for its seasonal component, which is independent of the prediction step. Hence, the predictions made may deviate immediately if the data exhibits frequent fluctuation at the initial prediction steps. The predictions of ARIMA tend~to~converge to the traffic mean, hence the almost flat long-term behavior.

Further, we compute the average and standard deviation of the NRMSE attained by both traditional and deep learning-based approaches over 11 prediction instances in Milan, 7 of these performed during weekdays and 4 over weekends, with different number of prediction steps employed. For each prediction instance, \textbf{most approaches forecast traffic consumption at all locations across the city for over 60 time steps for all instances}. We only employ AE+LSTM for one-step predictions, as this does not support longer forecasting. We summarise these results in Table \ref{Tab:mi_rpd} where we also include the traditional SVM for completeness. Observe that \textbf{\mbox{D-STN} performs best in all cases}, attaining superior performance to that of other neural network based schemes (i.e. Conv-LSTM, 3D-ConvNet, and MLP) used individually, \textbf{which confirms the effectiveness of combining such elements through multiple fusion layers}, as we propose. In addition D-STN achieves NRMSEs up to 60\% and 38\% smaller than those obtained with HW-ExpS and respectively ARIMA, and as expected SVM is inferior to the MLP. Our approach further outperforms the AE+LSTM solution in terms of one-step prediction by 26\%, which confirms the superior performance of our architecture.

\begin{table}[htb]
\centering
\begin{tabular}{c|cccc}
Method  & 1-step        & 10-step             & 30-step & 60-step        \\
\hline
STN     & \textbf{0.19$\pm$0.02} & 0.29$\pm$0.05 &   0.51$\pm$0.12 & 0.83$\pm$0.14 \\
D-STN   & \textbf{0.19$\pm$0.02} & \textbf{0.28$\pm$0.04} &   \textbf{0.48$\pm$0.09} & \textbf{0.71$\pm$0.18}\\
HW-ExpS & 0.33$\pm$0.03 & 0.51$\pm$0.08 &   0.96$\pm$0.01  &1.79$\pm$0.45\\
ARIMA   & 0.20$\pm$0.04 & 0.39$\pm$0.15 &   0.77$\pm$0.31& 1.00$\pm$0.27\\
MLP   & 0.23$\pm$0.02 & 0.38$\pm$0.03 &    0.67$\pm$0.13 &0.96$\pm$0.22\\
ConvLSTM   & 0.23$\pm$0.02 & 0.39$\pm$0.05 &    0.95$\pm$0.24 &1.49$\pm$0.24\\
3D-ConvNet  & 0.20$\pm$0.02 & 0.37$\pm$0.09 &   0.95$\pm$0.30 &1.64$\pm$0.34\\
SVM  & 0.39$\pm$0.16 & 0.46$\pm$0.11 &   0.62$\pm$0.14 &0.95$\pm$0.19\\
AE+LSTM & 0.24$\pm$0.05 &  -- & --   & -- \\
\end{tabular}
\vspace*{0.5em}
\caption{\label{Tab:mi_rpd} NRMSE (mean$\pm$std) comparison between different predictors over the Milan dataset. Eleven prediction instances triggered at different times of day are used to compute statistics in each case.}

\end{table}

\begin{table}[!b]
\begin{tabular}{c|cccc}
Method  & 1-step        & 10-step       & 30-step  & 60-step      \\
\hline
STN     & 0.47$\pm$0.03 & 0.66$\pm$0.10 &   \textbf{0.76$\pm$0.11} & \textbf{0.85$\pm$0.07}\\
D-STN    & 0.47$\pm$0.03 & 0.68$\pm$0.11 &  0.78$\pm$0.11 &0.86$\pm$0.08\\
HW-ExpS & 0.57$\pm$0.07 & 0.76$\pm$0.08 &   1.28$\pm$0.16  &2.19$\pm$0.49\\
ARIMA   &  \textbf{0.38$\pm$0.09}  &  \textbf{0.61$\pm$0.24}    & 0.95$\pm$0.34& 1.30$\pm$0.35\\
MLP   & 0.67$\pm$0.08 & 0.85$\pm$0.05 &   0.85$\pm$0.08 &0.86$\pm$0.07\\
ConvLSTM   & 0.52$\pm$0.05 & 0.72$\pm$0.07 &   0.80$\pm$0.10  &0.89$\pm$0.08\\
3D-ConvNet   & 0.48$\pm$0.04 & 0.67$\pm$0.07 &  0.77$\pm$0.09 &0.84$\pm$0.07\\
SVM  & 0.40$\pm$0.04 & 0.55$\pm$0.02 &   0.92$\pm$0.15 &1.63$\pm$0.55\\
AE+LSTM & 0.50$\pm$0.06 & -- &  --  & -- \\
\end{tabular}
\vspace*{0.5em}
\caption{\label{Tab:tre_rpd} NRMSE (mean$\pm$std) comparison between different predictors over the Trentino dataset. Eleven prediction instances triggered at different times of day are used to compute statistics in each case.}

\end{table}

We now turn attention to the performance of all schemes on the Trentino dataset. Recall that the deep learning based prediction schemes were trained on the Milan data, whilst HW-ExpS, ARIMA, and AE+LSTM had to be retrained to capture the traffic features of Trentino. We first examine the volume of traffic predicted by our (D-)STN proposals and the other traditional approaches 
in the lower sub-plots of Fig.~\ref{Fig:exa_1}. Note that the traffic volume is relatively small (8 Mbytes consumed on average every 10 minutes) and consumption patterns during weekdays are very similar. Observe that the \textbf{output of STN follows very closely the real measurements, while the performance of D-STN is nearly identical} (curves overlap).  On the contrary, the ARIMA predictor does not follow the ground truth, but instead yields a nearly constant output that is close to the average of the measurements. HW-ExpS substantially underestimates the volume of traffic (except when triggered at 18h). 

\begin{figure*}[!t]
\begin{center}
\includegraphics[width=1\textwidth]{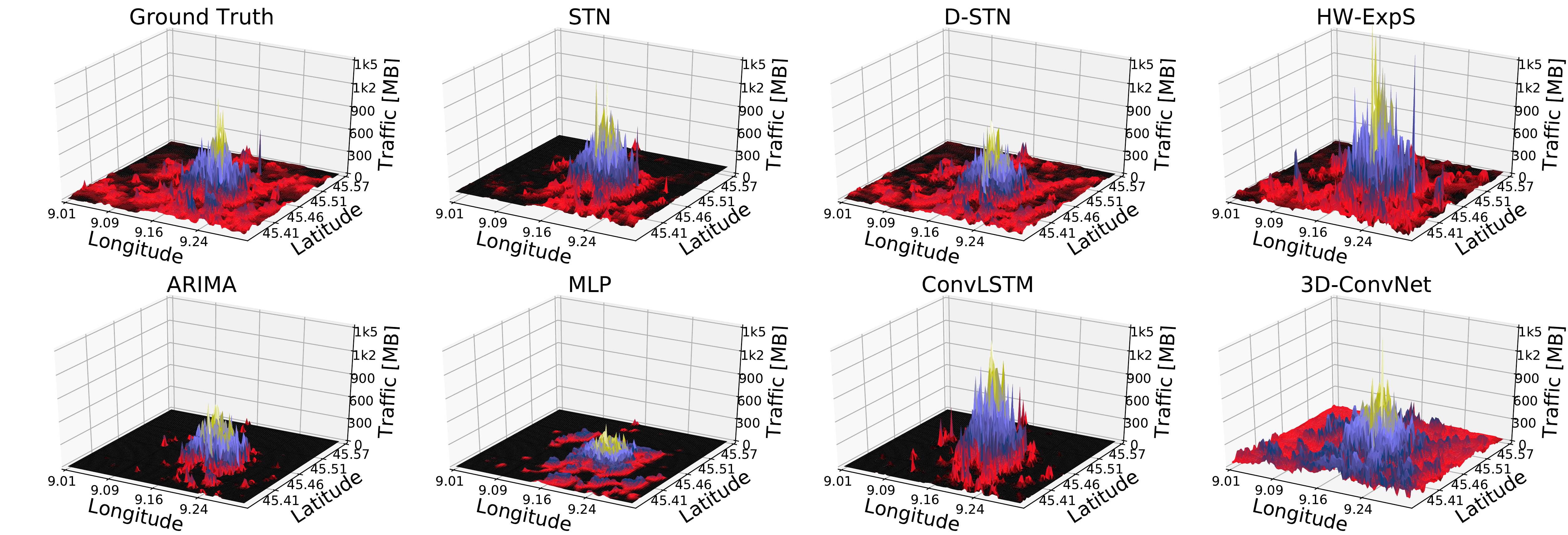}
\end{center}
\caption{\label{Fig:mi_global} Snapshots of network-wide predictions made after 10 hours by the proposed (D-)STN, and existing deep learning based and traditional forecasting approaches in Milan, on 24\textsuperscript{th} Dec 2013.}
\end{figure*}

\begin{figure}[htb]
\centering\includegraphics[width=\columnwidth]{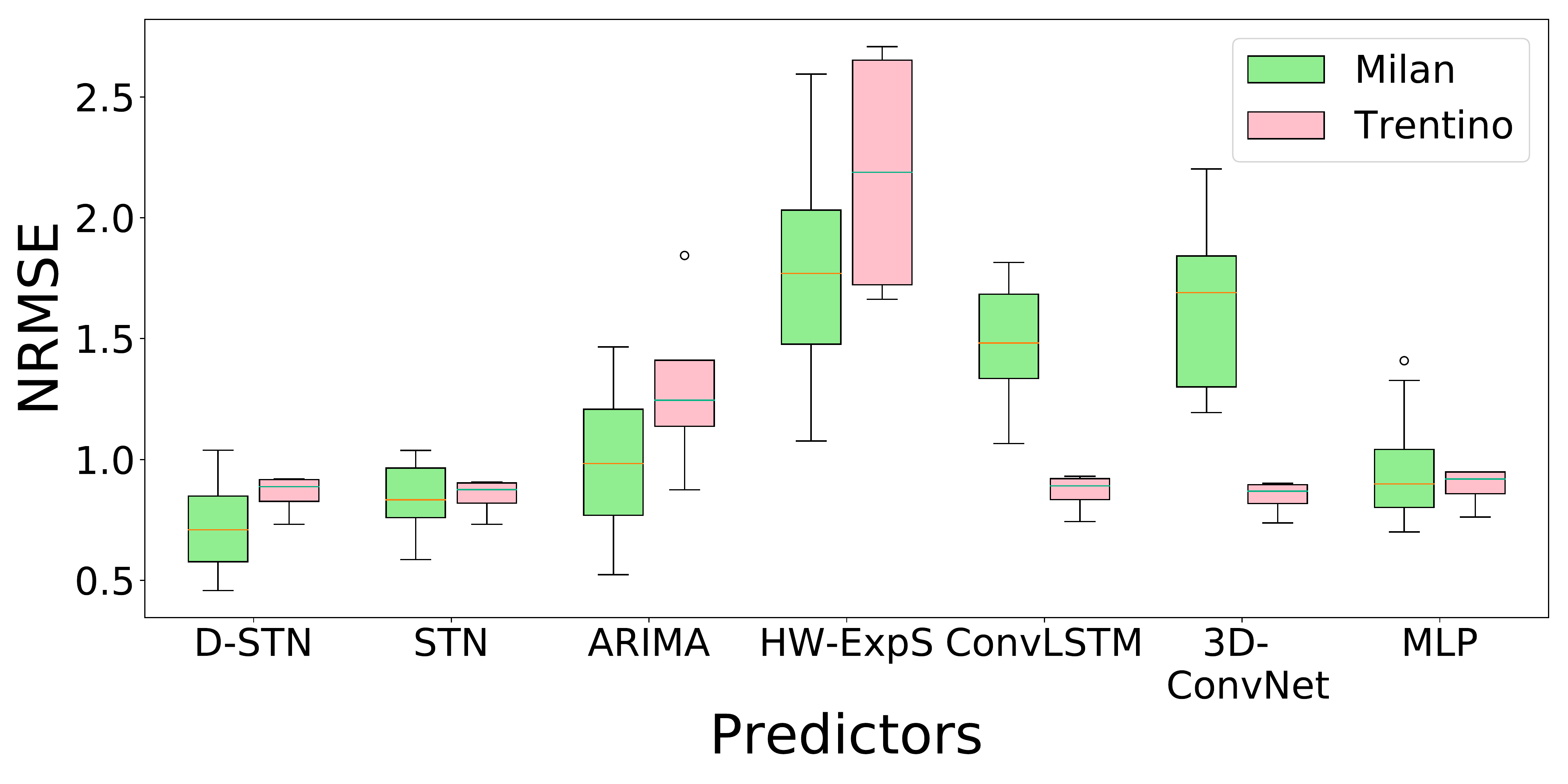}
\caption{\label{Fig:stats} Long-term (10h) prediction statistics (quartiles of NRMSE) for the proposed (D-)STN, conventional, and other deep learning based approaches in Milan and Trentino.}
\end{figure}

We further compute averages and standard deviation of the mean over 11 prediction instances and examine the performance over prediction durations of different lengths (1 to 60 steps). We summarize these results in Table~\ref{Tab:tre_rpd}, where we observe that on average ARIMA is slightly more accurate in short-term predictions as the volume of traffic remains low, while \textbf{STN performs substantially better in the long-term (61\% and 35\% lower NRMSE than HW-ExpS and ARIMA)}. We note that since the OTS is employed over the Milan dataset, in this case D-STN may be unnecessarily sophisticated, given the light and periodic nature of the traffic. 
Specifically, each data point corresponds to a measurement in a $235$m $\times 235$m cell in Milan, while in Trentino the granularity is relative to $1,000$m $\times 1,000$m squares. Further, data consumption in Milan over 10 minutes is almost 8 times higher than in Trentino. However, the performance of D-STN remains very close to that of the STN in Trentino. Short-term, our proposals yield again lower NRMSEs than AE+LSTM. Nonetheless, as we did not have to retrain (D-)STN with the Trentino dataset, the superior performance of our approach also confirms its excellent generalization abilities.

To give additional perspective on the value of employing our proposals for long-term mobile traffic forecasting, in Fig.~\ref{Fig:mi_global} we show snapshots of the predictions made by all approaches considered, across the \emph{entire} network. These are taken at instances that are 10h after measurements have been suspended in the city of Milan. Observe that \textbf{D-STN delivers the best prediction among all approaches}. STN works well at city center level, while slightly underestimating the traffic volume in the surrounding areas. HW-ExpS significantly overestimates the traffic volume, particularly in the central city. The other approaches (ARIMA, MLP, ConvLSTM, and 3D-ConvNet), although successfully capturing the spatial profile of mobile traffic, under-/over-estimate the traffic volume. 

We provide further evidence of the superior performance of the proposed (D-) STN methods in forecasting long-term data traffic consumption, by investigating the distribution of the long-term predictions (60 steps) of all methods considered when these are triggered at different times of the day and during both weekdays and weekends. To this end, we show in Fig.~\ref{Fig:stats} with box-and-whisker plots the quartiles of the NRMSE (min, max, median, 1\textsuperscript{st} and 3\textsuperscript{rd} quartiles) for all approaches in both Milan and Trentino areas. Observe that both STN and D-STN make consistently good predictions, as their median is the lowest in all cases. \rv{In addition, the performance of D-STN is statistically significant over STN in the Milan dataset, as the error box plot has much lower upper and lower bounds.  However, the difference between quartiles is small in the Trentino.}

Not least, it is important to note that linear time series predictors require longer observations to forecast the traffic volume. In particular the 50-day long \textbf{data series used for ARIMA and HW-ExpS are 600 times longer than those employed by the proposed deep learning approaches}. Moreover these techniques cannot be reused without prior re-training in each area. In contrast, \textbf{once trained with one dataset, \mbox{(D-)STN} can be easily re-used in other locations and provides excellent performance, even if the geographical layouts and traffic volume profiles differ}. 

\rv{We also note that our model is not perfect, as it sometime fails to predict traffic spikes. These spikes are mostly caused by noise in the collected data and equipment failures, which are hardly predictable. Other traffic peaks caused by human mobility, can be however, predicted well by our model by capturing the spatio-temporal correlations. We further stress that the traffic spikes can be more predictable if similar patterns have occurred in the training set, and therefore collecting more data for training becomes important to capture richer traffic patterns.}

\subsection{The Spatio-Temporal Impact\label{subsec: gradient}}
Lastly we examine the cross-spatial and cross-temporal correlations between data traffic that is provided as input to the predictors we propose. 

\begin{figure}[htb]
\begin{center}
\includegraphics[width=\columnwidth]{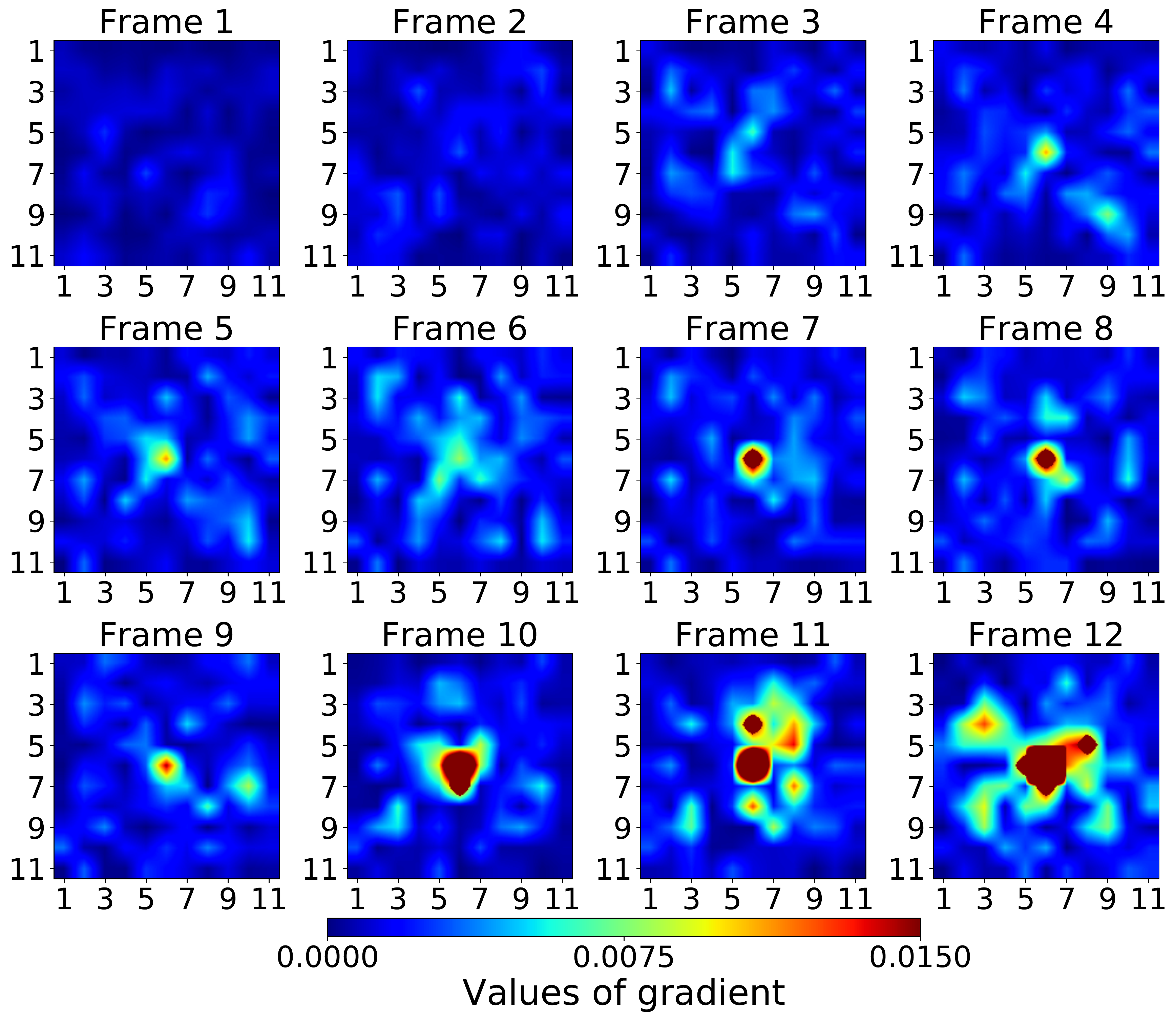}
\end{center}
\caption{\label{Fig:gradient00} Gradient of the loss function over the input for one cell represented as heat maps.}
\end{figure}

The overall model is essentially a complex non-linear function of the input $\mathbf{F}_k$. The relationship between the prediction and measurement observations can be approximated by the first-order term of its Taylor expansion, i.e.

\begin{equation}
\begin{aligned}
\mathcal{M}(\Theta;\mathbf{F}_k) \approx w_{\mathbf{F}_k}^T \cdot \mathbf{F}_k+ b,
\end{aligned}
\end{equation}
\noindent where $w_{\mathbf{F}_k}$ denotes the gradient of the model output $\mathcal{M}(\Theta;\mathbf{F}_k)$ over input $\mathbf{F}_k$ and $b$ is a constant. 

The point-wise contribution of each input data point in~(\ref{eq:matrix}) to the predicted future traffic can be estimated by the $w(\mathbf{F}_k)$ \cite{li2015visualizing},~i.e.
\begin{equation}\label{gra_map}
\begin{aligned}
w_{\mathbf{F}_k} = \frac{\partial\mathcal{M}(\Theta;\mathbf{F}_k) }{\partial  \mathbf{F}_k}.
\end{aligned}
\end{equation}
Evaluating the above should give insights into both the number of temporal steps and measurement points adjacent to a location of interest, required for accurate prediction.

Therefore in Fig.~\ref{Fig:gradient00} we plot as a heat map the gradient of the loss function over the input (\ref{gra_map}) for one cell. Each frame corresponds to one input sampled at a time instance between $t-S+1$ and $t$, to predict the traffic at $t+1$. The $(x,y)$ square in each such frame shows the extent to which the traffic in the cells adjacent to the center, where the target lies, influences the prediction (a color closer to red corresponds to a stronger correlation).

Unsurprisingly, the center of each input accounts for the majority of contributions to the value predicted at $t+1$. The cross-temporal correlation is stronger as the frame index approaches $t$ and we can see that the oldest frame (frame~1) provides limited useful information, since all gradients are close to zero. 
On the other hand, spatial components have a major contribution to the predictions, especially those in the frames closer to $t+1$, i.e. frames 9 to 12 (see the brighter squares around the center of each frame). This is indeed reasonable, since users move in clusters and the volume of traffic consumed is more likely to shift to adjacent cells over short intervals, due to their mobility. Further, we observe that cells at the boundaries of the squares we consider (i.e. $11\times 11 \times 12$) yield small gradient values, which supports the dependence assumption we make in (\ref{eq.indepedance}). 
We believe that taking into consideration such spatial information is key to the superior prediction accuracy attained by STN, as compared to other single time series predictive methods (e.g. HW-ExpS and ARIMA).

We conclude that, by exploiting the spatio-temporal features of mobile traffic, the proposed D-STN and STN schemes substantially outperform conventional traffic forecasting techniques (including HW-ExpS and ARIMA) as they attain up to 61\% lower prediction errors (NRMSE), do not require to be trained to capture the specifics of a particular city, and operate with up to 600 times shorter measurement intervals, in order to make long-term (up to 10h) predictions. (D-)STN further outperform other deep learning based approaches, such as ConvLSTM, 3D-ConvNet, MLP, and SVM.

\section{Summary}
Forecasting traffic in cellular networks is becoming increasingly important to dynamically allocate network resources and timely respond to exponentially growing user demand. This task becomes particularly difficult as network deployments densify and the cost of accurate monitoring steepens. In this chapter, we proposed a Spatio-Temporal neural Network (STN), which is a precise traffic forecasting architecture that leverages recent advances in deep neural networks and overcomes the limitations of prior forecasting techniques. We introduced an Ouroboros Training Scheme (OTS) to fine tune the pre-trained model. Subsequently, we proposed a Double STN (D-STN), which employs a light-weight mechanism for combining the STN output with historical statistics, thereby improving long-term prediction performance. Experiments conducted with publicly available 60-day long traffic measurements collected in the city of Milan and the Trentino region demonstrate the proposed \mbox{(D-)STN} provide up to 61\% lower prediction errors as compared to widely employed ARIMA and HW-ExpS methods, while requiring up to 600 times shorter ground truth measurement durations.

\rv{We further recognize that Gaussian Processes have potential to resolve the mobile traffic forecasting, by automatically selecting a combination of kernels to fit time series data \cite{steinruecken2019a}. This can further model the prediction uncertainty to better support traffic engineering for mobile operators. Future work will focus on comparing and combining a seasonal Gaussian Process with our D-STN, to further enhance forecasting performance.}

%% file: chap4.tex
\chapter{Deep Learning Driven Mobile Traffic Forecasting on Geospatial Point Clouds\label{chap:fore2}}


Digital services continue to diversify and demand often conflicting performance guarantees (\eg low latency vs. high throughput vs. high reliability). Precision traffic engineering thus becomes increasingly important, as operators must be able to allocate resources intelligently for each service and anticipate individual future demands. Network slicing is a first step towards addressing these challenges, enabling to logically isolate network infrastructure on a per-service basis. Allocating sufficient resources to a certain slice however requires precise measurements and detailed real-time analysis of mobile traffic~\cite{bega2019deepcog}. Achieving this is computationally expensive~\cite{naboulsi2016large} and relies heavily on specialized equipment (\eg measurement probes~\cite{keysight}). 


To achieve precise mobile traffic forecasting in support of network slicing, this chapter attacks the network-wide traffic prediction over a broad range of mobile services at the antenna level. Unlike previous forecasting on the city grids, antennas are usually non-uniformly distributed over a region. Forecasting over such scattered antennas poses a different point-cloud stream forecasting problem. Point-cloud stream forecasting seeks to predict the future values and/or locations of data streams generated by a geospatial point cloud $\mathcal{S}$, given sequences of historical observations \cite{shi2018machine}.
Unlike traditional spatio-temporal forecasting on grid-structural data (\eg precipitation nowcasting \cite{xingjian2015convolutional}, or video frame prediction \cite{wang2018predrnn++}), point-cloud stream forecasting needs to operate on geometrically scattered sets of points, which are irregular and unordered, and encapsulate complex spatial correlations. While vanilla Long Short-term Memories (LSTMs) have modest abilities to exploit spatial features \cite{xingjian2015convolutional}, convolution-based recurrent neural network (RNN) models (\eg ConvLSTM \cite{xingjian2015convolutional} and PredRNN++ \cite{wang2018predrnn++}) are limited to modeling grid-structural data, and are therefore inappropriate for handling scattered point-clouds. Leveraging the location information embedded in such irregular data sources, so as to learn important spatio-temporal features, is in fact challenging.

\begin{figure}[h]
\centering
\centering
\includegraphics[width=\columnwidth, trim={1cm 0.5cm 1cm 1cm}]{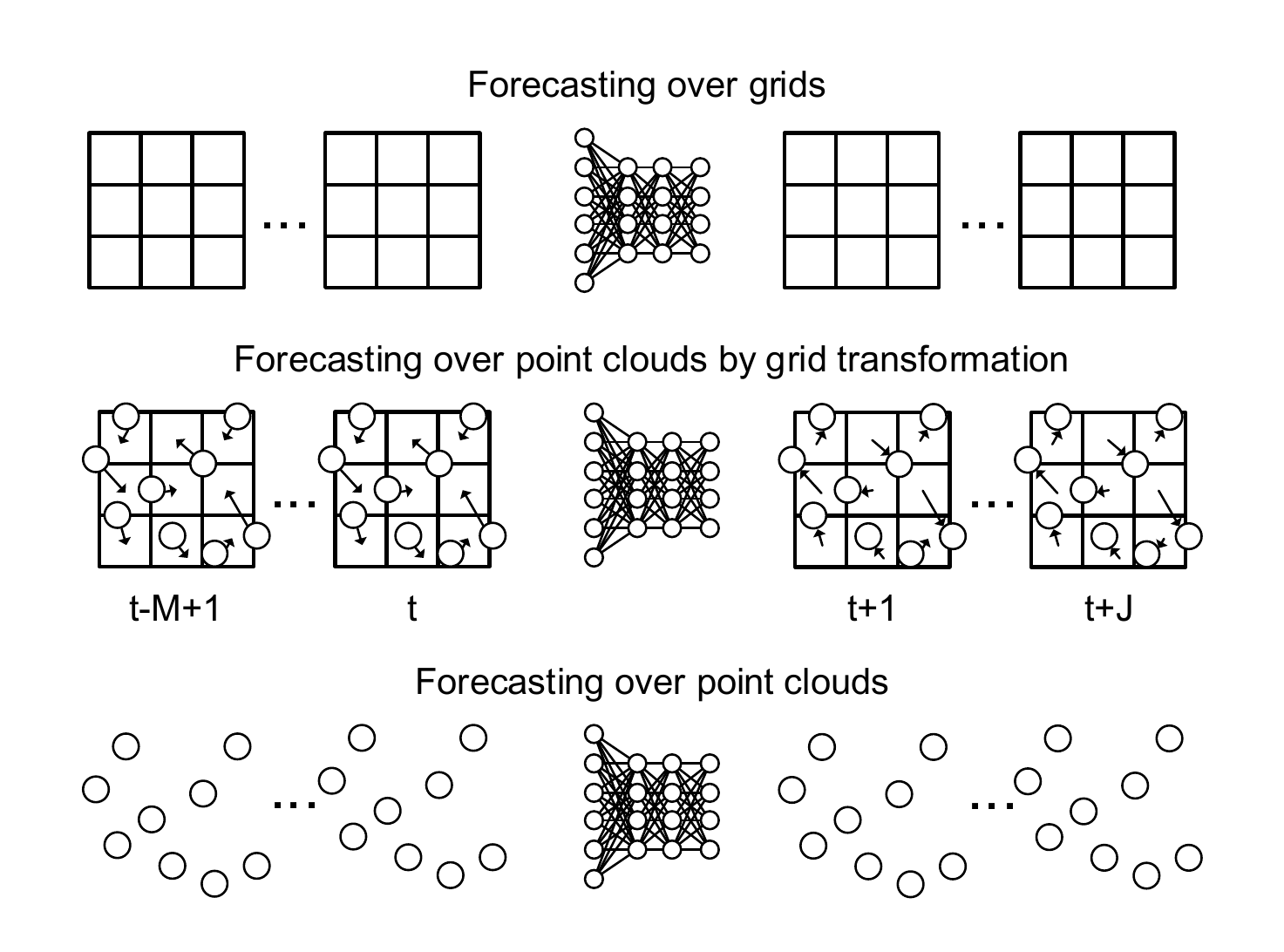}
\caption{Different approaches to geospatial data stream forecasting: predicting over input data streams that are inherently grid-structured, \eg video frames using ConvLSTMs (top); mapping of point-cloud input to a grid, \eg mobile network traffic collected at different antennas in a city, to enable forecasting using existing neural network structures (middle); forecasting directly over point-cloud data streams using historical information (as above, but without pre-processing), as proposed in this chapter (bottom).
\label{fig:forecast}}
\end{figure}

Existing approaches that tackle the point-cloud stream forecasting problem can be categorized into two classes, both of which bear significant shortcomings: \emph{(i)} methods that transform point-clouds into data structures amenable to processing with mature solutions (\eg grids \cite{Patr1907:Multi}, see Fig.\ref{fig:forecast}); and \emph{(ii)} models that ignore the exact locations of each data source and inherent spatial correlations (\eg \cite{liang2018geoman}). The transformations required by the former not only add data preprocessing overheads, but also introduce spatial displacements, which distorts relevant correlations among points \cite{Patr1907:Multi}. On the other hand, the latter are largely location-invariant, while recent literature suggests spatial correlations should be revisited over time, to suit series prediction tasks~\cite{shi2017deep}. In essence, \emph{overlooking dynamic spatial correlations will lead to modest forecasting performance}.

Beyond learning from sequential data, deep neural networks for feature learning on spatial-only point cloud are advancing rapidly. PointNet performs feature learning and maintains input permutation invariance~\cite{qi2017pointnet}. PointNet++ upgrades this structure by hierarchically partitioning point-clouds and performing feature extraction on local regions~\cite{qi2017pointnet++}. VoxelNet employs voxel feature encoding to limit inter-point interactions within a voxel~\cite{zhou2018voxelnet}. This effectively projects cloud-points onto sub-grids, which enables feature learning. Li \etal generalize the convolution operation on point-clouds and employ $\mathcal{X}$-transformations to learn the weights and permutations for the features \cite{li2018pointcnn}. Through this, the proposed PointCNN leverages spatial-local correlations of point clouds, irrespective of the order of the input. Notably, although these architectures can learn spatial features of point-clouds, they are designed to work with static data, thus have limited ability to discover temporal dependencies.

\textbf{Contributions.} In this chapter, we introduce Convolutional Point-cloud LSTMs (CloudLSTMs), a new branch of recurrent neural network models tailored to geospatial point-cloud stream forecasting. The CloudLSTM builds upon a Dynamic Point-cloud Convolution (\dc) operator, which takes raw point-cloud streams (both data time series and spatial coordinates) as input, and performs dynamic convolution over these, to learn spatio-temporal features over time, irrespective of the topology and permutations of the point-cloud. This eliminates the data preprocessing overheads mentioned above and circumvents the negative effects of spatial displacement. The proposed CloudLSTM takes into account the locations of each data source and performs \textit{dynamic positioning} at each time step, to conduct a deformable convolution operation over point-clouds \cite{dai2017deformable}. This allows revising the spatial and temporal correlations, and the configuration of the data points over time, and guarantees that the location-variant property is met at different steps. Importantly, the \dc operator is flexible, as it can be easily plugged into existing neural network models for different purposes, such as RNNs, LSTMs, sequence-to-sequence (Seq2seq) learning \cite{sutskever2014sequence}, and attention mechanisms \cite{luong2015effective}.

We perform antenna-level forecasting of data traffic generated by mobile services \cite{zhang2018long, bega2019deepcog} as a case study, experimenting with metropolitan-scale mobile traffic measurements collected in two European cities for 38 popular mobile apps. This represents an important application of geospatial point-cloud stream forecasting. We combine our CloudLSTM with Seq2seq learning and an attention mechanism, then undertake a comprehensive evaluation on both datasets. The results obtained demonstrate that our architecture can deliver precise long-term mobile traffic forecasting, outperforming eight different baseline neural network models in terms of four performance metrics, without any data preprocessing requirements. To the best of knowledge, \rv{\emph{the proposed CloudLSTM is the first dedicated neural architecture for spatio-temporal forecasting that operates directly on point-cloud streams}.}

\section{Convolutional Point-cloud LSTM}

Next, we describe in detail the concept and properties of forecasting over point cloud-streams. We then introduce the \dc operator, which is at the core of our proposed CloudLSTM architecture. Finally, we present CloudLSTM and its variants, and explain how to combine CloudLSTM with Seq2seq learning and attention mechanisms, to achieve precise forecasting over point-cloud streams.

\subsection{Forecasting over Point-cloud Streams \label{sec:fore}}
We formally define a point-cloud containing a set of $N$ points, as $\mathcal{S} = \{p_1, p_2, \cdots, p_N\}$. Each point $p_n \in \mathcal{S}$ contains two sets of features, \ie $p_n = \{\nu_n, \varsigma_n \}$, where $\nu_n = \{v_n^1, \cdots, v_n^H\}$ are value features (\eg mobile traffic measurements) of $p_n$, and $\varsigma_n = \{c_n^1, \cdots, c_n^L\}$ are its $L$-dimensional coordinates. At each time step $t$, we may obtain $U$ different channels of $\mathcal{S}$ by conducting different measurements\footnote{These resemble the RGB channels in images.} denoted by $\mathcal{S}_t^\upsilon = \{\mathcal{S}_t^1, \cdots, \mathcal{S}_t^U\}, \: \: \mathcal{S}_t^\upsilon\in \mathbb{R}^{U\times N\times (H + L)}$. We can then formulate the $J$-step point-cloud stream forecasting problem, given $M$ observations, as:
\begin{equation}
\widehat{\mathcal{S}}_{t+1}^\upsilon, \cdots, \widehat{\mathcal{S}}_{t+J}^\upsilon = \argmax_{\mathcal{S}_{t+1}^\upsilon, \cdots, \mathcal{S}_{t+J}^\upsilon}p(\mathcal{S}_{t+1}^\upsilon, \cdots, \mathcal{S}_{t+J}^\upsilon|\mathcal{S}_{t}^\upsilon, \cdots, \mathcal{S}_{t-M+1}^\upsilon).
\end{equation}
Note that, in some cases, each point's coordinates may be unchanged, since the data sources are deployed at fixed locations. \rv{In this chapter, we focus specifically on the mobile traffic volume collected at the antenna level, \ie predicting the $J$ step future mobile traffic consumption in a region covered by a scattered antenna set.} An ideal point-cloud stream forecasting model should embrace five key properties, similar to other point-cloud applications and spatio-temporal forecasting problems~\cite{qi2017pointnet, shi2017deep}:\\
\emph{(i)} \textbf{Order invariance:} A point cloud is usually arranged without a specific order. Permutations of the input points should not affect the output of the forecasting \cite{qi2017pointnet}.\\
\emph{(ii)} \textbf{Information intactness:} The output of the model should have exactly the same number of points as the input, without losing any information, $\ie N_\mathrm{out} = N_\mathrm{in}$.\\
\emph{(iii)} \textbf{Interaction among points:}  Points in $\mathcal{S}$ are not
isolated, thus the model should be able to capture local dependencies among neighboring points and allow interactions \cite{qi2017pointnet}.\\
\emph{(iv)} \textbf{Robustness to transformations:} The model should be robust to  correlation-preserving transformation operations on point-clouds, \eg scaling and shifting \cite{qi2017pointnet}.\\
\emph{(v)} \textbf{Location variance:} The spatial correlations among points may change over time. Such dynamic correlations should be revised and learnable during training \cite{shi2017deep}.

In what follows, we introduce the Dynamic Point-cloud Convolution (\dc) operator as the core module of the Cloud-LSTM, and explain how \dc satisfies the aforementioned properties.

\subsection{Dynamic Convolution over Point Cloud}
The Dynamic Point-cloud Convolution operator (\dc) absorbs the concept of ordinary convolution over grids, which takes $U_1$ channels of 2D tensors as input, and outputs $U_2$ channels of 2D tensors of smaller size (if without padding). Similarly, the \dc takes $U_1$ channels of a point-cloud $\mathcal{S}$, and outputs $U_2$ channels of a point-cloud, but with the same number of elements as the input, to ensure the \emph{information intactness} property \emph{(ii)} discussed previously. For simplicity, we denote the $i^{th}$ channel of the input set as $\mathcal{S}^i_{in}$ and the $j^{th}$ channel of the output as $\mathcal{S}^j_\mathrm{out}$. Both $\mathcal{S}^i_{in}$ and $\mathcal{S}^j_\mathrm{out}$ are 2D tensors, of shape $(N,(H\!+\!L))$ and $(N,(H\!+\!L))$ respectively.

We also define $\mathcal{Q}_n^{\mathcal{K}}$ as a subset of points in $\mathcal{S}^i_{in}$, which includes the $\mathcal{K}$ nearest points with respect to $p_n$ in the Euclidean space, \ie $\mathcal{Q}_n^{\mathcal{K}} = \{p_n^1,\cdots,p_n^k,\cdots,p_n^\mathcal{K}\}$, where $p_n^k$ is the $k$-{th} nearest point to $p_n$ in the set $\mathcal{S}^i_{in}$. Note that $p_n$ itself is included in $\mathcal{Q}_n^{\mathcal{K}}$ as an anchor point, \ie $p_n \equiv  p_n^1$. Recall that each $p_n \in \mathcal{S}$ contains $H$ value features and $L$ coordinate features, \ie $p_n = \{\nu_n, \varsigma_n \}$, where $\nu_n = \{v_n^1, \cdots, v_n^H\}$ and $\varsigma_n = \{c_n^1, \cdots, c_n^L\}$. Similar to the vanilla convolution operator, for each $p_n$ in $\mathcal{S}^i_{in}$, the \dc sums the element-wise product over all features and points in $\mathcal{Q}_n^{\mathcal{K}}$, to obtain the values and coordinates of a point $p_n'$ in $\mathcal{S}^j_\mathrm{out}$. Note that we assume the value features are related to their positions at the previous layer/state, to better exploit the dynamic spatial correlations. Therefore, we aggregate coordinate features $c(p_n^k)_i^l$ when computing the value features $v_{n,j}^{h'}$. The mathematical expression of the \dc is thus:
\begin{align*}
v_{n,j}^{h'} &= \sum_{i\in U_\mathrm{in}}\sum_{p_n^k \in \mathcal{Q}_n^{\mathcal{K}}} \Big(\sum_{h \in H}w_{i,j}^{h,h',k} v(p_n^k)_i^h+\sum_{l \in L}w_{i,j}^{(H+l),h',k}  c(p_n^k)_i^l\Big) + b_j,\\
c_{n,j}^{l'} &=\sigma\Bigg(\sum_{i\in U_\mathrm{in}}\sum_{p_n^k \in \mathcal{Q}_n^{\mathcal{K}}} \Big(\sum_{h \in H}w_{i,j}^{h,l',k} v(p_n^k)_i^h+\sum_{l \in L}w_{i,j}^{(H+l),l',k}  c(p_n^k)_i^l\Big) + b_j\Bigg),
\end{align*}
\vspace*{-0.5cm}
\begin{align}
\mathcal{S}^j_\mathrm{out} &= (p_1',\cdots, p_N') 
\label{eq:cloudcnn}\\ 
&= 
\left \{  \{\nu_1', \varsigma_1' \}, \cdots, \{\nu_N', \varsigma_N' \} \right \} \nonumber \\ 
&= 
\Bigg(  \Big((v_1^{1'}, \cdots, v_1^{H'}), (c_1^{1'}, \cdots, c_1^{L'})\Big), \cdots, \Big((v_N^{1'}, \cdots, v_N^{H'}), (c_N^{1'}, \cdots, c_N^{L'})\Big) \Bigg). \nonumber
\end{align}
In the above, we define learnable weights $\mathcal{W}$ as 5D tensors with shape $(U_\mathrm{in}, \mathcal{K}, (H+L), (H+L), U_\mathrm{out})$. The weights are shared across different anchor points in the input map. Each element $w_{i, j}^{m, m', k} \in \mathcal{W}$ is a scalar weight for the $i$-{th} input channel, $j$-{th} output channel, $k$-{th} nearest neighbor of each point corresponding to the $m$-{th} value and coordinate features for each input point, and $m'$-th value and coordinate features for output points.  Similar to the convolution operator, we define $b_j$ as a bias for the $j$-{th} output map. In the above,  $h$ and $h'$ are the $h$\textsuperscript{(}$'$\textsuperscript{)}-th value features of the input/output point set. Likewise, $l$ and $l'$ are the $l$\textsuperscript{(}$'$\textsuperscript{)}-th coordinate features of the input/output. $\sigma(\cdot)$ is the sigmoid function, which limits the range of predicted coordinates to $(0, 1)$, to avoid outliers.  Before feeding them to the model, the coordinates of raw point-clouds are normalized to $(0, 1)$ by $\varsigma = (\varsigma - \varsigma_{\min})/(\varsigma_{\max} - \varsigma_{\min})$, on each dimension. This improves the transformation robustness of the operator. 

Note that The $\mathcal{K}$ nearest points can indeed vary for each channel at each location. The reason is that the channels in the point-cloud dataset may represent different types of measurements. For example, channels in the mobile traffic dataset are related to the traffic consumption of different mobile apps. The spatial correlations will vary between different measurements (channels), due to human mobility. For instance, more people may use Facebook at a social event, but YouTube traffic may be less significant in this case. This will be reflected by the data consumption of each mobile service. We want these spatial correlations to be learnable, so we do not fix the $\mathcal{K}$ nearest neighbours across channels, but encourage each channel to find the best neighbour set. This is also a contribution of the CloudLSTM, which helps improve the forecasting performance, which will be demonstrated later.

Here, we show that the normalization of the coordinates features enables transformation invariance with shifting and scaling. The shifting and scaling of a point can be represented as:
\begin{equation}
\varsigma' = A\varsigma + B,
\end{equation}
where $A$ and $B$ are a positive scaling coefficient and respectively an offset. By normalizing the coordinates, we have:
\begin{equation}
\begin{aligned}
\varsigma'&= \frac{\varsigma' - \varsigma'_{\min}}{\varsigma'_{\max} - \varsigma'_{\min}}\\
&= \frac{(A\varsigma + B) - (A\varsigma_{\min} +B)}{(A\varsigma_{\max} +B) - (A\varsigma_{\min} +B)}\\
&= \frac{A(\varsigma - \varsigma_{\min})}{A(\varsigma_{\max} - \varsigma_{\min})}\\
&= \frac{\varsigma - \varsigma_{\min}}{\varsigma_{\max} - \varsigma_{\min}}.
\end{aligned}
\end{equation}
This implies that, by using normalization, the model is invariant to shifting and scaling transformations.

\begin{figure}[h]

\centering
\centering\includegraphics[width=\columnwidth, trim={1cm 1cm 1cm 1cm}]{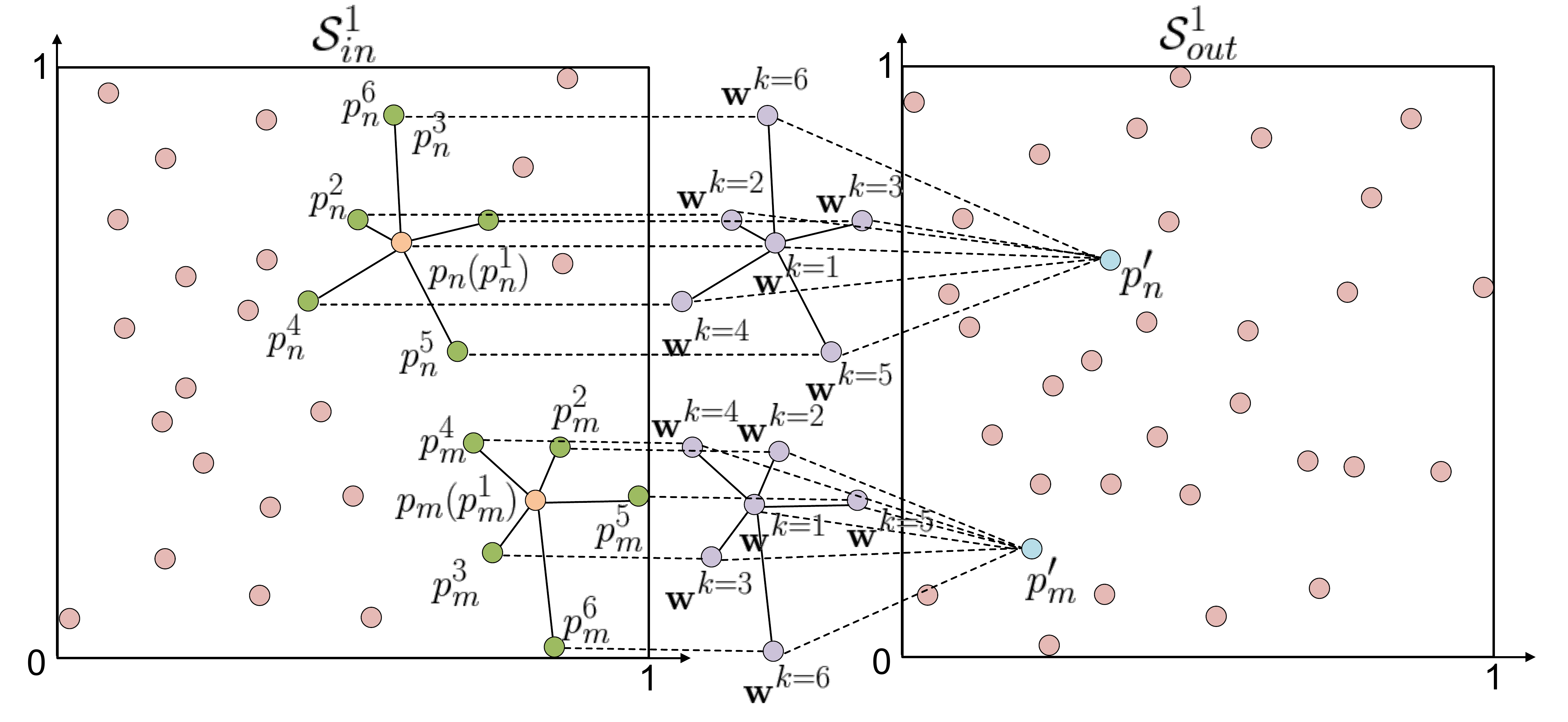}
\caption{Illustration of the \dc operator, with a single input channel and $\mathcal{K}=6$ neighbors. For every $p \in \mathcal{S}_{in}^1$, \dc weights its $\mathcal{K}$ neighboring set $\mathcal{Q}_n^{\mathcal{K}} = \{p_n^1,\cdots,p_n^6\}$ to produce values and coordinate features for $p_n' \in \mathcal{S}_\mathrm{out}^1$. Here, each $\mathbf{w}^k$ is a set of weights $w$ with index $k$ (\ie $k$-{th} nearest neighbor) in Eq.~\ref{eq:cloudcnn}, shared across different $p$. 
\label{fig:dc}} 
\end{figure}

We provide a graphical illustration of \dc in Fig.~\ref{fig:dc}. For each point $p_n$, the \dc operator weights its $\mathcal{K}$ nearest neighbors across all features, to produce the values and coordinates in the next layer. Since the permutation of the input neither affects the neighboring information nor the ranking of their distances for any $\mathcal{Q}_n^{\mathcal{K}}$, \dc is a symmetric function whose output does not depend on the input order. This means that the property \emph{(i)} discussed in Sec.~\ref{sec:fore} is satisfied. Further, \dc is performed on every point in set $\mathcal{S}^i_{in}$ and produces exactly the same number of features and points for its output; property \emph{(ii)} is therefore naturally fulfilled. In addition, operating over a neighboring point set, irrespective of its layout, allows to capture local dependencies and improve the robustness to global transformations (\eg shifting and scaling). The normalization over the coordinate features further improves the robustness to those transformations. This enables to meet the desired properties \emph{(iii)} and \emph{(iv)}. More importantly, \dc learns the layout and topology of the cloud-point for the next layer, which changes the neighboring set $\mathcal{Q}_n^{\mathcal{K}}$ for each point at output \smash{$\mathcal{S}^j_\mathrm{out}$}. This enables the ``location-variance'' (property \emph{(v)}), allowing the model to perform dynamic positioning tailored to each channel and time step. This is essential in spatio-temporal forecasting neural models, as spatial correlations change over time \cite{shi2017deep}. \dc can be efficiently implemented using simple 2D convolution, by reshaping the input map and weight tensor, which can be parallelized easily in existing deep learning frameworks. 

We assume a batch size of 1 for simplicity. Recall that the input and output of \dc, $\mathcal{S}'_{in}$ and $\mathcal{S}_\mathrm{out}$, are 3D tensors with shape $(N, (H+L), U_1)$ and $(N, (H+L), U_2)$, respectively. Note that for each $p_n$ in $\mathcal{S}^i_{in}$, we find the set of top $\mathcal{K}$ nearest neighbors $\mathcal{Q}_n^\mathcal{K}$. Combining these, we transform the input into a 4D tensor $\mathcal{S}^{i'}_{in}$, with shape $(N, \mathcal{K}, (H+L), U_1)$. To perform \dc over $\mathcal{S}^{i'}_{in}$, we split the operator into the following steps in Algorithm~\ref{alg:dc}, which enables to translate the \dc into a standard convolution operation, which is highly optimized by existing deep learning frameworks.
\begin{algorithm}[h]
  \caption{Efficient algorithm for \dc implementation using the 2D convolution operator \label{alg:dc}}
  \begin{algorithmic}[1]
  \color{black}
    \Inputs{$\mathcal{S}'_{in}$, with shape $(N, \mathcal{K}, (H+L), U_1)$.}
    \Initialize{The weight tensor $\mathcal{W}$.}
    \State{Reshape the input map $\mathcal{S}^{i'}_{in}$ from shape $(N, \mathcal{K}, (H+L), U_1)$ to shape $(N, \mathcal{K}, (H+L)\times U_1)$}
    \State{Reshape the weight tensor $\mathcal{W}$ from shape $(U_1, \mathcal{K}, (H+L), (H+L), U_2)$ to shape $(1, \mathcal{K}, U_1 \times (H+L), U_2 \times (H+L))$}
    \State{Perform 2D convolution $\mathcal{S}_\mathrm{out} = Conv(\mathcal{S}^{i'}_{in}, \mathcal{W})$ with step 1 without padding. $\mathcal{S}_\mathrm{out}$ becomes a 3D tensor with shape $(N, 1, U_2 \times (H+L))$}
    \State{Reshape the output map $\mathcal{S}_\mathrm{out}$ to $(N, (H+L), U_2)$}
    \State{Apply the sigmoid function $\sigma (\cdot)$ to the coordinates feature in $\mathcal{S}_\mathrm{out}$}
    \end{algorithmic}
\end{algorithm}

\textbf{Complexity Analysis.} We study the complexity of \dc by separating the operation into two steps: \emph{(i)} finding the neighboring set $\mathcal{Q}_n^{\mathcal{K}}$ for each point $p_n \in \mathcal{S}$, and \emph{(ii)} performing the weighting computation in Eq.~\ref{eq:cloudcnn}. We discuss the complexity of each step separately. For simplicity and without loss of generality, we assume the number of input and output channels are both 1. For step \emph{(i)}, the complexity of computing a point-wise Euclidean distance matrix is $O(L\cdot N^2)$, while finding $\mathcal{K}$ nearest neighbors for one point has complexity $O(N\log \mathcal{K})$, if using heapsort \cite{schaffer1993analysis}. As such, \emph{(i)} has complexity $O(L\cdot N^2+N\log \mathcal{K})$. For step \emph{(ii)}, it is easy to see from Eq.~\ref{eq:cloudcnn} that the complexity of computing one feature of the output $p'_n$ is $O((H+L)\cdot \mathcal{K})$. Since each point has $(H+L)$ features and the output point set \smash{$\mathcal{S}^j_\mathrm{out}$} has $N$ points, the overall complexity of step \emph{(ii)} becomes $O(N\cdot \mathcal{K} \cdot (H+L)^2)$. This is equivalent to the complexity of a vanilla convolution operator, where both the input and output have $(H+L)$ channels, and the input map and kernel have $N$ and $\mathcal{K}$ elements, respectively. This implies that, compared to the convolution operator whose inputs, outputs, and filters have the same size, \dc introduces extra complexity by searching the $\mathcal{K}$ nearest neighbors for each point.

\rv{\textbf{Relations with PointCNN \cite{li2018pointcnn} and Deformable Convolution \cite{dai2017deformable}.}} The \dc operator builds upon the PointCNN \cite{li2018pointcnn} and deformable convolution neural network (DefCNN) on grids \cite{dai2017deformable}, but introduces several variations tailored to point-cloud structural data. PointCNN employs the $\mathcal{X}$-transformation over point clouds, to learn the weight and permutation on a local point set using multilayer perceptrons (MLPs), which introduces extra complexity. This operator guarantees the order invariance property, but leads to information loss, since it performs aggregation over points. In our \dc operator, the permutation is maintained by aligning the weight of the ranking of distances between point $p_n$ and $\mathcal{Q}_n^{\mathcal{K}}$. Since the distance ranking is unrelated to the order of the inputs, the order invariance is ensured in a parameter-free manner without extra complexity and loss of information.

Further, the \dc operator can be viewed as the DefCNN \cite{dai2017deformable} over point-clouds, with the differences that \emph{(i)} DefCNN deforms weighted filters, while \dc deforms the input maps; and \emph{(ii)} DefCNN employs bilinear interpolation over input maps with a set of continuous offsets, while \dc instead selects $\mathcal{K}$ neighboring points for its operations. Both DefCNN and \dc have transformation modeling flexibility, allowing adaptive receptive fields on convolution. \rv{We note that to the best of our knowledge, \dc is the first convolutional operator for point cloud stream processing, which makes the major contribution of this chapter.}

\subsection{The CloudLSTM Architecture \label{sec:cloudlstm}}
\begin{figure*}[htb]
\centering\includegraphics[width=1.03\columnwidth]{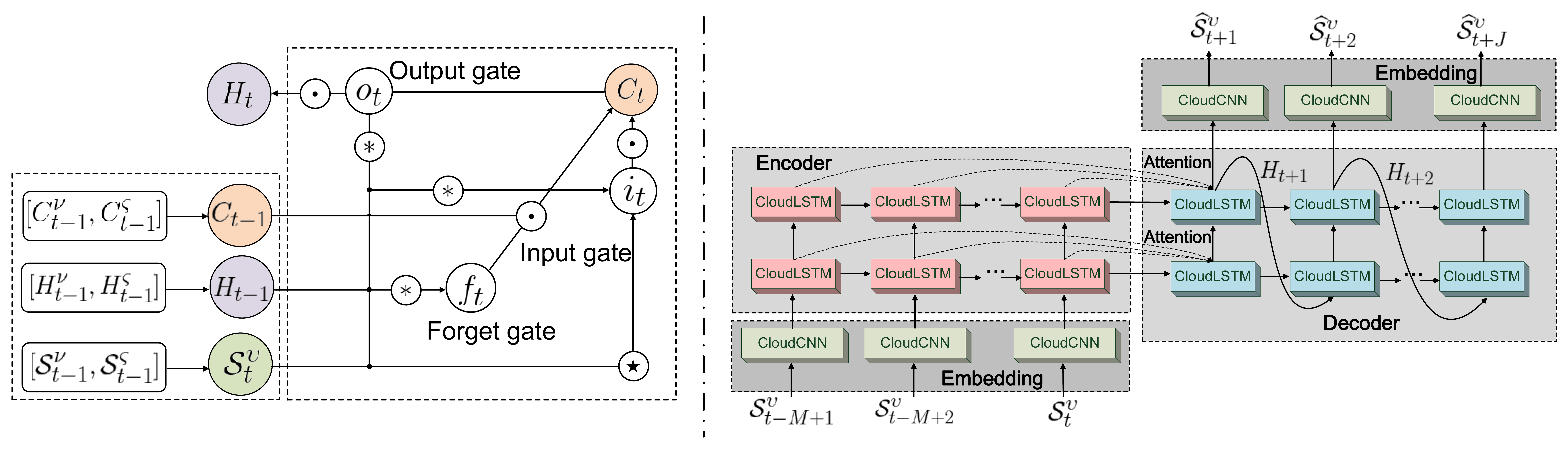}
\caption{The inner structure of the CloudLSTM cell (left) and the overall Seq2seq CloudLSTM architecture (right). We denote by $(\cdot)^{\nu}$ and $(\cdot)^{\varsigma }$ the value and coordinate features of each input, while these features are unified for gates.\label{fig:cloudlstm}} 
\end{figure*}

The \dc operator can be plugged straightforwardly into LSTMs, to learn both spatial and temporal correlations over point-clouds. We formulate the Convolutional Point-cloud LSTM (CloudLSTM) as:
\begin{equation}
\begin{aligned}
&i_t = \sigma(\mathcal{W}_{si}\oast \mathcal{S}_{t}^\upsilon + \mathcal{W}_{hi}\oast H_{t-1} + b_i),\\
&f_t = \sigma(\mathcal{W}_{sf}\oast \mathcal{S}_{t}^\upsilon + \mathcal{W}_{hf}\oast H_{t-1}+b_f),\\
&C_t = f_t\odot C_{t-1} + i_t \odot \tanh(\mathcal{W}_{sc}\ostar \mathcal{S}_{t}^\upsilon + \mathcal{W}_{hc}\ostar H_{t-1}+b_c),\\
&o_t = \sigma(\mathcal{W}_{so}\oast \mathcal{S}_{t}^\upsilon + \mathcal{W}_{ho}\oast H_{t-1} +b_o),\\
&H_t = o_t\odot \tanh(C_t).
\end{aligned}
\label{eq:cloudlstm}
\end{equation}
Similar to ConvLSTM \cite{xingjian2015convolutional},  $i_t$, $f_t$, and $o_t$, are input, forget, and output gates respectively. $C_t$ denotes the memory cell and $H_t$ is the hidden states. Note that $i_t$, $f_t$, $o_t$, $C_t$, and $H_t$ are all point cloud representations. $\mathcal{W}$ and $b$ represent learnable weight and bias tensors. In Eq.~\ref{eq:cloudlstm}, `$\odot$' denotes the element-wise product, `$\ostar$' is the \dc operator formalized in Eq.~\ref{eq:cloudcnn}, and `$\oast$' a simplified \dc that removes the sigmoid function in Eq.~\ref{eq:cloudcnn}. The latter only operates over the gates computation, as the sigmoid functions are already involved in outer calculations (first, second, and fourth expressions in Eq.~\ref{eq:cloudlstm}). We show the structure of a basic CloudLSTM cell in the left subplot of Fig.~\ref{fig:cloudlstm}.

We combine our CloudLSTM with Seq2seq learning \cite{sutskever2014sequence} and the soft attention mechanism \cite{luong2015effective}, to perform forecasting, given that these neural models have been proven to be effective in spatio-temporal modeling on grid-structural data (\eg \cite{xingjian2015convolutional, zhang2018attention}). 

\noindent \textbf{Sequence-to-sequence learning} is a technique widely-applied to recurrent neural network (RNN) for machine translation tasks \cite{sutskever2014sequence}. This structure employs a RNN model to encode the input into a low-dimensional tensor. The decoder is another RNN model that decodes the tensor encoded into a sequence. The sequence-to-sequence architecture provides an end-to-end mapping from different sequences, which is particular well-suited for the mobile traffic forecasting problem, since essentially we seek to infer a sequence (\ie future mobile traffic demand) from another sequence (\ie mobile traffic measurements observation).

\noindent \textbf{Soft attention mechanism} We denote the $j$-{th} and $i$-{th} states of the encoder and decoder as $H^{j}_{en}$ and $H^{i}_{de}$. The context tensor for state $i$ at the encoder can be represented as:
\begin{equation}
\begin{aligned}
c_i &= \sum _{j\in M} a_{i,j}H^{j}_{en} = \frac{e^{i,j}}{\sum _{j\in M} e_{i,j}},
\end{aligned}
\end{equation}
where $e_{i,j}$ is a score function, which can be selected among many alternatives. In this chapter, we choose $e_{i,j} = \mathbf{v}_a^T\tanh (\mathbf{W}_a\ast [H^{j}_{en};H^{i}_{de}])$. Here $[\cdot;\cdot]$ is the concatenation operator and $\ast$ is the convolution function. Both $\mathbf{W}_a$ and $\mathbf{v}_a$ are learnable weights. The $H^{i}_{de}$ and context tensor are concatenated into a new tensor for the following operations.

We show the overall Seq2seq CloudLSTM in the right subplot of Fig.~\ref{fig:cloudlstm}. The architecture incorporates an encoder and a decoder, which are different stacks of CloudLSTMs. The encoder encodes the historical information into a tensor, while the decoder decodes the tensor into predictions. The states of the encoder and decoder are connected using the soft attention mechanism via a context vector \cite{luong2015effective}.
Before feeding the point-cloud to the model and generating the final forecasting, the data is processed by Point Cloud Convolutional (CloudCNN) layers, which perform the \dc operations. Their function is similar to the word embedding layer in natural language processing tasks \cite{mikolov2013distributed}, which helps translate the raw point cloud into tensors and \emph{vice versa}. In this study, we employ a two-stack encoder-decoder architecture, and configure 36 channels for each CloudLSTM cell, as we found that further increasing the number of stacks and channels does not improve the performance significantly.

Beyond CloudLSTM, we also explore plugging the \dc into vanilla RNN and Convolutional GRU, which leads to a new Convolutional Point-cloud RNN (CloudRNN) and Convolutional Point-cloud GRU (CloudGRU), as formulated by the following equations respectively:\\
\begin{equation}
\begin{aligned}
&\textbf{CloudRNN:}\\
&h_t = \sigma(\mathcal{W}_{sh}\oast \mathcal{S}_{t}^\upsilon + \mathcal{W}_{sy}\oast y_{t-1} + b_h),\\
&y_t = \sigma(\mathcal{W}_{yh}\oast h_t + b_y)
\end{aligned}
\label{eq:cloudrnn}
\end{equation}
\begin{equation}
\begin{aligned}
&\textbf{CloudGRU:}\\
&z_t = \sigma(\mathcal{W}_{sz}\oast \mathcal{S}_{t}^\upsilon + \mathcal{W}_{hz}\oast H_{t-1} + b_z),\\
&r_t = \sigma(\mathcal{W}_{sr}\oast \mathcal{S}_{t}^\upsilon + \mathcal{W}_{hr}\oast H_{t-1}+b_r),\\
&H_t' = \tanh(r_t \odot \mathcal{W}_{h'z}\oast H_{t-1} + \mathcal{W}_{x'z}\oast \mathcal{S}_{t}^\upsilon)\\
&H_t = (1-z_t) \odot H_t' + z_t \odot H_{t-1}
\end{aligned}
\label{eq:cloudlgru}
\end{equation}
The CloudRNN and CloudGRU share a similar Seq2seq architecture with CloudLSTM, except that they do not employ the attention mechanism. \rv{They are employed as baselines to compare different sequential structures for mobile traffic forecasting. Note that the attention mechanism is not used in the CloudRNN and CloudGRU.}  We compare their performance in the following section.

\section{Experiments}

To evaluate the performance of our architectures, we employ antenna-level mobile traffic forecasting as a case study and experiment with two large-scale mobile traffic datasets. We use the proposed CloudLSTM to forecast future mobile traffic consumption at scatter-distributed antennas in the regions of interest. We provide a comprehensive comparison with 12 baseline deep learning models, over four performance metrics. All models considered in this study are implemented using the open-source Python libraries TensorFlow~\cite{tensorflow2015-whitepaper} and TensorLayer~\cite{tensorlayer}. We train all architectures with a computing cluster with two NVIDIA Tesla K40M GPUs. We optimize all models by minimizing the mean square error (MSE) between predictions and ground truth, using the Adam optimizer \cite{kingma2015adam}.

Next, we first introduce the dataset employed in this study, then discuss the baseline models used for comparison, the experimental settings, and the performance metrics employed for evaluation. Finally, we report on the experimental results and provide visualizations that reveal further insights.

\subsection{Dataset and Preprocessing \label{sec:dataset_paris}}
We conduct experiments using large-scale multi-service datasets collected by a major operator in two large European metropolitan areas during 85 consecutive days. The data consists of the volume of traffic generated by devices associated to each of the 792 and respectively 260 antennas in the two target areas. The antennas are non-uniformly distributed over the urban regions, thus they can be viewed as 2D point clouds over space. Their locations are fixed across the measurements period.

\begin{figure*}[h]
\centering\includegraphics[width=1\columnwidth]{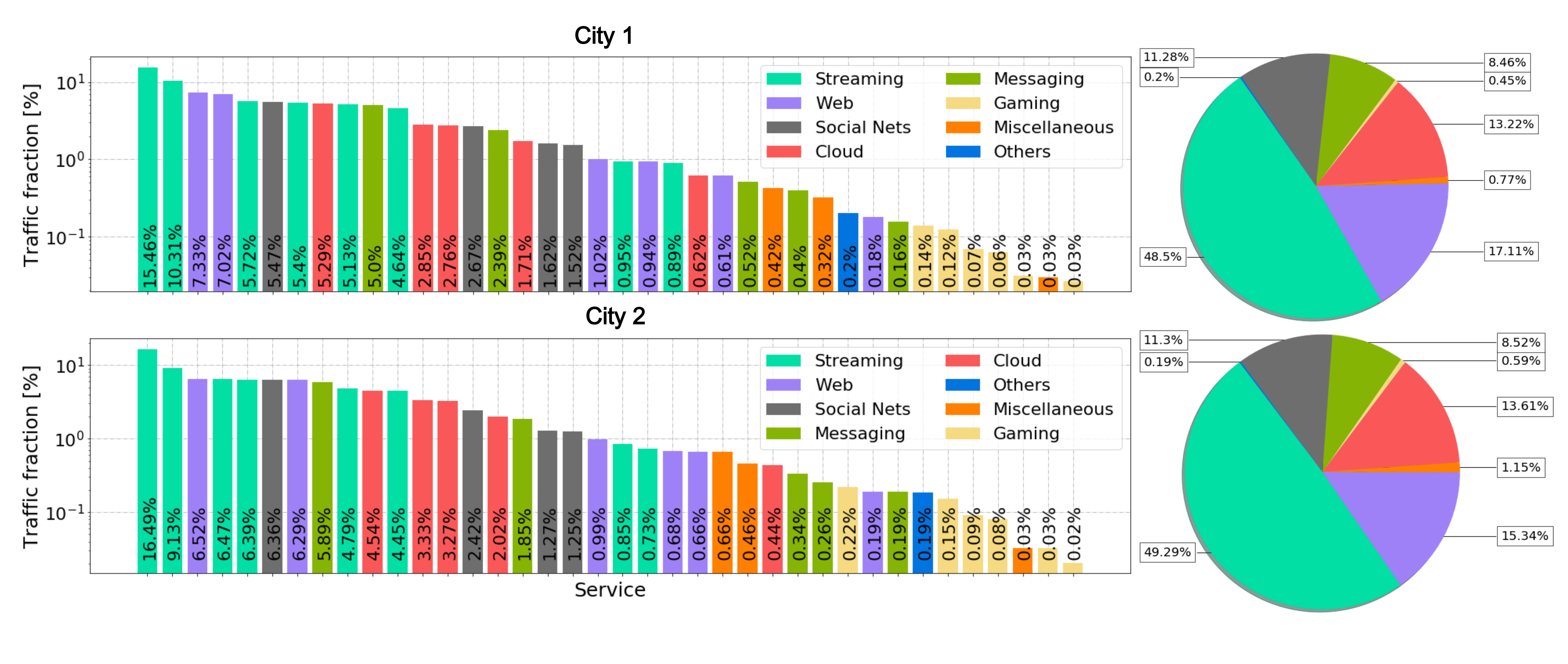}
\caption{Fraction of the total traffic consumed by each mobile service (left) and each service category (right) in the considered set. \label{fig:traffic-stats0}} 
\end{figure*}
\begin{figure*}
\centering\includegraphics[width=1\columnwidth]{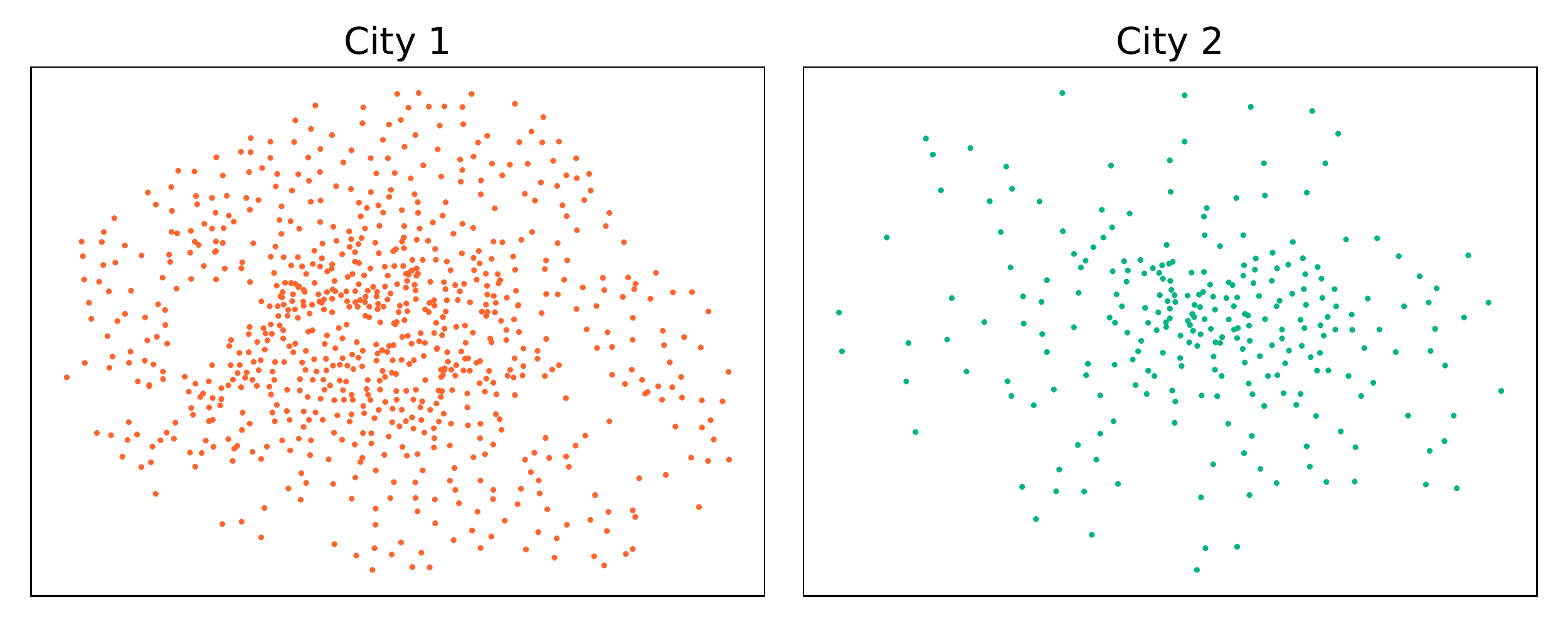}
\caption{The anonymized locations of the antenna set in both cities.} \label{fig.city_locs}
\end{figure*}

The measurement data is collected via traditional flow-level deep packet inspection at the packet gateway (PGW). Proprietary traffic classifiers are used to associate flows to specific services. \rv{Due to data protection and confidentiality constraints, we do not disclose the name of the operator, the target metropolitan regions, or the detailed operation of the classifiers.} For similar reasons, we cannot name the exact mobile services studied. We show the anonymized locations of the antennas sets in both cities in Fig.~\ref{fig.city_locs}.

As a final remark on data collection, we stress that all measurements were carried out under the supervision of the competent national privacy agency and in compliance with applicable regulations. In addition, the dataset we employ for our study only provides mobile service traffic information accumulated at the antenna level, and does not contain personal information about individual subscribers. This implies that the dataset is fully anonymized and its use for our purposes does not raise privacy concerns.

As already mentioned, the set of services $\mathcal{S}$ considered in our analysis comprises 38 different services. An overview of the fraction of the total traffic consumed by each service and each category in both cities throughout the duration of the measurement campaign is in Fig.\,\ref{fig:traffic-stats0}. The left plot confirms the power law previously observed in the demands generated by individual mobile services. Also, streaming is the dominant type of traffic, with five services ranking among the top ten. This is confirmed in the right plot, where streaming accounts for almost half of the total traffic consumption. Web, cloud, social media, and chat services also consume large fractions of the total mobile traffic, between 8\% and 17\%, whereas gaming only accounts for 0.5\% of the demand.

At each antenna, the traffic volume is expressed in Megabytes and aggregated over 5-min intervals, which leads to 24,482 traffic snapshots. These snapshots are gathered independently for each of 38 different mobile services, selected among the most popular mobile services, including video streaming, gaming, messaging, cloud services, social networking, \etc\, Due to data protection and confidentiality constraints, we do not disclose the identity of the mobile operator, and we do not provide information about the exact location of the data collection equipment, or the names of the mobile services considered. The data collection procedure was conducted under the supervision of the competent national privacy agency, which complies with regulations. The dataset is fully anonymized, as it only comprises service traffic aggregated at the antenna level, without unveiling personal information.%

Before feeding to the models the traffic measurements for each mobile service, these are transformed into different input channels of the point-cloud $\mathcal{S}$. All coordinate features $\varsigma$ are normalized to the $(0,1)$ range. In addition, for the baseline models that require grid-structural input (\ie CNN, 3D-CNN, ConvLSTM and PredRNN++), the point clouds are transformed into grids \cite{Patr1907:Multi} using the Hungarian algorithm~\cite{kuhn1955hungarian}, as required. The ratio of training, validation, and testing sets is $8$:$1$:$1$.

\subsection{Benchmarks and Performance Metrics}
We compare the performance of our proposed CloudLSTM with a set of baseline models, as follows. MLP \cite{Goodfellow-et-al-2016}, CNN \cite{krizhevsky2012imagenet}, and 3D-CNN \cite{ji20133d} are frequently used as benchmarks in mobile traffic forecasting (\eg \cite{bega2019deepcog, zhang2018long}). DefCNN learns the shape of the convolutional filters and has similarities with the \dc operator proposed in this study \cite{dai2017deformable}. PointCNN \cite{li2018pointcnn} performs convolution over point-clouds and has been employed for point-cloud classification and segmentation. LSTM is an advanced RNN frequently employed for time series forecasting \cite{hochreiter1997long}. CloudCNN is an original benchmark we introduce, which stacks the proposed \dc operator over multiple layers for feature extraction from point-clouds. 
\rv{PointLSTM is another original benchmark we design, obtained by replacing the cells in ConvLSTM with the $\mathcal{X}$-Conv operator employed by PointCNN, which provides a fair term of comparison for other Seq2seq architectures.} While ConvLSTM \cite{xingjian2015convolutional} can be viewed as a baseline model for spatio-temporal predictive learning, the PredRNN++ is the state-of-the-art architecture for spatio-temporal forecasting on grid-structural data and achieves the best performance in many applications \cite{wang2018predrnn++}. Beyond these models, we also compare the CloudLSTM with two of its variations, \ie CloudRNN and CloudGRU, which were introduced in Sec.~\ref{sec:cloudlstm}.
We show in Table~\ref{tab:model2} the detailed configuration along with the number of parameters for each model considered in this study. \rv{We note that the hyper-parameters of all models are optimized based on our experience, as well as on the performance obtained on the validation set during the training.}
\begin{table}[h!]
\centering
\caption{The configuration of all models considered in this study. \label{tab:model2}}
\begin{tabular}{|c|C{10cm}|}
\hline
\textbf{Model}                       & \textbf{Configuration}  \\ \hline
MLP                         &  Five hidden layers, 500 hidden units for each layer                               \\ \hline
CNN                         &  Eleven 2D convolutional layers, each applies 108 channels and $3\times 3$ filters, with batch normalization and ReLU functions.                               \\ \hline
3D-CNN                      &  Eleven 3D convolutional layers, each applies 108 channels and $3\times 3\times 3$ filters, with batch normalization and ReLU functions.                               \\ \hline
DefCNN                      &  Eleven 2D convolutional layers, each applies 108 channels and $3\times 3$ filters, with batch normalization and ReLU functions. Offsets are predicted by separate convolutional layers                              \\ \hline
PointCNN                    &  Eight $\mathcal{X}$-Conv layers, with K, D, P, C=[9, 1, -1, 36]                              \\ \hline
CloudCNN                    &  Eight \dc layers, , with 36 channels and $\mathcal{K}=9$                              \\ \hline
LSTM                        &  2-stack Seq2seq LSTM, with 500 hidden units                               \\ \hline
ConvLSTM                    &  2-stack Seq2seq ConvLSTM, with 36 channels and $3\times 3$ filters                               \\ \hline
PredRNN++                   &  2-stack Seq2seq PredRNN++, with 36 channels and $3\times 3$ filters                               \\ \hline
PointLSTM                  &  2-stack Seq2seq PointLSTM, with K, D, P, C=[9, 1, -1, 36]                               \\ \hline
CloudRNN                    & 2-stack Seq2seq CloudRNN, with 36 channels and $\mathcal{K}=9$                                \\ \hline
CloudGRU                    & 2-stack Seq2seq CloudGRU, with 36 channels and $\mathcal{K}=9$                                \\ \hline
CloudLSTM ($\mathcal{K}=3$) & 2-stack Seq2seq CloudLSTM, with 36 channels and $\mathcal{K}=3$                                \\ \hline
CloudLSTM ($\mathcal{K}=6$) & 2-stack Seq2seq CloudLSTM, with 36 channels and $\mathcal{K}=6$                                \\ \hline
CloudLSTM ($\mathcal{K}=9$) & 2-stack Seq2seq CloudLSTM, with 36 channels and $\mathcal{K}=9$                                \\ \hline
Attention CloudLSTM         & 2-stack Seq2seq CloudLSTM, with 36 channels, $\mathcal{K}=9$ and soft attention mechanism                               \\ \hline
\end{tabular}
\end{table}

We quantify the accuracy of the proposed CloudLSTM, in terms of Mean Absolute Error (MAE) and Root Mean Square Error (RMSE)\@. Since the mobile traffic snapshots can be viewed as ``urban images''~\cite{liu2015urban}, we also select Peak Signal-to-Noise Ratio (PSNR) and Structural Similarity Index (SSIM) \cite{hore2010image} to quantify the fidelity of the forecasts and their similarity with the ground truth, as suggested by relevant recent work \cite{zhang2017zipnet}. These are defined as:
\begin{equation}
\begin{aligned}
\mathrm{MAE}(t) =  \frac{1}{|N|\cdot |H|} \sum_{n \in \mathcal{N}} \sum_{h \in \mathcal{H}}|\widehat{v}_n^h(t) - v_n^h(t)|.\label{eq:mae}
\end{aligned}
\end{equation}
\begin{equation}
\begin{aligned}
\mathrm{RMSE}(t) =  \sqrt{\frac{1}{|N|\cdot |H|} \sum_{n \in \mathcal{N}} \sum_{h \in \mathcal{H}}||\widehat{v}_n^h(t) - v_n^h(t)||^2}.\label{eq:rmse}
\end{aligned}
\end{equation}
\begin{align}
 \text{PSNR}(t) = &\ 20 \log v_{\max}(t)\ - 10 \log \frac{1}{|N|\cdot |H|} \sum_{n \in \mathcal{N}} \sum_{h \in \mathcal{H}}||\widehat{v}_n^h(t) - v_n^h(t)||^2.\label{eq:psnr}
\end{align}
\begin{equation}
\begin{aligned}
&\text{SSIM}(t) =  
&\frac{\left(2\ \widehat{v}_n^h(t)\ \mu_v(t) + c_1\right)\left(2\ \text{\sc cov} (v_n^h(t),\widehat{v}_n^h(t)) + c_2\right)}
{\left({\widehat{v}_n^h(t)}^2\ \mu_v(t)^2  + c_1\right)\left( \text{\sc var} (v_n^h(t)) \text{\sc var}(\widehat{v}_n^h(t) +c_2 \right)},\label{eq:ssim}
\end{aligned}
\end{equation}
where $\mu_v(t)$ and $v_{\max}(t)$ are the average and maximum traffic recorded for all services, at all antennas and time instants of the test set. $\text{\sc var}(\cdot)$ and $\text{\sc cov}(\cdot)$ denote the variance and covariance, respectively. Coefficients $c_1$ and $c_2$ are employed to stabilize the fraction in the presence of weak denominators. Following standard practice, we set $c_1 = (k_1L)^2$ and $c_2 = (k_2L)^2$, where $L = 2$ is the dynamic range of float type data, and $k_1 = 0.1$, $k_2 = 0.3$. We optimize all architectures using the MSE loss function:
\begin{equation}\label{eq:mse}
    \mathrm{MSE}(t) = \frac{1}{|N|\cdot |H|} \sum_{n \in \mathcal{N}} \sum_{h \in \mathcal{H}}||\widehat{v}_n^h(t) - v_n^h(t)||^2.
\end{equation}
Here $\widehat{v}_n^h$ is the mobile traffic volume forecast for the $h$-{th} service at antenna $n$ at time $t$, and $v_n^h$ is its corresponding ground truth.

We employ all neural networks to forecast city-scale future mobile traffic consumption for up to 30 mins, given consecutive 30-min measurements sampled every 5 minutes. That is all models take as input 6 snapshots ($M=6$) and forecast following 6 traffic volume snapshots ($J=6$). For RNN-based models, \ie LSTM, ConvLSTM, PredRNN++, CloudLSTM, CloudRNN, and CloudGRU, we extend the number of prediction steps to $J=36$ (3 hours), to evaluate their long-term performance.

\subsection{Result and Visualization}
We perform 6-step forecasting for 4,888 instances across the test set, and report in Table~\ref{tab:metric} the mean and standard deviation (std) of each metric. We also investigate the effect of different numbers of neighboring points considered (\ie $\mathcal{K}=3, 6, 9$), as well as the influence of the attention mechanism.

\begin{table}[]\scriptsize
\centering
\caption{The mean$\pm$std of MAE, RMSE, PSNR, and SSIM across all models considered, evaluated on two datasets collected in different cities for mobile traffic forecasting. \label{tab:metric}}
\vspace*{1em}
\begin{tabular}{|>{\hspace{-0.2pc}}c<{\hspace{-2.2pt}}|>{\hspace{-0.2pc}}c<{\hspace{-2.2pt}}>{\hspace{-0.2pc}}c<{\hspace{-2.2pt}}>{\hspace{-0.2pc}}c<{\hspace{-2.2pt}}>{\hspace{-0.2pc}}c<{\hspace{-2.2pt}}|>{\hspace{-0.2pc}}c<{\hspace{-2.2pt}}>{\hspace{-0.2pc}}c<{\hspace{-2.2pt}}>{\hspace{-0.2pc}}c<{\hspace{-2.2pt}}>{\hspace{-0.2pc}}c<{\hspace{-2.2pt}}|}
\hline
\multirow{2}{*}{\textbf{Model}}                                                 & \multicolumn{4}{c|}{\textbf{City 1}}                                                               & \multicolumn{4}{c|}{City 2}                                                                        \\ \cline{2-9} 
                                                                                & \textbf{MAE}           & \textbf{RMSE}          & \textbf{PSNR}           & \textbf{SSIM}          & \textbf{MAE}           & \textbf{RMSE}          & \textbf{PSNR}           & \textbf{SSIM}          \\ \hline
MLP                                                                             & 4.79$\pm$0.54          & 9.94$\pm$2.56          & 49.56$\pm$2.13          & 0.27$\pm$0.12          & 4.59$\pm$0.59          & 9.44$\pm$2.45          & 50.30$\pm$2.28          & 0.33$\pm$0.14          \\
CNN                                                                             & 6.00$\pm$0.62          & 11.02$\pm$2.09         & 48.93$\pm$1.60          & 0.25$\pm$0.12          & 5.30$\pm$0.51          & 10.05$\pm$2.06         & 49.97$\pm$1.87          & 0.32$\pm$0.14          \\
3D-CNN                                                                          & 4.99$\pm$0.57          & 9.94$\pm$2.44          & 49.74$\pm$2.13          & 0.33$\pm$0.14          & 5.21$\pm$0.48          & 9.97$\pm$2.03          & 50.13$\pm$1.85          & 0.37$\pm$0.16          \\
DefCNN                                                                          & 6.76$\pm$0.81          & 11.72$\pm$2.57         & 48.43$\pm$1.82          & 0.16$\pm$0.08          & 5.31$\pm$0.51          & 9.99$\pm$2.13          & 49.84$\pm$1.87          & 0.32$\pm$0.14          \\
PointCNN                                                                        & 4.95$\pm$0.53          & 10.10$\pm$2.46         & 49.43$\pm$2.06          & 0.27$\pm$0.12          & 4.75$\pm$0.56          & 9.55$\pm$2.32          & 50.17$\pm$2.16          & 0.35$\pm$0.15          \\
CloudCNN                                                                        & 4.81$\pm$0.58          & 9.91$\pm$2.81          & 49.93$\pm$2.21          & 0.29$\pm$0.11          & 4.68$\pm$0.52          & 9.39$\pm$2.22          & 50.31$\pm$2.03          & 0.36$\pm$0.14          \\
LSTM                                                                            & 4.20$\pm$0.66          & 9.58$\pm$3.17          & 50.47$\pm$3.29          & 0.36$\pm$0.10          & 4.32$\pm$1.64          & 9.17$\pm$3.03          & 50.79$\pm$3.26          & 0.42$\pm$0.12          \\
ConvLSTM                                                                        & 3.98$\pm$1.60          & 9.25$\pm$3.10          & 50.47$\pm$3.29          & 0.36$\pm$0.10          & 4.09$\pm$1.59          & 8.87$\pm$2.97          & 51.10$\pm$3.33          & 0.42$\pm$0.12          \\
PredRNN++                                                                       & 3.97$\pm$1.60          & 9.29$\pm$3.12          & 50.43$\pm$3.30          & 0.36$\pm$0.10          & 4.07$\pm$1.56          & 8.87$\pm$2.97          & 51.09$\pm$3.34          & 0.42$\pm$0.12          \\
PointLSTM                                                                       & 4.63$\pm$0.45          & 9.47$\pm$2.55          & 50.02$\pm$2.26          & 0.34$\pm$0.14          & 4.56$\pm$0.54          & 9.26$\pm$2.43          & 50.52$\pm$2.35          & 0.37$\pm$0.15          \\ \hline
CloudRNN ($\mathcal{K}=9$)                                                      & 4.08$\pm$1.66          & 9.19$\pm$3.17          & 50.45$\pm$3.23          & 0.32$\pm$0.12          & 4.08$\pm$1.65          & 8.74$\pm$3.03          & 51.10$\pm$3.26          & 0.39$\pm$0.14          \\
CloudGRU ($\mathcal{K}=9$)                                                      & 3.79$\pm$1.59          & 8.90$\pm$3.11          & 50.73$\pm$3.29          & 0.39$\pm$0.10          & 3.90$\pm$1.57          & 8.47$\pm$2.96          & 51.40$\pm$3.33          & 0.45$\pm$0.12          \\ \hline
CloudLSTM ($\mathcal{K}=3$)                                                     & 3.71$\pm$1.63          & 8.87$\pm$3.11          & 50.76$\pm$3.30          & 0.39$\pm$0.10          & 3.86$\pm$1.51          & 8.42$\pm$2.94          & 51.45$\pm$3.32          & 0.46$\pm$0.11          \\
CloudLSTM ($\mathcal{K}=6$)                                                     & 3.72$\pm$1.63          & 8.91$\pm$3.13          & 50.72$\pm$3.29          & 0.38$\pm$0.10          & 3.84$\pm$1.59          & 8.46$\pm$2.96          & 51.43$\pm$3.33          & 0.45$\pm$0.12          \\
CloudLSTM ($\mathcal{K}=9$)                                                     & 3.72$\pm$1.62          & 8.88$\pm$3.11          & 50.75$\pm$3.29          & 0.39$\pm$0.10          & 3.89$\pm$1.55          & 8.46$\pm$2.96          & 51.41$\pm$3.32          & 0.46$\pm$0.11          \\ \hline
\begin{tabular}[c]{@{}c@{}}Attention\\ CloudLSTM ($\mathcal{K}=9$)\end{tabular} & \textbf{3.66$\pm$1.64} & \textbf{8.82$\pm$3.10} & \textbf{50.78$\pm$3.21} & \textbf{0.40$\pm$0.11} & \textbf{3.79$\pm$1.57} & \textbf{8.43$\pm$2.96} & \textbf{51.46$\pm$3.33} & \textbf{0.47$\pm$0.11} \\ \hline
\end{tabular}
\vspace*{-1em}
\end{table}
Observe that RNN-based architectures in general obtain superior performance, compared to CNN-based models and the MLP. In particular, our proposed CloudLSTM, and its CloudRNN, and CloudGRU variants outperform all other banchmark architectures, achieving lower MAE/RMSE and higher PSNR/SSIM on both urban scenarios. This suggests that the \dc operator learns features over geospatial point-clouds more effectively than vanilla convolution and PointCNN. Among our approaches, CloudLSTM performs better than CloudGRU, which in turn outperforms CloudRNN.

Interestingly, the forecasting performance of the CloudLSTM seems fairly insensitive to the number of neighbors ($\mathcal{K}$). It is therefore worth using a small $\mathcal{K}$ in practice, to reduce model complexity, as this does not compromise the accuracy significantly. Further, we observe that the attention mechanism improves the forecasting performance, as it helps capturing better dependencies between input sequences and vectors in decoders. This effect has also been confirmed by other NLP tasks.

Note that we conduct our experiments using strict variable-controlling methodology, i.e., only changing one factor while keep the remaining the same. Therefore, it is easy to study the effect of each factor. For example, taking a look at the performance of LSTM, ConvLSTM, PredRNN++, PointLSTM and CloudLSTM, which employ dense layers, and CNN, PointCNN and D-Conv as core operators but using LSTM as the RNN structure, it is clear that the D-Conv contributes significantly to the performance improvements. Further, by comparing CloudRNN, CloudGRU and CloudLSTM, it appears that CloudRNN $\ll$ CloudGRU $<$ CloudLSTM. Similarly, by comparing the CloudLSTM and Attention CloudLSTM, we see that the effect of the attention mechanism is not very significant. Therefore, we believe the core operator $>$ RNN structure $>$ attention, ranked by their contribution. 

\begin{figure*}[htb]
\centering\includegraphics[width=1.03\columnwidth, trim={1cm 1cm 1cm 1cm}]{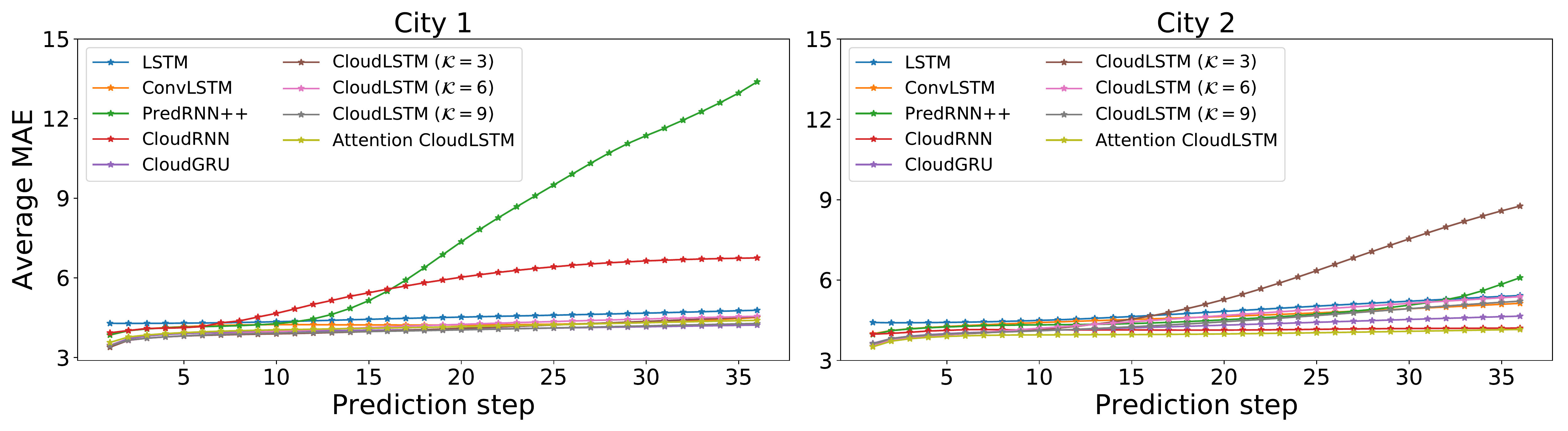}
\caption{MAE evolution wrt.\ prediction horizon achieved by RNN-based models on both cities. \label{fig:mae}} 
\end{figure*}

\noindent \textbf{Long-term Forecasting Performance.} We extend the prediction horizon to up to $J=36$ time steps (\ie 3 hours) for all RNN-based architectures, and show their MAE evolution with respect to this horizon in Fig.~\ref{fig:mae}. Note that the input length remains unchanged, \ie 6 time steps. In city 1, observe that the MAE does not grow significantly with the prediction step for most models, as the curves flatten. This means that these model are reliable in terms of long-term forecasting. As for city 2, we note that low $\mathcal{K}$ may lead to poorer long term performance for CloudLSTM, though not significant before step 20. This provides a guideline on choosing  $\mathcal{K}$ for different forecast length required.

\begin{figure*}[htb]
\centering\includegraphics[width=1.03\columnwidth]{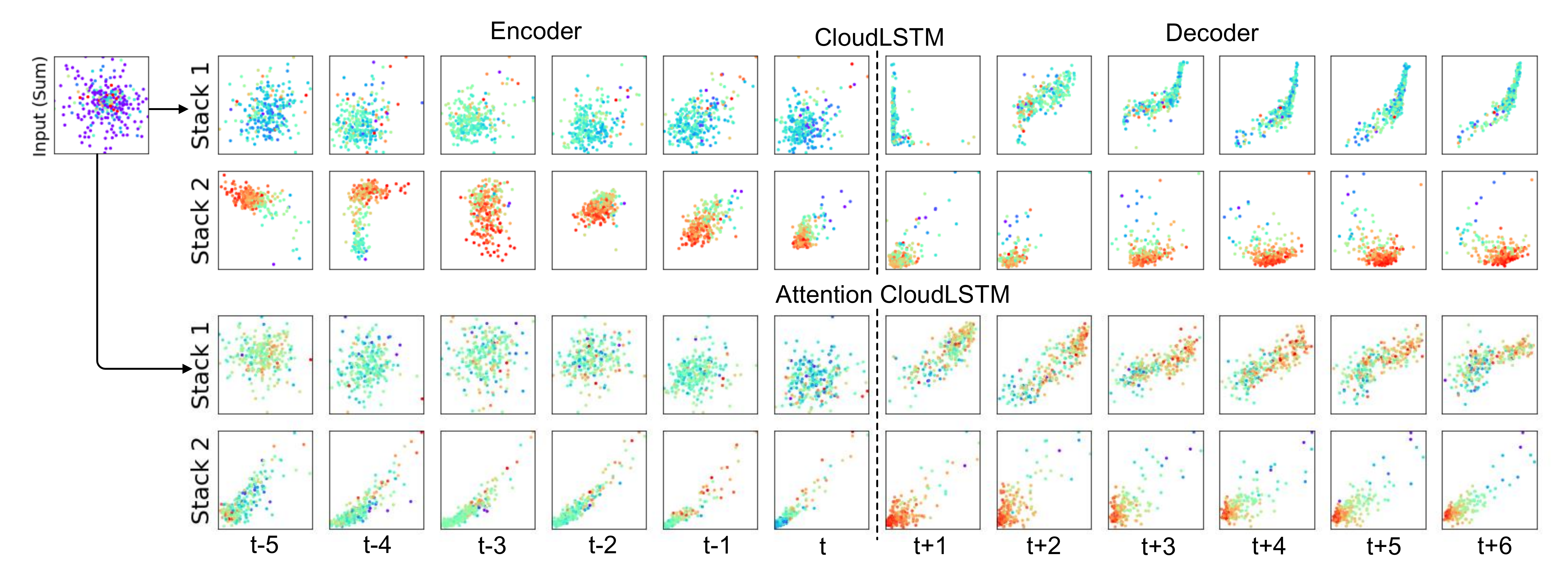}
\caption{The scatter distributions of the value and coordinate features of the hidden state in $H_t$ for CloudLSTM and Attention CloudLSTM. Values and coordinates are averaged over all channels. \label{fig:exa}} 
\end{figure*}

\noindent \textbf{Visualization.} We complete the evaluation of the mobile traffic forecasting task by visualizing the hidden features of the CloudLSTM, which provide insights into the knowledge learned by the model. In Fig.~\ref{fig:exa}, we show an example of the scatter distributions of the hidden state in $H_t$ of CloudLSTM and Attention CloudLSTM at both stacks, along with the first input snapshots. The first 6 columns show the $H_t$ for encoders, while the rest are for decoders. The input data snapshots are samples selected from City 2 (260 antennas/points). Recall that each $H_t$ has 1 value features and 2 coordinate features for each point, therefore each scatter subplot in Fig.~\ref{fig:exa} shows the value features (volume represented by different colors) and coordinate features (different locations), averaged over all channels. Observe that in most subplots, points with higher values (warm colors) tend to aggregate into clusters and have higher densities. These clusters exhibit gradual changes from higher to lower values, leading to comet-shape assemblages. This implies that points with high values also come with tighter spatial correlations, thus CloudLSTMs learn to aggregate them. This pattern becomes more obvious in stack~2, as features are extracted at a higher level, exhibiting more direct spatial correlations with respect to the output.

\subsection{Augmenting Forecasting with Seasonal Information}
\vspace*{-0.5em}
We notice that seasonal information exists in the mobile traffic series, which can be further exploited to improve the forecasting performance. However, directly feeding the model with data spanning multiple days is infeasible, since, e.g., a 7-day window corresponds to a 2016-long sequence as input (given that data is sampled every 5 minutes) and it is very difficult for RNN-based models to handle such long sequences. In addition, by considering the number of mobile services (38) and antennas (792), the input for 7 days would have 60,673,536 data points. This would make any forecasting model extremely large and therefore impractical for real deployment.

To capture seasonal information more efficiently, we concatenate the 30 minute-long sequences (sampled every 5 minutes) with a sub-sampled 7-day window (sampled every 2h). This forms an input with length 90 (6 + 84). We conduct experiments on a randomly selected subset (100 antennas) of the mobile traffic dataset (City 1), and show the forecasting performance without and with seasonal information (7-day window) in Table~\ref{tab:seasonal}. By incorporating the seasonal information, the performance of most forecasting models is boosted. This indicates that the periodic information is learnt by the model, which helps reduce the prediction errors. However, the concatenation increases the length of the input, which also increases the model complexity. Future work will focus on a more efficient way to fuse the seasonal information, with marginal increase in complexity.

\begin{table*}[t]
\centering
\caption{The mean$\pm$std of MAE and RMSE across all models considered without/with seasonal information, evaluated on a subset of antennas in City 1 for  mobile traffic forecasting. \label{tab:seasonal}}
\begin{tabular}{|c|cc|cc|}
\hline
\multirow{2}{*}{Model}                                                          & \multicolumn{2}{c|}{30 Minutes Window}                     & \multicolumn{2}{c|}{30 Minutes + 7 Days Window}                   \\ \cline{2-5} 
                                                                                & \textbf{MAE}            & \textbf{RMSE}            & \textbf{MAE}           & \textbf{RMSE}           \\ \hline
MLP                                                                             & 4.86$\pm$0.51       & 10.30$\pm$2.52          & 4.93$\pm$0.53       & 11.30$\pm$2.2         \\
CNN                                                                             & 6.10$\pm$0.59         & 11.12$\pm$2.04         & 5.98$\pm$0.60         & 10.52$\pm$2.11          \\
3D-CNN                                                                          & 5.01$\pm$0.51         & 9.89$\pm$2.54         & 4.82$\pm$0.54         & 9.49$\pm$2.36          \\
DefCNN                                                                          & 6.79$\pm$0.89         & 11.92$\pm$2.47         & 6.40$\pm$0.83         & 11.55$\pm$2.33          \\
PointCNN                                                                        & 5.01$\pm$0.55         & 10.22$\pm$2.40         & 4.88$\pm$0.51         & 9.86$\pm$2.31          \\
CloudCNN                                                                        & 4.79$\pm$0.53         & 9.94$\pm$2.75          & 4.63$\pm$0.50         & 9.57$\pm$2.66          \\
LSTM                                                                            & 4.24$\pm$0.64         & 9.67$\pm$3.23          & 4.04$\pm$0.67         & 9.28$\pm$3.03            \\
ConvLSTM                                                                        & 4.10$\pm$1.61          & 9.28$\pm$3.11          & 3.82$\pm$1.54          & 8.87$\pm$3.00          \\
PredRNN++                                                                       & 3.94$\pm$1.62          & 9.31$\pm$3.10          & 3.61$\pm$1.55          & 8.95$\pm$2.92          \\
PointLSTM                                                                       & 4.63$\pm$0.41          & 9.44$\pm$2.46        & 4.44$\pm$0.41          & 9.00$\pm$2.32          \\ \hline
CloudRNN ($\mathcal{K}=9$)                                                      & 4.14$\pm$1.67          & 9.18$\pm$3.13          & 3.99$\pm$1.62          & 8.88$\pm$2.93          \\
CloudGRU ($\mathcal{K}=9$)                                                      & 3.77$\pm$1.58          & 8.95$\pm$3.08          & 3.42$\pm$1.53          & 8.53$\pm$2.88          \\ \hline
CloudLSTM ($\mathcal{K}=3$)                                                     & 3.69$\pm$1.61         & 8.83$\pm$3.09            & 3.38$\pm$1.57         & 8.40$\pm$2.86          \\
CloudLSTM ($\mathcal{K}=6$)                                                     & 3.68$\pm$1.60          & 8.87$\pm$3.10          & 3.39$\pm$1.54          & 8.43$\pm$2.76          \\
CloudLSTM ($\mathcal{K}=9$)                                                     & 3.69$\pm$1.61          & 8.86$\pm$3.12           & 3.40$\pm$1.56          & 8.46$\pm$2.77          \\ \hline
\begin{tabular}[c]{@{}c@{}}Attention\\ CloudLSTM ($\mathcal{K}=9$)\end{tabular} & \textbf{3.60$\pm$1.59} & \textbf{8.77$\pm$3.06}  & \textbf{3.22$\pm$1.55} & \textbf{8.36$\pm$2.66}          \\ \hline
\end{tabular}
\end{table*}

\subsection{Service-wise Evaluation}
\begin{figure*}[t]
\centering\includegraphics[width=1\columnwidth]{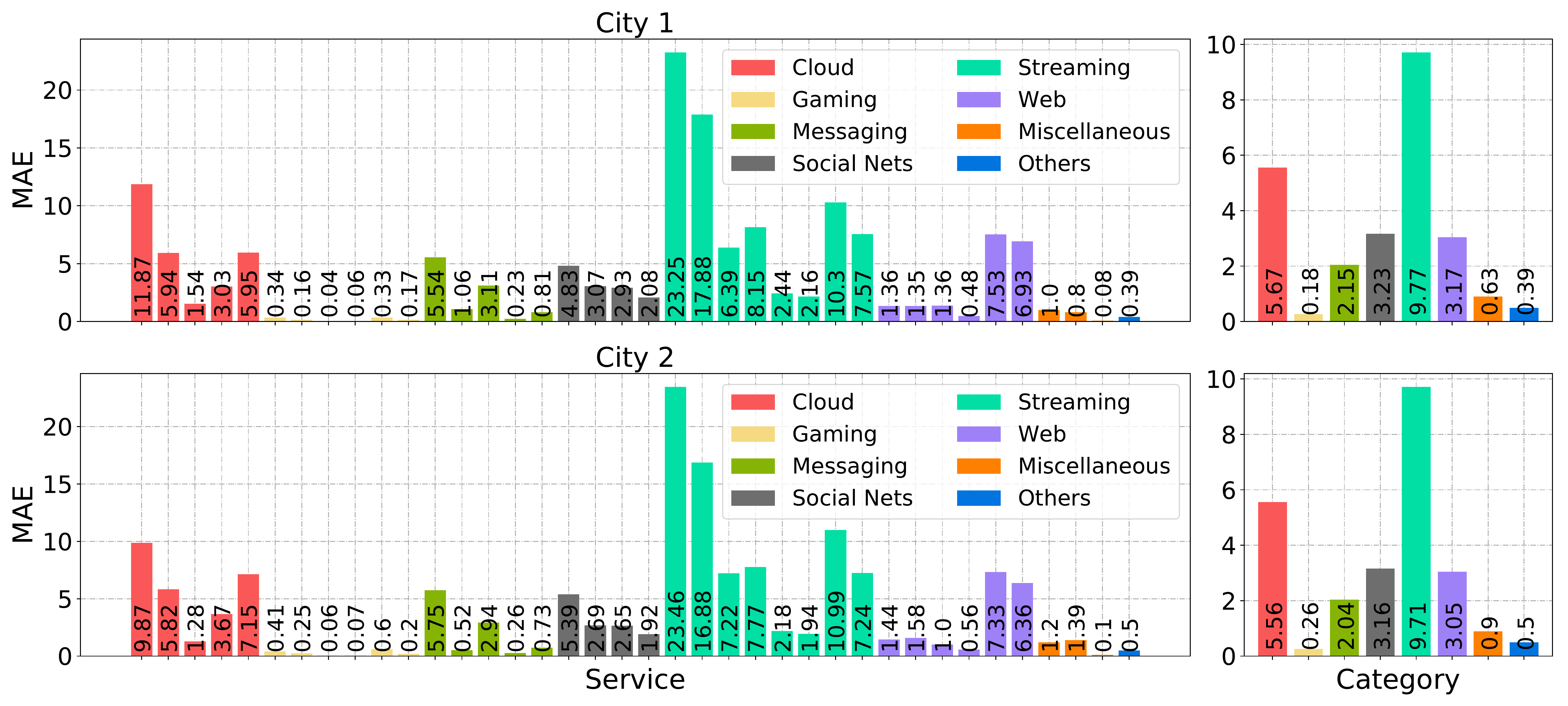}
\caption{Service-level MAE evaluation on both cities for the Attention CloudLSTMs, averaged over 36 prediction steps.\label{fig:both_mae}} 
\end{figure*}
Finally, we dive deeper into the performance of the proposed Attention CloudLSTMs, by evaluating the forecasting accuracy for each individual service, averaged over 36 steps. To this end, we present the MAE evaluation on a service basis (left) and category basis (right) in Fig.~\ref{fig:both_mae}. Observe that the attention CloudLSTMs obtain similar performance over both cities at the service and category level. Jointly analyzing with Fig.~\ref{fig:traffic-stats0}, we see that services with higher traffic volume on average (\eg streaming and cloud) also yield higher prediction errors. This is because their traffic evolution exhibits more frequent fluctuations, which introduces higher uncertainty, making the traffic series more difficult to predict.

\section{Summary}
In this chapter, we attacked the multi-service mobile traffic forecasting problem to support resource management for network slicing. We introduce CloudLSTM, a dedicated neural model for spatio-temporal forecasting tailored to point-cloud data streams. The CloudLSTM builds upon the \dc operator, which performs convolution over point-clouds to learn spatial features while maintaining permutation invariance. The \dc simultaneously predicts the values and coordinates of each point, thereby adapting to changing spatial correlations of the data at each time step. \dc is flexible, as it can be easily combined with various RNN models (\ie RNN, GRU, and LSTM), Seq2seq learning and attention mechanisms. We employ antenna-level mobile traffic forecasting as a case study, where we show that our proposed CloudLSTM achieves state-of-the-art performance on large-scale datasets collected in two major European cities. We believe the CloudLSTM gives a new perspective on point-cloud stream modeling, and it can be easily extended to higher dimension point-clouds, without requiring changes to the model.

%% file: chap5.tex
\chapter{Deep Learning Driven Mobile Traffic Super Resolution\label{chap:mtsr}}

In this chapter, we propose an original mobile traffic `super-resolution' technique that drastically reduces the complexity of measurements collection and analysis, \rev{characteristic to purely RAN based tools}, while gaining accurate insights into \emph{fine-grained mobile traffic patterns} at city scale.  Mobile traffic measurement collection and monitoring currently rely on dedicated probes deployed at different locations within the network infrastructure \cite{naboulsi2016large}. Unfortunately, this equipment \emph{(i)}~either acquires only coarse information about users' position, the start and end of data sessions, and volume consumed \rev{(which is required for billing)}, while roughly approximating location throughout sessions --  Packet Data Network Gateway (PGW) probes; \emph{(ii)} or has small geographical coverage, e.g. Radio Network Controller (RNC) \rev{or \mbox{eNodeB}} probes that need to be densely deployed, \rev{require to store tens of gigabytes of data per day at each RNC~\cite{Xu:2011}, but cannot independently quantify data consumption. In addition, context information recorded is often stale, which renders timely inferences difficult. Mobile traffic prediction within narrowly localized regions is vital for precision traffic-engineering and can further benefit the provisioning of emergency services. Yet} this remains costly, as it involves non-negligible overheads associated with the transfer of reports~\cite{naboulsi2016large}, substantial storage capabilities, and intensive off-line post-processing. In particular, combining different information from a large number of such specialized equipment is required, e.g. user position obtained via triangulation \cite{averin2010locating}, data session timings, traffic consumed per location, etc. \rev{Overall this is a challenging endeavour that cannot be surmounted in real-time by directly employing measurements from independent RNC probes.}

In order to simplify the analysis process, mobile operators make simple assumptions about the distribution of data traffic consumption across cells. For instance, it is frequently assumed users and traffic are uniformly distributed, irrespective of the geographical layout of coverage areas~\cite{lee2014spatial}. Unfortunately, such approximations are usually highly inaccurate, as traffic volumes exhibit considerable  disparities between proximate locations \cite{wang2015understanding}. Operating on such simplified premises can lead to deficient network resources allocations and implicitly to modest end-user quality of experience. \rev{Alternative coarse-grained measurement crowdsourcing~\cite{kang2017enhance} remains impractical for traffic inference purposes.}

Our objective is to precisely infer narrowly localized traffic consumption from aggregate coarse data recorded by a limited number of probes \rev{(thus reducing deployment costs) that have arbitrary granularity}. Achieving this goal is challenging, as small numbers of probes summarize traffic consumption over wide areas and thus only provide `low-resolution' snapshots that cannot capture distributions correlated with human mobility. Further, measurement instrumentation is non-uniformly deployed, as coverage area sizes depend on the population density~\cite{Gramaglia:2015}. Such diversity creates considerable ambiguity about the traffic consumption at sub-cell scale and multiple `high-resolution' snapshots could be matched to their aggregate counterparts. Inferring the precise traffic distribution is therefore hard.

\begin{figure}[t]
\begin{center}
\includegraphics[width=\columnwidth]{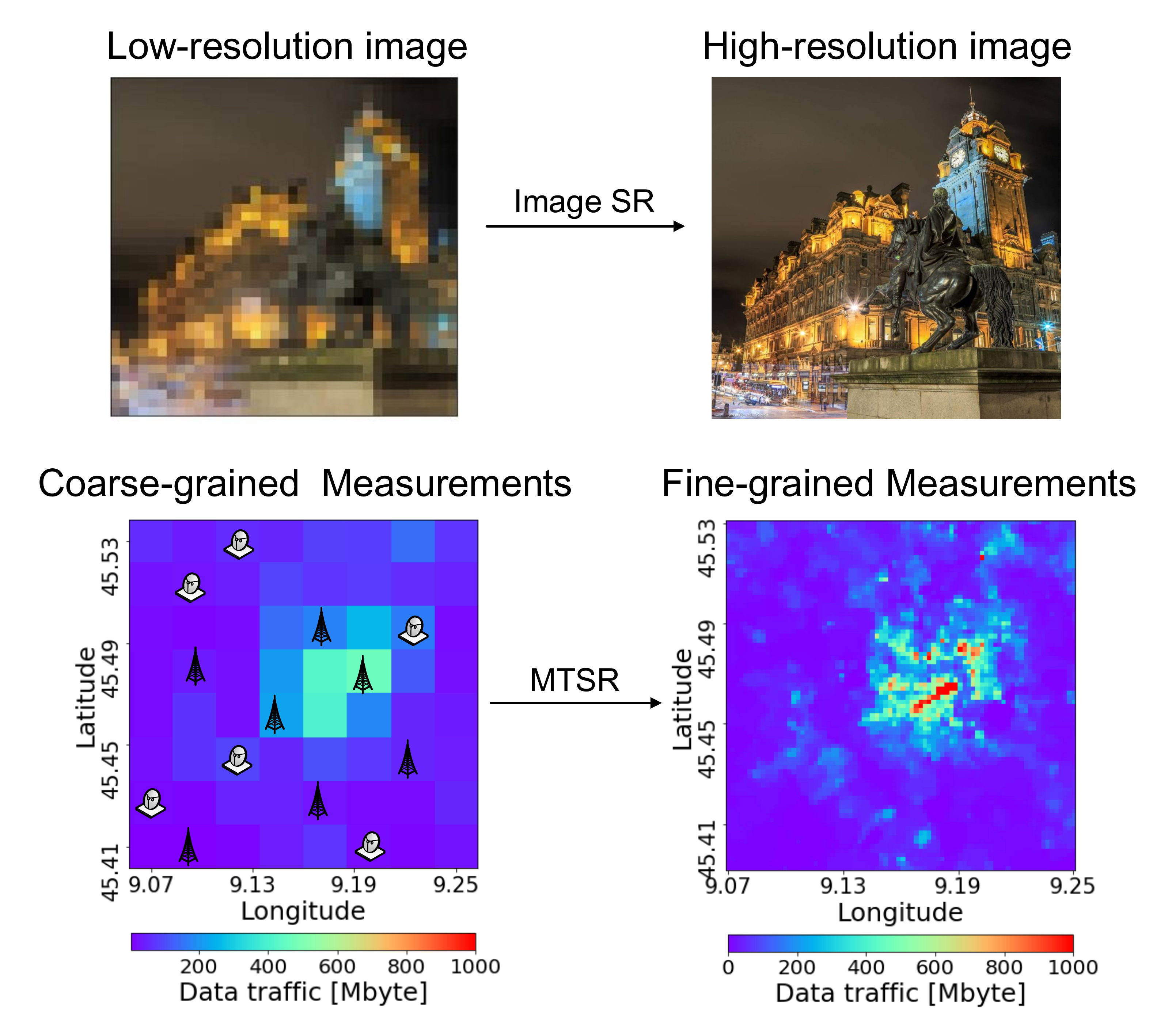}
\end{center}
\caption{\label{Fig:sr} Illustration of the image super-resolution (SR) problem (above) and the underlying principle of the proposed mobile traffic super-resolution (MTSR) technique (below). Figure best viewed in color.}

\end{figure}

\textbf{Drawing inspiration from image processing.} Spatio-temporal correlations we observe between mobile traffic patterns prompt us to represent these as tensors that highly resemble images (cross-spatial relations) or videos (cross-temporal relations). It becomes apparent that a problem similar to the one we tackle exists in the image processing field, where images \rev{with small number} of pixels are enhanced to high-resolution. There, a super-resolution (SR) imaging approach mitigates the multiplicity of solutions by constraining the solution space through prior information \cite{simpkins2012introduction}. This inspires us to employ image processing techniques to learn end-to-end relations between low- and high-resolution mobile traffic snapshots. We illustrate the similarity between the two SR problems in Fig.~\ref{Fig:sr}. Recent achievements in GPU based computing \cite{tensorflow2015-whitepaper} have led to important results in image classification \cite{krizhevsky2012imagenet}, while different neural network architectures have been successfully employed to learn complex relationships between low- and high-resolution images \cite{dong2016image,yu2016ultra}. Thus we recognise the potential of exploiting deep learning models to achieve reliable mobile traffic super-resolution (MTSR) and make the following \textbf{key contributions:}

\begin{enumerate}[topsep=5pt]
 \item We propose a novel Generative Adversarial neural Network (GAN) architecture tailored to MTSR, to infer fine-grained mobile traffic patterns, from aggregate measurements collected by network probes. Specifically, in our design high-resolution traffic maps are obtained through a \emph{generative model} that outputs approximations of the real traffic distribution. This is trained with a \emph{discriminative model} that estimates the probability a sample snapshot comes from a fine-grained ground truth measurements set, rather than being produced by the generator. 

 \item We construct the generator component of the GAN using an original deep zipper network (ZipNet) architecture. This upgrades a sophisticated ResNet model \cite{he2016deep} with a set of additional `skip-connections', without introducing extra parameters, while allowing gradients to backpropagate faster through the model in the training phrase. The ZipNet further introduces 3D upscaling blocks to jointly extract spatial and temporal features that exist in mobile traffic patterns. Our design accelerates training convergence and we demonstrate it outperforms the baseline Super-Resolution Convolutional Neural Network (SRCNN) \cite{dong2016image} in terms of prediction accuracy.

 \item To stabilise the adversarial model training process, and prevent model collapse or non-convergence problems (common in GAN design), we introduce an empirical loss function. The intuition behind our choice is that the generator can be optimized and complexity reduced, if using a loss function for model training that is insensitive to model structure and hyper-parameters configuration.

 \item We propose a data processing and augmentation procedure to handle the insufficiency of training data and ensure the neural network model does not turn significantly over-fitted. Our approach crops the original city-wide mobile data `snapshots' to smaller size windows and repeats this process with different offsets to generate extra data points from the original ones, thereby maximizing the usage of datasets available for training.
 
 \item We conduct experiments with a publicly available real-world mobile traffic dataset and demonstrate the proposed ZipNet (-GAN) precisely infer fine-grained mobile traffic distributions with up to 100$\times$ higher granularity as compared to standard probing, irrespective to the coverage and the position of the probes. Importantly, our solutions outperform existing traditional and deep-learning based interpolation methods, as we achieve up to 78\% lower reconstruction errors, 40\% higher fidelity of reconstructed traffic pattern, and improve the structural similarity by 36.4$\times$.
\end{enumerate}

We believe the proposed ZipNet(-GAN) techniques can
be deployed at a gateway level to effectively reduce the complexity and enhance the quality of mobile traffic analysis, by simplifying the networking infrastructure, underpinning intelligent resource management, and overcoming service congestion in popular `hot spots'. 

\rv{\section{Problem Formulation}}
\label{sec:problem2}
The objective of MTSR is to infer city-wide fine-grained mobile data traffic distributions, using sets of coarse-grained measurements collected by probes deployed at different locations. We formally define low-resolution traffic measurements as a spatio-temporal sequence of data points $\mathcal{M}^L = \{D_1^L,D_2^L,\ldots,D_T^L\}$, where $D_t^L$ is a snapshot at time $t$ of the mobile traffic consumption summarized over the entire coverage area and in practice partitioned into $V$ cells (possibly of different size), i.e. $D_t^L = \{l_t^1, \ldots, l_t^V \}$. Here $l_t^v$ represents the data traffic consumption in cell $v$ at time $t$.

We denote $\mathcal{M}^H = \{D_1^H,D_2^H,\ldots,D_T^H\}$ the high-resolution mobile traffic measurement counterparts (which are traditionally obtained via aggregation and post-processing), where $D_t^H$ is a mobile traffic consumption snapshot at time $t$ over $I$ sub-cells, i.e. $D_t^H = \{h_t^1, \ldots, h_t^I \}$. Here $h_t^i$ denotes the data traffic volume in sub-cell $i$ at time $t$. Dissimilar to $\mathcal{M}^L$, we work with sub-cells of the same size and shape, therefore $D_t^H$ points have the same measuring granularity. Note that both $\mathcal{M}^L$ and $\mathcal{M}^H$ measure the traffic consumption in the same area and for the same duration.

From a machine learning prospective, the MTSR problem is to infer the most likely current fine-grained mobile traffic consumption, given previous $S$ observations of coarse-grained measurements. Denoting this sequence $F_t^S = \{D_{t-S+1}^{L},\\\ldots,D_t^L\}$, MTSR solves the following optimization problem:
\begin{equation}
\label{eq:problem}
\begin{aligned}
\widetilde{D}_t^H &:= \mathop{\arg\max}\limits_{D_t^H} p\left(D_t^H|F_t^S\right),
\end{aligned}
\end{equation}
where $\widetilde{D}_t^H$ denotes the solution of the prediction. Both $D_t^H$ and $F_t^S$ are high-dimensional, since they represent different traffic measurements across a city. To precisely learn the complex correlation between $D_t^H$ and $F_t^S$, in this work we propose to use a Generative Adversarial Network (GAN), which will model the conditional distribution $p\left(D_t^H|F_t^S\right)$. As we will demonstrate, the key advantage of employing a GAN structure is that it will not only minimize the mean square error between predictions and ground truth, but also yield remarkable fidelity of the high-resolution inferences made.

\section{Performing MTSR via ZipNet-GAN}
In what follows we propose a deep-learning approach to tackle the MTSR problem using GANs. This is a novel unsupervised learning framework for generation of artificial data from real distributions through an adversarial training process~\cite{goodfellow2014generative}. In general, a GAN is composed of two neural network models, a generator $\mathcal{G}$ that learns the data distribution, and a discriminative model $\mathcal{D}$ that estimates the probability that a data sample came from the real training data rather than from the output of $\mathcal{G}$. We first give a brief overview of general GAN operation, then explain how we adapt this structure to the MTSR problem we aim to solve.


\subsection{Essential Background}
As with all neural networks, the key to their performance is the training process that serves to configure their parameters. When training GANs, the generator $\mathcal{G}$ takes as input a noisy source (e.g. Gaussian or uniformly distributed) $z\sim P_n(z)$ and produces an output $\hat{x}$ that aims to follow a target unknown data distribution (e.g. pixels in images, voice samples, etc.). On the other hand, the discriminator $\mathcal{D}$ randomly picks data points generated by $\mathcal{G}$, i.e. $\hat{x}\sim \mathcal{G}(z)$, and others sampled from the target distribution $x\sim P_r(x)$, and is trained to maximize the probabilities that $\hat{x}$ is fake and $x$ is real. In contrast, $\mathcal{G}$ is trained to produce data whose distribution is as close as possible to $P_r(x)$, while maximizing the probability that $\mathcal{D}$ makes mistakes. This is effectively a two-player game where each model is trained iteratively while fixing the other one.
The joint objective is therefore to solve the following minimax problem \cite{goodfellow2014generative}:
\begin{equation}
\label{eq:minmax2}
\begin{aligned}
\min\limits_{\mathcal{G}}\max\limits_{\mathcal{D}} \mathbb{E}_{x\sim P_r(x)}[\log \mathcal{D}(x)]+\mathbb{E}_{z\sim P_n(z)}[\log(1- \mathcal{D}(\mathcal{G}(z)))].
\end{aligned}
\end{equation}
Once trained, the generator $\mathcal{G}$ is able to produce artificial data samples from the target distribution given noisy input~$z$. 

\begin{figure}[t]
 \includegraphics[width=\columnwidth]{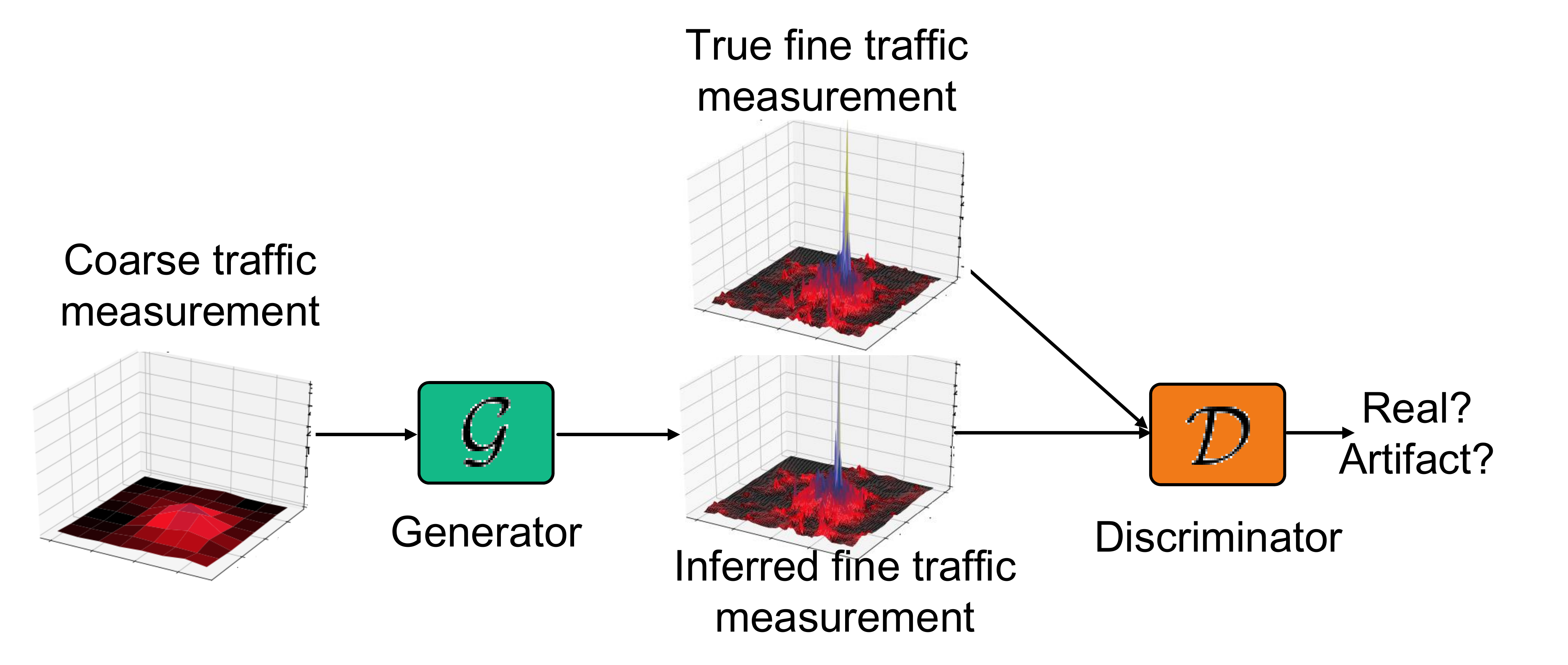}
 \caption{GAN operating principle in MTSR problem. Only the generator is employed in prediction phase.}
 \label{fig:pipeline}

\end{figure}

In the case of our MTSR problem, the input of $\mathcal{G}$ is sampled from the distribution of coarse-grained measurements $p(F_t^S)$, instead of traditional Gaussian or uniform distributions. Our objective is to understand the relations between $D_t^H$ and $F_t^S$, i.e. modeling $p\left(D_t^H|F_t^S\right)$. $\mathcal{D}$ is trained to discriminate whether the data is a real fine-grained traffic measurement snapshot, or merely an artifact produced by $\mathcal{G}$. We summarize this principle in Fig.~\ref{fig:pipeline}.

\begin{figure*}[t]
\begin{center}
\includegraphics[width=1\columnwidth]{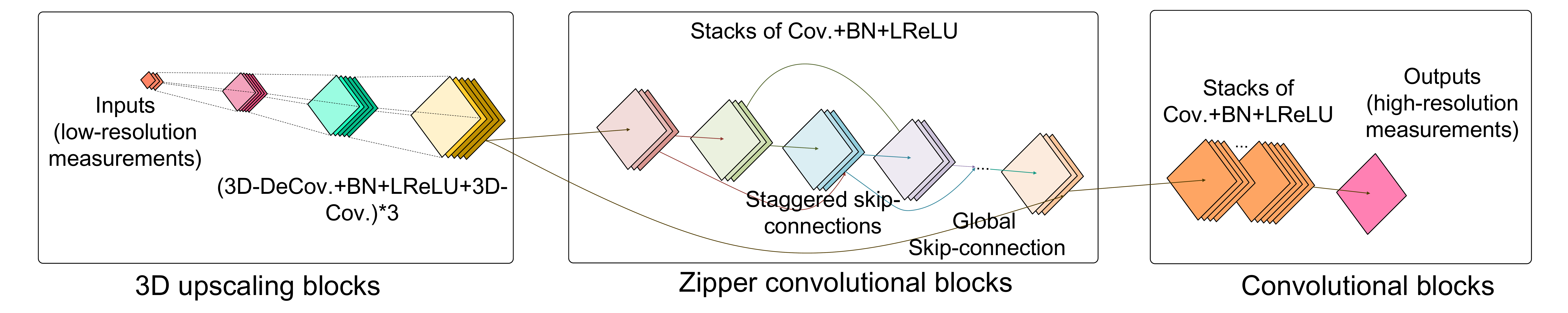}
\end{center}
\caption{\label{Fig:zipnet} The deep Zipper Network (ZipNet) generator architecture, consisting of 3D upscaling, zipper convolutional, and standard convolutional blocks. Conv. and deconv. denote convolutional and deconvolutional layers.}
\end{figure*}

\subsection{The ZipNet-GAN Architecture}
Recall that our GAN is composed of a generator $\mathcal{G}$ that takes low-resolution measurements $F_t^S$ as input and reconstructs their high-resolution counterparts  $D_t^H$, and a discriminator $\mathcal{D}$ whose task is relatively light (learning to discriminate samples and minimizing the probability of making mistakes).

In general the generator has a complicated structure, which is required for the reconstruction of data points with unknown distributions. In our MTSR problem, we want to capture the complex correlations between $F_t^S$ and $D_t^H$. To this end, we propose Zipper Network (ZipNet), an original deep neural network architecture. This upgrades a ResNet model \cite{he2016deep} with additional `skip-connections', without introducing extra parameters, while allowing gradients to backpropagate faster through the model, and accelerate the training process. The overall ZipNet architecture is illustrated in Fig.~\ref{Fig:zipnet}. To exploit important historical traffic information, we introduce 3D upscaling blocks that extract the spatio-temporal features specific to mobile traffic patterns.
The proposed ZipNet comprises three main components, namely:

$\bullet$ \textbf{3D upscaling blocks} consisting of a 3D deconvolutional layer \cite{noh2015learning}, three 3D convolutional layers \cite{ji20133d}, a batch normalization (BN) layer \cite{ioffe2015batch}, and a Leaky ReLU (LReLU) activation layer \cite{maas2013rectifier}. 3D (de)convolutions are employed to capture spatial (2D) and temporal relations between traffic measurements. The deconvolution is a transposed convolutional layer widely employed for image upscaling. 3D convolutional layers enhance the model representability. BN layers normalize the output of each layer and are effective in training acceleration~\cite{ioffe2015batch}. LReLUs improve the model non-linearity and their output is of the form:
\begin{equation}\label{lrelu}
\begin{aligned}
\text{LReLu($x$)}  = \left\{\begin{matrix}
x, \hfill x\geqslant 0;\\ 
\alpha x, \hfill x<0,
\end{matrix}\right.
\end{aligned}
\end{equation}
where $\alpha$ is a small positive constant (e.g. 0.1). The objective of such upscaling blocks is to up-sample the input $F_t^S$ into tensors that have the same spatial resolution as the desired output $D_t^h$. The number of upscaling blocks increases with the resolution of the input (from 1 to 3).  \rv{Upscaling blocks are frequently used to increase the data dimension for image super resolution, whereas we introduce an extra dimension to process the temporal domain.} \textbf{These 3D upscaling blocks are key to jointly extracting spatial and temporal features specific to mobile traffic.} 

$\bullet$ \textbf{Zipper convolutional blocks (core)}, which pass the output of the 3D upscaling blocks, through 24 convolutional layers, BN layers, and LReLU activations, with staggered and global skip connections. These blocks hierarchically extract different features and construct the input for the next convolutional component, \rv{and also are the purpose-built components we tailored for mobile traffic data processing.} 
We give further details about this block in the rest of this sub-section.

\begin{figure}[b]
\begin{center}
\includegraphics[width=\columnwidth]{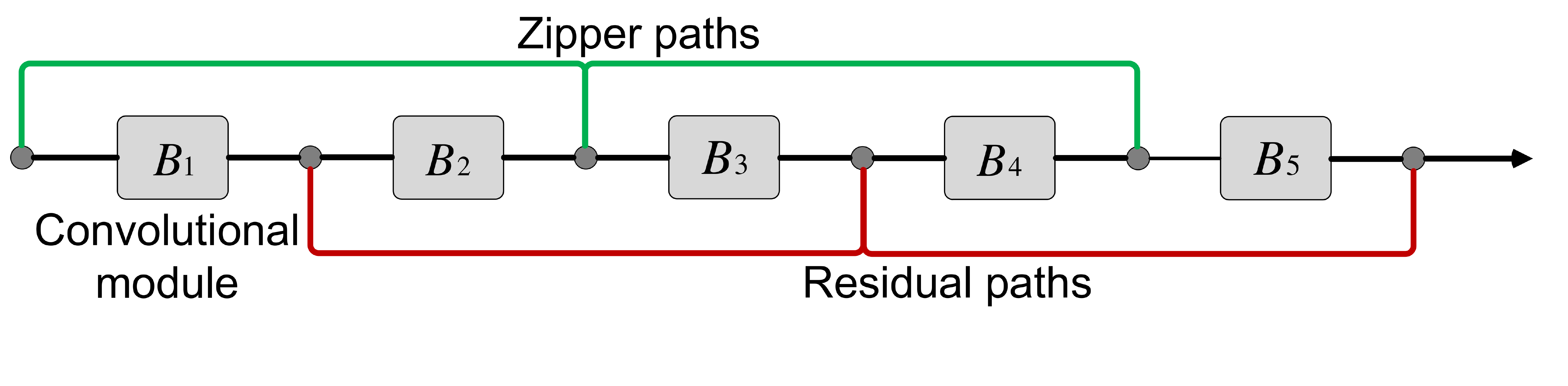}
\end{center}
\caption{\label{Fig:zip} 5 Zipper convolutional blocks; each module $B$ includes a convolutional layer, a BN layer and a LReLU activation. Circular nodes represent additions.} 
\end{figure}

$\bullet$ \textbf{Convolutional blocks} that summarize the features distilled by the Zipper blocks and make the final prediction decisions. The block consists of three standard convolutional layers, BN layers, and LReLU activations, without skip connections. Each layer is configured with more feature maps as compared to the previous block, in order to provide sufficient features for the mobile traffic final prediction.

The proposed ZipNet is a complex architecture that has over 50 layers. In general, accuracy grows with the depth (which is precisely what we want to achieve), but may also degrade rapidly, if the network becomes too deep. To overcome this problem, we use shortcut connections at different levels between the comprising blocks, as illustrated in Fig.~\ref{Fig:zip} --this resembles a zipper, hence the naming. The structure can be regarded as an extension to residual networks (ResNets) previously used for image recognition~\cite{he2016deep}, where the zipper connections significantly reduce the convergence rate and improve the model's accuracy, without introducing extra parameters and layers. 

The fundamental module $B$ of the zipper convolutional block comprises a convolutional layer, a BN layer, and an activation function. Staggered skip connections link every two modules, and a global skip connection performs element-wise addition between input and output (see Fig.~\ref{Fig:zipnet}) to ensure fast backpropagation of the gradients. The skip connections also build an ensemble of networks with various depths, which has been proven to enhance the model's performance~\cite{veit2016residual}. Further, replacing layer-wise connections (as e.g. in \cite{huang2016densely}) with light-weight staggered shortcuts reduces the computational cost. \rev{Compared to the original ResNets, extra zipper paths act as an additional set of residual paths, which reinforce the ensembling system and alleviate the performance degeneration problem introduced by deep architectures. The principle behind ensembling systems is collecting a group of models and voting on their output for prediction (e.g. using a random forest tool). Extra zipper paths increase the number of candidate models in the voting process, which improves the robustness of the architecture and contributes to superior performance.} 

\begin{figure}[t]
\begin{center}
\includegraphics[width=\columnwidth]{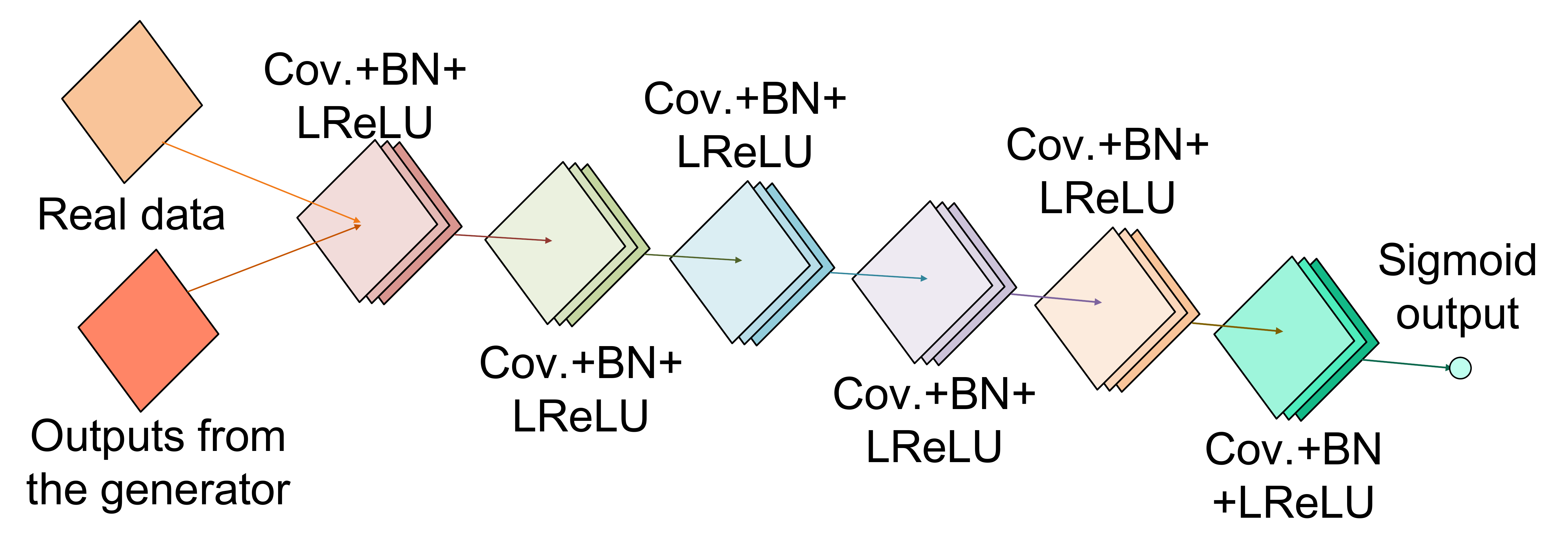}
\end{center}
\caption{\label{Fig:discriminator} Architecture of the discriminator $\mathcal{D}$ we use in ZipNet-GAN, based on the VGG-net structure~\cite{simonyan2014very}.}
\end{figure}

Finally, the discriminator $\mathcal{D}$ accepts simultaneously the predictions made by the generator and fine-grained ground truth measurements, and minimizes the probability of misjudgement. In our design $\mathcal{D}$ follows a simplified version of a VGG-net neural network architecture \cite{simonyan2014very} routinely used for imaging applications, and consists of 6 convolutional blocks (convolutional layer + BN layer + LReLU), as illustrated in Fig.~\ref{Fig:discriminator}. The number of feature maps doubles every other layer, 
such that the final layer will have sufficient features for accurate decision making. The final layer employs a sigmiod activation function, which constrains the output to a probability range, i.e. $(0, 1)$.

The remaining task in the design of the proposed neural network is to configure the loss functions employed by the generator and discriminator, which is key to the adjustment of the weights within the different layers.

\subsection{Designing the Loss Functions}
The MTSR is a supervised learning problem, hence directly deploying the traditional adversarial training process is inappropriate. This is because the data generated by $\mathcal{G}$ in traditional GANs is stochastically sampled from the approximated target distribution, whereas in MTSR we expect the output of the generator to follow as close as possible the distribution of real fine-grained measurements. That is, individual data points $\mathcal{G}(F_t^S)$ produces should be close to their corresponding ground truth. This requires to minimize \emph{(i)}~the divergence between the distributions of the real data $p(D_t^H)$ and the generated data $p(\mathcal{G}(F_t^S))$, and \emph{(ii)}~the Euclidean distance between individual predictions and their corresponding ground truth. 

$\text{L}_2$ loss functions commonly used in neural networks to solve the optimization problem given in (\ref{eq:problem}) only fulfill the second objective above, but do not guarantee that the generator's output will follow the real data distribution. This may lead to failure to match the fidelity expected of high-resolution predictions~\cite{ledig2016photo}. Therefore we add another term to (\ref{eq:problem}), which requires to design different loss functions, while changing the original problem to
\begin{equation}
\label{eq:new_problem}
\begin{aligned}
\widetilde{D}_t^H &= \mathop{\arg\max}\limits_{D_t^H} p\left(D_t^H|F_t^S\right) \cdot \mathcal{D}(\mathcal{G}(F_t^S)),
\end{aligned}
\end{equation}
where recall that $ \mathcal{D}(\mathcal{G}(F_t^S))$ is the probability that $\mathcal{D}$ predicts whether data is sampled from the real distribution. Minimizing this new term $\mathcal{D}(\mathcal{G}(F_t^S))$ aims to `fool' the discriminator, so as to minimize the divergence between real and generated data distributions, i.e. objective \emph{(i)} above, and remedy the aforementioned fidelity problem. 

The discriminator $\mathcal{D}$ works as a binary classifier, which is trained to configure the discriminator's parameters $\Theta_\mathcal{D}$ by maximizing the following loss function: 
\begin{equation}
\label{eq:dloss}
\begin{aligned}
{\hat{\mathcal{L}}(\Theta_\mathcal{D}) =  \sum_{t=S}^T \log \mathcal{D}(D_t^h) + \log (1-\mathcal{D}(\mathcal{G}(F_t^S))).}
\end{aligned}
\end{equation}

The more challenging part is the design of the loss function employed to train the generator. Following the common hypothesis used in regression problems, we assume that the prediction error $\epsilon$ follows a zero-mean Gaussian distribution, with diagonal covariance matrix $\sigma ^2 \mathbf{I}$ \cite{Goodfellow-et-al-2016}, i.e.
\begin{equation}
\begin{aligned}
\epsilon \sim N(\epsilon|\mathbf{0}, \sigma ^2 \mathbf{I}).
\end{aligned}
\end{equation}
Then a fine-grained data point $D_t^H$ can be approximated with respect to the corresponding prediction error as following a multivariate Gaussian distribution with a mean that depends on $F_t^S$ and a diagonal covariance matrix $\sigma ^2 \mathbf{I} $.
This allows us to rewrite the conditional distribution in~(\ref{eq:new_problem}) as 
\begin{equation}
\label{eq:gau}
\begin{aligned}
p\left(D_t^H|F_t^S\right) \sim N(D_t^H| \mathcal{G}(F_t^S), \sigma ^2 \mathbf{I}).
\end{aligned}
\end{equation}
Substituting (\ref{eq:gau}) in~(\ref{eq:new_problem}), and adopting maximum likelihood estimation, the problem (\ref{eq:new_problem}) is equivalent to minimizing the following loss function, in order to configure the parameters $\Theta_\mathcal{G}$ of the generator:
\begin{equation}
\label{eq:loss}
\begin{aligned}
\mathcal{L}(\Theta_\mathcal{G}) &= \frac{1}{T} \sum_{t = S}^{T} \left[||D_t^H -  \mathcal{G}(F_t^S)||^2 - 2\sigma^2 \log \mathcal{D}(\mathcal{G}(F_t^S))\right].
\end{aligned}
\end{equation}
Recall that $\sigma^2$ is the variance of the Gaussian distribution of $\epsilon$, which can be considered as a trade-off weight between the mean square error (MSE) $||D_t^H -  \mathcal{G}(F_t^S)||^2$ and the adversarial loss $\log \mathcal{D}(\mathcal{G}(F_t^S))$. The same loss function is used in \cite{yu2016ultra} for image SR purposes, while the weight $\sigma^2$ is manually set. 

However, \textbf{we find that the training process is highly sensitive to the configuration of $\sigma^2$}. Specifically, the loss function does not converge when $\sigma^2$ is large, while the discriminator rapidly reach an optimum if $\sigma^2$ is small, which in turn may lead to model collapse \cite{arjovsky2017towards} and overall poor performance. 
To solve this problem, we propose an alternative loss function in which we replace the $\sigma^2$ term with the MSE. This yields
\begin{equation}
\label{eq:newloss}
\begin{aligned}
\hat{\mathcal{L}}(\Theta_\mathcal{G}) =  \frac{1}{T} \sum_{t = S}^{T} (1- 2 \log \mathcal{D}(\mathcal{G}(F_t^S)))\cdot ||D_t^H -  \mathcal{G}(F_t^S)||^2.
\end{aligned}
\end{equation}
In the above, the MSE term forces the predicted individual data points to be close to their corresponding ground truth, while the adversarial loss works to minimize the divergence between two distributions. Our experiments suggest that this new loss function significantly stabilises the training process, as the model never collapses and the process converges fast. In what follows, we detail the training procedure.

\subsection{Training the ZipNet-GAN}
To train the ZipNet-GAN we propose for solving the MTSR problem, we employ  Algorithm~\ref{tra_gan2}, which takes a Stochastic Gradient Descent (SGD) approach and we explain the steps involved next. Recall that the purpose of training is to configure the parameters of the neural network components, $\Theta_\mathcal{G}$ and $\Theta_\mathcal{D}$. We work with the Adam optimizer \cite{kingma2014adam}, which yields faster convergence as compared to traditional SGD. 

We train $\mathcal{G}$ and $\mathcal{D}$ iteratively, each time for $n_\mathcal{G}$ and $n_\mathcal{D}$ sub-epochs, by fixing the parameters of one and configuring the others', and vice-versa, until their loss functions converge (line~\ref{ln:bigloop2}). 
At every sub-epoch (lines~\ref{ln:loopd2}, \ref{ln:loopg}), $\mathcal{G}$ and~$\mathcal{D}$ randomly sample $m$ low-/high-resolution traffic measurement pairs (lines \ref{ln:sampling11}, \ref{ln:sampling22}) and compute the gradients $g_\mathcal{G}$ and $g_\mathcal{D}$ (lines~\ref{ln:gradn}, \ref{ln:grad2}) to be used in the optimization (lines~\ref{ln:adam11}, \ref{ln:adam22}). 

\begin{algorithm}[t]
  \caption{The GAN training algorithm for MTSR.}
  \label{tra_gan2}
  \begin{algorithmic}[1]
    \Inputs{Batch size $m$, low-/high-res traffic measure-\\ments $\mathcal{M}^L$ and $\mathcal{M}^H$, learning rate $\lambda$, genera-\\\mbox{tor and discriminator sub-epochs, $n_\mathcal{G}$ and $n_\mathcal{D}$.}}
    \Initialize{Generative and discriminative models, $\mathcal{G}$~and $\mathcal{D}$,  parameterized by $\Theta_\mathcal{G}$ and $\Theta_\mathcal{D}$.\\
    Pre-trained $\mathcal{G}$ by minimizing (\ref{eq:mse}).}\label{ln:initial}
    \While{$\Theta_\mathcal{G}$ and $\Theta_\mathcal{D}$ not converge}\label{ln:bigloop2}
    \For{$e_\mathcal{D} = 1$ to $n_\mathcal{D}$} \label{ln:loopd2}
        	\State{Sample low-/high-res traffic measurement pairs}\label{ln:sampling11}
        	\Statex{\hspace*{3em}$\{F_t^S, D_t^H\}_{t=1}^m$, $F_t^S \in\mathcal{M}^L, D_t^H \in \mathcal{M}^H$.}
               \State{$g_\mathcal{D} \leftarrow \Delta_{\Theta_\mathcal{D}}[\frac{1}{m}\sum_{i=1}^m \log \mathcal{D}(D_t^H) +$}\label{ln:gradn}
               \Statex{\hspace*{6em}$+ \frac{1}{m}\sum_{i=1}^m \log (1-\mathcal{D}(\mathcal{G}(F_t^S)))]$.}               
               \State{$\Theta_\mathcal{D}\leftarrow \Theta_\mathcal{D} + \lambda \cdot \text{Adam}(\Theta_\mathcal{D}, g_\mathcal{D})$.} \label{ln:adam11}
	\EndFor
        \For{$e_\mathcal{G} = 1$ to $n_\mathcal{G}$}\label{ln:loopg}
        	\State{Sample low-/high-res traffic measurement pairs}\label{ln:sampling22}
        	\Statex{\hspace*{3em}$\{F_t^S, D_t^H\}_{t=1}^m$, $F_t^S \in\mathcal{M}^L, D_t^H \in \mathcal{M}^H$.}
               \State{$g_\mathcal{G} \leftarrow \Delta_{\Theta_\mathcal{G}}\frac{1}{m} \sum_{t = 1}^{m} (1- 2 \log \mathcal{D}(\mathcal{G}(F_t^S)))\cdot$}\label{ln:grad2}
               \Statex{\hspace{6em}$\cdot||D_t^H -  \mathcal{G}(F_t^S)||^2.$}               
               \State{$\Theta_\mathcal{G}\leftarrow \Theta_\mathcal{G} - \lambda \cdot \text{Adam}(\Theta_\mathcal{G}, g_\mathcal{G})$.}\label{ln:adam22}
	\EndFor   
    \EndWhile
    \end{algorithmic}
\end{algorithm}

The key to this training process is that $\mathcal{G}$ and $\mathcal{D}$ make progress synchronously. At early training stages, when $\mathcal{G}$ is poor, $\mathcal{D}$ can reject samples with high confidence as they are more likely to differ from the real data distribution. An ideal $\mathcal{D}$ can always find a decision boundary that perfectly separates true and generated data points, as long as the overlapping measure support set of these two distributions is null. This is highly likely in the beginning and can compromise the training of $\mathcal{G}$.
To overcome this issue, we initialize the generator by minimizing the following MSE until convergence
\begin{equation}
\label{eq:mse3}
\begin{aligned}
\text{MSE}(\Theta_\mathcal{G}) =  \frac{1}{T} \sum_{t = S}^{T} ||D_t^H -  \mathcal{G}(F_t^S)||^2.
\end{aligned}
\end{equation}
Note that at this stage the initialized $\mathcal{G}$ (line~\ref{ln:initial}) could be directly deployed for the MTSR purpose, since the MSE based initialization is equivalent to solving (\ref{eq:problem}). However, this only minimizes the point-wise Euclidean distance, and the further training steps in Algorithm~\ref{tra_gan2} are still required to ensure the predictor does not lose accuracy. In our experiments, we set $n_\mathcal{G}$ and $n_\mathcal{D}$ to 1, and learning rate $\lambda$ to $0.0001$.

Next we discuss the processing and augmentation of the dataset we employ to train and evaluate the performance of the proposed ZipNet(-GAN).

\section{\mbox{Data Processing \& Augmentation}}
\begin{figure}[t]
\centering
\includegraphics[width=2.83in]{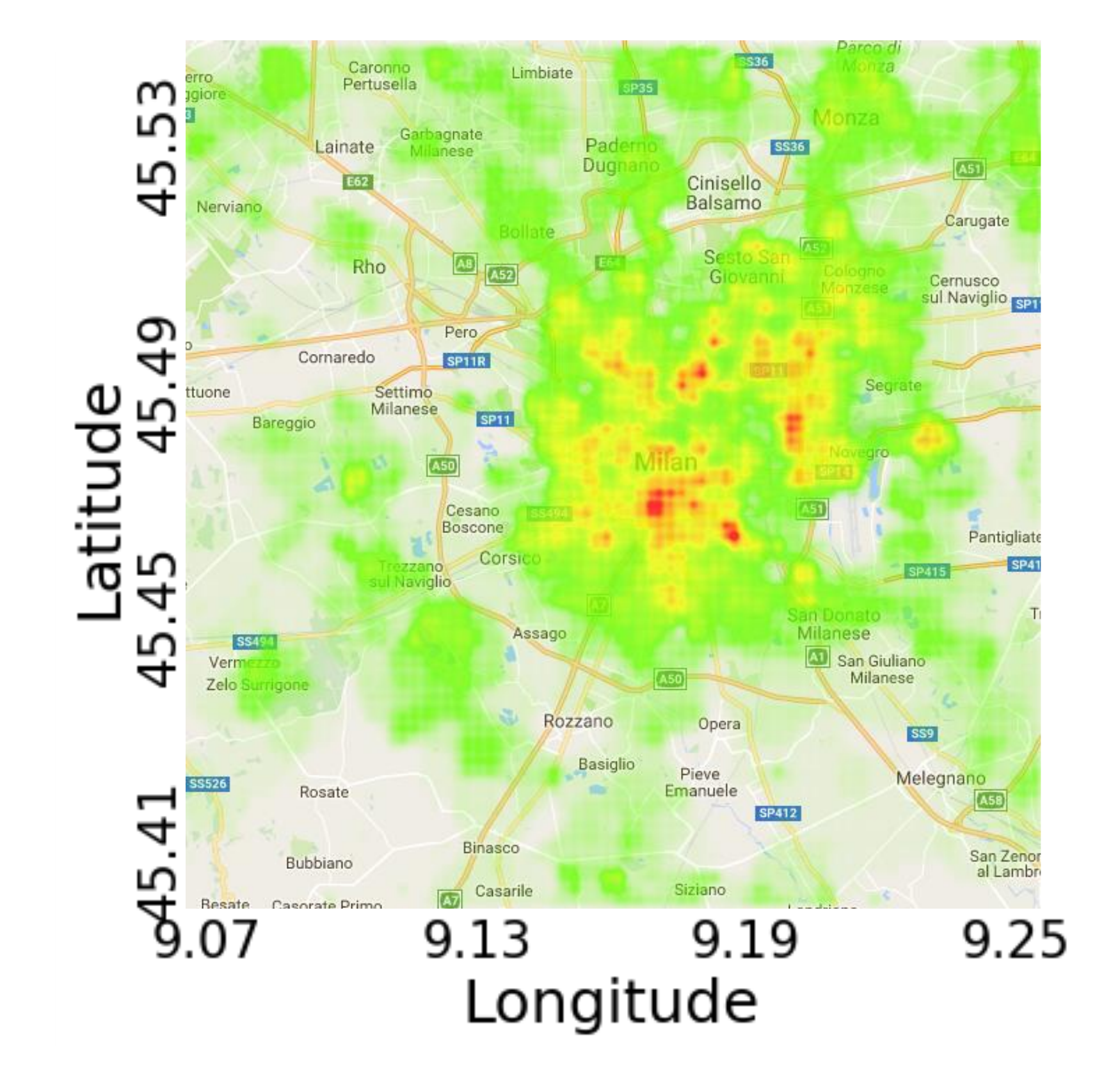}
\includegraphics[width=2.83in]{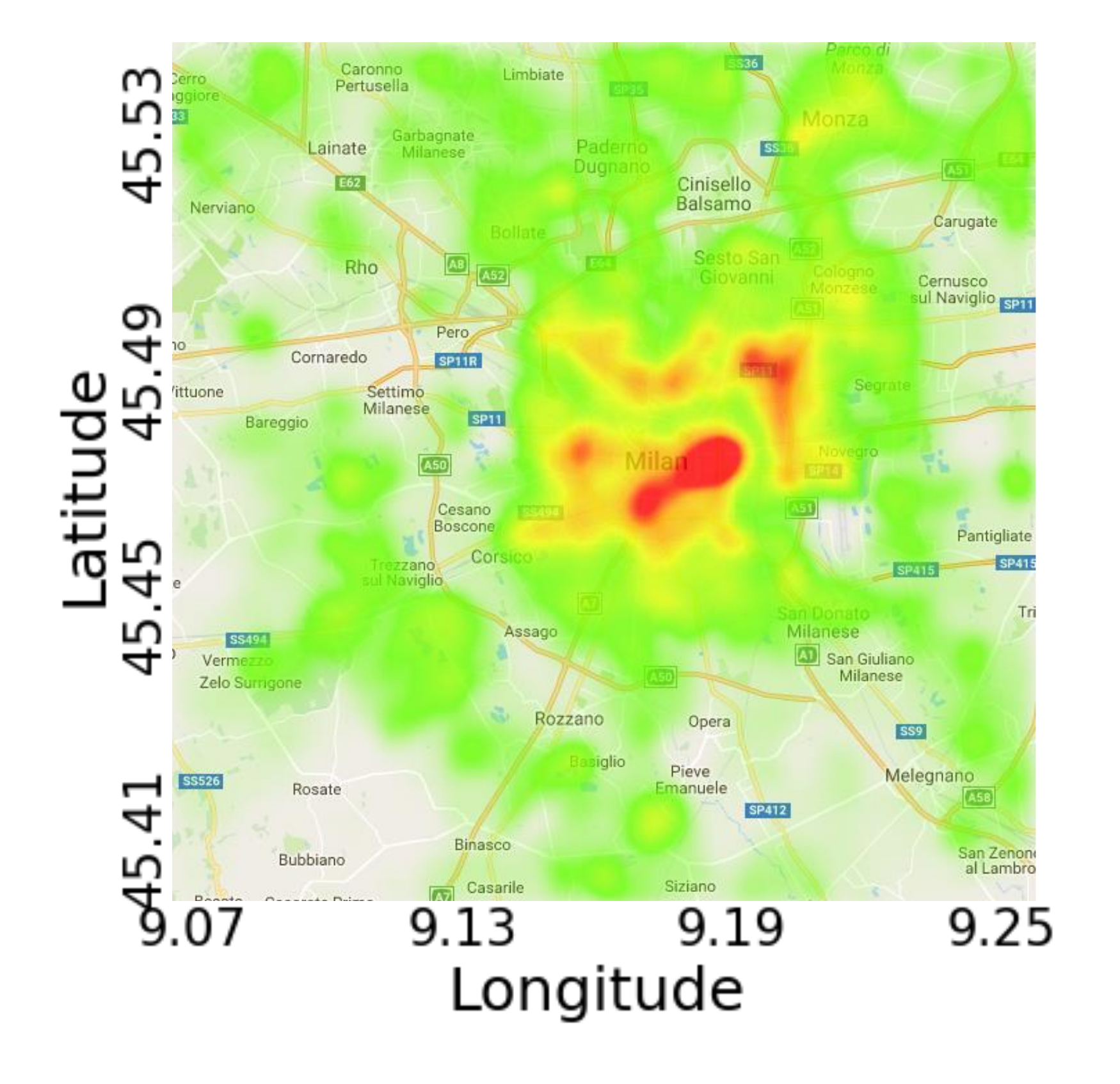}

\caption{Spatial distribution of mobile data traffic consumption in Milan during off-peak (left) and peak (right) times. Traffic consumption per 10-minute interval varies between 20 MB (green) to 5,496 MB (red). Figure best viewed in color.}
\label{Fig:topology}
\end{figure}

To train and evaluate the ZipNet(-GAN) architecture, we conduct experiments with a publicly available real-world mobile traffic dataset released through Telecom Italia's Big Data Challenge~\cite{barlacchi2015multi}, as we discuss in Sec.~\ref{sec:grid_dataset}. We show snapshots of the traffic's spatial distribution for both off-peak and peak times in Fig.~\ref{Fig:topology}. 

\begin{figure}[t]
\begin{center}
\includegraphics[width=\columnwidth]{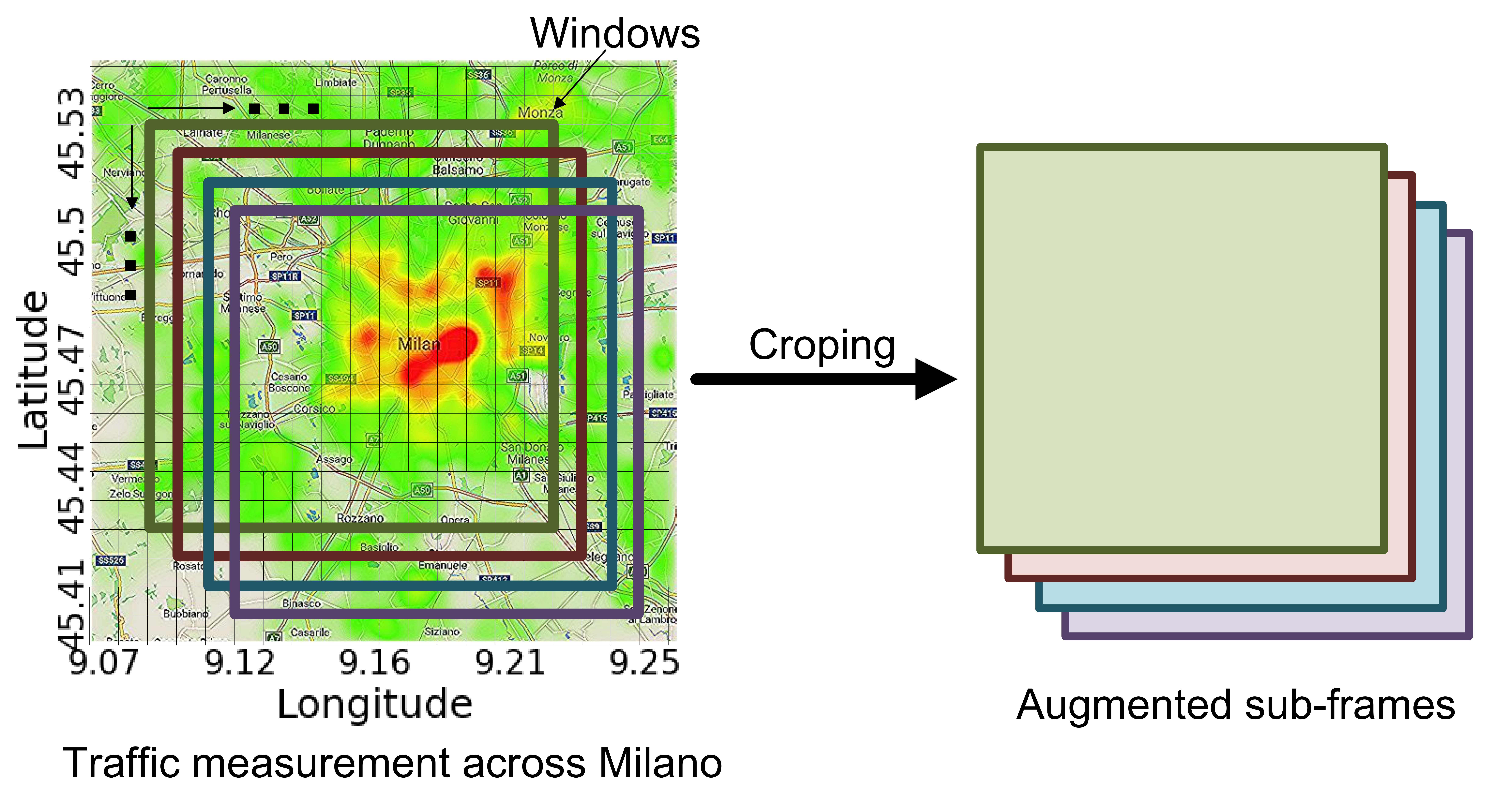}
\end{center}
\caption{\label{Fig:aug}Illustration of the data augmentation technique we propose to enable precise training in view of performing MTSR. Figure best viewed in color.}
\end{figure}
Overall the dataset includes $T=8,928$ snapshots of fine-grained mobile traffic measurements. This number appears sufficient when using traditional machine learning tools, though it is grossly insufficient for the purpose of training a complex neural network, as in our case. This is due to the highly dimensional parameter set, which can turn the model over-fitting if improperly configured through training with small datasets. To overcome this problem, we augment the Milan dataset by cropping each snapshot to multiple `windows' of smaller size, each with different offsets (1 cell increment in every dimension). We illustrate this augmentation process in Fig.~\ref{Fig:aug}. 

We work with `sub-frames' of $80\times 80$ sub-cells, thereby producing 441 new data points from every snapshot. Note that the initial size of model's prediction matrix is also $80\times 80$ and we employ a moving average filter to construct the final predictions across the original grid (i.e. $100\times 100$).

\section{Experiments}
In this section we first describe briefly the implementation of the proposed ZipNet(-GAN), then evaluate its performance in terms of prediction accuracy (measured as Normalized Mean Root Square Error -- NMRSE), fidelity of the inferences made (measured through Peak Signal-to-Noise Ratio -- PSNR), and similarity between predictions made and ground truth measurements (structural similarity index  -- SSIM). We compare the performance of ZipNet(-GAN) with that of Uniform and Bicubic interpolation techniques~\cite{carlson1985monotone}, Sparse Coding method (SC)~\cite{yang2010image}, Adjusted Anchored Neighboring Regression (A+) \cite{timofte2014a+}, and Super-Resolution Convolutional Neural Network (SRCNN) previously used in imaging applications~\cite{dong2016image}.

\subsection{Implementation}
We implement the proposed ZipNet(-GAN), the SRCNN, the Uniform and Bicubic interpolation methods using the open-source Python libraries TensorFlow~\cite{tensorflow2015-whitepaper} and TensorLayer~\cite{tensorlayer}.\footnote{\rev{The source-code of our implementation is available at \texttt{\url{https://github.com/vyokky/ZipNet-GAN-traffic-inference}}}.} We train the models using a GPU cluster comprising 19 nodes, each equipped with 1-2 NVIDIA TITAN X and Tesla K40M computing accelerators (3584 and respectively 2280 cores). To evaluate their prediction performance, we use one machine with one NVIDIA GeForce GTX 970M GPU for acceleration (1280 cores). SC and A+ super-resolution techniques are implemented using Matlab. In what follows, we detail four MTSR instances with different granularity, which we use for comparison of the different SR methods.

\subsection{Different MTSR Granularity}
In practice measurement probes are deployed at different locations (e.g. depending on user density) and often have different coverage in terms of number of cells. We approximate the coverage of a probe by a square area consisting of $r_f = n_f\times n_f$ sub-cells, and refer to $n_f$ as an upscaling factor. The smaller the $n_f$, the higher the granularity of measurement. Given the heterogeneous nature of cellular deployments, we construct four different MTSR instances with different granularity, as summarized in Table~\ref{probes}.

\begin{table}[htb]
\centering
\begin{tabular}{|c|p{8cm}|c|c|}
\hline
Instance &  Configuration & $n_f$ &$r_f$ \\
& & (Avg.) & (Avg.)\\ \hline
Up-2     & Probes cover $2\times 2$ sub-cells                                               & 2                                & 4                                \\ \hline
Up-4     & Probes cover $4\times 4$ sub-cells                                               & 4                                & 16                               \\ \hline
Up-10    & Probes cover $10\times 10$ sub-cells                                             & 10                               & 100                              \\ \hline
Mixture  & 7\% of probes cover $10\times 10$ sub-cells, 44\% cover $4\times 4$, and 49\% cover $2\times 2$ sub-cells. & 4                                & 16                               \\ \hline
\end{tabular}
\caption{Configuration of MTSR instances with different granularity considered.}
\label{probes}

\end{table}

\begin{figure}[t]
\begin{center}
\includegraphics[width=\columnwidth]{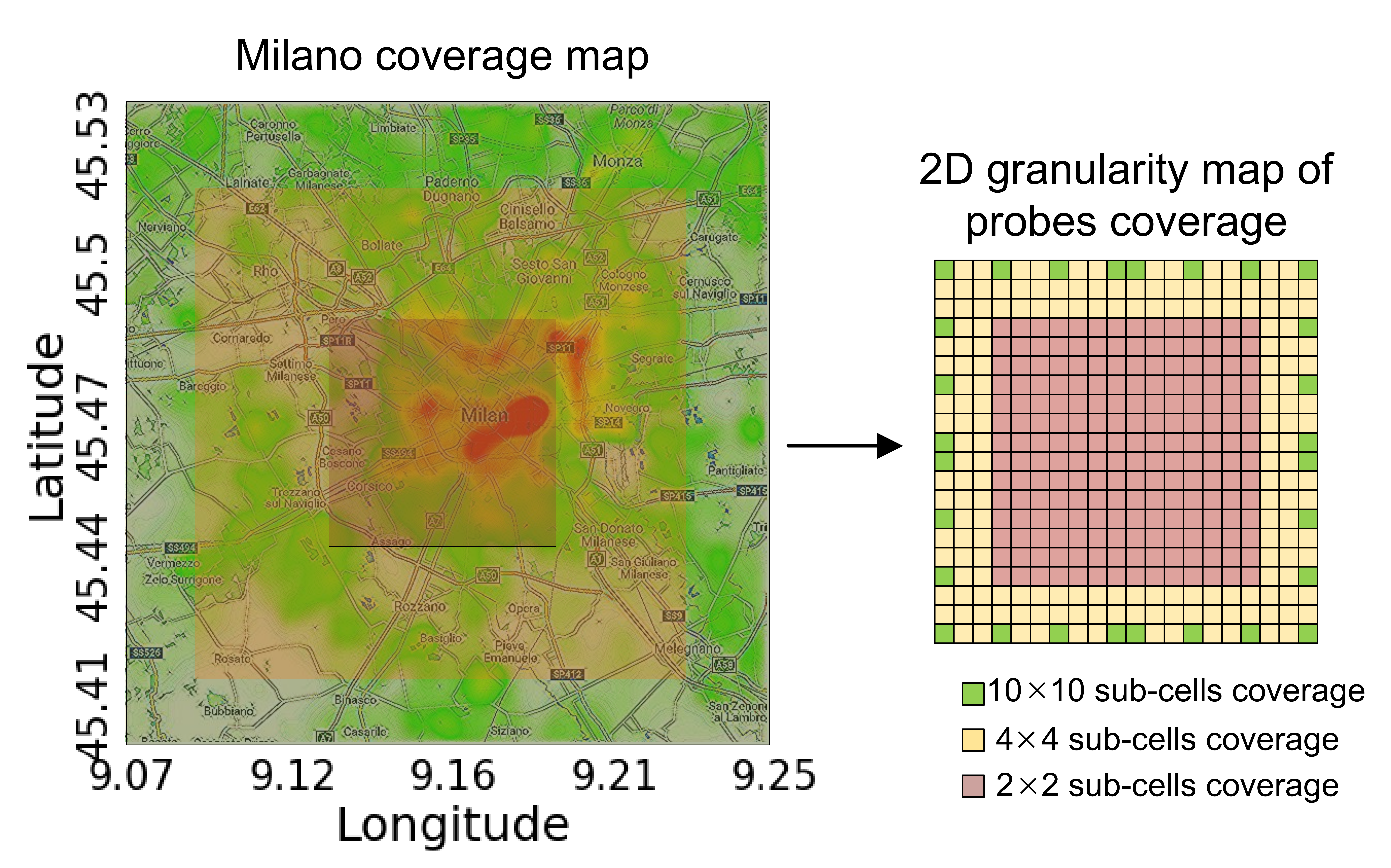}
\end{center}
\caption{\label{Fig:mix} Cellular deployment map for Milan served by a mixture of probes with different coverage (left), and the corresponding measurement granularity projected onto a 2D map (right). Best viewed in color.}

\end{figure}

The first three instances assume that all probes are uniformly distributed in Milan and have the same coverage, i.e. 4, 16, and 100 sub-cells respectively. The fourth instance corresponds to a mixture of probes, each with different coverage, as we illustrate in Fig.~\ref{Fig:mix}. Specifically, we consider three types of probes covering 2$\times$2, 4$\times$4, and respectively 10$\times$10 sub-cells. 
In this instance, more probes serve the city center (red area), each with smaller coverage and thus finer granularity. Fewer probes are assumed in surrounding areas, covering larger regions (the yellow and green areas). \rev{We compute aggregate measurements collected by such dissimilar probes, each aggregate corresponding to the sum over the covered cells. We then project these points onto a square that becomes the input to our model, as shown on the right in Fig.~\ref{Fig:mix}.} Therefore, for each traffic snapshot we can obtain 400 measurement points that can be viewed as a 2D square (right side of~Fig.~\ref{Fig:mix}), which we use for training and performing inferences. Note that although the average $n_f$ of the mixture instance is 4, its structure differs significantly from that of the `up-4' instance. Our evaluation will reveal that such differences will affect the accuracy of the MTSR outcome. 

We train all deep learning models with data collected between 1 Nov 2013 and 10 Dec 2013 (40 days), validate their configuration with measurements observed over the following 10 days, and finally evaluate their performance using the averages aggregated by the different probes (different granularity) between 20--30 Dec 2013. Prior to training, all data is normalized by subtraction of the mean and division by the standard deviation. The coarse-grained input measurement data (i.e. $\mathcal{M}^L$) is generated by averaging the traffic consumption over cells' coverage.

\subsection{Performance Evaluation}

\noindent \textbf{Evaluation Metrics:}  We quantify the accuracy of the proposed ZipNet(-GAN), comparing against that of existing super-resolution methods, in terms of Normalized Root Mean Square Error (NRMSE, see Eq.(\ref{eq:nrmse})), Peak Signal-to-Noise Ratio (PSNR, see Eq.(\ref{eq:psnr})), and Structural Similarity Index (SSIM, see Eq.(\ref{eq:ssim})). 


\vspace*{0.5em}
\noindent \textbf{Techniques for comparison:} \rv{Since MTSR is an original technique, there exist no previous solutions to tackle this problem. We select the most relevant methods as baselines for comparison.} We summarize the performance of the proposed ZipNet(-GAN) against a range of existing interpolation or image super resolution techniques, including Uniform interpolation, Bicubic interpolation, Sparse Coding based methods (SC) \cite{yang2010image}, Adjusted Anchored Neighboring Regression (A+) \cite{timofte2014a+}, and Super-Resolution Convolutional Neural Network (SRCNN) \cite{dong2016image}. Uniform interpolation is routinely used by operators and assumes that the data traffic volume is uniformly distributed across cells \cite{lee2014spatial}. The Bicubic interpolation is a popular non-parametric tool frequently used to enhance the resolution of images \cite{carlson1985monotone}. SC and A+ are commonly used as benchmarks in image super-resolution evaluation. 
Finally, SRCNN is a benchmark deep learning architectures that comprises three convolutional layers. We note that in our evaluation ZipNet is a simplified version of the ZipNet-GAN, which is purely trained with the mean square error (\ref{eq:mse3}), without the help of the discriminator.

\begin{figure}[h!]
\begin{center}
\includegraphics[width=\columnwidth]{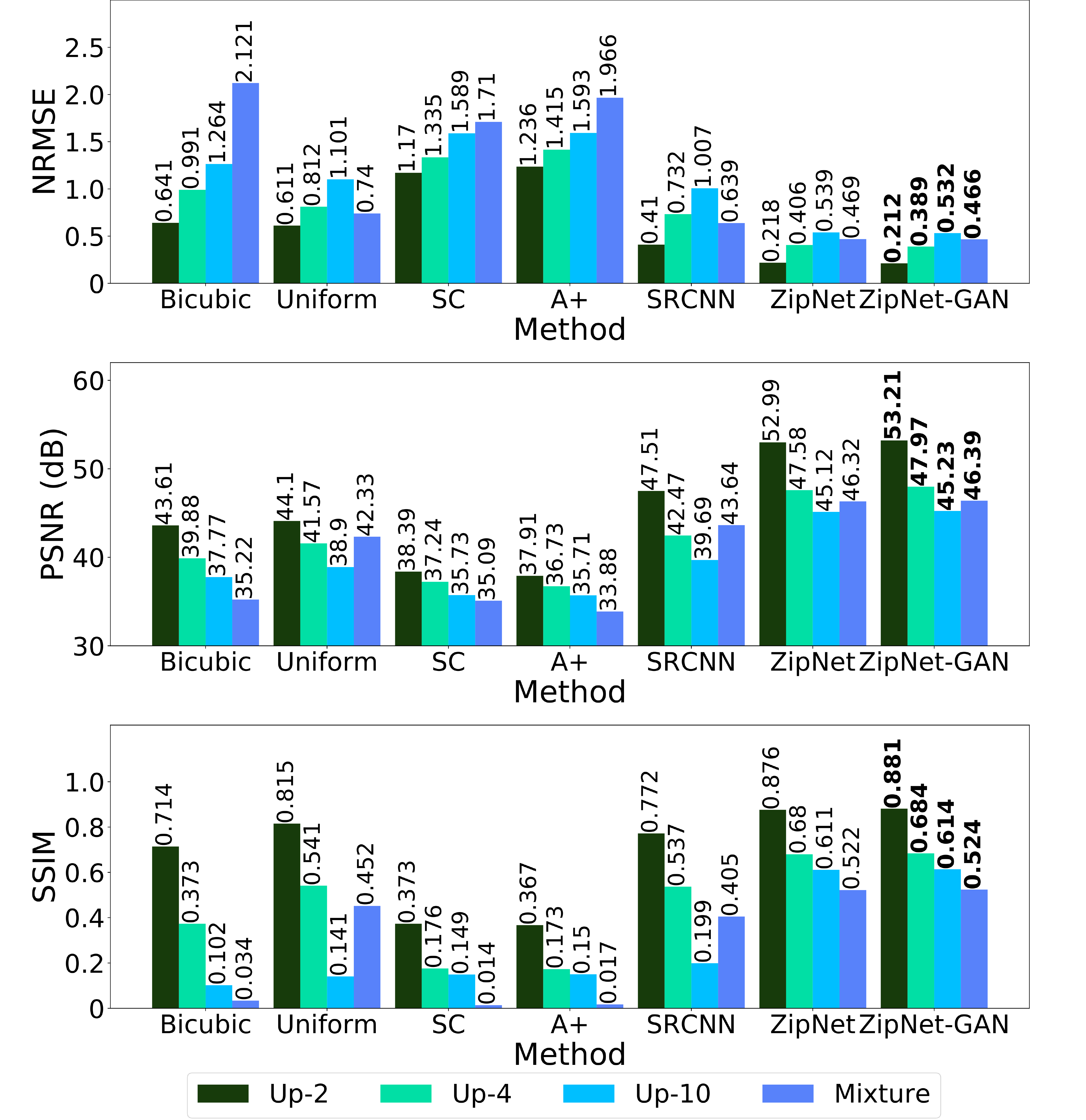}
\end{center}
\caption{\label{Fig:metric0} Comparison of inference accuracy of the proposed ZipNet(-GAN) and existing SR techniques in terms of NRMSE (above), PSNR (middle), SSIM (below). Four upscaling instances (different color bars). Experimental results with the Milan dataset~\cite{barlacchi2015multi}.}
\end{figure}

\noindent\textbf{Assessing the quality of inferences:}
We summarize the performance of the proposed ZipNet(-GAN) and that of existing SR techniques in terms of the aforementioned metrics in Fig.~\ref{Fig:metric0}, where we use different bars for each of the four MTSR instances considered for inference (2, 4, and 10 upscaling, and upscaling mixture). The bars correspond to averages for inferences made over 10 days (i.e. 1440 snapshots).

Observe that \textbf{ZipNet-GAN achieves the best performance for all MTSR instances}, outperforming traditional super resolution schemes. 
In particular, although SC and A+ work well in image SR, their performance is inferior to that of simple Uniform and Bicubic interpolation techniques when performing MTSR. Our intuition is that the mobile traffic data has substantially different spatial structure and scale, as compared to imaging data. Therefore traditional SR approaches are unable to capture accurately the relations between low- and high- resolution traffic `frames'. On the other hand, SRCNN works acceptably with low upscaling MTSR instances (i.e. `up-4' and `up-2'), but performs poorly when processing the `up-10' instances. Unlike Uniform, A+, and Bicubic, the proposed ZipNet-GAN achieves an up to 65\%, 76\% and respectively 78\% smaller NRMSE. The ZipNet-GAN further attains the highest PSNR and SSIM among all approaches, namely up to 40\% higher PSNR and a remarkable 36.4$\times$ higher SSIM. Although perhaps more subtle to observe, the ZipNet-GAN is more accurate than the ZipNet (i.e. without employing a discriminator), which confirms the effectiveness of the GAN approach.

\begin{figure*}[htb]
\begin{center}
\includegraphics[width=\columnwidth]{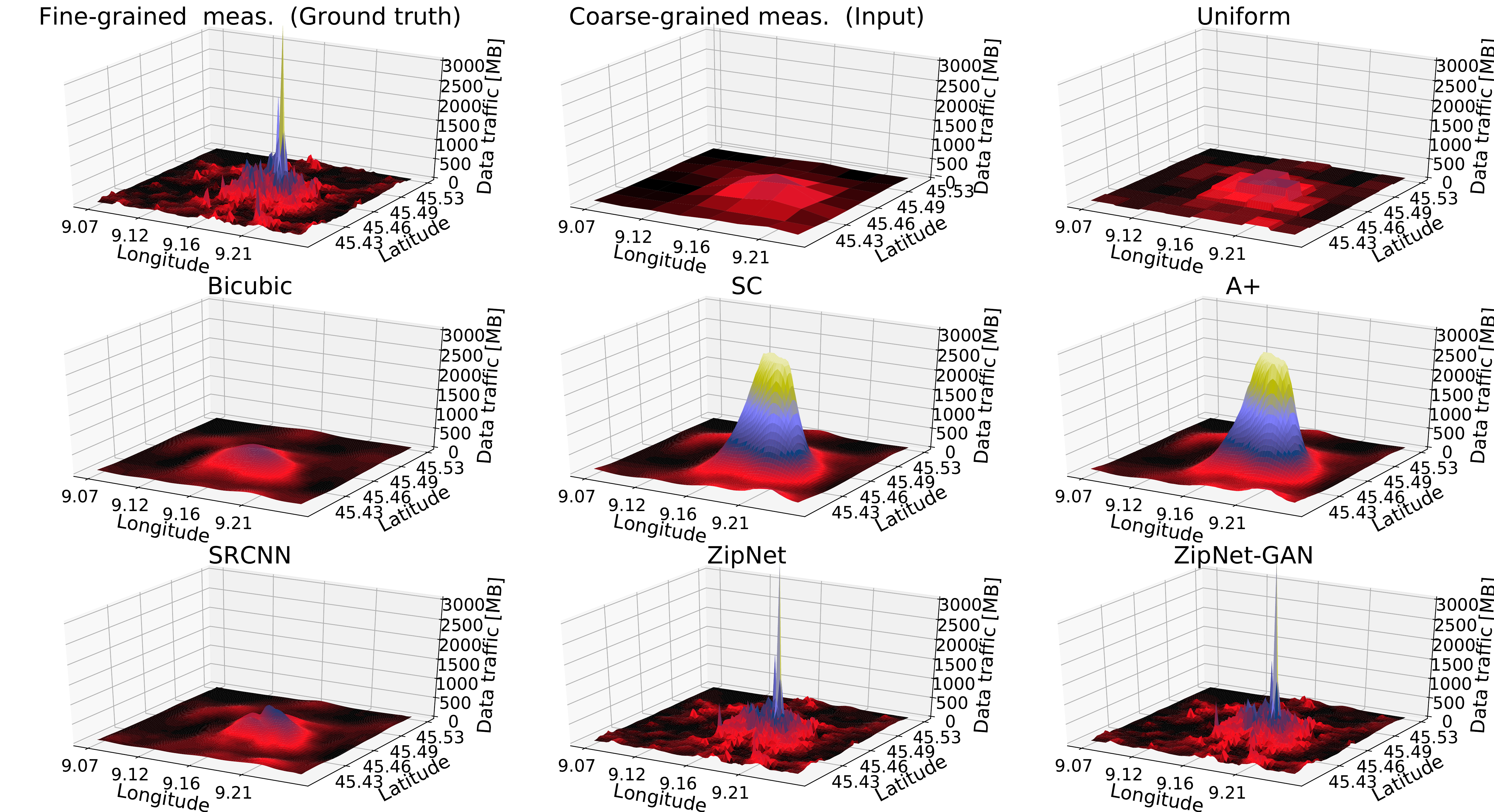}
\end{center}
\caption{\label{Fig:d10} Snapshots of ground truth, input data, and the \emph{`up-10' MTSR instance} predicted by the proposed ZipNet (-GAN), existing interpolation methods and image SR techniques, using data collected in Milan on 21\textsuperscript{st} Dec 2013.}
\end{figure*}
\begin{figure*}[htb]
\begin{center}
\includegraphics[width=\columnwidth]{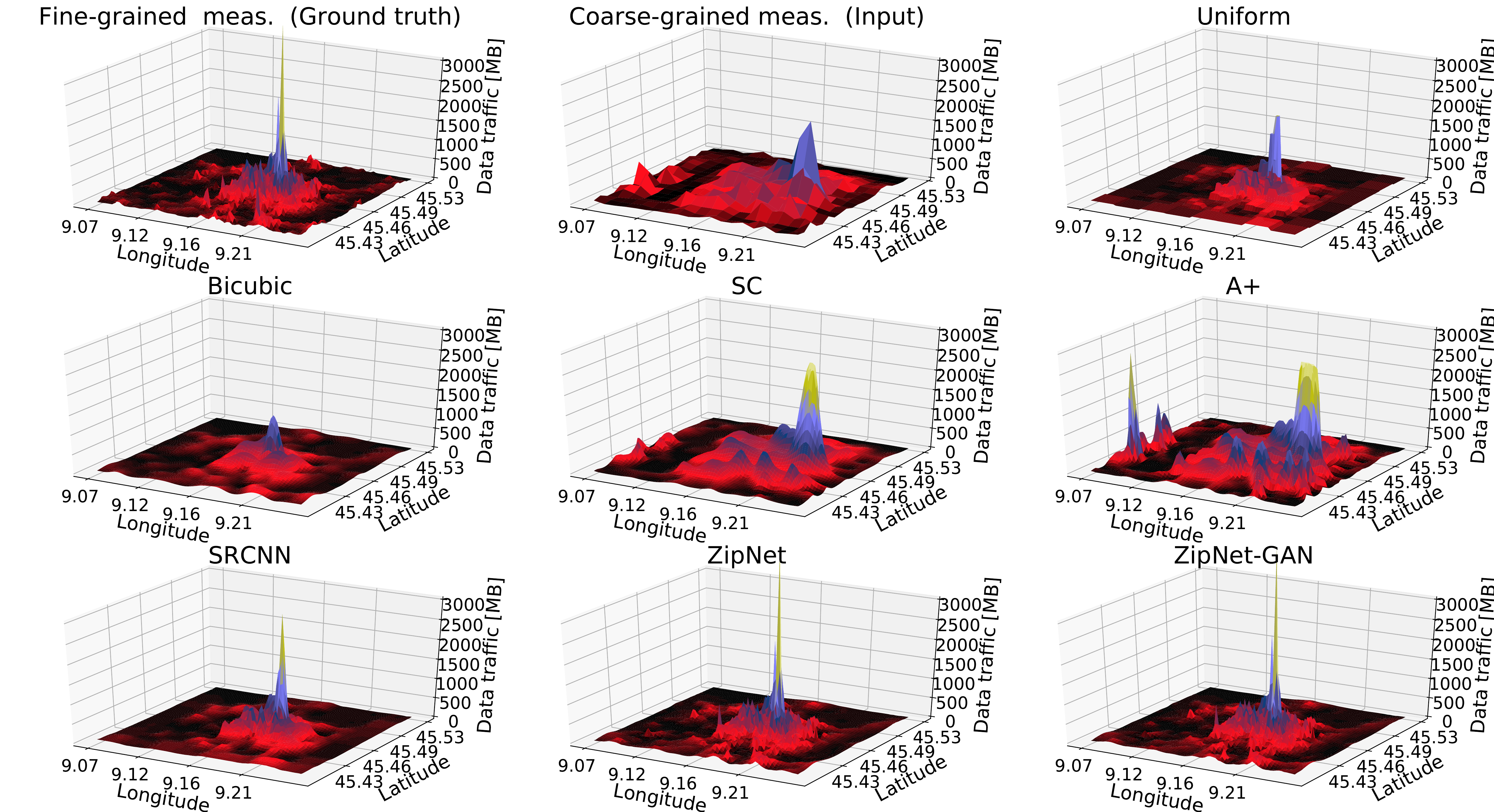}
\end{center}
\caption{\label{Fig:gau}  Snapshots of ground truth, input data, and the \emph{mixture MTSR instance} predicted by the proposed ZipNet(-GAN), existing interpolation and image SR techniques, using data collected in Milan on 21\textsuperscript{st} Dec 2013.}
\end{figure*}

Taking a closer look at Fig.~\ref{Fig:metric0}, observe that the prediction accuracy of all approaches drops as the upscaling factor $n_f$ grows. This is indeed reasonable, since a larger $n_f$ corresponds to a greater degree of aggregation and thus detail information loss, which will pose more uncertainty to the models. Further, although the `up-4' and mixture instances have the same average upscaling factor, the proposed ZipNet(-GAN) operate somewhat better with the former. Our intuition is that the mixture instance distorts the spatial correlation of the original measurements, as shown in Fig.~\ref{Fig:mix}. However, given the subtle performance differences, MTSR with probes of dissimilar coverage and granularity remains feasible with the proposed ZipNet-GAN.

We now delve deeper into the performance of all methods consider for MSTR and examine the behavior of each with individual snapshots. To this end, Figs.~\ref{Fig:d10} and~\ref{Fig:gau} provide additional perspectives on the value of employing the proposed ZipNet(-GAN) to infer mobile traffic consumption with fine granularity. Each figure shows snapshots of the predictions made by all approaches, when working with the `up-10' (Fig.~\ref{Fig:d10}) and the mixture MTSR (Fig.~\ref{Fig:gau}) instances. In particular, in Fig.~\ref{Fig:d10} we observe that ZipNet(-GAN) deliver remarkably accurate predictions with a 99\% reduction in term of measurement points required, as the texture and details are almost perfectly recovered. In contrast, the Uniform, Bicubic, SC, A+ and SRCNN techniques, although improve the resolution of data traffic `snapshots', lose significant details and deviate considerable from the ground truth measurements (upper left corner). 

Turning attention to Fig.~\ref{Fig:gau}, observe that the input exhibits some spatial distortion (top centre plot), as the probes aggregating measurements have different coverage and locations. Despite this, the ZipNet(-GAN) still capture well spatial correlations and continues to perform very accurately (two plots in the bottom right corner). In contrast, the Uniform and Bicubic interpolations, although capture some spatial distribution mobile traffic profiles, significantly under-estimate the traffic volume in the city centre. In addition, the predictions made by SC and A+ exhibit significant distortions and yield inferior performance, demonstrating their image SR capabilities cannot be mapped directly to MTSR tasks. Lastly, the SRCNN approach, which employs deep learning, works acceptably in areas with low traffic intensity, but largely underestimates the traffic volume in the city center.

\subsection{The Benefits of GAN}
\begin{figure*}[htb]
\begin{center}
\includegraphics[width=\columnwidth]{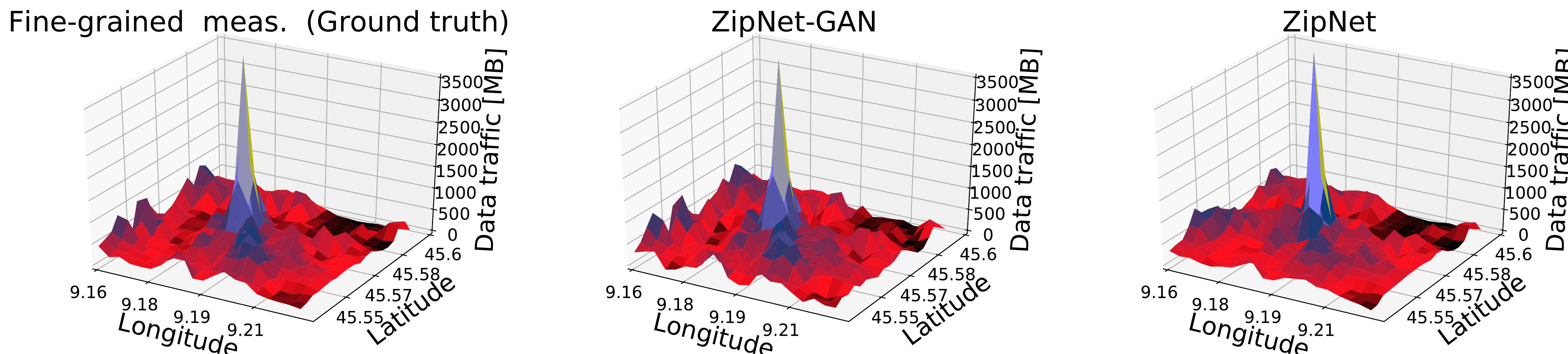}
\end{center}
\caption{\label{Fig:zoom} \rev{Zoom snapshots of ground truth and fine-grained traffic consumption predicted by the proposed ZipNet(-GAN) in an up-10 instance in central Milan.}}
\end{figure*}

\rev{In image SR, GAN architectures improve the fidelity of high-resolution output, making the image more photo-realistic. Here we show the GAN plays a similar role in MTSR. To this end, in Fig. \ref{Fig:zoom} we present zoom snapshots of the predictions made by ZipNet and ZipNet-GAN. These offer a clear visual perception of the inference improvements of GAN in central parts of the city.}
\rev{Indeed, observe that including a GAN in the architecture improves the prediction fidelity, as it minimizes the divergence between real and predicted data distributions, although this does not necessarily enhance overall accuracy. Note that the additional accuracy does not come at the cost of increased complexity, since the adversarial training is fast in terms of convergence and the discriminator will be abandoned in the inference phase.}

\subsection{\rv{Robustness to Abnormal Traffic}}
\begin{figure*}[htb]
\begin{center}
\includegraphics[width=\columnwidth]{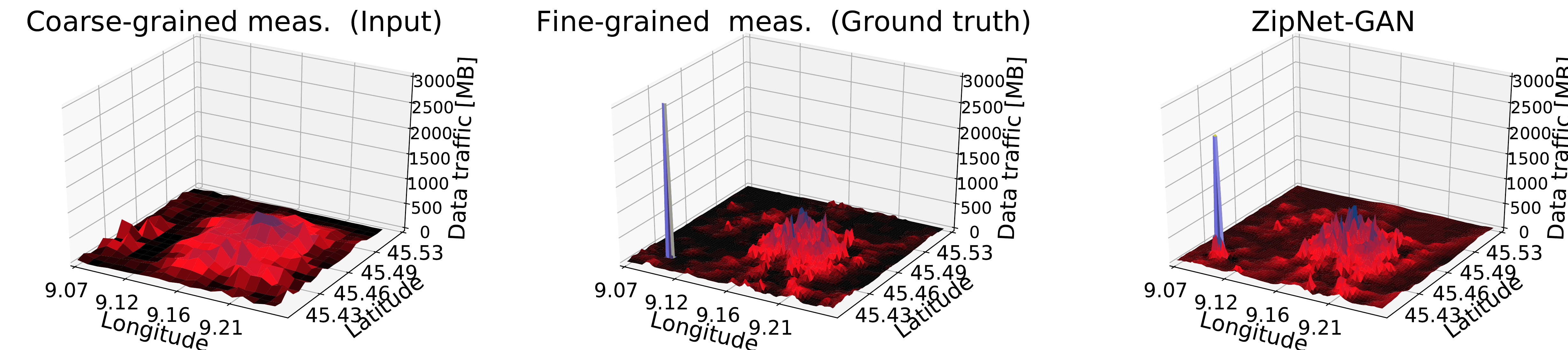}
\end{center}
\caption{\label{Fig:anomaly} \rev{Behavior in the presence of anomalies: snapshots of coarse-grained measurements fed as input (left), ground truth (middle) and fine-grained traffic patterns predicted by the proposed ZipNet-GAN (right) in a mixture instance.}}
\end{figure*}

\rev{To evaluate the robustness of our solution in the presence of traffic anomalies, we artificially add such traffic patterns to the test dataset and investigate the behavior of our proposal \rv{with the best inferred accuracy, \ie ZipNet-GAN.} Specifically, we introduce abrupt traffic demands in suburban areas, which can be regarded as occurrences of social events (e.g. concert, football match, etc.), as seen in the bottom left corner of the second sub-plot in Fig.~\ref{Fig:anomaly}. Although such anomalies do not occur in the training set, the proposed ZipNet-GAN still successfully identifies the locations of abnormal traffic, given averaged and smoothed inputs (first sub-plot). This implies that, to some extent, our proposal can perform as an anomaly detector operating only with coarse measurements.} 

\subsection{Impact of Cross-Temporal Relations}
We conclude our experiments by examining the impact cross-temporal correlations between traffic measurements provided as input have on the ZipNet-GAN architecture we propose. To this end, we \emph{(i)} compare the MTSR accuracy with input of different temporal lengths $S$, and \emph{(ii)}~compute the absolute value of first-order derivatives of the loss function employed over input. The magnitude of these gradients are a good approximation of the sensitiveness of final prediction decisions to changes of the input~\cite{li2015visualizing}. 

\begin{figure}[ht]
\begin{center}
\includegraphics[width=\columnwidth]{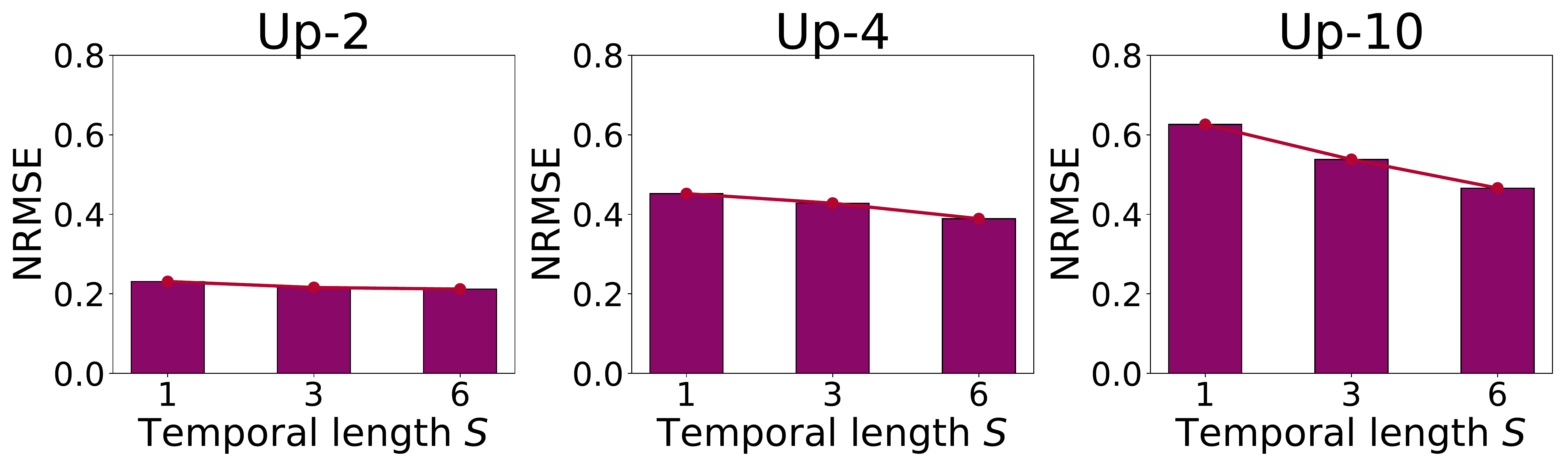}
\end{center}
\caption{\label{Fig:temporal} NRMSE comparison for three MTSR instances, with different temporal length $S$ of the input.}
\end{figure}

\noindent \textbf{NRMSE with different length inputs:} We feed the proposed ZipNet-GAN with input of temporal length $S \in \{1, 3, 6\}$ snapshots, and illustrate in Fig.~\ref{Fig:temporal} the NRMSE attained with the three homogeneous MTSR instances considered. Observe that the prediction error drops with the increase of the number of snapshot we provide to our model in all instances, which indicates that earlier observations provide valuable insights toward inferring real fine-grained data traffic consumption. Additionally, the historical measurements play a more significant role with the increase of $n_f$ -- in the `up-10' instance the error between predictions made with $S=1$ and $S=6$ sequence lengths is much larger than in the same case for the other instances. This brings important implications to operators assessing the trade-offs between the length of inputs (which affects model complexity) and prediction accuracy.

\begin{figure}[t]
\begin{center}
\includegraphics[width=\columnwidth]{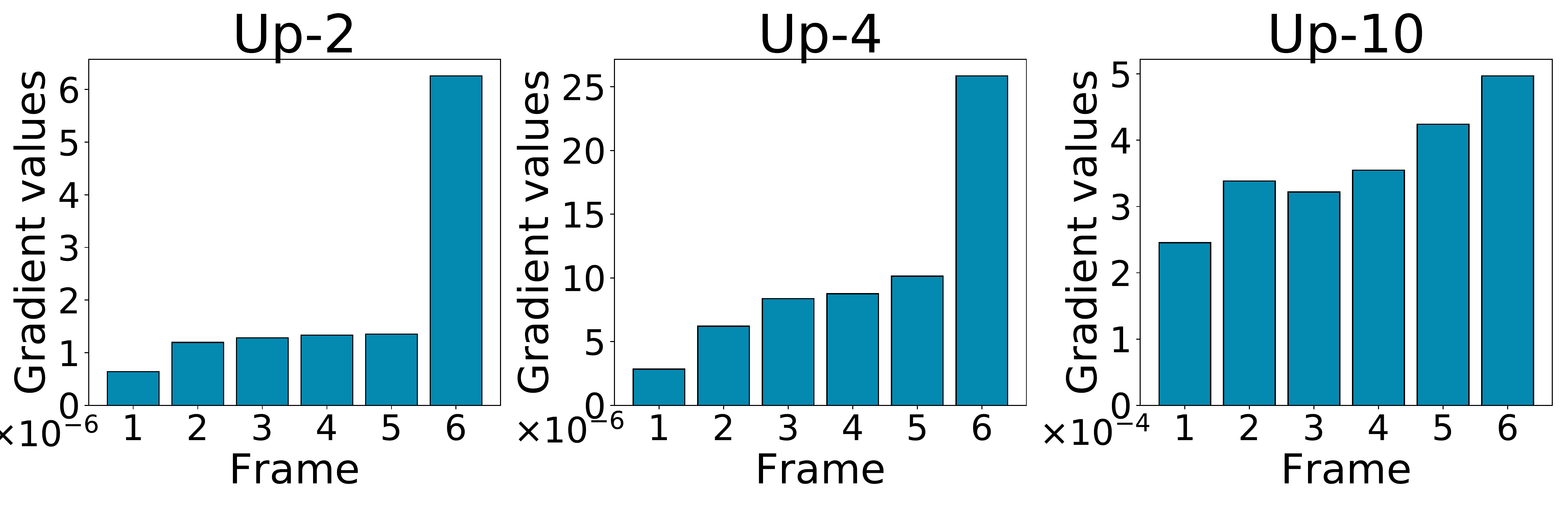}
\end{center}
\caption{\label{Fig:gradient} Mean magnitude of the gradient of the loss function $\text{\normalfont L}(F_t^S)$ over inputs ${F_t^S}$, with different MTSR instances. Averages computed over the entire network across all test data. }
\end{figure}

\noindent \textbf{Magnitudes of gradients:} The impact of the input's temporal dimension can also be evaluated from the gradients' perspective. The loss function $\text{L}$ is essentially a complex non-linear of the input $F_t^S$. However, this can be approximated by the first-order term of its Taylor expansion, i.e.

\begin{equation}
\begin{aligned}
\text{L}(F_t^S) \approx w(F_t^S)^T \cdot F_t^S+ b,
\end{aligned}
\end{equation}
\noindent where $w(F_t^S)$ is the gradient of the loss function $\text{L}(F_t^S)$ over input $F_t^S$ and $b$ is a bias. Computing the absolute value of the gradient $w(F_t^S)^T$, i.e.

\begin{equation*}
\frac{\partial\text{L}(F_t^S)}{\partial  F_t^S} 
= \frac{\partial }{\partial F_t^S}  \left[||D^H_t -  \mathcal{G}(F_t^S)||^2(1- 2 \log \mathcal{D}(\mathcal{G}(F_t^S)))\right],
\end{equation*}
\noindent should give insights into the number of temporal steps required for accurate prediction with different MTSR instances. Therefore in Fig.~\ref{Fig:gradient} we plot the average magnitude of the gradient of the loss function over all inputs ${F_t^S}$, for all three homogeneous MTSR instances. 
Observe that the most recent `frame' (i.e. frame 6) yields the largest gradient for all instances, as we expect. This means that the current measurement `snapshot' provides the most valuable information for the model to reconstruct the fine-grained counterpart. Further, the contribution of historical measurements (i.e. frames 1 to 5) increases with the upscaling factor (from 2 to 10), which suggests that historical information becomes more significant when less spatial information is available, which is consistent with the insight gained earlier.  

We conclude that, by exploiting the potential of deep neural networks and generative adversarial networks, the proposed ZipNet-GAN scheme can infer with high accuracy fine-grained mobile traffic patterns, and substantially outperforms traditional interpolation methods and image SR techniques, as it attains up to 78\% lower prediction errors (NRMSE) and 36.4$\times$ higher structural similarity (SSIM). \rev{By employing a high performance GPU, training ZipNet-GAN takes 2-3 days, which we believe is negligible, considering the long-term costs of post-processing alternatives. Given the reasonable training time, ZipNet-GAN can be easily ported to different urban layouts after retraining for the target regions.}

\section{Discussion}
Obtaining city-scale fine-grained mobile traffic maps from coarse aggregates collected by probes brings important~\mbox{benefits} to a range of applications critical to the urban realm. Namely,
\begin{itemize}[topsep=5pt]
 \item \textbf{Agile network resource management:} our approach enables infrastructure simplification and reduces OpEx. This is because base station controllers (BSC) are shared by increasing numbers of BSs, and already handle radio quality measurements and handovers. Adding further measurement capabilities, co-ordination of triangulation, and real-time processing of multi-modal measurements, poses scalability concerns, which may eventually impact on end user QoE. The proposed ZipNet-GAN shifts the burden to a single dedicated cluster that only requires to be trained infrequently and can perform coarse- to fine- grained measurement transformations, fed by summary statistics already available at selected probes.
 
 \item \textbf{Events localization \& response:} popular social events (e.g. concerts, fairs, sports, etc.) or emergency situations (fires, riots, or terrorist acts) exhibit high spatial similarity in terms of mobile data traffic consumption, which can serve as an indicator of density/size of the crowds involved. Accurate and timely localization of such instances through inference of data traffic can help overcome network congestion in hot spots, but also appropriately provision policing and emergency services. 
 
 \item \textbf{Context-based business:} accurate MTSR can also aid small businesses that can offer promotional products depending on spontaneous circumstances, e.g. families gathering in the park on a sunny day. Likewise university PR services can disseminate relevant information during highly-attended open/graduation days.

\end{itemize}

Therefore, we argue ZipNet-GAN is not limited to network infrastructure related tasks, e.g. precision traffic engineering or anomalous traffic detection \cite{roughan2012spatio}, but could also serve civil applications such as  events planning or even transportation. \rev{Importantly, once trained the proposed technique can continuously perform inferences on live streams, unlike post-processing approaches that only work off-line and thus have limited utility.}

\section{Summary}
Precision large-scale mobile traffic analytics is increasingly vital for operators, to provision network resources agilely and keep up with end-user application requirements. Extracting fine-grained mobile traffic patterns is however complex and expensive, as this relies on specialized equipment, substantial storage capabilities, and intensive post-processing. In this chapter we proposed a mobile traffic super resolution (MTSR) technique that overcomes these issues and infers mobile traffic consumption with fine granularity, given only limited coarse-grained measurement collected by probes. Our solution consists of a Generative Adversarial Network (GAN) architecture combined with an original deep Zipper Network (ZipNet) inspired by image processing tools. The proposed ZipNet-GAN is tailored to the specifics of the mobile networking domain and achieves reliable network-wide MTSR.
Specifically, experiments we conduct with a real-world mobile traffic dataset collected in Milan, demonstrate the proposed schemes predict narrowly localized traffic consumption, while improving measurement granularity by up to 100$\times$, irrespective of the position and coverage of the probes. The ZipNet(-GAN) reduce the prediction error (NRMSE) of existing interpolation and super-resolution approaches by~78\%, and achieve up to 40\% higher fidelity (PSNR) and 36.4$\times$ greater structural similarity (SSIM). \rv{Future work will focus on evaluating our proposal with other traffic datasets, \eg the one collected in Trentino.}

%% file: chap6.tex
\chapter{Deep Learning Driven Mobile Traffic Decomposition\label{chap:decompose}}

In this chapter, we introduce an original mobile traffic decomposition (MTD) technique, which breaks down the aggregated mobile traffic consumption into service-wise demand, to support the resource allocation for network slicing \cite{li2018delmu}. Strong service differentiation needs are already emerging in current deployments, where, \eg live video streaming must coexist with on-line gaming or shared cloud services. An important instrument that 5G mobile communication infrastructures will leverage to answer such diverse necessities is a much-improved flexibility in the management of resources. 
This will be realized via novel solutions for the virtualization of network functions at both the edge and core of the network, including dynamic spectrum allocation~\cite{bogucka2015dynamic}, baseband processing~\cite{nikaein}, scheduling~\cite{anand2017joint}, or task containerization~\cite{taleb2016}. 


Network operators remain in charge of performing management and orchestration (MANO) of resources dedicated to each slice, through platforms like ONOS~\cite{onos} or OSM~\cite{osm}. A critical operation in MANO is the anticipatory provisioning of isolated capacity (\eg spectrum, computation, storage, or transport) to each slice.
This requires knowledge of the total demand generated by each mobile service over horizons of minutes, which is the common resource reallocation periodicity supported by state-of-the-art Virtual Infrastructure Manager (VIM) and Network Function Virtualization (NFV) architectures~\cite{gil-herrera16,sciancalepore17}.

Unfortunately, mobile operators do not have direct visibility of service-level traffic. The current common practices for extracting such information combine two steps, as follows:
\noindent($i$) \textit{Deep Packet Inspection} (DPI) allows collecting IP traffic header metadata, \eg by sniffing on the GPRS Tunneling Protocol user plane (GTP-U) via probes tapping into the interfaces of the Packet Data Network Gateway~\mbox{(PGW)} \cite{shafiq15,marquez17};
\noindent($ii$) \textit{Flow-level classification} then leverages such metadata to identify the service generating each flow. Nowadays, classifiers mostly rely on cleartext hostnames in DNS queries, or revealing fields in the TLS handshake, like the Server Name Indication (SNI), due to the increasing adoption of traffic encryption~\cite{trevisan18}.

{Yet, running DPI at line rate and at scale is computationally expensive, while the surge in mobile data traffic and rapid growth of transport link speeds beyond the Terabit-per-second barrier exacerbate the problem. Further, while on-line classification of streaming data is currently an open research challenge, recent proposals for DNS over TLS~\cite{dot} or encrypted SNI in TLS 1.3~\cite{esni} will make classifiers based on DNS and TLS ineffective, and force the adoption of even more complex fingerprinting techniques~\cite{rezaei19}. Overall, there is a real risk that current trends in mobile data usage, network architectures, and security will render scalable on-line flow-level classification a much more tangled and resource-intensive problem than it already is today.}

\begin{figure}[t]
\centering
\includegraphics[width=\columnwidth]{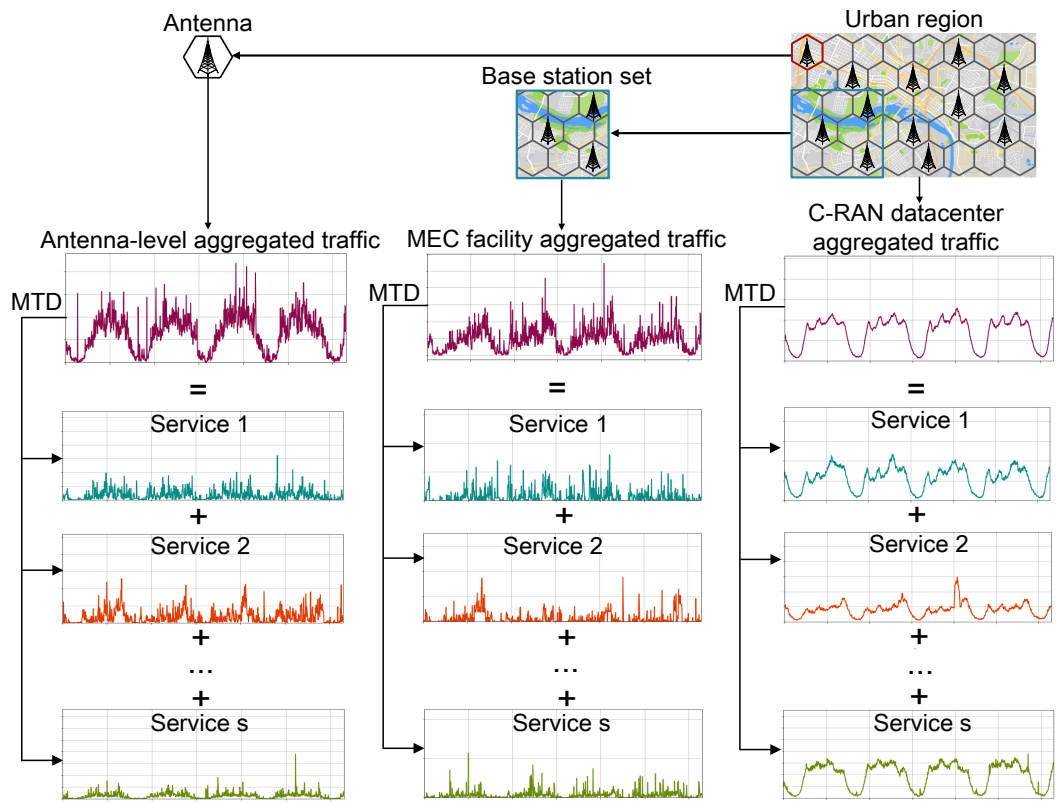}
\caption{Example of MTD at the level of individual antenna sector (left), set of base stations controlled by a single Mobile Edge Computing (MEC) facility (center), and set of Distributed Units (DU) associated with the same Cloud Radio Access Network (C-RAN) datacenter (right). At each level, the time series of aggregate mobile data traffic (top) is the sum of the time-varying demands associated to many different mobile services. The MTD problem consists of retrieving per-service time series from the aggregates. Note that the scales of the time series are different at each level.
\label{fig:mtd}}
\end{figure}

\textbf{Contributions.}
We propose an alternative approach to demand estimation for sliced network MANO. 
We target the inference of \textit{service-level} demands that are required for capacity provisioning to individual slices.
Our proposal builds on \emph{mobile traffic decomposition} (MTD), \ie the process of breaking down aggregate traffic time series into separate time series for each individual service. We illustrate 
the MTD concept by the example in Fig.\,\ref{fig:mtd}.
Performing MTD is technically challenging, since ($i$)~the decomposition of a single signal into multiple service-wise time series may have a multitude of solutions and yields inherent ambiguity; ($ii$)~complex spatial and temporal correlations exist in mobile traffic, and capturing these in order to resolve the ambiguity is non trivial; and ($iii$)~traditional decomposition techniques, {including factorial hidden Markov models~\cite{zhong2014signal} or neural networks for decomposition considered in other domains~\cite{zhang2018sequence},} are unsuitable for MTD, as they only work on single input time series, whereas in our case \emph{multiple input time series} are concurrently generated at different geographical locations.

To tackle these challenges and achieve effective and scalable MTD, we design a dedicated deep learning framework. We rely on deep learning due to its demonstrated effectiveness in discovering knowledge from time series under spatial correlations~\cite{zhang2018deep}, and in operating on large-scale mobile traffic in real time~\cite{zhang2018long}. This leads to the following contributions:

\begin{enumerate}
    \item We introduce the concept of mobile network traffic decomposition (MTD), and show that it can be a low-cost yet effective solution for the real-time inference of the demands generated by individual mobile services.
    \item We propose \name, a framework that solves the MTD problem effectively via a new class of deformable convolutional neural networks, named 3D-DefCNNs, which enhance traditional CNN
    structures~\cite{dai17dcn} and can extract complex spatio-temporal correlations from aggregate traffic data.
    \item We experiment with metropolitan-scale measurement data collected in an operational networks in a major European city; we show that \name makes inferences about per-service traffic consumption with errors below 1.2\%.
    \item We give a quantitative assessment of the practical feasibility of employing \name in operational networks, showing that the additional cost incurred by decomposition is small, yet the benefits are on par with having perfect knowledge of per-service demands.
\end{enumerate}{}

\begin{figure}[t]
\centering
\includegraphics[width=\columnwidth]{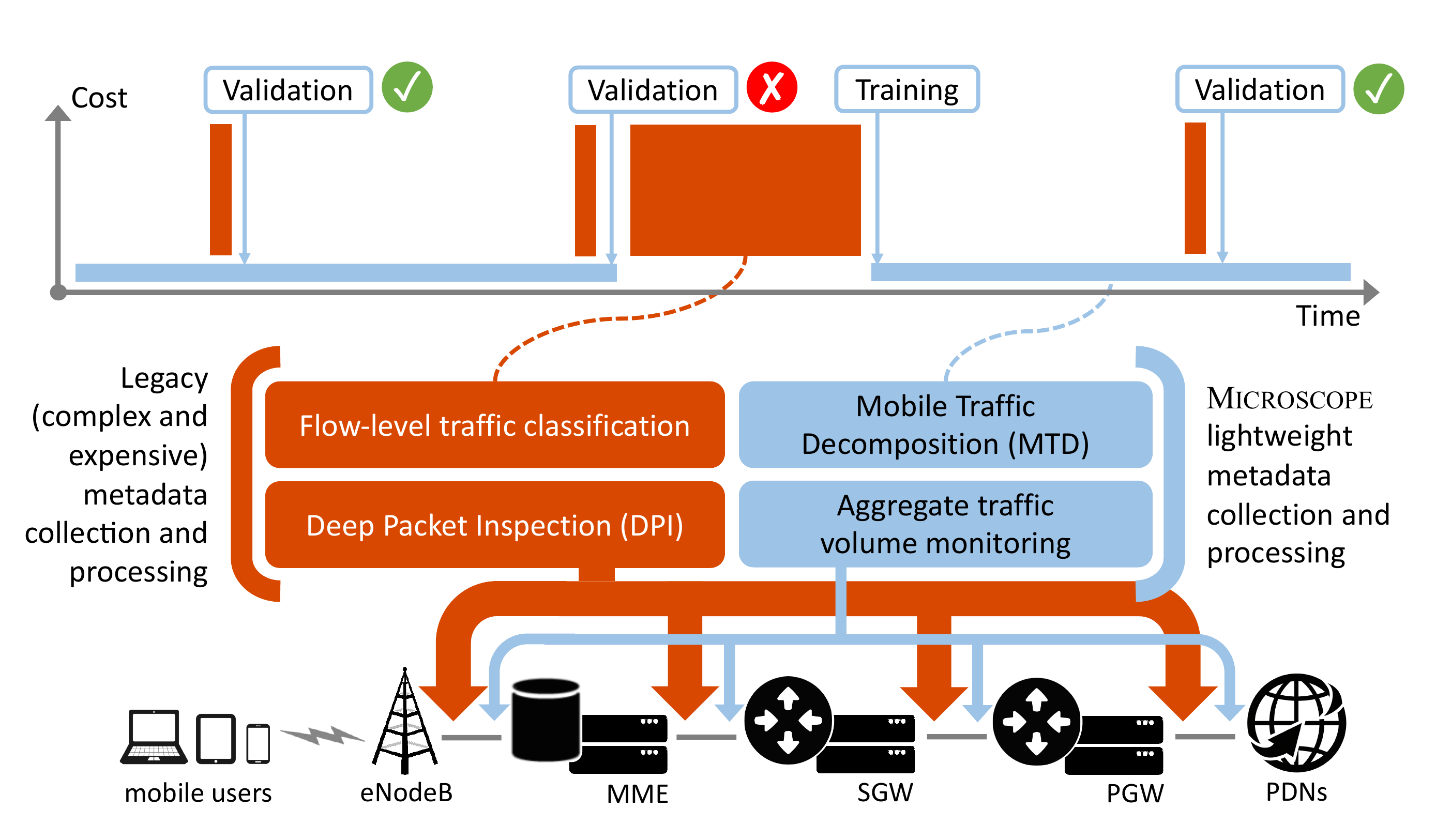}
\caption{{Hybrid service-level demand estimation framework for NSaaS capacity provisioning. The framework combines DPI and MTD, so that \name lightweight monitoring and decomposition are used most of the time, and costly flow-level classification approaches are only triggered when strictly necessary.}
\label{fig:system}}
\end{figure}

Ultimately, our results indicate that the MTD approach implemented by \name can drive service-level capacity provisioning in emerging NSaaS network management models.
\name can then be combined with legacy techniques based on DPI and flow-level classification into an efficient hybrid framework, as illustrated in Fig.\,\ref{fig:system}. By only requiring monitoring of the total traffic volume, \name offers a lightweight solution for per-service demand estimation that can be run persistently at low cost. The mechanics and popularity of mobile applications are however evolving continuously, which calls for periodic validation of the MTD solution: expensive DPI-based classification is thus still performed for short time periods to produce the necessary ground-truth data. Extensive flow-level classification is only required when the validation fails and re-training is necessary.
Nonetheless, major changes in the operation and adoption of mobile service usage occur over long timescales (\eg months), which lets us speculate that the latter would be a relatively infrequent event. Although we do not have sufficient data to demonstrate that this is the case, our results are promising in this sense, as they show that the MTD model yields excellent performance two weeks after training.

\section{Mobile Traffic Decomposition}
\label{sec:mtd}

We formalize the MTD problem in Sec.\,\ref{sub:problem}, and then outline the \name framework we propose to solve it in Sec.\,\ref{sub:microscope}.

\begin{figure*}[tb]
\begin{center}
\includegraphics[trim=10 20 15 30, clip, width=1.03\columnwidth]{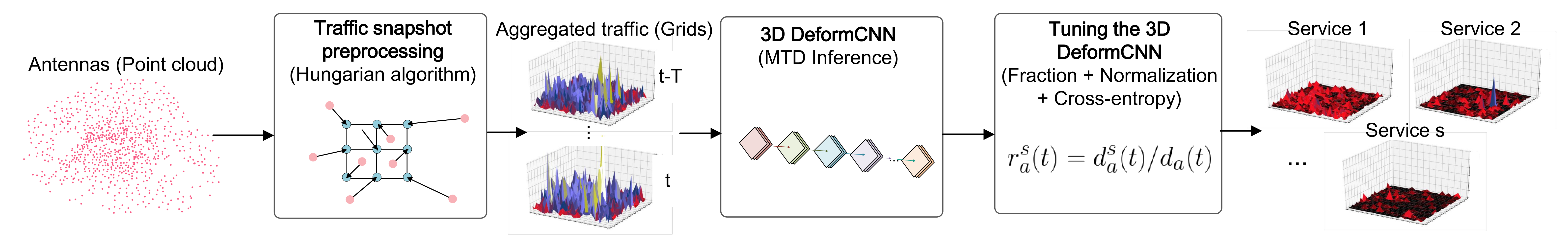}
\end{center}
\caption{\name framework for MTD. Three pipelined components perform
preprocessing of mobile traffic snapshots (left), actual traffic decomposition via a dedicated \mbox{3D-DefCNNs} (center), and fine-tuning (right).
\label{fig:microscope}}
\end{figure*}

\subsection{Problem Formulation}
\label{sub:problem}

Consider a geographical region where mobile network coverage is provided by a set $\mathcal{A}$ of antennas,
\footnote{{An eNodeB may serve users present in one or multiple sectors, each covered by a co-located antenna with a different azimuth. We refer to a single radio front-end unit as one `antenna' hereafter.}}
which accommodate traffic generated by a set $\mathcal{S}$ of mobile services. We denote by $d_a^s(t)$ the traffic demand (expressed in Mbps) accommodated by antenna $a\in\mathcal{A}$ for a specific service $s\in\mathcal{S}$ at time $t$. The sum of demands over all services gives the aggregate demand at antenna $a$ and time $t$, $d_a(t)=\sum_{s\in\mathcal{S}} d_a^s(t)$.
A mobile network traffic \textit{snapshot} is the set of demands recorded at all antennas in the target region at a specific time. This applies both to individual services, leading to a \textit{service snapshot} $D^s(t) = \left\{d_a^s(t) | a\in\mathcal{A}\right\}$, and to traffic aggregates over all services, obtaining an \textit{aggregate snapshot} $D(t) = \left\{d_a(t) | a\in\mathcal{A}\right\}$.

The MTD problem is formally defined as that of inferring the service snapshots $D^s(t)$ of all services $s\in\mathcal{S}$ at current time $t$, by only knowing the aggregate snapshots up to $T$ previous time instants, \ie $\left\{D(t\hspace*{-2pt}-\hspace*{-2pt}T\hspace*{-2pt}+\hspace*{-2pt}1),\dots,D(t)\right\}$. If we define as $\mathcal{D}^s(t) = \left\{D^s(t) | s\in\mathcal{S}\right\}$ the set of current service snapshots, the solution to the MTD problem is 
\begin{equation}
    \widetilde{\mathcal{D}}^s(t) := \mathop{\arg\max}\limits_{\mathcal{D}^s(t)} \; p\big(\:\mathcal{D}^s(t) \: | \: \left\{D(t\hspace*{-2pt}-\hspace*{-2pt}T\hspace*{-2pt}+\hspace*{-2pt}1),\dots,D(t)\right\}\big),
\label{eq:mtd}
\end{equation}
where $\widetilde{\mathcal{D}}^s(t)$ denotes the estimated current traffic demands disaggregated over the service set $\mathcal{S}$ for all antennas in $\mathcal{A}$, and $p(\cdot)$ is the probability of the argument. The MTD results are restricted by two obvious constraints in the system \ie

\begin{flalign}
& \tilde{d}_a^s(t) \geq 0, \hspace*{42pt} \forall a\in\mathcal{A}, \forall s\in\mathcal{S}, \forall t, \label{eq:const_pos} \\
& \sum_{s\in\mathcal{S}} \tilde{d}_a^s(t) = d_a(t), \hspace*{10pt} \forall s\in\mathcal{S}, \forall t.\label{eq:const_sum}
\end{flalign}
The expression in (\ref{eq:const_pos}) enforces that all estimated traffic demands are positive, while (\ref{eq:const_sum}) implies that the sum of the per-service traffic must be equal to the aggregate traffic. The constraints hold for all antennas and at all time instants.

We stress that \rv{MTD is a supervised learning problem,} but fundamentally different from prediction, as it assumes knowledge of the aggregate measurement data at the current time instant $t$, and aims at inferring information relative to that same time instant.
In fact, MTD is a one-to-many problem that seeks to decompose one aggregate traffic measurement into the underlying $|\mathcal{S}|$ per-service snapshots: $\mathcal{D}^s(t)$ includes $|\mathcal{S}|$ snapshots, each featuring the same cardinality as $D(t)$.
Iteratively solving this problem over time ultimately leads to the reconstruction of per-service demand time series at each antenna from the aggregate traffic, as originally illustrated in Fig.\,\ref{fig:mtd}.

\subsection{\name in a Nutshell}
\label{sub:microscope}

\name is a novel machine learning framework that is specifically designed for MTD. Here, we provide an overview of the framework, discussing the functionality and integration of the modules it comprises; the following sections give full details about the implementation of each component.

As shown in Fig.\,\ref{fig:microscope}, there are three main elements in \name. The first is a \textit{traffic snapshot preprocessing} block, which receives the current aggregate mobile network traffic measurement data $D(t)$ and converts them into a format that is suitable for the following analysis. Specifically, this component limits the spatial distortion of antenna locations as they are fed to the neural network, by solving an opportune association problem. 
This preprocessing is critical to generalizing the framework, since it allows \name to accommodate any antenna deployment layout with minimum loss of geographical information. Details are in Sec.\,\ref{sec:input}.

The second component implements an original \textit{3D deformable convolutional neural network} (\mbox{3D-DefCNN}), capable of learning abstract spatio-temporal cross-service correlations that are unique to mobile traffic~\cite{wang2015,furno:2017} and which are needed to solve the MTD problem. The network takes as input preprocessed measurement data corresponding to the aggregate traffic recorded during most recent $T$ time instants. Its internal structure also allows to compensate for remaining spatial distortions of antenna locations, and dynamically adapts the weights of contributions from past aggregate snapshots. Details are in Sec.\,\ref{sec:defcnn}--\ref{sec:3dcnn-all}.

The third component is concerned with \textit{fine tuning the neural network} via a suitable output normalization and adoption of a loss function that fits well MTD. Specifically, we enforce that the inferred mobile network traffics of individual services satisfy the constraint that their sum matches the total traffic consumption, for each antenna $a\in\mathcal{A}$. We further propose to train the neural network by minimizing a cross-entropy function that yields higher accuracy when solving the MTD problem.
The final outputs are the estimated service snapshots $\widetilde{\mathcal{D}}^s(t)$ as per (\ref{eq:mtd}). Details are in Sec.\,\ref{sec:tuning}.

\section{Mobile Network Traffic Dataset and Preprocessing}
\label{sec:data}

\rv{We conduct our experiments on real-world 3G/4G mobile network traffic collected by a major operator in a large European metropolitan area during 85 consecutive days, as we discuss in Sec.~\ref{sec:dataset_paris}. Note that we only test our proposal on one city, that which has a larger set of antennas and relatively richer traffic behaviors. Evaluations on a different city will be a part of our future work.} Each mobile network traffic snapshot consists of the demand accumulated over a 5-minute time interval at 792 different antenna sectors,\footnote{As we find measurements to be sometimes incomplete, we perform MTD over antennas where active traffic flows exist at least 90\% of the time. This filters out 88 decommissioned or sporadically enabled antennas. The demand measured at each antenna accounts for both uplink and downlink directions.} for several tens of services separately.

The neural network is trained on data collected by the operator in the first 51 days (60\%), validated with the measurements gathered in the following 17 days (20\%), and tested on the traffic observed during the last 17 days of the dataset (20\%). In these conditions, the training process converges at the tenth epoch, taking around 48 hours in total. Performing MTD inference requires less than 1 second per instance.

\subsection{Data Collection and Overview}


\begin{figure}[tb]
\begin{center}
\includegraphics[width=\columnwidth]{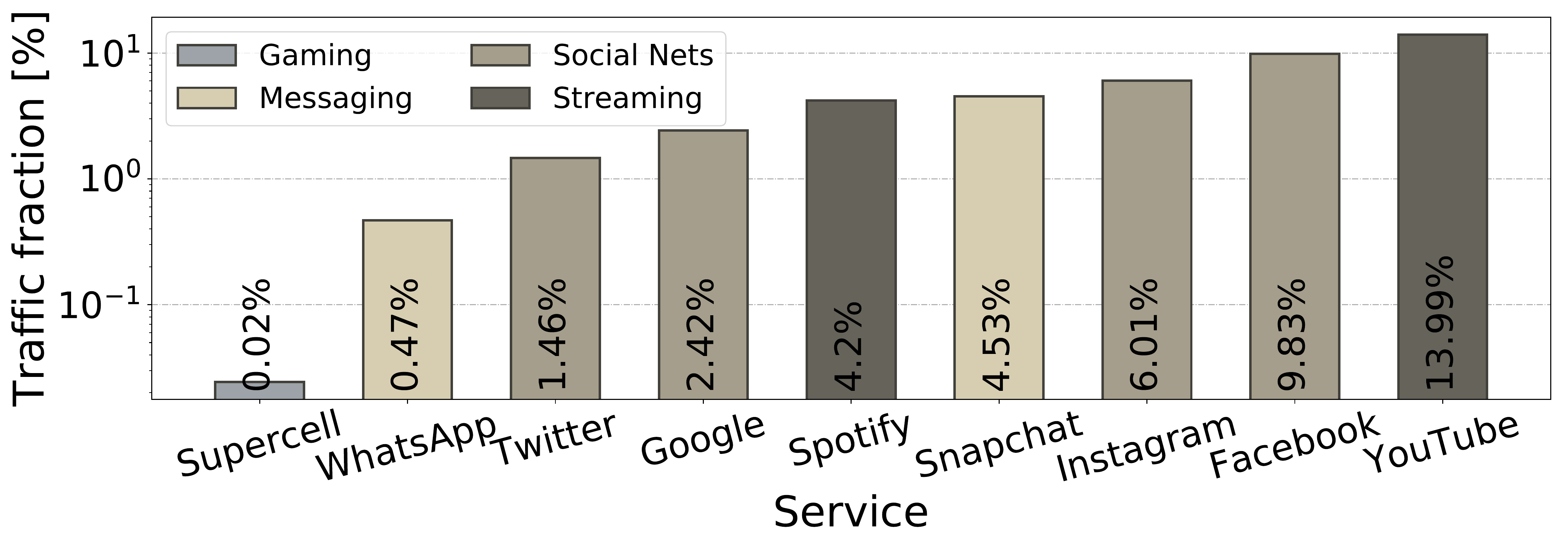}
\end{center}
\caption{\label{fig:mean_app} Overview of the traffic generated by services in the set $\mathcal{S}$ considered in our study.}
\label{fig:traffic-stats}
\end{figure}

The set of services $\mathcal{S}$ considered in our analysis comprises tens of different services. However, we confine the evaluation of \name to a representative set of services that encompass a variety of application types and are reasonable candidates to be allocated dedicated network slices. Such services are \textit{heavy hitters}, \ie generate sizeable amounts of network traffic, and have Quality of Service (QoS) requirements. %
Then, $\mathcal{S}$ includes mobile games (Clash Royale and Clash of Clans, grouped under the Supercell label), messaging apps (Snapchat and WhatsApp), social media (Facebook, Twitter, Instagram), video (YouTube) and audio (Spotify) streaming platforms, as well as Google services. Fig.~\ref{fig:traffic-stats} illustrates the fraction of total traffic induced by each service: owing to the well-known Zipfian distribution of the demand across services~\cite{shafiq11,marquez17}, the nine selected services are responsible for more than 42\% of the total mobile data traffic in the target region. Interestingly, the diverse nature of applications in $\mathcal{S}$ leads to dissimilar temporal dynamics of the corresponding demands~\cite{shafiq2012,marquez17}, allowing us to test our MTD approach in a realistic scenario where heterogeneous services coexist.

\subsection{Traffic Snapshots Preprocessing}
\label{sec:input}

\begin{figure*}[tb]
\begin{center}
\includegraphics[width=\columnwidth]{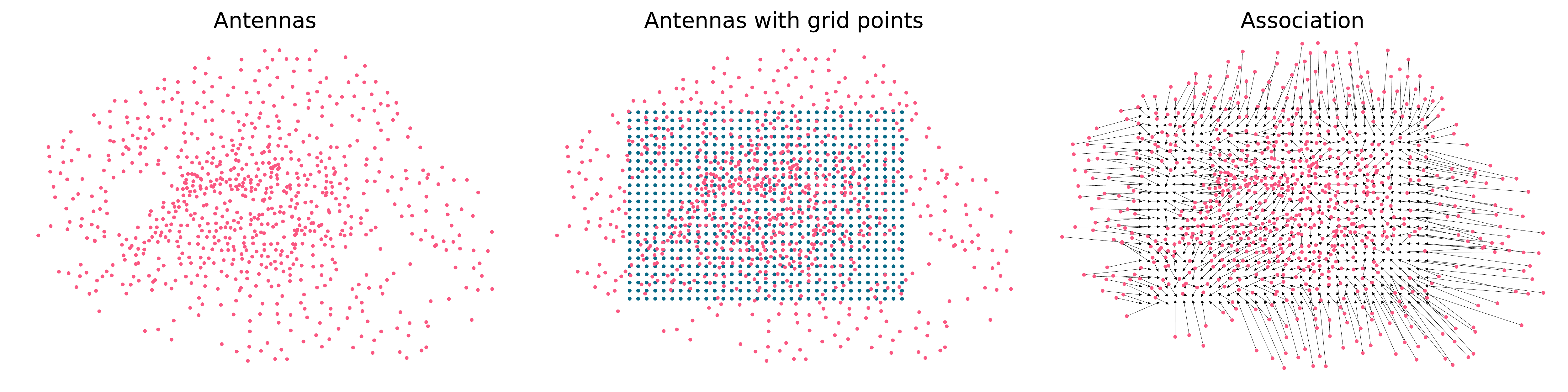}
\end{center}
\caption{\label{fig:antenna} Traffic snapshots preprocessing for the metropolitan deployment considered. Selected antennas positions (left), antennas and constructed grid points (center), and the result of antenna-to-point mapping, where each arrow denotes a transition from the original antenna position to the assigned point on the grid (right).
}
\end{figure*}

The core of \name is a specialized type of convolutional neural network (CNN), and thus requires an input in matricial form. In our case, an input matrix describes an aggregate snapshot $D(t)$ of mobile network traffic, as explained in Sec.\,\ref{sub:problem}. The input format creates the problem of mapping antennas to matrix elements, \ie points on a regular grid. As a matter of fact, and as later detailed in Sec.\,\ref{sec:nn}, the adjacency of elements in the matrix has an impact on the operation of the CNN, and should reflect in the best way possible the geographical proximity of the antennas.  
However, the deployment of the antennas can be highly irregular: as an example, Fig.\,\ref{fig:antenna} (left) shows the uneven antenna distribution in the metropolitan region we consider for our experiments. While this is expected, as the antenna positions are driven by the area topography and varied density of subscriber~\mbox{presence}, it also makes the mapping problem not trivial, since a non-uniform distribution must be transformed into a regular grid.

One solution to reconciling real-world antenna deployments with the input requirement of CNNs would be that of spatial supersampling~\cite{barlacchi15}. This involves approximating the coverage area of each antenna (\eg via a Voronoi tessellation), assuming the traffic recorded at each antenna to be uniformly distributed within the associated coverage area, and superposing a (possibly dense) grid to the resulting continuous traffic map. Then, each matrix element is easily filled with the traffic in the underlying map. However, we discard this option, since ($i$) it introduces strong unrealistic assumptions about the coverage and spatial distribution of users, and ($ii$) in the case of dense grids, it artificially increases the size of the matrices and the computational cost of the CNN.

Instead, we solve the mapping problem by constructing a regular grid that has the same number of points as the number of antennas and performing a one-to-one antenna-to-point association. Below we discuss the design of the regular grid and the solution to the mapping problem.

\subsubsection{Regular Grid Design}

To preserve spatial correlations within mobile network traffic, a properly designed grid should ($i$) overlap with the coverage of the antennas as much as possible, and ($ii$) only introduce small spatial displacements after the mapping procedure. Then, given the target set $\mathcal{A}$ of antennas, our grid design follows the logic below.
\begin{enumerate}[leftmargin=1.5em, topsep=0.25em]
    \item We project the locations of all antennas to an Euclidean space, and determine the extreme values on $x$ and $y$ axes, \ie $x_{\min}, x_{\max}, y_{\min}, y_{\max}$, obtainig a rectangular representation of the coverage region with vertices $(x_{\min}, y_{\min})$, $(x_{\min}, y_{\max})$, $(x_{\max}, y_{\min})$, $(x_{\max}, y_{\max})$;
    \item We compute the aspect ratio $r_a$ of such a rectangular area, as 
    $\rho = (x_{\max}-x_{\min})/(y_{\max}-y_{\min})$;
    \item We dimension the grid so that it reflects such an aspect ratio, with
    $n_r=\sqrt{|\mathcal{A}|/{\rho}}$ rows and $n_c= n_r/\rho$ columns;
    \item We assign geographical coordinates to the regular grid, by first superposing the grid to the matching-ratio antenna positions, and then scaling the inter-point distance by a factor $\kappa$ to account for the denser antenna deployment typical of central urban areas. 
\end{enumerate}
\rv{This is an original grid design strategy we propose for the MTD purpose.} The center plot of Fig.\,\ref{fig:antenna} portrays the regular grid points obtained by means of the technique above in the case of the antenna deployment shown in the left plot of the same figure. Here, we employed $\kappa=0.25$, and obtained a $33\times 24$ grid. 

\subsubsection{Antenna-to-point Mapping}
\label{sec:hungarian}

We aim to minimize the overall displacement when performing the one-to-one association between antennas and grid points. We define the displacement $c_{a,p}$ as the Euclidean distance between the geographical positions of antenna $a$ and grid point $p$. We construct a cost matrix $C := \{c_{a,p}\}_{|\mathcal{A}|\times |\mathcal{A}|}$ to represent the distances between antennas and grid points. We further define a binary matrix $X := \{x_{a,p}\}_{|\mathcal{A}|\times |\mathcal{A}|}$, where $x_{a,p} = 1$, if and only if the antenna $a$ is assigned to the grid point $p$. Hence, we formulate the association problem as:

\begin{flalign}
\underset{x}{\text{minimize}}
& \hspace*{10pt}\sum_{a\in\mathcal{A}} \sum_{p=1}^{|\mathcal{A}|} c_{a,p} \cdot x_{a,p}, \label{eq:opt_obj}\\
\text{subject to}
& \hspace*{10pt}\sum_{a\in\mathcal{A}} x_{a,p} = 1, \; \forall p \in \left[1,|\mathcal{A}|\right], \label{eq:opt_c1}\\
& \hspace*{12pt}\sum_{p=1}^{|\mathcal{A}|} \hspace*{1pt} x_{a,p} = 1, \; \forall a\in\mathcal{A},\label{eq:opt_c2}
\end{flalign}
where constraints enforce that one antenna will be assigned to only one grid point, and vice versa.

The expressions in (\ref{eq:opt_obj})--(\ref{eq:opt_c2}) define an assignment problem that is efficiently solved via the Hungarian algorithm~\cite{kuhn1955hungarian}, which has four major steps. Namely:
\begin{enumerate}
\setlength\itemindent{2em}
    \item[\textbf{Step 1:}] For each row of the cost matrix $C$, subtract the lowest value $c_i^{\min}$ from each element in that row, \ie 
        \begin{equation*}
        c_{i,j} = c_{i,j} - c_i^{\min}, \; \forall i \in \overline{1, V}, \forall j \in \overline{1, V}.
        \end{equation*}
    \item[\textbf{Step 2:}] For each column of the cost matrix $C$, subtract the lowest value $c_j^{\min}$ from each element in that column, \ie 
        \[
        c_{i,j} = c_{i,j} - c_j^{\min}, \; \forall j \in \overline{1, V}, \forall i \in \overline{1, V}.
        \]
    \item[\textbf{Step 3:}] Find the minimum number of lines $l_{\min}$ required to cover the zero elements in matrix $C$, both vertically and horizontally. An optimum (the zero-element location) is reached and the algorithm terminates when $l_{\min} = V$; otherwise, continue with the next step, if $l_{\min} < V$.
    \item[\textbf{Step 4:}] Find the smallest value $c_{\min}$ that is not covered by a line at Step 3. Subtract $c_{\min}$ from all uncovered elements, and add $c_{\min}$ to all elements that are covered twice. Repeat Steps 3--4 until an optimum is found.
\end{enumerate}

The algorithm has a polynomial complexity $O(|\mathcal{A}|^3)$, hence runs efficiently in practical cases.
In the case of the antenna deployment we consider in our experiments, in Fig.\,\ref{fig:antenna}, it returns the mapping in the right plot. We observe that only antennas on the outskirts of the urban area are subject to significant spatial shifts, whereas the movement is limited in all other cases; this results in an average displacement of $1.22$~km overall and geographical correlations are maintained in the regular grid; significant displacements are restricted to the borders, and most points incur a shift below $500$ m.

\subsection{Displacement Assessment}

Although minimized by the approach presented above, some spatial displacement is unavoidable, due to cell coverage heterogeneity. We further investigate the level of displacement induced by our proposed solution, by considering the representative case of our reference large metropolis scenario.

\begin{figure}[tb]
\begin{center}
\includegraphics[width=\columnwidth]{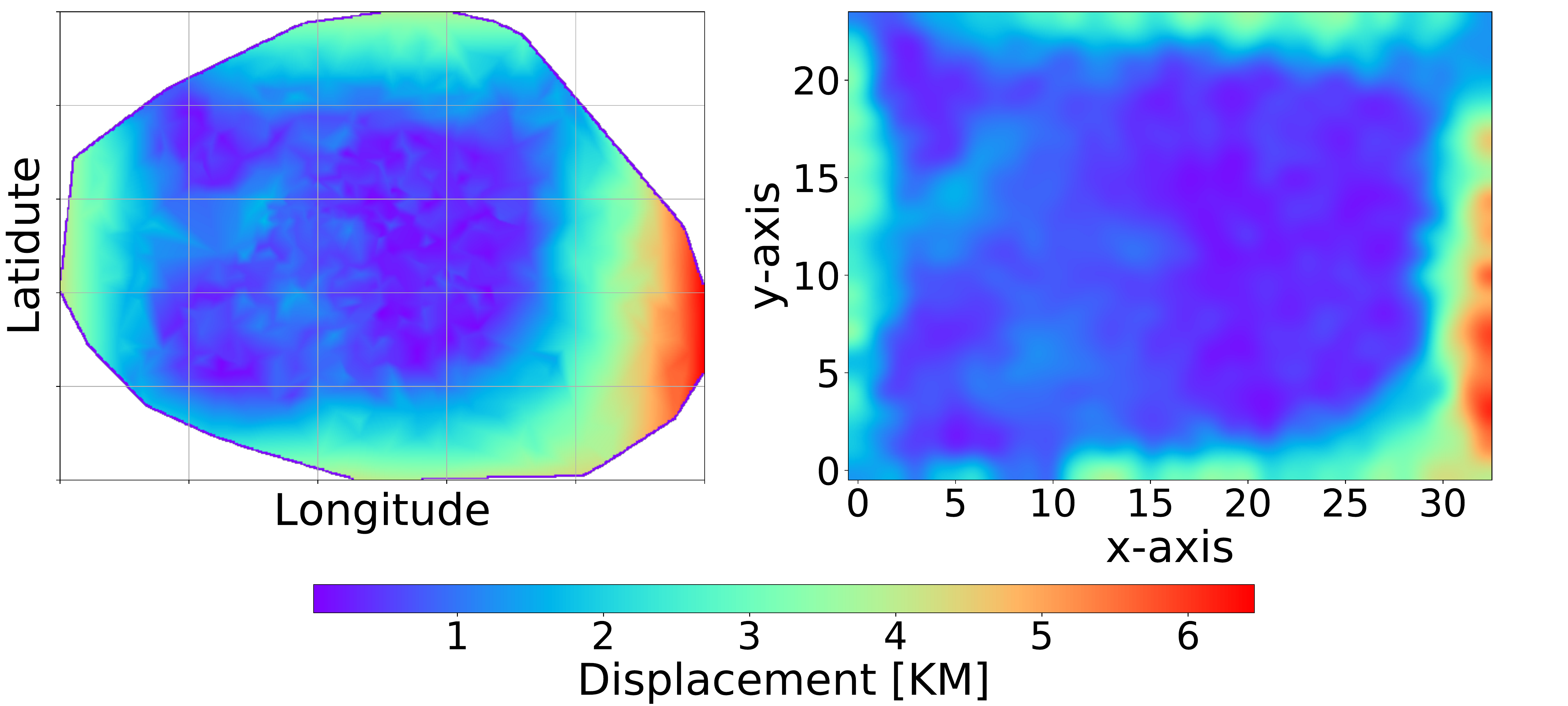}
\end{center}
\caption{\label{fig:displacement} Heatmaps of the spatial displacement for grid mapping based on the original antenna locations (left), and projection on the regular grid space (right).}
\end{figure}

Fig.\,\ref{fig:displacement} shows heatmaps of the displacement incurred by antennas in the target region upon mapping to the regular grid. In the left plot, we observe that only antennas in the urban area outskirts are forced to significant spatial shifts, whereas the movement is limited in all other cases; this results in an average displacement of $1.22$ km overall. The right plot makes it even more clear that geographical correlations are maintained in the regular grid, where significant displacement are restricted to the borders, and most points incur a shift below $500$ m.
Hence, in our reference scenario, the proposed preprocessing of traffic snapshots generates an input matrix for the neural network that preserves substantial spatial relationships in the mobile network demands.

\section{3D-Deformable convolutional neural network}
\label{sec:nn}

The core of \name is a new deep learning architecture, named \emph{3D deformable convolutional neural network}, or \mbox{3D-DefCNN}. The \mbox{3D-DefCNN} is specifically designed for MTD: it compensates for the spatial displacement in network traffic snapshots, discovers spatio-temporal correlations in aggregate traffic, and exploits them for decomposition. 

\subsection{Deformable CNN}
\label{sec:defcnn}
Our 3D-DefCNN design is an enhancement of the deformable convolutional neural network (\mbox{DefCNN})~\cite{dai17dcn} originally proposed for computer vision application, such as image classification~\cite{zhu2018deformable}, object detection and semantic segmentation~\cite{qi2017deformable}. 
Whilst classic convolutional neural networks (CNNs) apply fixed geometric transformations to a 2D space, DefCNN architectures perform deformable convolution operations over the same type of input. This allows flexible transformations that can compensate for distortions in the 2D input space.

As discussed in Sec.\,\ref{sec:input}, we adopt a traffic snapshot preprocessing that mitigates but cannot completely remove the displacement of antenna locations into a regular grid. Therefore, aggregate snapshots recorded in the 2D geographical space are distorted, with a risk that spatial correlations in the original data are misrepresented in the input matrix.
DefCNN can reduce such a risk, by enabling convolutional filters to access any element of the input matrix. These connections are dynamically tailored to the input, and can be learned jointly with the other synapses via gradient descent.

\begin{figure*}[htb]
\begin{center}
\includegraphics[width=1.03\columnwidth]{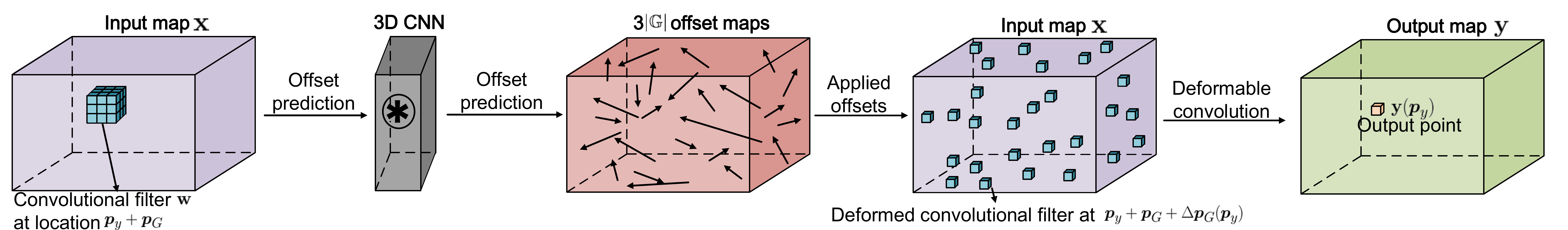}
\end{center}
\caption{\label{fig:deform} Graphical illustration of 3D-Deformable convolutional operation. 
}
\end{figure*}
The approach is equivalent to letting the neural network re-organize the 2D spatial structure of the input data. In our context, this allows identifying and exploiting any geographical correlation inherent to the mobile network traffic that may be lost during the snapshot preprocessing phase.

\subsection{3D-Deformable Convolution}
\label{sub:3D Def-conv}

Legacy DefCNNs perform deformable convolution over 2D input matrices. In MTD, the data fed to the network includes time, thus it is three-dimensional. Specifically, the input consists of the aggregate snapshot set $\left\{D(t\hspace*{-2pt}-\hspace*{-2pt}T\hspace*{-2pt}+\hspace*{-2pt}1),\dots,D(t)\right\}$ (Sec.\,\ref{sub:problem}), which is preprocessed into $T$ subsequent 2D matrices (Sec.\,\ref{sec:hungarian}).
Hence, we extend the deformable convolution operation to the temporal dimension. In doing so, we account for the fact that not all last $T$ input snapshots have the same relevance to MTD at the current time instant, and weight in an adaptive manner the different 2D input matrices.

To achieve our objective, we combine the DefCNN model with 3D convolution, a technique previously used for action recognition~\cite{ji20133d}, which operates over both time and a  bidimensional space. Essentially, 3D convolution performs summation over an input map $\mathbf{x}$ weighted by a filter matrix $\mathbf{w}$, which results
in an output map $\mathbf{y}$.\footnote{In action recognition tasks, $\mathbf{x}$ and $\mathbf{y}$ are four-dimensional tensors: the first three are the spatio-temporal dimensions; the fourth is the RGB channel dimension. This allows defining dedicated filters on each channel. In our case, we employ a single filter shared across all channels of the input. Our neural network will still produce multiple \emph{channels} throughout the hidden layers, which do not have a direct physical meaning, but rather provide intermediate outputs that aid extracting abstract features. The final layer making predictions has however one output channel for each mobile service.}
Each convolutional filter $\mathbf{w}$ has a receptive field $\mathbb{G}$, which defines the spatio-temporal extent of connectivity between different locations in the input. As an example, the receptive field of a $3\times 3\times 3$ convolutional filter can be defined as $\mathbb{G} = \{(-1, -1, -1), (-1, -1, 0),$ $\ldots, (1, 1, 0), (1, 1, 1)\}$. For each location $\boldsymbol{p}_{y}$ of the output $\mathbf{y}$, the 3D convolution performs the following calculation:
\begin{equation}\label{eq:conv2}
    \mathbf{y}(\boldsymbol{p}_y) = \sum_{\boldsymbol{p}_G \in \mathbb{G}} \mathbf{w}(\boldsymbol{p}_G)\cdot \mathbf{x}(\boldsymbol{p}_y + \boldsymbol{p}_G),
\end{equation}
where $\boldsymbol{p}_G$ denotes all positions in $\mathbb{G}$.
For instance, if $\boldsymbol{p}_y = (1, 2, 3)$ and $\boldsymbol{p}_G = (-1, 1, 0)$, then $\mathbf{x}(\boldsymbol{p}_y + \boldsymbol{p}_G$) is the value of $\mathbf{x}$ at index $(0, 3, 3)$.

Our proposed 3D-DefCNN modifies the operations above by introducing a set of additional offsets $\{\Delta \boldsymbol{p}_G | G = 1,\dots,|\mathbb{G}| \}$. The actual $\Delta \boldsymbol{p}_G$ is selected from the set by an additional 3D CNN, which takes $\mathbf{x}$ as input. Our model then extends (\ref{eq:conv2}) as:
\begin{equation}\label{eq:dfconv}
    \mathbf{y}(\boldsymbol{p}_y) = \sum_{\boldsymbol{p}_G \in \mathbb{G}} \mathbf{w}(\boldsymbol{p}_G)\cdot \mathbf{x}(\boldsymbol{p}_y + \boldsymbol{p}_G + \Delta \boldsymbol{p}_G(\boldsymbol{p}_y)).
\end{equation}
Note, for each location $\boldsymbol{p}_y$, we apply different $\Delta \boldsymbol{p}_G(\boldsymbol{p}_y)$. This requires the offset map to have the same shape as $\mathbf{x}$. Overall, the 3D CNN will output $3|\mathbb{G}|$ offset maps, each corresponding to the offset applied to location $\boldsymbol{p}_G$ in one \mbox{dimension}.\footnote{We need $|\mathbb{G}|$ offsets for each dimension, thus  we construct $3|\mathbb{G}|$ offsets in the 3D architecture, \ie $2|\mathbb{G}|$ for spatial and $|\mathbb{G}|$ for temporal dimensions.}

We illustrate the principle of 3D deformable convolution in Fig.\,\ref{fig:deform}.
As mentioned above, we first apply a 3D-CNN structure (3 layers) onto the original convolutional filter (which is compact) to predict $3|\mathbb{G}|$ offset maps $\Delta \boldsymbol{p}_G(\boldsymbol{p}_y)$. These offset maps essentially seek to alter the position of the filter elements, in order to scan, at each step, locations in the input that are not necessarily adjacent. The offsets are learned and shared across the different input channels. Subsequently, we apply these offsets to the original convolutional filter, to construct a deformed convolutional filter. Finally, by performing convolution between input and deformed filter, we obtain the output at location $\boldsymbol{p}_y$ via (\ref{eq:dfconv}). Since 3D deformable convolution operations are fully differentiable, the offsets to be applied can be learned through standard back-propagation. As such, 3D-DefCNNs grant convolutional filters complete freedom to query any location in the input maps. This significantly improves the model flexibility and enables to adapt to any spatial displacement (spatial deformation) and diverse importance of historical data (temporal deformation).

\begin{figure}[htb]
\centering
\includegraphics[width=\columnwidth]{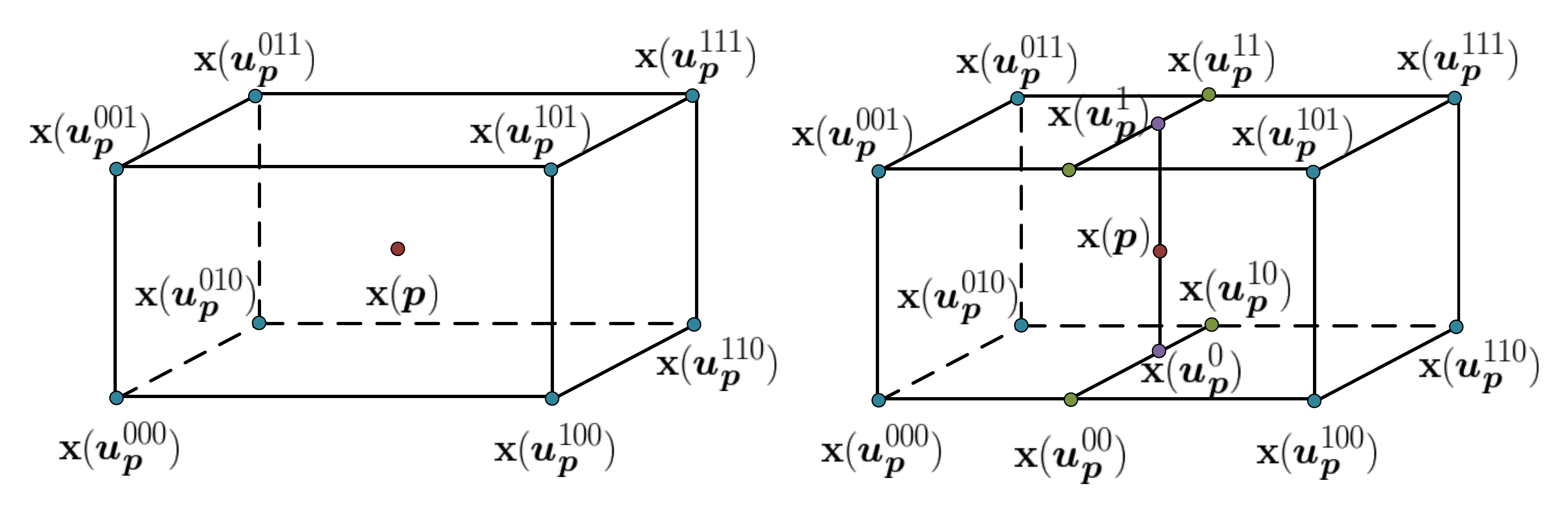}
\caption{Trilinear interpolation: compute 
$\{\mathbf{x}(\boldsymbol{u}_{\boldsymbol{p}}^{00}),$  
$\mathbf{x}(\boldsymbol{u}_{\boldsymbol{p}}^{01}), \mathbf{x}(\boldsymbol{u}_{\boldsymbol{p}}^{10}), \mathbf{x}(\boldsymbol{u}_{\boldsymbol{p}}^{11})\}$ via linear interpolation, then compute $\{\mathbf{x}(\boldsymbol{u}_{\boldsymbol{p}}^{0}), \mathbf{x}(\boldsymbol{u}_{\boldsymbol{p}}^{1})\}$, and finally obtain the value of $\mathbf{x}(\boldsymbol{p})$.}
\label{fig:tri}
\end{figure}

A complication introduced by equation (\ref{eq:dfconv}) is that the sample positions $\boldsymbol{p} = \boldsymbol{p}_y + \boldsymbol{p}_G + \Delta \boldsymbol{p}_G$ can be fractional.
Indeed, while $\boldsymbol{p}_G$ and $\boldsymbol{p}_y$ are indices of $\mathbb{G}$ and $\mathbf{y}$, and thus integer values, $\Delta \boldsymbol{p}_G$ may not be integer, as it is estimated by a CNN. 
We compute the deformed input in (\ref{eq:dfconv}) as:
\begin{equation}
	\mathbf{x}(\boldsymbol{p}_y + \boldsymbol{p}_G + \Delta \boldsymbol{p}_G) =
	\mathbf{x}(\boldsymbol{p}) =
	\sum_{\boldsymbol{u}_{\boldsymbol{p}} \in \boldsymbol{U}_{\boldsymbol{p}}} \textsl{Tri}(\boldsymbol{U}_{\boldsymbol{p}}, \boldsymbol{p})\cdot \mathbf{x}(\boldsymbol{u}_{\boldsymbol{p}}),
\end{equation}
where $\boldsymbol{U}_{\boldsymbol{p}} = \{\boldsymbol{u}_{\boldsymbol{p}}^{000}, \boldsymbol{u}_{\boldsymbol{p}}^{001},
\dots,\boldsymbol{u}_{\boldsymbol{p}}^{111}\}$ are the locations of the eight input samples around $\boldsymbol{p}$, and $\textsl{Tri}(\cdot)$ denotes the \mbox{trilinear} interpolation operator~\cite{rajon2003marching}, whose principle we show in~Fig.\,\ref{fig:tri}.

\begin{figure*}[t!]
\begin{center}
\includegraphics[width=1.03\columnwidth]{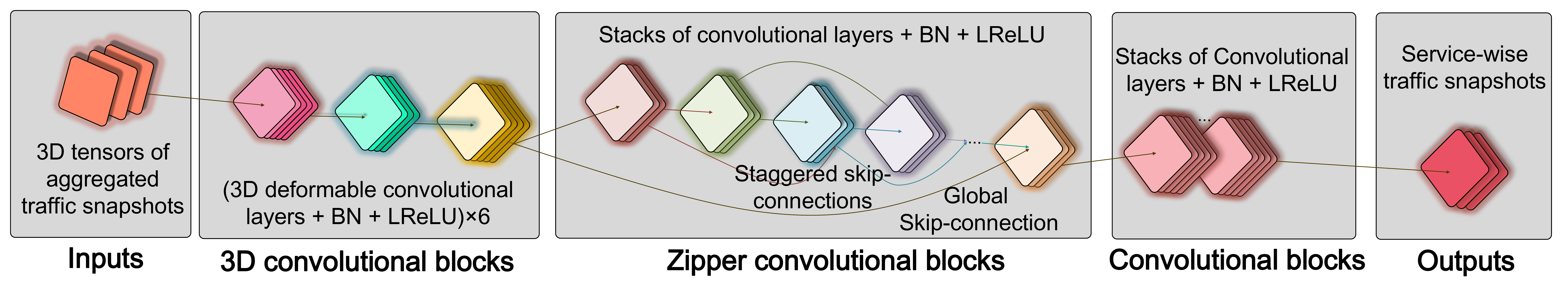}
\end{center}
\caption{\label{Fig:arc} The overall structure of our 3D-DefCNN, consisting of 3D-Deformable convolutional blocks (left), zipper convolutional blocks (middle), and standard 2D convolutional blocks (right).}
\end{figure*}

\subsection{Overall \mbox{3D-DefCNN} Structure}
\label{sec:3dcnn-all}
We integrate the 3D deformable convolutional operations presented above in the complete \mbox{3D-DefCNN} structure shown in Fig.\,\ref{Fig:arc}. Our design encompasses three major components: ($i$) 3D-Deformable convolutional blocks, ($ii$)~zipper convolutional blocks, and ($iii$) standard 2D convolutional blocks.

The 3D-Deformable convolutional blocks consist of stacks of 3D-Deformable convolutional layers, batch normalization (BN) layers~\cite{ioffe2015batch}, and leaky rectified linear unit (LReLU) activation layers~\cite{maas2013rectifier}. As detailed in Sec.\,\ref{sub:3D Def-conv}, 3D deformable convolutions are employed to mitigate the spatial displacements and perform adaptive weighting over historical observations; in addition, they extract important spatio-temporal patterns in mobile network traffic. BN layers perform normalization over a batch of output of each layer. This effectively reduces output's variance and can significantly accelerate the model training. LReLUs perform as activation functions. They improve the model non-linearity and representability, which enables the model to extract even more abstract features.

The zipper convolutional blocks receive the output of the 3D-Deformable convolutional blocks and are responsible for feature extraction. The structure of these blocks is inspired by that of a deep zipper network (ZipNet). The rationale is that the ZipNet, originally proposed for mobile network traffic super-resolution~\cite{zhang2017zipnet}, works demonstrably well in extracting spatio-temporal correlations hidden in this type of measurement data.
More precisely, global and multiple skip connections are employed within these blocks, to perform effective residual learning~\cite{he2016deep}, which is known to make the model more robust by constructing an ensembling system of neural networks with different depths~\cite{veit2016residual}. Skip connections also significantly smoothen the loss surface, which enables faster convergence of the model training~\cite{li2017visualizing}.

Upon processing by the zipper convolutional blocks, the mobile network traffic data is transformed into highly abstracted representations, ready for final MTD inference. This is performed by standard 2D convolutional blocks. Compared to the previous blocks, these are configured with a larger number of feature maps, so as to provide sufficient information for the inference process. The last layer of this block has $|\mathcal{S}|$ channels, \ie feature maps, each corresponding to the decomposed traffic volume of an individual mobile service.

\subsection{Tuning the Neural Network}
\label{sec:tuning}
To train the proposed 3D-DefCNN to solve the problem in~(\ref{eq:mtd}), we explore three different methods, namely ($i$) regression, ($ii$) ratio prediction trained with Mean Square Error (MSE), and ($iii$) ratio prediction trained with Cross-Entropy (CE).
\begin{enumerate}[leftmargin=1.5em]
    \item The \emph{Regression} method trains neural networks with a loss function over the traffic volume, which directly aims at minimizing the difference between $\widetilde{d}_a^s(t)$ and $d_a^s(t)$, \ie
    \begin{equation}\label{eq:mse2}
        \mathrm{L}(t) = \frac{1}{|\mathcal{S}|\cdot |\mathcal{A}|} \sum_{s \in \mathcal{S}} \sum_{a \in \mathcal{A}}||\widetilde{d}_a^s(t) - d_a^s(t)||^2.
    \end{equation}
    Due to model imperfections, the output obtained with this approach may violate the constraints (\ref{eq:const_pos}) and (\ref{eq:const_sum}).
    
    \item Instead, the ratio prediction using Mean Square Error (MSE) method seeks to infer the fraction of traffic consumed by each service, relative to the corresponding aggregate, \ie $r_a^s(t) = d_a^s(t) / d_a(t)$. Clearly, $\sum_{s\in\mathcal{S}} r_a^s(t) = 1$, $\forall a\in\mathcal{A}, \forall s\in\mathcal{S}, \forall t$.
    An equivalent transformation can be introduced for the estimated service demand by means of a softmax function:
    \begin{equation}
    	\tilde{r}_a^s(t) = \frac{\text{exp}(\tilde{\chi}_a^s(t))}{\sum_{s\in\mathcal{S}} \text{exp}(\tilde{\chi}_a^s(t))}.
    	\label{eq:softmax}
    \end{equation}
    Here $\tilde{\chi}_a^s(t)$ denotes the intermediate (unnormalized) output of the neural network. In essence, by the above we map a vector output with elements of arbitrary values onto a probability vector where the elements are in the (0,1) range~\cite{bishop}. The expression of $\tilde{r}_a^s(t)$ in (\ref{eq:softmax}) satisfies both (\ref{eq:const_pos}) and (\ref{eq:const_sum}). Moreover, following training this approach allows for an immediate derivation of the mobile network traffic, as $\tilde{d}_a^s(t) = \tilde{r}_a^s(t) \cdot d_a(t)$. Hence, we train the neural network to enforce $\tilde{r}_a^s(t)$ to be close to $r_a^s(t)$, by minimizing the following loss function:
    \begin{equation}\label{eq:mse}
        \mathrm{MSE}(t) = \frac{1}{|\mathcal{S}|\cdot |\mathcal{A}|} \sum_{s \in \mathcal{S}} \sum_{a \in \mathcal{A}}||\widetilde{r}_a^s(t) - r_a^s(t)||^2.
    \end{equation}
    We refer to this approach as $MSE$ in the following.
    
    \item Alternatively, we can uses the \emph{cross-entropy (CE)} for training, which is formally defined as 
    \begin{equation}\label{eq:ce}
        \mathrm{CE}(t) = \frac{1}{|\mathcal{A}|} \sum_{a\in\mathcal{A}} \sum_{s\in\mathcal{S}} -r_a^s(t)\log(\tilde{r}_a^s(t)),
    \end{equation}
    for the service snapshot estimated at time $t$. The overall CE for a given time span $\mathcal{T}$ can then be computed as the average over all time instants $t\in\mathcal{T}$.
    The expression of CE in (\ref{eq:ce}) is minimized when the estimated and actual traffic demand ratios match, \ie $\tilde{r}_a^s(t) = r_a^s(t), \forall a\in\mathcal{A}, \forall s\in\mathcal{S}, \forall t$.
This method makes the assumption that the traffic fraction follows a multinomial distribution, and minimizing the loss (\ref{eq:ce}) is equivalent to minimizing the \emph{Kullback--Leibler} (KL) divergence between the model and target distribution \cite{murphy2012machine}.  We opt for this function since, as reported in~\cite{golik2013cross}, training a neural network with CE usually finds a better optimum than other loss functions when the output is normalized, as in our case. This will be later confirmed by our comparative performance evaluation in Sec.\ref{sub:comp}.
\end{enumerate}

\section{Experiments\label{Sec:exp}}

We implement \name using the open-source Python libraries TensorFlow~\cite{tensorflow2015-whitepaper} and TensorLayer~\cite{tensorlayer}. We train and evaluate the framework on a high-performance computing cluster with two NVIDIA Tesla K40M GPUs with 2280 cores. The model parameters are optimized using the popular Adam stochastic gradient descent-based optimizer~\cite{kingma2015adam}.

In the following, we first introduce in Sec.\,\ref{sec:metric} the metrics used to assess the accuracy of \name, and present the benchmarks against which we test it. We then perform in Sec.\,\ref{sec:mtd_antenna} a comprehensive comparative evaluation of MTD performance on mobile network traffic recorded at the antenna level, which corresponds to end-to-end NSaaS settings where each network slice enjoys dedicated spectrum~\cite{marquez2018should}. In Sec.\,\ref{sec:cluster}, we investigate how such performance vary at different network levels, including MEC facilities, C-RAN or core network datacenters where network slices are also to be implemented~\cite{marquez2018should}. Finally, we comment on the complexity of the different models used in \name and discuss accuracy--complexity trade-offs in Sec.~\ref{sec:complex}.

\subsection{Performance Metrics and Benchmarks\label{sec:metric}}
We evaluate the performance of \name, along with that of all benchmarks, by means of the  mean absolute error (MAE, see Eq.(\ref{eq:mae})) and Normalized MAE (NMAE).
The NMAE normalizes MAE by the range of ground truth traffic measurements:
\begin{equation}
\hspace*{-10pt}\mathrm{NMAE}(t) = \frac{1}{|\mathcal{S}|\cdot |\mathcal{A}|} \sum_{s \in \mathcal{S}} \hspace*{-2pt}\sum_{a \in \mathcal{A}}\frac{|\widetilde{d}_a^s(t) - d_a^s(t)|}{\max_t d_a^s(t) - \min_t d_a^s(t)}.
\label{eq:nmae}
\end{equation}

In both (\ref{eq:mae}) and (\ref{eq:nmae}), the set of antennas $\mathcal{A}$ shall be replaced by the set of facilities or datacenters, depending on the network level at which MTD is performed. 

\begin{table}[tb]
\centering
\footnotesize
\caption{Benchmark neural networks configuration.}
\label{tab:configure}
\begin{tabular}{|l|L{11cm}|}
\hline
\textbf{Class}  & \textbf{Configuration}                                                                                                                            \\ \hline
MLP          & 5 hidden layers, 1,000 hidden units each                                                                                      \\ \hline
CNN          & Fig.\,\ref{Fig:arc}, without 3D convolutional block                                                                   \\ \hline
LSTM          & 3 stacks, 500 units for each stack                                                                   \\ \hline
ConvLSTM          & 3 stacks, 108 channels for each stack                                                                   \\ \hline
ZipNet       & Fig.\,\ref{Fig:arc}, where 3D deformable convolutional layers in the first block are replaced by legacy 3D convolutional layers \\ \hline
DefCNN    & Same as ZipNet, but convolutional layers in the last block are replaced by 2D deformable convolutional layers                  \\ \hline
\end{tabular}
\end{table}

We compare our proposed 3D-DefCNNs against six deep learning architectures, whose configurations are summarized in Tab.~\ref{tab:configure}. \rv{Note that since there is no existing solution for MTD, these baselines are what we find to be most relevant for comparison purposes.} The multi-layer perceptron (MLP) is the simplest neural network class~\cite{Goodfellow-et-al-2016} and we consider it as a baseline solution. Convolutional neural networks (CNNs) are commonly used for imaging applications~\cite{krizhevsky2012imagenet}. Long Short-Term Memory (LSTM) is frequently exploited for modeling sequential data \cite{hochreiter1997long}, \eg speech signals and natural language. Convolutional LSTM (ConvLSTM) is a dedicated model for spatio-temporal data forecasting \cite{xingjian2015convolutional}. Deep zipper networks (ZipNets) were originally proposed for mobile traffic super resolution tasks with remarkable results~\cite{zhang2017zipnet}. 2D deformable CNN (DefCNN) extend the transformation ability of CNNs and perform very well in many computer vision tasks~\cite{dai17dcn}.

\subsection{MTD at radio access \label{sec:mtd_antenna}}
\label{sub:comp}

At radio access, network slicing can provide very high QoS guarantees by isolating spectrum or even dedicated antenna sites for specific mobile services~\cite{rost17,sallent18}. To fulfill this vision, resource management decisions need to be made at the individual antenna level, which in turn calls for fine-grained traffic usage information for different services as close as possible to the user. Our initial scenario for evaluation of MTD is thus the access network, with decomposition carried out on traffic recorded at each antenna sector separately.

\subsubsection{Comparative analysis}

\begin{figure}[tb]
\begin{center}
\includegraphics[width=\columnwidth]{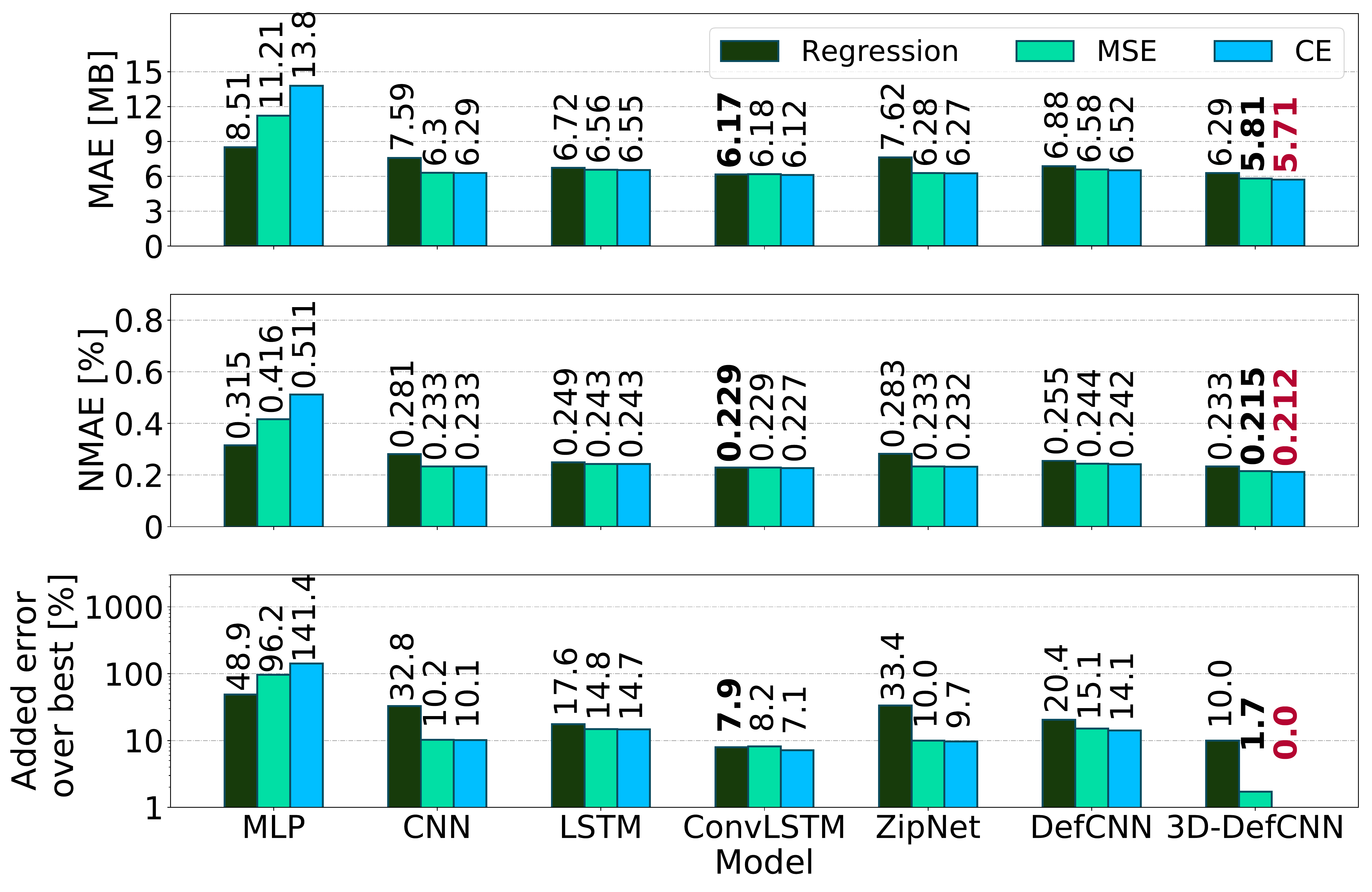}
\end{center}
\caption{\label{Fig:metric} MTD performance in terms of MAE (top), NMAE (middle), and additional error with respect to the best solution (bottom), across all combinations of neural networks in Sec.\,\ref{sec:metric} and loss functions in Sec.\,\ref{sec:tuning}. Results are averaged over all services in $\mathcal{S}$.
}
\end{figure}

Fig.\,\ref{Fig:metric} gives an overview of the comparative performance evaluation in the RAN setting, juxtaposing the MTD results attained by \name 
with those obtained by the benchmark neural network architectures above,
under all training methods listed in Sec.\,\ref{sec:tuning}.
We emphasize in bold the error yielded by the best model with each loss function, while the overall best performance is highlighted in red. By leveraging the \mbox{3D-DefCNN} deep learning architecture trained using a CE loss function, \name achieves the lowest MAE and NMAE.
A closer look at Fig.\,\ref{Fig:metric} reveals that training with the CE loss function on traffic demand ratios leads to better estimates than using Regression or MSE. While this holds in the vast majority of cases, the improvement brought by CE over the benchmark loss functions is especially consistent for the \mbox{3D-DefCNN} architecture, where it allows performance gains up to 10\% in terms of MAE over the competing loss functions. This confirms that the normalization and its combination with CE indeed improve the MTD accuracy of \name. 

Concerning the neural network architectures, MLP delivers the poorest MTD performance among all approaches considered, as it lacks blocks capable of extracting spatial features. {As such, the added estimation error introduced by MLP with respect to \mbox{3D-DefCNN} can be as high 141\%.} In contrast, LSTM and ConvLSTM can effectively model the temporal correlations inherent to mobile traffic, which leads to much improved results. Among convolutional architectures, Def-CNN trained with CE achieves lower error as compared to traditional CNN and ZipNet structures. Still, the \mbox{3D-DefCNN} we propose consistently outperforms all other solutions, {and yields error reductions in the range 7--33\%.}

\begin{figure}[tb]
\begin{center}
\includegraphics[width=1.03\columnwidth]{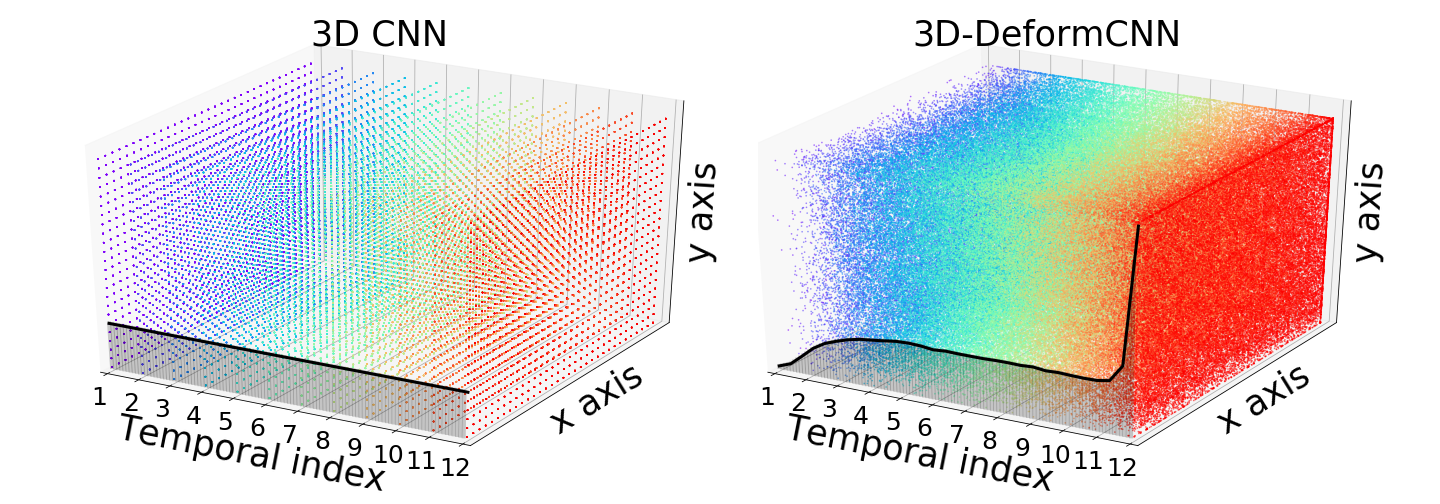}
\end{center}
\caption{\label{fig:offset_st} Spatio-temporal distribution of positions of one tensor input visited by the filter of a traditional 3D convolutional layer (left), and the filter learned by our 3D deformable convolutional layer (right).
}

\end{figure}

The results suggest that: ($i$) both spatial and temporal features are significant to the MTD performance; ($ii$) the spatial deformation operations help reorganizing the geographical displacements in the aggregate snapshots provided as input; ($iii$) incorporating temporal deformation into the model enables scanning of historical data with frequencies that are tailored to their importance. The second point is equivalent to an \textit{attention-like mechanism}~\cite{yu2018generative}, where relevant information is sampled more frequently.
We provide an intuitive visualization of this last point in Fig.\,\ref{fig:offset_st}, where we compare the spatio-temporal distribution of the locations of one tensor input (\ie a set of consecutive aggregate snapshots) visited by a legacy 3D convolutional layer (employed by ZipNet and DefCNN) and by our novel 3D deformable convolutional layer (adopted by \name), when both are used as the first layer of the network. Both plots also show the density of the number of sampled points projected onto the temporal dimension ($x=0$) as black solid lines subtending a shaded area. The structure shown on the left is not flexible, leading to a regular sampling in space and time, which results into a uniform distribution of points over the input matrices. Instead, the 3D deformable convolutional layer learns which positions of the input tensor are the most relevant: as a result, it samples more frequently recent input matrices, deeming that fresher snapshots are more relevant to the estimation.

\begin{figure}[tb]
\begin{center}
\includegraphics[width=\columnwidth]{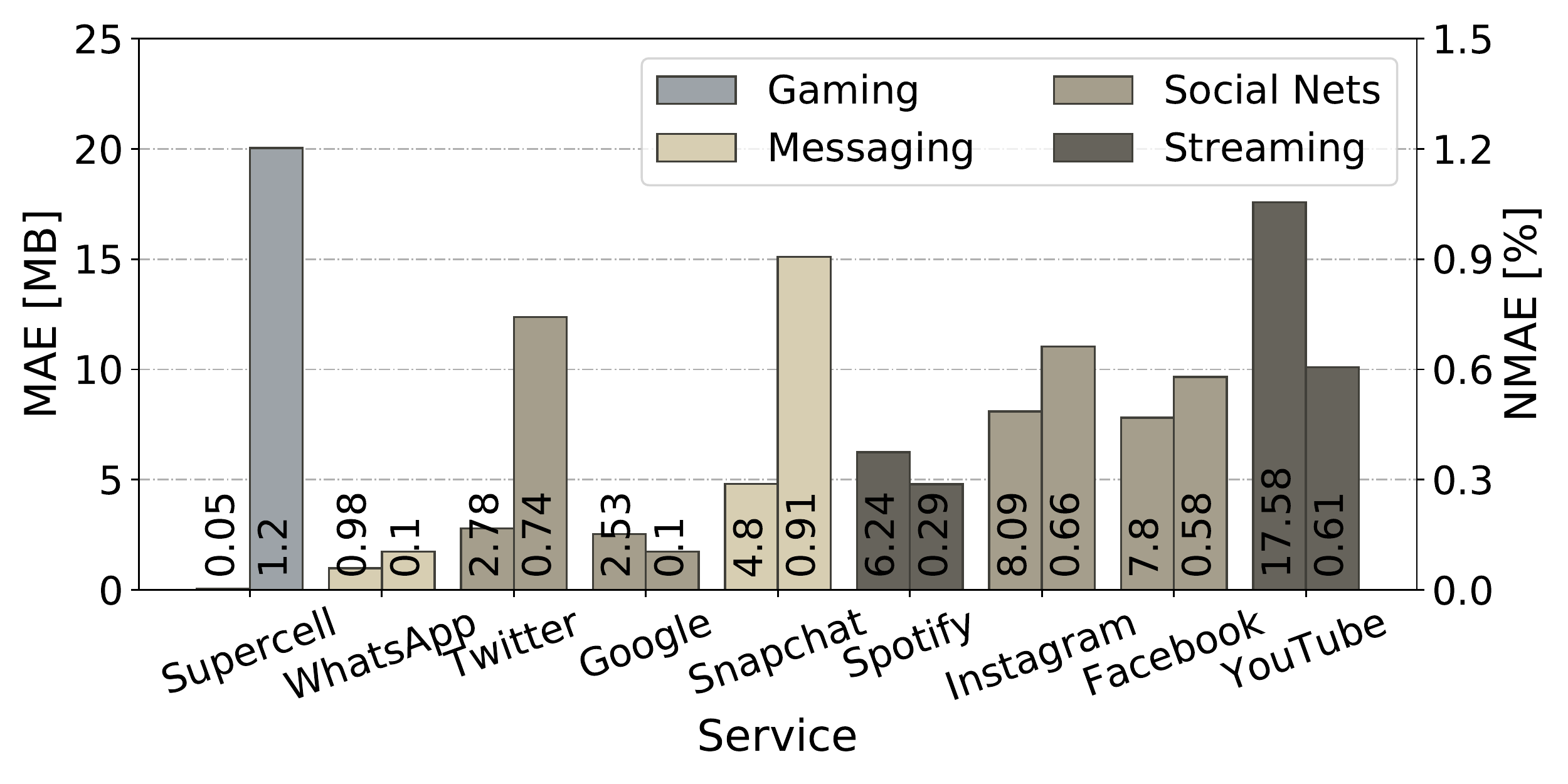}
\end{center}
\caption{\label{fig:mae_s} \name performance, in terms of MAE (left bars) and NMAE (right bars) for all services in $\mathcal{S}$. Results are averaged over the whole test period.
}
\end{figure}

\subsubsection{Service-level performance of \name}

Having clarified that \name outperforms all considered benchmarks, we shift our attention to the MTD performance offered by our framework on a per-service basis. Fig.\,\ref{fig:mae_s} offers a breakdown of MAE (left bars) and NMAE (right bars) across the nine services in our reference set $\mathcal{S}$, when using \name. There exists some apparent variability in the estimation quality across services. \emph{E.g.}, streaming services (\ie YouTube and Spotify), which consume a large fraction of the total traffic, are also subject to higher MAE. However, this does not necessarily lead to high relative error: in fact, their NMAE is as low as 0.61\% and 0.29\%, respectively. In contrast, MTD works very well for Supercell games traffic, due to the smaller volume of data generated by such apps. The associated NMAE is the highest recorded among all services, yet it stays at a very reasonable 1.2\%. Overall, the MTD performance of \name is remarkable, as the inference errors are typically well below 1\% for all services.

\begin{figure}[tb] 
\begin{center}
\includegraphics[width=\columnwidth]{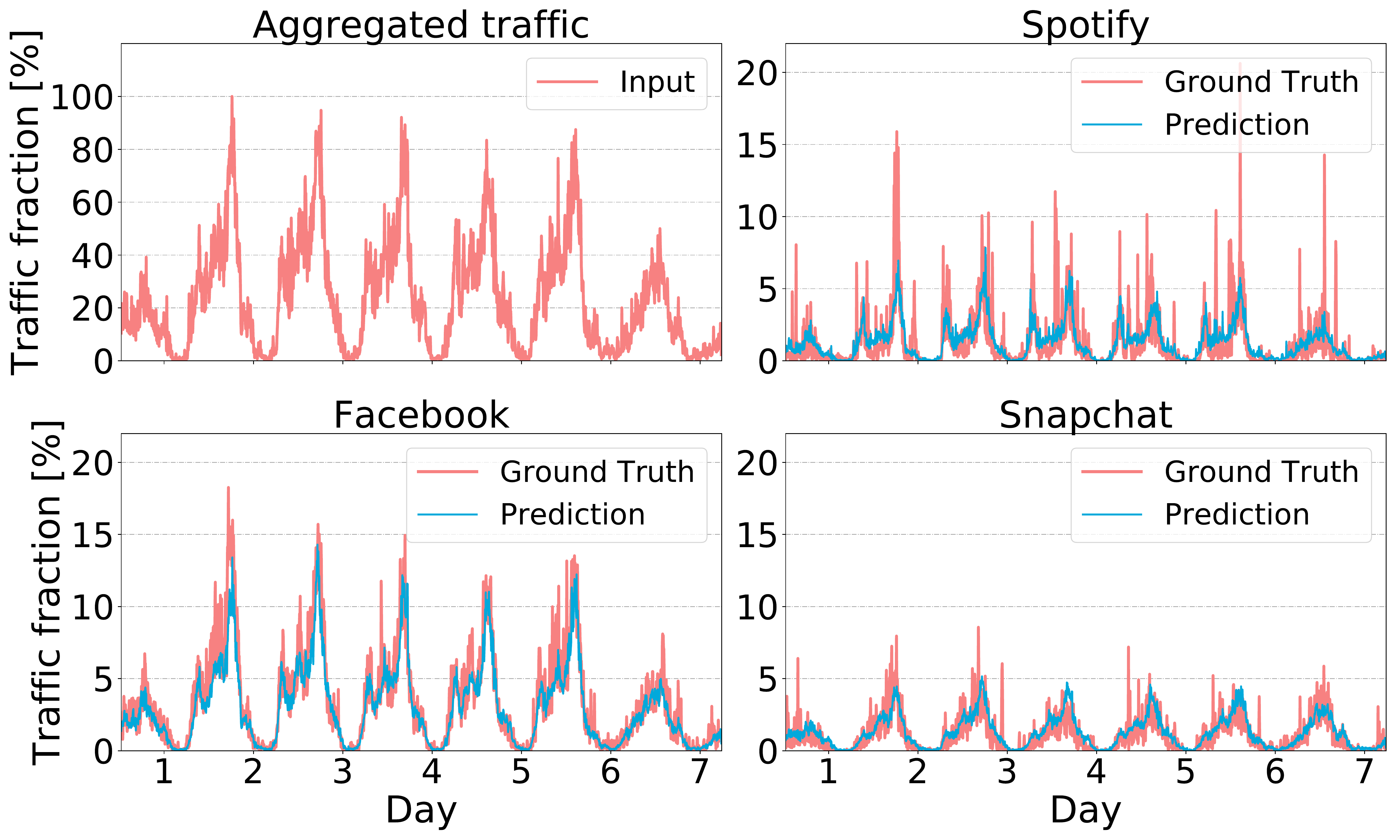}
\end{center}
\caption{\label{Fig:exa_a} Example of MTD on the aggregate traffic recorded at one antenna (top left) and decompositions for three representative services by \name. All traffic is normalized to the aggregate activity peak.}
\end{figure}

An illustrative example of the MTD quality granted by our framework is in Fig.\,\ref{Fig:exa_a}, which shows the inferred time series of the demand for three representative services, \ie Spotify, Facebook, and Snapchat, based on the sole input provided by the aggregate traffic (top left subplot). The result focuses on one test week at a random single antenna. Observe that although the Spotify service exhibits frequent fluctuations, our framework still captures well the overall traffic profile (top right). As for Facebook and Snapchat, the predicted traffic is very close to the ground truth. This confirms that \name yields precise MTD, irrespective of service type.

\subsection{MTD at network datacenters \label{sec:cluster}}
\label{sub:clust}

Network slicing heavily relies on the capability of the operator to dynamically orchestrate virtualized functions at the edge and core of their infrastructures~\cite{ngmn16,gil-herrera16,rost17}. Similarly to the radio access case, this requires efficient data-driven resources management, which could be fueled by service-level demands inferred via MTD. In order to assess the flexibility of \name in serving heterogeneous edge and core network scenarios, we consider three use cases: ($i$) fifty MEC facilities deployed at the edge, aggregating traffic from 10-20 antennas each; ($ii$) thirty C-RAN datacenters, each providing MAC-layer functionalities for 20-40 antennas; ($iii$) ten core network datacenters implementing, \eg Serving Gateway (S-GW) functions and accommodating traffic generated by 50 or more antennas each. We model the deployment of such diverse network entities according to a strategy of load balancing and latency reduction.

\begin{figure}[tb] 
\begin{center}
\includegraphics[width=1\columnwidth]{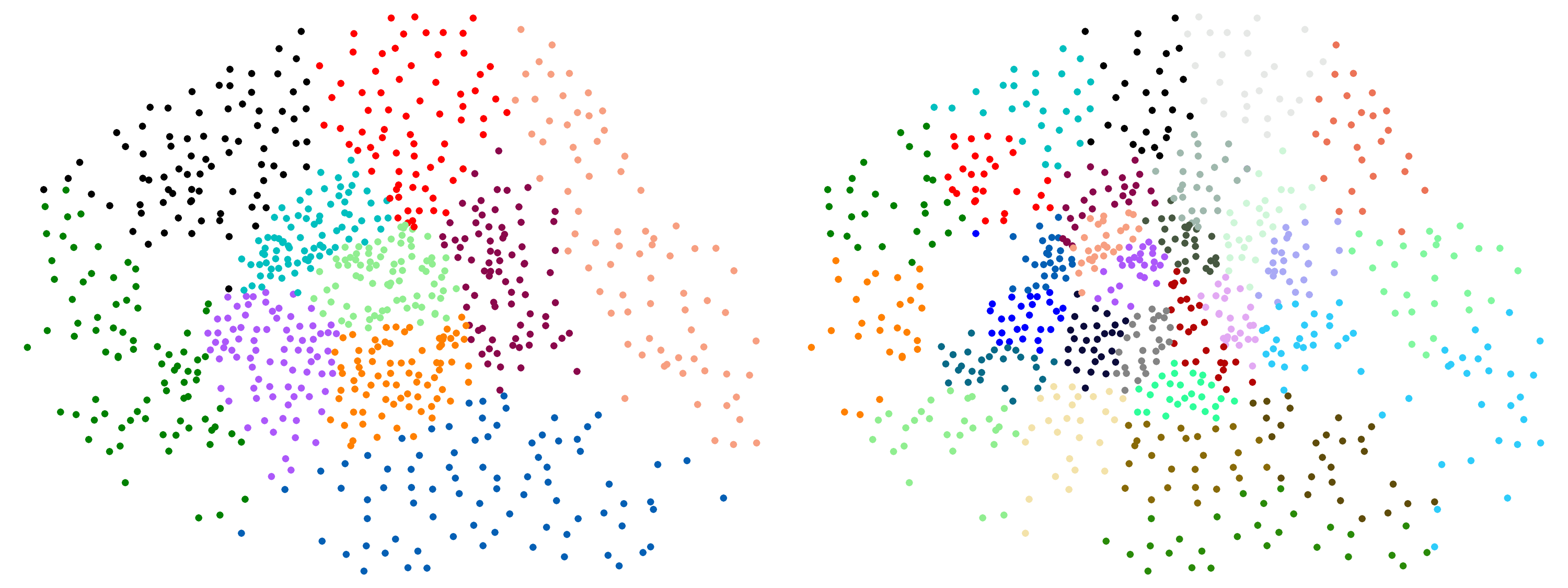}
\end{center}
\caption{\label{Fig:cluster} Assignments of antennas in our reference metropolitan scenario to ten Core datacenters (left) and thirty C-RAN datacenters (right), fulfilling load balancing and latency requirements. Different colors denote clusters of antennas associated to a same datacenter. Figure best viewed in color.}
\end{figure}

In order to decide on the associations between antennas and MEC facilities, C-RAN and core datacenters, we run the balanced graph $k$-partitioning algorithm proposed in \cite{marquez2018should} over the Delaunay triangulation graph~\cite{lee1980two} of the 792 antenna locations. This creates a fixed number of antenna clusters (10, 30 or 50, for MEC, C-RAN and core datacenters, respectively) that serve comparable traffic loads, while minimizing the latency (\ie distance) with respect to associated antennas. Examples of partitions returned by the algorithm via the Karlsruhe Fast Flow Partitioning (KaFFPa) heuristic~\cite{sandersschulz2013} are provided in Fig.\,\ref{Fig:cluster} for the core and C-RAN datacenter scenarios.

Although the architectural settings above are arbitrary, and do not necessarily reflect the future organization of the Network Virtualization Function (NVF)-compliant mobile network in the target region, they provide a reasonable approximation of what one could expect, and allow us to investigate the performance of \name in diverse virtualized network scenarios.
In all these cases, we run \name using the geographic centroid of each cluster as the location associated to the traffic served by each datacenter.

We also consider different time resolutions for the management of NSaaS, from 5 mins to 1 h. The rationale is that Virtual Network Functions (VNF) and their associated resources at datacenter level are likely to be reconfigured over longer timescales than at radio access. Hence, varying the temporal granularity offers a more complete analysis of the system.

Results are summarized in Tab.~\ref{Tab:mae_cluster}, in terms of MAE and NMAE. The minimum recorded error for each combination of network level and time resolution is highlighted in bold. The key takeaway is that \name still outperforms all other architectures as long as MTD is performed with higher frequency (\ie 5 to 30 minutes) or at MEC facilities and C-RAN datacenters. Instead, when mobile traffic is accumulated in large volumes, by considering hourly aggregates or demands at network core datacenters, LSTM yields the lowest errors.
Indeed, ($i$) coarse temporal granularities lead to more regular time series, and generally easier to decompose, whereas ($ii$) considering combined at core datacenters diminishes the impact of spatial correlations between locations. Under these settings, the complexity of \name becomes unnecessary, while LSTM thrives by avoiding looking for complicate interactions that are in fact absent in the input data. For the same reason, the performance of CNN-based models, including ConvLSTM, CNN, ZipNet, and DefCNN, improves as we move from the core to the edge of the network, where the impact of spatial correlations increases. Overall, \name proves the best solution for MTD in the majority of settings, which confirms the flexibility of our framework.

\begin{table}[h!]
\caption{MAE (top, in MB), NMAE (middle, in \%) and additional error with respect to \name (bottom, also in \%) returned by all models over different combinations of network level and time resolution. Each cell shows the mean MTD error at core datacenters (left), C-RAN datacenters (middle) and MEC facilities (right). \label{Tab:mae_cluster}}
\centering
\begin{tabular}{|l|c|c|c|c|}
\hline
\textbf{Model} & 5 mins    & 10 mins    & 30 mins     & 1 h          \\ \hline
MLP            & 116/53/38 & 213/96/69  & 602/263/175 & 5022/673/401 \\ \hline
CNN            & 152/64/43 & 290/119/79 & 842/321/206 & 1617/589/306 \\ \hline
LSTM           & \textbf{106}/52/38 & \textbf{191}/93/67  & \textbf{483}/236/163 & \textbf{895}/\textbf{414}/\textbf{283}  \\ \hline
ConvLSTM       & 129/55/39 & 268/104/71 & 837/319/199 & 1026/611/345 \\ \hline
ZipNet         & 149/60/41 & 281/110/75 & 699/262/185 & 1313/524/344 \\ \hline
DefCNN         & 127/53/37 & 228/96/67  & 632/244/156 & 1315/491/295 \\ \hline
\name          & 124/\textbf{49}/\textbf{34} & 231/\textbf{89}/\textbf{62}  & 696/\textbf{230}/\textbf{155} & 1194/451/292 \\ \hline
\end{tabular}\\

\vspace*{1em}
\begin{tabular}{|>{\hspace{-0.3pc}}l<{\hspace{-3.6pt}}|>{\hspace{-0.3pc}}c<{\hspace{-3.7pt}}|>{\hspace{-0.3pc}}c<{\hspace{-3.7pt}}|>{\hspace{-0.3pc}}c<{\hspace{-3.7pt}}|>{\hspace{-0.3pc}}c<{\hspace{-3.7pt}}|}
\hline
\textbf{Model} & 5 mins         & 10 mins        & 30 mins        & 1 h            \\ \hline
MLP            & 1.99/1.58/1.23 & 1.96/1.58/1.62 & 1.91/1.84/1.61 & 8.37/2.56/1.99 \\ \hline
CNN            & 2.61/1.90/1.38 & 2.67/1.97/1.86 & 2.67/2.25/1.88 & 2.70/2.24/1.92 \\ \hline
LSTM           & \textbf{1.82}/1.56/1.21 & \textbf{1.76}/1.54/1.58 & \textbf{1.53}/1.65/1.50 & \textbf{1.49}/\textbf{1.57}/\textbf{1.40} \\ \hline
ConvLSTM       & 2.22/1.64/1.26 & 2.47/1.71/1.68 & 2.65/2.23/1.82 & 2.71/2.32/1.71 \\ \hline
ZipNet         & 2.56/1.77/1.31 & 2.59/1.82/1.76 & 2.22/1.84/1.69 & 2.19/1.99/1.70 \\ \hline
DefCNN         & 2.18/1.58/1.20 & 2.10/1.59/1.60 & 2.00/1.71/1.43 & 2.19/1.87/1.46 \\ \hline
\name          & 2.13/\textbf{1.47}/\textbf{1.08} & 2.12/\textbf{1.48}/\textbf{1.47} & 2.21/\textbf{1.61}/\textbf{1.42} & 1.99/1.71/1.45 \\ \hline
\end{tabular}\\
\vspace*{1em}

\begin{tabular}{|l|>{\hspace{-0.3pc}}c<{\hspace{-3.5pt}}|>{\hspace{-0.3pc}}c<{\hspace{-3.5pt}}|>{\hspace{-0.3pc}}c<{\hspace{-3.5pt}}|>{\hspace{-0.3pc}}c<{\hspace{-3.5pt}}|}
\hline
\textbf{Model} & 5 mins                         & 10 mins                       & 30 mins                       & 1 h                                  \\ \hline
MLP            & -6.5/8.2/11.8                  & -7.8/7.9/11.3                 & -13.5/14.3/12.9               & 320.6/49.2/37.3                            \\ \hline
CNN            & 22.6/30.6/26.5                 & 25.5/33.7/27.4                & 21.0/39.6/32.9                & 35.4/30.6/4.8                              \\ \hline
LSTM           & \textbf{-14.5}/6.1/11.8        & \textbf{-17.3}/4.5/8.1        & \textbf{-30.6}/2.6/5.2        & \textbf{-25.0}/\textbf{-8.2}/\textbf{-3.1} \\ \hline
ConvLSTM       & 4.0/12.2/14.7                  & 16.0/16.9/14.5                & 20.3/38.7/28.4                & -14.1/35.5/18.2                            \\ \hline
ZipNet         & 20.2/22.4/20.6                 & 21.6/23.6/21.0                & 0.4/13.9/19.4                 & 10.0/16.2/17.8                             \\ \hline
DefCNN         & 2.4/8.2/8.8                    & -1.3/7.9/8.1                  & -9.2/6.1/0.6                  & 10.0/8.9/1.0                               \\ \hline
\end{tabular}

\end{table}

\subsection{Complexity Analysis \label{sec:complex}}

\begin{figure}[tb] 
\begin{center}
\includegraphics[width=1\columnwidth]{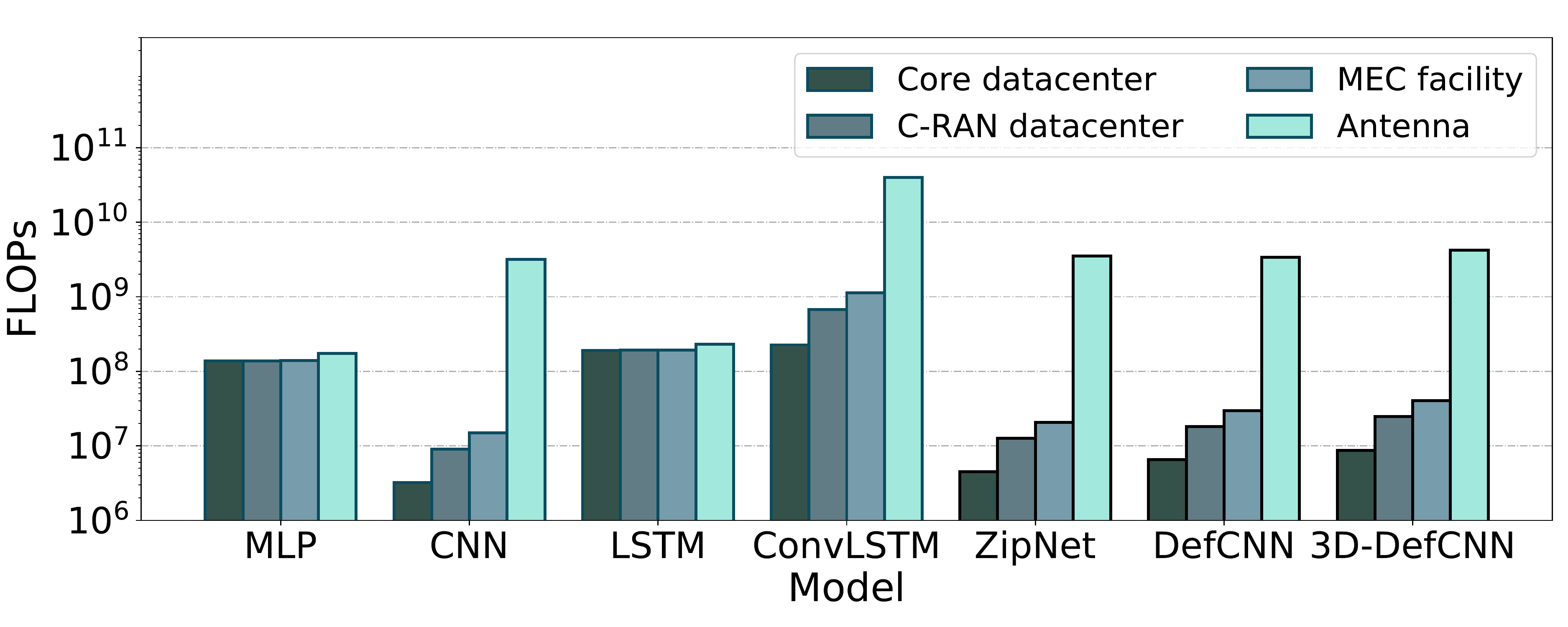}
\end{center}
\caption{Evaluation of FLOPs of different neural networks models across all considered network levels. \label{Fig:complexity}}
\end{figure}

We also evaluate the models' complexity in terms of floating point operations (FLOPs) per inference instance, a frequently employed measure of the complexity of neural networks \cite{molchanov2017pruning}. The number of FLOPs is computed by counting the number of mathematical operation or assignments that involve floating-point numbers. Observe in Fig.~\ref{Fig:complexity} that models which include convolution operations, \ie CNN, ConvLSTM, ZipNet, DefCNN, and 3D-DefCNN are more sensitive to spatial granularity and as a result their complexity grows significantly with the size of the input. In contrast, as MLP mostly employs dot product operation with fixed hidden layers, its complexity is lower, but so is its overall MTD accuracy.

An interesting remark is that, although LSTM yields the best accuracy in some cases, as indicated in Tab.~\ref{Tab:stat1}, it entails the highest complexity in MTD tasks at core, C-RAN and MEC levels. The complexities of CNN-based models surpass that of LSTM only at antenna-level MTD. Observe that deformable operations do not introduce significant additional complexity compared to plain CNN. Instead, their advantage lies withing the important accuracy gains. Overall, 3D-DefCNN requires around $4\times 10^9$ FLOPs per inference instance for antenna-level MTD, which can be easily handled in real-time even by a modern CPU: an Intel Core i7 980 XE CPU can execute $1.076 \times 10^{11}$ FLOPs per second, which exceeds the requirements of MTD at antenna level.

\section{Implications for Operators}

We complete our evaluation of the performance of \name by assessing the viability of the MTD it offers in practical case studies. Specifically, we take the perspective of the mobile network operator, and assess the incurred costs when \name is used to determine the capacity allocated to each network slice, solely based on aggregate traffic information. Such costs directly stem from estimation errors in the decomposed demands, which may entail Service-Level Agreements (SLAs) violations (if the estimate is below the reference) or overprovisioning of unnecessary capacity (if the estimate is above the reference).

We thus re-train \name to perform MTD with a recently proposed loss function that specifically aims at minimizing monetary costs for network operators~\cite{bega2019deepcog}:
\begin{equation}
    \mathrm{L}(t) = \frac{1}{|\mathcal{S}|\cdot |\mathcal{A}|} \sum_{s \in \mathcal{S}} \sum_{a \in \mathcal{A}}\mathcal{L}(\Delta d_a^s(t)),
    \label{eq:deepcog}
\end{equation}
where
\begin{eqnarray}
\mathcal{L}(\Delta d_a^s(t)) = 
\begin{cases}
\alpha - \epsilon \cdot \Delta d_a^s(t), &\Delta d_a^s(t)<0\cr 
\alpha - \frac{1}{\epsilon} \Delta d_a^s(t), &0<\Delta d_a^s(t)\leq \alpha\epsilon \cr 
\Delta d_a^s(t) -\alpha\epsilon , &\Delta d_a^s(t)>\alpha\epsilon.
\end{cases}
\end{eqnarray}
Here $\Delta d_a^s(t) = \widetilde{d}_a^s(t) - d_a^s(t)$, $\alpha$ controls the cost of an SLA violation, and $\epsilon$ is a small constant. The expression accounts for the cost of SLA violations (quasi constant, and intervening for $\Delta d_a^s(t) \leq \alpha\epsilon$) and overprovisioning (growing as additional resources are erroneously reserved, when $\Delta d_a^s(t) > \alpha\epsilon$), so that \name can perform MTD by trying to balance them. We configure $\alpha = 1$ and $\epsilon = 0.01$, as suggested in \cite{bega2019deepcog}.

We test \name with this configuration in three case studies: ($i$)~MTD at MEC facilities on traffic aggregated at every 30 minutes; ($ii$)~MTD at C-RAN datacenters on traffic aggregated at every 30 minutes; and ($iii$)~MTD at core datacenters on traffic aggregated at every hour. Exclusively in the last scenario, we choose LSTM as the neural network architecture, according to the performance results in Sec.\ref{Sec:exp}.

The costs incurred to the operator in the datacenter scenarios ($i$)--($iii$) above are listed in Tab.~\ref{Tab:stat}. The table expresses costs in MB/s, which can then be translated into actual monetary values based on the operator's price of the capacity needed to accommodate one MB/s of traffic at each network level. For overprovisioning, the cost maps to proper additional MB/s of allocated capacity beyond what strictly required; for SLA violations, each infringement is translated into MB/s as per (\ref{eq:deepcog}), where one violation has the same cost as allocating additional capacity to cover $\alpha$ times the peak demands for the considered mobile service traffic.
In both cases, results are reported as the total over all MEC (respectively, C-RAN and network core) nodes in the target metropolitan region.

\begin{table}[tb]
\caption{Total SLA violation cost and overprovisioning cost determined by MTD at different network levels. \label{Tab:stat}}
\centering
\begin{tabular}{|l|c|c|}
\hline
\textbf{Use case} & \textbf{SLA violation [MB/s (\%)]} & \textbf{Overprovisioning [MB/s (\%)]} \\ \hline
MEC facility                   & 247.95 (136.31)                  & 203.38 (111.80)                                        \\ \hline
C-RAN datacenter                    & 25.82 (14.20)                  & 106.59 (58.60)                                      \\ \hline
Core datacenter              & 15.54 (8.55)                   & 61.71 (33.92)                                        \\ \hline
\end{tabular}
\end{table}

At C-RAN and core datacenters, \name carries percent costs in the range from 8\% to 58\%, computed with respect to the true demand. These are in fact comparable to the equivalent costs for capacity allocation to network slices at the same network levels, incurred when the operator has perfect knowledge of the traffic demand associated to each mobile service: in these conditions, by using a state-of-the-art one-step predictor, costs are up to 30\% for SLA violations and up to 18\% for overprovisioning~\cite{bega2019deepcog}. When pushing \name at MEC facilities, the performance degrades sensibly, and the stronger fluctuations in per-service traffic make a MTD-based estimation of the demand generated by each service less suitable to capacity allocation.


When MTD is performed at antenna level, we translate the extra resources required to support the unnecessary capacity overprovisioned at each antenna in terms of additional subcarriers/spectrum~\cite{Throughput}. From this, we compute the cost in CPU time, based on experimental models obtained with open LTE stacks such as OpenAirInterface~\cite{gringoli2018performance}. We summarize our findings in Tab.~\ref{Tab:stat1}. When \name performs MTD at  antenna level, just 6 Mbps of additional throughput is needed per antenna, which yields a 3~MHz spectrum cost and requires 7.5\% additional CPU time with respect to the case where perfect knowledge of service traffic is available.

\begin{table}[t]\footnotesize
\caption{Mean additional costs of antenna-level MTD. \label{Tab:stat1}}
\centering
\small
\begin{tabular}{|c|c|c|c|}
\hline
\textbf{Throughput} & \textbf{Subcarriers} & \textbf{Spectrum cost} & \textbf{CPU time} \\ \hline
6.114 Mbps        & 110           & 3 MHz                  & 7.5\%             \\ \hline
\end{tabular}

\end{table}

Given the positive results obtained in the majority of the scenarios considered, we believe that MTD can substantially reduce the cost of service-level demand estimation in many practical cases, with minimal impact to the quality of the resource allocation.


\section{Summary}
In this chapter, we introduced \name, a dedicated framework for aggregate Mobile Traffic Decomposition (MTD) into service-level demands, intended to assist capacity allocation to network slices.
By hinging on an original 3D-DefCNN structure, \name decomposes traffic aggregates into service-level consumption time series with inference errors below 1.2\%, in large-scale and heterogeneous network deployments. In doing so, it also outperforms a wide range of other deep learning approaches.

\name can perform MTD inference in real-time, and we show that a resource allocation based on decomposition yields affordable costs for the operator. As a result, MTD via \name provides a means to limit the need for extensive deep packet inspection (DPI) in traffic collected at different levels of the network infrastructure. We believe that it is a promising novel approach that can complement the DPI technique.

%% file: chap7.tex
\chapter{Attacking and Defending Deep Learning-based Intrusion Detection Systems\label{chap:tita}}

Network Intrusion Detection (NID) aims at identifying malicious traffic flows, so as to protect computers, networks, servers, and data from attacks, unauthorized access, modification, or destruction \cite{buczak2015survey}. Given the unprecedented growth in the data traffic volume transiting both wired and wireless infrastructure, NID is becoming increasingly important to ensure system/service availability and protect individuals' safety and privacy online. As new cyberattacks proliferate, traditional intrusion detection methodologies that rely on pattern matching (\eg IP address and Port number) and classification are losing their effectiveness \cite{zhou2019evaluation}. In this context, deep learning-based solutions are gaining traction, as they rely less often on deep packet inspection (hence raising fewer privacy concerns) and have better generalization abilities. However, due to their complex structures, DNN also suffer from limited interpretability, which inevitably raises important questions: \emph{Is deep learning a truly reliable option for NID? Is there any ``Achilles heel'' that can be exploited to compromise the alleged high detection accuracy?} Answering these questions is crucial to guaranteeing the reliability of Network Intrusion Detection System (NIDS).


Unfortunately, Deep Neural Networks (DNNs) have been proven vulnerable to adversarial examples \cite{nasr2019comprehensive} or blakdoor attacks~\cite{ccs2} in several applications \cite{236248, ccs1}, whereby they can be fooled by dedicated unnoticeable perturbations introduced in the input \cite{yuan2019adversarial}, which interfere with the correctness of the inferences made. Since such perturbations are very subtle, they are often extremely difficult to detect. Deep learning-based NIDS may be also susceptible to such adversarial manipulations. Attackers unaware of the properties of an NIDS (\ie black-box system) could generate adversarial samples by repeatedly changing small subsets of the traffic features, and make ``queries'' to the NIDS. After each query, the attacker obtains some feedback (\eg ACK), which indicates the success or failure of the attack attempt.\footnote{Note the distinction between adversarial attacks and network attacks, which would typically work in tandem. The former aims to compromise classification (including that of malicious traffic) performed by a DNN, whilst the latter targets specific weaknesses that NIDS usually seek to mitigate.} Based on this feedback, the attacker can adjust perturbations on selected time-based features (\eg intervals between each two packets) of the traffic, or introduce new ones, without changing its essence, until succeeding in bypassing the NIDS~\cite{ibitoye2019threat, usama2019adversarial}. By this approach, malicious flows could then be disguised into benign traffic and compromise their targets~\cite{kuppa2019black}, while remaining undetected even by NIDS previously thought to be highly accurate.

\begin{figure}[t]
\centering
\includegraphics[trim=20 20 20 10, width=\columnwidth]{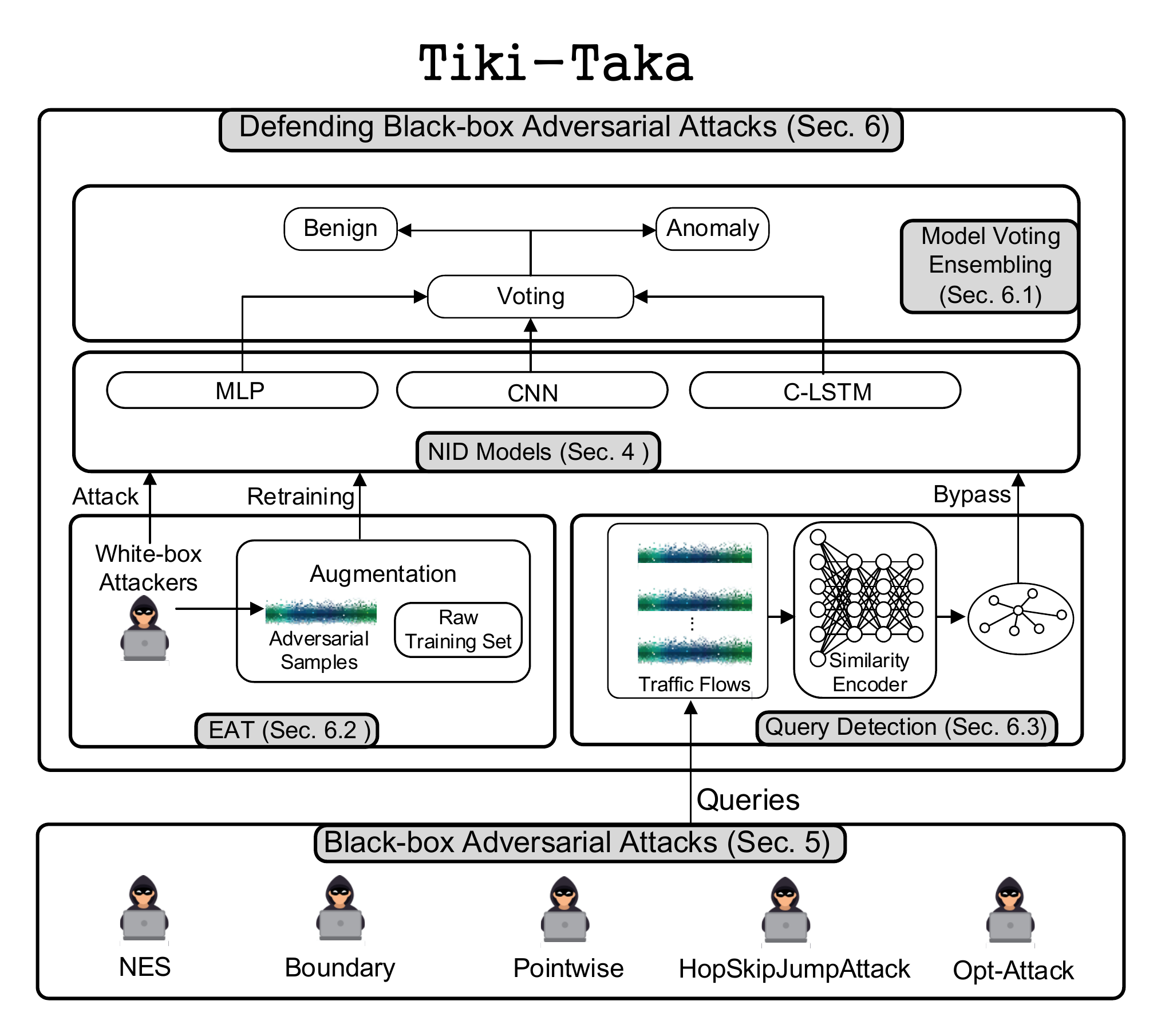}
\caption{The \tita framework for crafting and defending against adversarial attacks towards NIDS.
\label{fig:tikitaka}}
\end{figure}

To mitigate the aforementioned issues, in this chapter we investigate the robustness of state-of-the-art deep learning NID models against different adversarial mechanisms. We consider attacks in black-box and decision-based settings, and test the effectiveness and efficiency of each attack in two detection scenarios: one-to-all and one-to-one, \ie aiming to discriminate malicious vs. benign traffic, and respectively identify precise types of attacks. To defend against those threats, we propose three different solutions, which effectively reduce the success rate of each attack to a large extent. This enables more robust and reliable NID and leads to the following contributions:

\begin{enumerate}
    \item We implement three types of DNN architectures based on state-of-the-art NID models, \ie MLP, CNN and LSTM, and perform NID on the realistic cyber defense dataset CSE-CIC-IDS2018 \cite{sharafaldin2018toward} in both one-to-all and one-to-one scenarios. The NID models implemented achieve over 98.7\% detection accuracy based on a limited set of traffic features, which matches the performance of previously reported NIDS implementations (details in Sec.~\ref{sec:nid}).
    \item To demonstrate the vulnerabilities of deep NIDS, we employ five state-of-the-art black-box attack strategies (\ie \nes, \bd, \hj, \pw and \opt) to generate adversarial samples under black-box and decision-based settings. These samples can bypass the pre-trained NID models within traffic features domain constraints (\ie features that are not amendable or can only vary within some range), in both detection scenarios. We conduct a comprehensive evaluation on the effectiveness of each adversarial attack, and provide an in-depth analysis of their characteristics (Sec.~\ref{sec:bba}).
    \item We propose three defense mechanisms to strengthen deep learning based NIDS against the aforementioned attacks, namely model voting ensembling \cite{abbasi2017robustness}, ensembling adversarial training \cite{tramer2018ensemble}, and adversarial query detection \cite{chen2019stateful}. Each defense method can either operate individually or jointly with others. Experiments show that our defense methods can drastically reduce the success rate of each attack, which significantly improves the robustness of the NID models (see Sec.~\ref{sec:def}).
\end{enumerate}
We name our attack--defense framework \tita\footnote{\tita is a football tactic that encourages short and fast passing of the ball and tackling on the spot when losing possession. We use this is to metaphorize the frequent queries passed to the intrusion detector in the attack process, and with the detector subsequently regaining control through defenses.} and illustrate the workflow of our methodology in Fig.~\ref{fig:tikitaka}. \rv{To the best of our knowledge, we are the first to study 
defense mechanisms against adversarial attacks targeting NID systems.} 

\section{Related Work}
DNN are increasingly used  or NID purposes, as they help minimize the effort of feature engineering, and operate with high detection accuracy \cite{buczak2015survey}. However, recent research suggests that there exist loopholes that can degrade their high performance, as perturbation added to their input can trigger traffic misclassification~\cite{heaven2019why, zhang2018deep}. Thus, defending deep learning based approaches from adversarial samples becomes a crucial issue for network security.

\subsection{Attacking Deep Learning Driven NIDS}
The majority of existing methods that employ adversarial samples to compromise classifiers target image applications (\eg \cite{li2019nattack, moon2019parsimonious}). Research on evading deep learning driven NIDS is scarce. Wang \etal employ four sets of white-box attack algorithms designed for image classification, to bypass MLP-based intrusion detectors trained on the NSL-KDD dataset \cite{wang2018deep22222}. Their experiments suggest that these attack algorithms are transferable to the NID domain and the MLP detectors are vulnerable to adversarial samples. However, since attackers usually do not have access to the neural model underpinning targeted NIDS, such settings are only useful to NIDS designers to assess their robustness \cite{kuppa2019black}.

Yang \etal generate adversarial samples in black-box settings \cite{yang2018adversarial} using three types of approaches, namely surrogate models \cite{carlini2017towards}, Zeroth Order Optimization (ZOO)-based \cite{chen2017zoo}, and Generative Adversarial Model (GAN)-based \cite{arjovsky2017wasserstein}. These methods can decrease the detection rate of the MLP-based detector, thus becoming a threat to NIDS. Kuppa \etal consider a more realistic situation, performing black-box attacks against different deep learning based detectors under decision-based and query-limited setting \cite{kuppa2019black}. By learning and approximating the distribution of benign and anomalous samples, these methods can evade NID with high success rate.

\subsection{Defending from Adversarial Samples}
There exist a range of strategies to defend deep learning models from adversarial examples. Commonly used methods include Network Distillation \cite{papernot2016distillation}, Adversarial Training \cite{tramer2018ensemble}, Adversarial Detecting \cite{lu2017safetynet}, Input Reconstruction \cite{defensegan}, Classifier Robustifying \cite{abbasi2017robustness}, Network Verification  \cite{katz2017reluplex}, and an ensemble of them \cite{meng2017magnet,yuan2019adversarial}, which counteract adversarial samples either reactively or proactively. 

Network distillation methods employ a student neural network to learn knowledge from a more complex teacher network. With this approach, the student network generalizes better and becomes more robust to the adversarial samples. Adversarial training retrains the neural networks by augmenting the original training set with adversarial samples, such that they can better defend against those inputs with subtle feature perturbations. Input reconstruction reduces the effectiveness of the perturbations by recovering the original input. Classifier robustifying employs various approaches (\eg model ensembling) to improve the robustness of the original classifier. Network verification uses an additional classifier to identify adversarial samples.

While these approaches are effective in the computer vision and natural language processing domains, none of them target defending against adversarial samples in NIDS, which is precisely the problem we tackle in this chapter.

\section{Dataset}
We conduct all experiments (\ie NID, black-box adversarial attacks, and with the proposed defenses) using the publicly available CSE-CIC-IDS2018 dataset~\cite{sharafaldin2018toward}. This encloses 14 types of artificially generated network intrusion traffic flows along with benign traffic. The network attack traffic can be categorized into seven classes, namely Brute-force, Heartbleed, Botnet, Denial of Service (DoS), Distributed Denial of Service (DDoS), Web attacks, and infiltration of the network from inside. Table~\ref{tab:dataset} summarizes the prevalence of these types of traffic. The infrastructure employed for attacks includes 50 machines, which attempt to intrude a victim network that consists of 420 end hosts and 30 servers. 

A total of 80 features of the traffic flows are extracted to perform intrusion detection and we filter 65 of them for the purpose of our work. The features selected can be grouped into 8 classes, specifically \emph{(a)} Forward Inter Arrival Time -- the time between two packets sent in the forward direction (mean, min, max, std); \emph{(b)} Backward Inter Arrival Time -- the time between two packets sent backwards (mean, min, max, std); \emph{(c)} Flow Inter Arrival Time -- the time between two packets sent in either direction (mean, min, max, std); \emph{(d)} Active-Idle Time -- the amount of time a flow was idle before becoming active (mean, min, max, std) and the amount of time a flow was active before becoming idle (mean, min, max, std); \emph{(e)} Flags based features -- the number of times the URG, PSH flags are set, both in the forward and backward direction; \emph{(f)} Flow characteristics -- bytes per second, packets per second, flow length (mean, min, max, std) and ratio between number of bytes sent downlink and uplink; \emph{(g)} Packet count with flags FIN, SYN, RST, PUSH, ACK, URG, CWE and ECE; \emph{(h)} Average number of bytes and packets sent in forward/backward directions in the initial window, bulk rate, and sub flows count. As `NaN' and `Inf' values exist in the dataset, we processed these 0 and the maximum values of a specific feature, respectively.

We train all deep learning models, implement and defend against the black-box adversarial attacks using the selected features. 

\begin{table}[t]
\centering
\caption{Statistics of the CSE-CIC-IDS2018 dataset employed. 
\label{tab:dataset}}
\begin{tabular}{|c|c|c|}
\hline
\textbf{Flow Type}       & \textbf{Number of Instances} & \textbf{Ratio} \\ \hline
Benign                   & 14,097,779                     & 83.6861\%      \\ \hline
Bot                      & 286,191                       & 1.6989\%       \\ \hline
DoS attack-SlowHTTPTest & 139,890                       & 0.8304\%       \\ \hline
DoS attacks-Hulk         & 461,912                       & 2.7420\%       \\ \hline
Brute Force-XSS         & 230                          & 0.0014\%       \\ \hline
SQL Injection            & 87                           & 0.0005\%       \\ \hline
Infiltration            & 161,934                       & 0.9613\%       \\ \hline
DoS attack-GoldenEye    & 41,508                        & 0.2464\%       \\ \hline
DoS attack-Slowloris    & 10,990                        & 0.0652\%       \\ \hline
Brute Force-Web          & 611                          & 0.0036\%       \\ \hline
FTP-Brute Force           & 193,360                       & 1.1478\%       \\ \hline
SSH-Brute Force           & 187,589                       & 1.1136\%       \\ \hline
DDoS attack-LOIC-UDP     & 1,730                         & 0.0103\%       \\ \hline
DDoS attack-HOIC         & 686,012                       & 4.0723\%       \\ \hline
DDoS attack-LOIC-HTTP   & 576,191                       & 3.4203\%       \\ \hline
All of the above attacks        & 2,748,235                      & 16.3139\%      \\ \hline
Total                    & 16,846,014                     & 100\%          \\ \hline
\end{tabular}
\end{table}

\section{Training Network Intrusion Detectors\label{sec:nid}}
\begin{figure}[]
\centering
\includegraphics[width=\columnwidth]{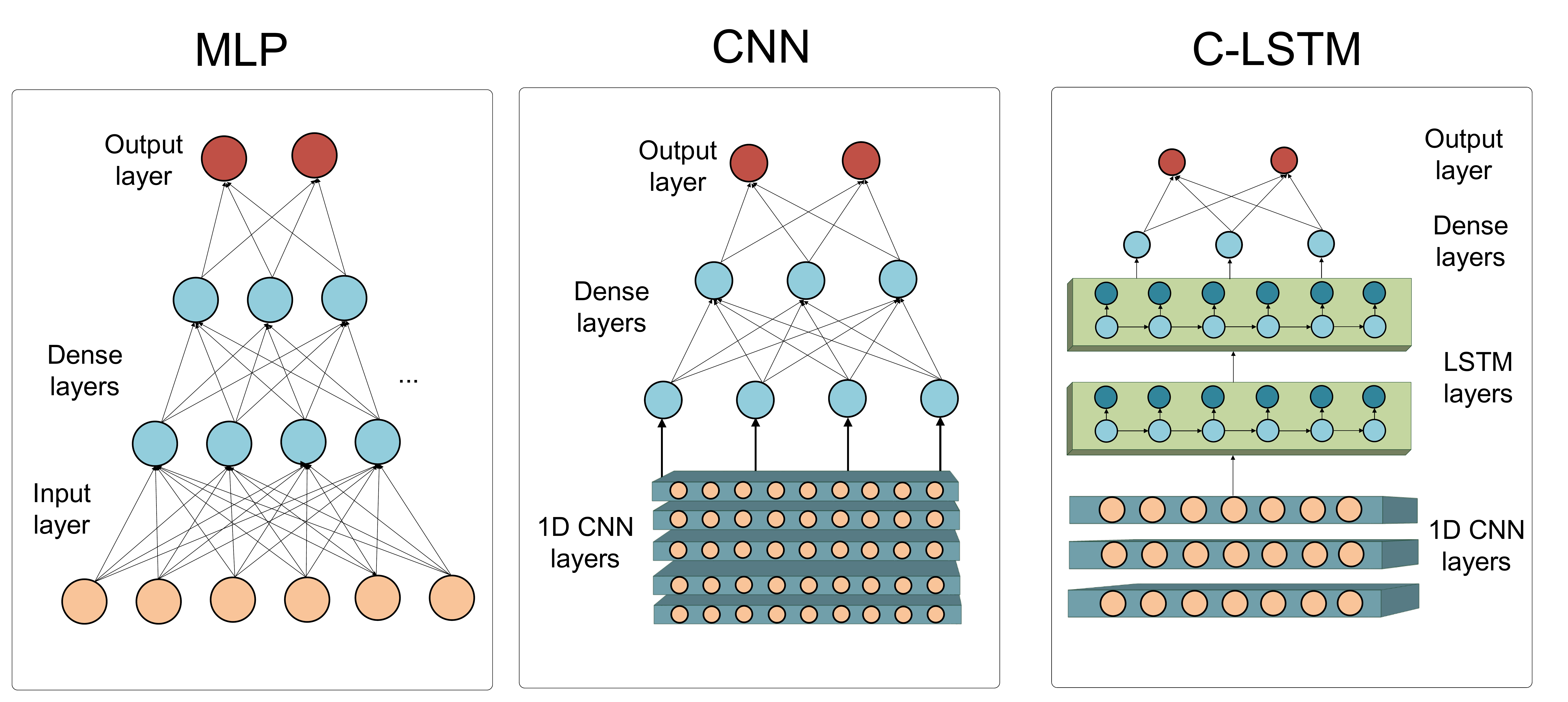}
\caption{Architectures of the deep learning based NID models employed in this study. 
\label{fig:model_nid}}
\end{figure}

Training accurate deep network intrusion detectors is the initial important step of the study, as \tita builds on the pre-trained NID models. To this end, we employ 3 well-known deep learning architectures, namely MLP \cite{faker2019intrusion}, CNN \cite{zhang2019pccn}, and CNN with LSTM layers, \ie C-LSTM \cite{zhang2019network}. These models are frequently used for NID purposes and have achieved notable performance. We illustrate the specific architectures of all models considered in Fig~\ref{fig:model_nid}. 

The MLP is the most simple deep learning architecture, which employs multiple stacks of fully-connected layers for features extraction. It is particularly suitable for handling traffic flows that have mixture types of features and ranges. In our study, we construct an MLP with 3 hidden layers. Each layer has 200 units, except for the last hidden layer, which has 400 units. CNN have good spatial perception abilities and have demonstrated remarkable precision in NID tasks \cite{zhang2019pccn}. In this work, we design a CNN with 10 one-dimensional CNN layers, each equipped with 108 filters, with filter size of 5. Lastly, we replicate the C-LSTM employed in \cite{zhang2019network}, with our C-LSTM operating on the features that characterize the traffic, instead of the raw flows. The C-LSTM combines CNN and LSTM structures to extract spatial and temporal features separately. Data will be first processed by a CNN with 5 hidden layers, then passed to a 2-layers LSTM for the final predictions. Each LSTM layer has 160 units. In this study, we perform NID, black-box adversarial attacks and defend against them based on these models, as we will detail in the next sections.

We consider NID in two different scenarios, namely \emph{(i)} one-to-all detection and \emph{(ii)} one-to-one detection. The one-to-all scenario groups all types of attack into a single ``anomaly'' class, which leads to a supervised binary classification problem. Instead, one-to-one detectors separate each network attack type (14 in total) into individual classes, and perform multi-class classification. In our study, the same neural network architectures are employed for both scenarios, except for changes in the final layers, as their number directly depends on the number of classes considered for identification. We train and validate all models using 80\% of the dataset and test on 20\% of it, as customary. All models are trained via minimizing the cross-entropy loss function through the Adam optimizer \cite{kingma2015adam}. Super-sampling is employed to reduce the effect of class imbalance between benign and malicious traffic, which is inherent to the dataset.

All models are trained and evaluated on a parallel computing cluster equipped with one or multiple Nvidia TITAN X, Tesla M40 or/and Tesla P100 GPUs. The neural models are implemented in Python using the Tensorflow \cite{tensorflow2015-whitepaper} and TensorLayer \cite{tensorlayer} packages.

\subsection{One-to-all NID}
We quantify the performance of the NIDS using four performance metrics, namely accuracy, precision, recall and F1 score, as shown in Table~\ref{tab:12a}. These metrics are frequently employed for the evaluation of binary classifiers, \rv{and they are computed over the positive (anomaly) class}.

\begin{table}[b]
\centering
\caption{The detection performance of MLP, CNN, and C-LSTM in the one-to-all scenario.\label{tab:12a} }
\begin{tabular}{|c|c|c|c|c|}
\hline
    & Accuracy & Precision & Recall & F1 score \\ \hline
MLP & 0.987    & 0.968     & 0.954  & 0.961    \\ \hline
CNN & 0.987    & 0.968     & 0.953  & 0.960    \\ \hline
C-LSTM & 0.987    & 0.967     & 0.952  & 0.960    \\ \hline
\end{tabular}
\end{table}
Observe that all models achieve high detection performance, as all F1 scores are above 0.960. In addition, the performance of the three models considered is similar, since the difference between the F1 scores attained by each never exceed 0.01. This matches the performance of state-of-the-art deep learning based NID solutions, thus the models can be considered ``reliable''. 

\subsection{One-to-one NID}
\begin{figure*}[]
\centering
\includegraphics[width=1.1\textwidth]{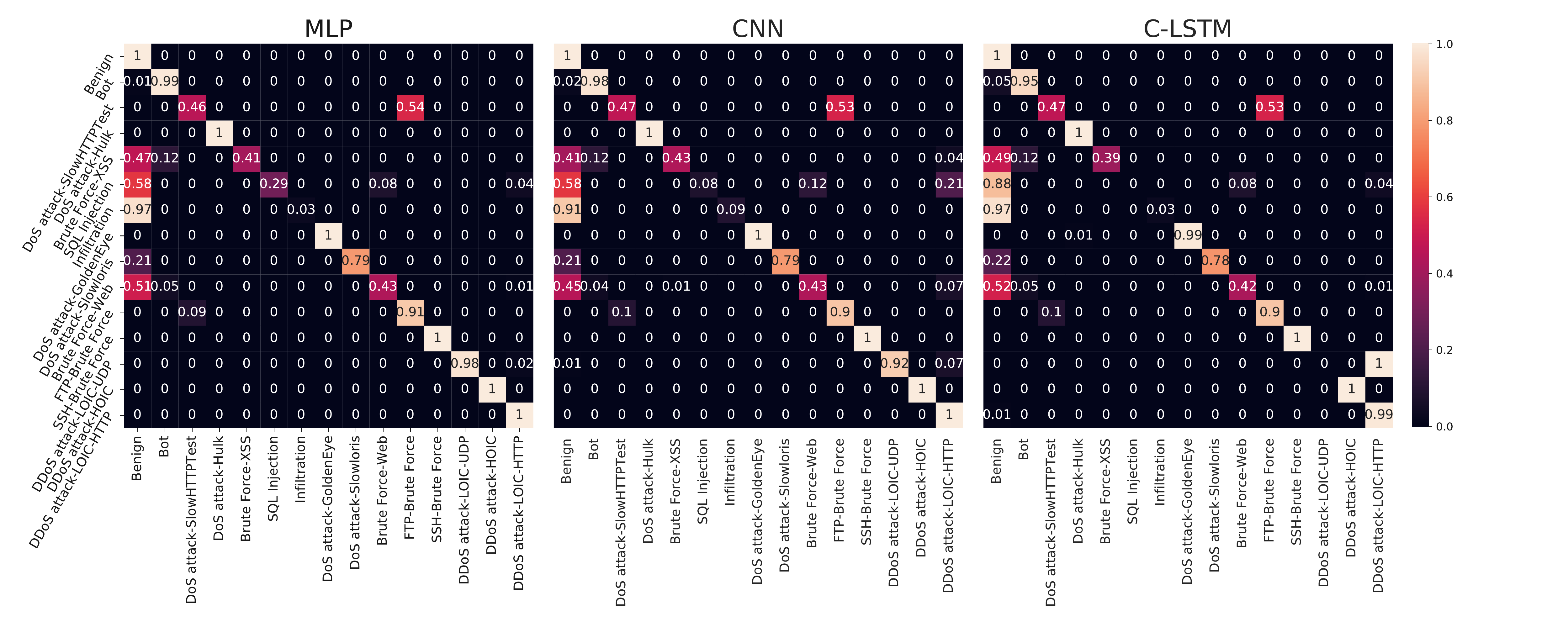}
\caption{Confusion matrices of the MLP, CNN, and C-LSTM models for the one-to-one NID. 
\label{fig:cm}}
\end{figure*}

One-to-one NIDS aim at classifying each traffic flow into 14 types of anomalies and benign. We employ the same neural network architectures to assess their detection performance also in this scenarios, this time resorting to normalized confusion matrices, as shown in Fig.~\ref{fig:cm}. The diagonal elements represent ratios of points for which the predicted label is equal to the true label, while off-diagonal elements are indicate misclassification ratios \cite{powers2011evaluation}. Therefore, the elements of each row sum up to 1. The higher the diagonal values of the confusion matrix, the higher the performance, indicating many correct predictions. 

Observe that all NID models achieve high detection accuracy for most types of anomalies, as diagonal values are close to 1. However, taking a closer look at the Brute Force-XSS, SQL Injection, Infiltration, and Brute Force-Web attacks, it appears the NID models usually fail to identify those and tend to misclassify them as ``benign''. In addition, all neural networks face difficulties in dealing with DoS attacks-SlowHTTPTest and FTP-BruteForce, as they mix them roughly 50/50. Further, the C-LSTM misclassifies almost all DDoS attack-HOIC traffic as DDoS attacks-LOIC-HTTP. This is perhaps less critical though, since both these attacks belong to the DDoS category. 
Overall, MLP, CNN and C-LSTM attain classification accuracies of 98.4\%, 98.3\%, and 98.3\% respectively, which are fairly close to the performance observed in the one-to-all scenario.

In what follows, we will demonstrate that although the NID performance of these solutions seems reliable in terms of detection accuracy, they can be easily compromised through a sequence of perturbations and queries without requiring to have information about the underlying models. 

\section{Black-box Attacks against NIDS\label{sec:bba}}
The objective of adversarial attacks is to generate adversarial samples, by adding small perturbations to the input given to NIDS, so as to cause misclassification. We denote by $x$ the input to a classifier (\ie features extracted from traffic flows), an adversarial sample as $x_{\textup{adv}} = x + \sigma_{x}$, and the targeted class as $y_{\textup{adv}}$. The objective of the adversarial attacks can be formulated as finding $x_{\textup{adv}}$ such that $||x_{\textup{adv}} - x||_\infty<\epsilon$ and $x_{\textup{adv}}$ is classified as $y_{\textup{adv}}$. Here, $\sigma_{x}$ is the perturbation added to the input and $\epsilon$ limits the scale of the perturbation~\cite{ilyas2018black}.

\subsection{Threat Model \label{sec:attack}}
Typical attacks against machine learning models can be categorized into three classes, namely \emph{(i)} white-box attacks, \emph{(ii)} grey-box attacks, and \emph{(iii)} black-box attacks. White-box and grey-box attacks assume the malicious actors have access to the training data or/and model structures. Such hypotheses apply in cases where system designers seek to improve the robustness of their NIDS, but rarely to external adversaries. Instead, potential hackers are forced to treat an NIDS as a black-box, since the details of a victim system's inner workings remain hidden and the only way in which the NIDS behavior can be learned is through a sequence of queries and the feedback received. This is also the practical threat model that we consider in this work.

\begin{figure}[]
\includegraphics[trim=40 1 1 1, width=\columnwidth]{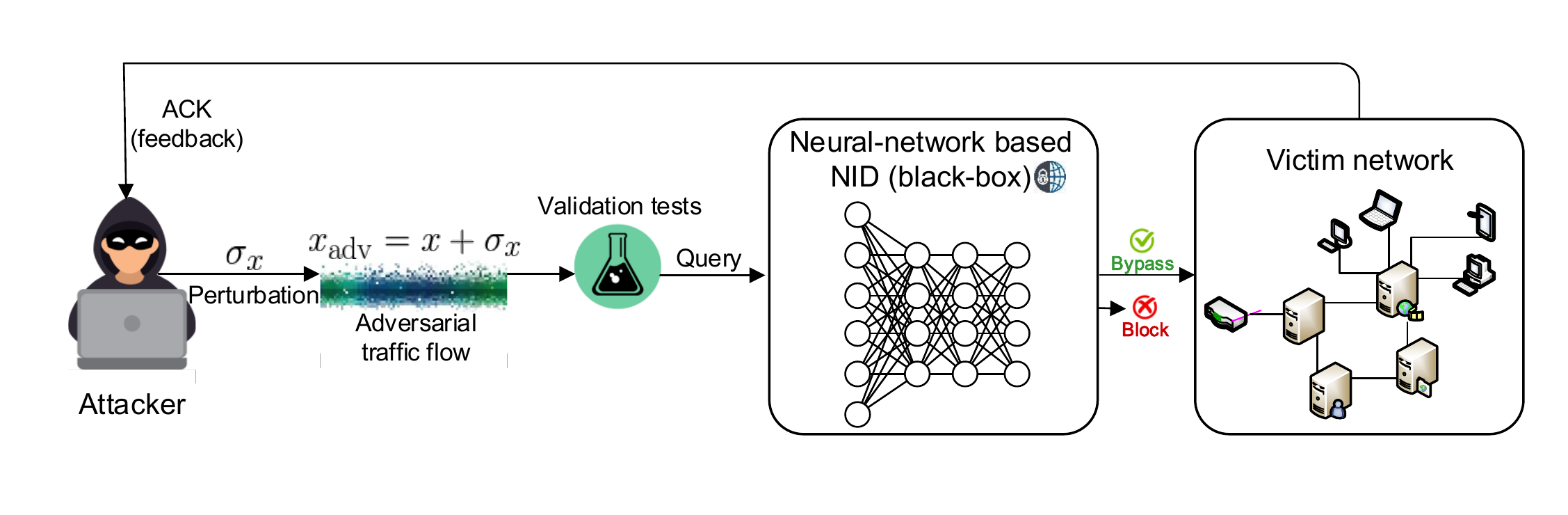}
\caption{An illustration of the black-box attack process against machine learning-based NID models.
\label{fig:bba}}
\end{figure}

Specifically, an attacker may send a traffic flow towards the target network, which will be first examined by a NID model. This is known as a query process. Subsequently, the attacker will receive implicit/explicit feedback from the model, \eg an ACK packet, which reflects whether the traffic flow was classified as an anomaly. Based on the feedback, the attacker can adjust and apply subtle perturbations to the malicious traffic flow, thereby producing adversarial samples that eventually may compromise the effectiveness of the NIDS, which will end up classifying malicious traffic as benign. On the other hand, the attacker may not have confidence about the exact decision class decided by the NIDS.
We illustrate this black-box attack process against NID models in Fig.~\ref{fig:bba}. 

\subsection{Domain Constraints\label{sec:constrains}}
Unlike adversarial attack against image classifiers, adversarial samples against NIDS must respect certain domain constrains \cite{kuppa2019black}, such that preserving the functionality and intactness of the samples when introducing perturbations $\sigma_{x}$. This means that \emph{(i)} only a subset of features are amendable; and \emph{(ii)} features of the adversarial samples do not violate the properties inherent to the original samples. To meet these requirements, we only select 22 time-based features to which we add perturbations, as also suggested in \cite{kuppa2019black}. These include \emph{(a)} Forward Inter Arrival Time -- the time between two packets sent forward direction (mean, min, max, std); \emph{(b)} Backward Inter Arrival Time -- the time between two packets sent backwards (mean, min, max, std);  \emph{(c)} Active-Idle Time, \emph{(d)} Average number of bytes and packets sent in forward and backward directions in the initial window or/and sub-flows. Features outside this subset remain unchanged during the attack process, which is inline with recent research confirming that adversarial samples can be constructed effectively by perturbing only a small subset of the input features \cite{su2019one}. In addition, 3 features (\ie flow duration, total time between two packets sent in the forward and backward directions) that may be affected by the perturbations are discarded from the NID model training, attacks, and defense mechanisms proposed.
 
In addition, we further impose that \emph{(i)} the Mean Absolute Percentage Error (MAPE) for each feature $k$ does not exceed 20\%, \ie $100\cdot |(x^{(k})-x^{(k)}_{\textup{adv}})/{x^{(k)}}|\leq 20\%$; \emph{(ii)} the perturbed features preserve the mean property (\eg the mean forward inter arrival time) plus/minus std features do not exceed their corresponding max and min features; \emph{(iii)} the sign of each perturbed sample remains the same as that of the original; and \emph{(iv)} if the std feature is zero, the corresponding mean, max, min and std features remain unchanged in the adversarial sample. Samples that violate these constrains will be automatically rejected and characterized as unsuccessful trials. These validation tests ensure that the generated adversarial samples preserve the originally intended functionality.

\subsection{Black-box Attack Methods}

We consider 5 state-of-the-art black-box attack approaches to generate adversarial samples and compromise the pretrained deep anomaly detectors. These include \emph{(i)} Natural Evolution Strategies (\nes) \cite{ilyas2018black}, \emph{(ii)} \bd Attack \cite{brendel2018decision}, \emph{(iii)}~\pw Attack \cite{schott2018towards}, \emph{(iv)} \hj~\cite{chen2019boundary++}, and \emph{(v)} \opt \cite{cheng2018queryefficient}, all of which were originally designed to attack image classifiers.  \\

\noindent \textbf{\nes \cite{ilyas2018black}} are black-box gradient estimation methods for machine learning models. Estimated gradients can be used for projected gradient descent (as used in white-box attacks) to construct adversarial examples. This approach does not require a surrogate network, thus it is more query-efficient and reliable in crafting adversarial examples for black-box attacks. Notably, \nes work well in decision-based settings, which makes them suitable for attacks against NID models. \\

\noindent \textbf{\bd Attack \cite{brendel2018decision}} is a method that generates adversarial examples in black-box and decision-based settings. It follows the decision boundary between adversarial and non-adversarial samples via rejection sampling. At each step, it employs constrained i.i.d. samples following a Gaussian distribution, starting from a large perturbation and successively reducing this until successful. This attack is highly flexible and can accommodate a set of adversarial criteria.\\

\noindent \textbf{\pw Attack \cite{schott2018towards}} is a simple decision-based attack method that greedily minimizes the $L_0$-norm between raw and adversarial samples. In image applications, it first introduces salt-and-pepper noise until misclassification, and then repeatedly iterates over each perturbed pixel, resetting it to the initial value if the perturbed image remains adversarial. We implement a similar approach to attack the NID models, but substitute the salt-and-pepper noise with additive Gaussian noise, to better suit network traffic.\\

\noindent \textbf{\hj \cite{chen2019boundary++}} is a hyperparameter-free, query-efficient attack method, which consists of three main steps, namely \emph{(i)} estimation of the gradient direction, \emph{(ii)}~step-size search via geometric progression, and \emph{(iii)} boundary search via a binary search approach. It is applicable to more complex settings, such as non-differentiable models or discrete input transformations, and achieves competitive performance against several defense mechanisms.\\

\noindent \textbf{\opt \cite{cheng2018queryefficient}} projects the decision-based attack into a continuous optimization problem and solves it via randomized zeroth-order gradient update. In particular, a Random Gradient-Free (RGF) method is employed to find appropriate perturbations and converge to stationary points. Since \opt does not rely on gradients, it can attack other non-differentiable classifiers besides neural networks, \eg Gradient Boosting Decision Trees.\\

We employ a modified version of the mean absolute percentage error to quantify the deviation between each unmodified sample $x$ and its adversarial version $x_{\textup{adv}}$, \ie
\begin{equation}\label{eq:MAPE}
    \mathbf{MAPE}(x, x_{\textup{adv}}) = \frac{100\%}{N}\sum_{k=1}^N\left |\frac{x^{(k)}-x^{(k)}_{\textup{adv}}}{x^{(k)}}\right |,
\end{equation}
where $N$ is the total number of perturbed features in $x$ and $x^{(k)}$, $x^{(k)}_{\textup{adv}}$ are the $k$\textsuperscript{th} perturbed features of the original and adversarial samples respectively. Smaller MAPE indicates higher similarity between the raw input $x$ and the adversarial sample $x_{\textup{adv}}$. 

\begin{table}[t]
\centering
\caption{Statistics of the dataset used to generate adversarial samples. \label{tab:adv_dataset}}
\begin{tabular}{|l|r|r|}
\hline
\textbf{Attack Type}     & \textbf{\begin{tabular}[c]{@{}c@{}}Number of \\ Instances\end{tabular}} & \textbf{Ratio [\%]} \\ \hline
Bot                      & 5,217                         & 10.434       \\ \hline
DoS attack-SlowHTTPTest & 5,217                         & 10.434       \\ \hline
DoS attack-Hulk         & 5,217                         & 10.434       \\ \hline
Brute Force-XSS         & 51                           & 0.102        \\ \hline
SQL Injection            & 24                           & 0.048        \\ \hline
Infiltration            & 5,217                         & 10.434       \\ \hline
DoS attack-GoldenEye    & 5,217                         & 10.434       \\ \hline
DoS attack-Slowloris    & 2,475                         & 4.950        \\ \hline
Brute Force-Web          & 117                          & 0.234        \\ \hline
FTP-Brute Force           & 5,217                         & 10.434       \\ \hline
SSH-Brute Force           & 5,217                         & 10.434       \\ \hline
DDoS attack-LOIC-UDP     & 383                          & 0.766        \\ \hline
DDoS attack-HOIC         & 5,217                         & 10.434       \\ \hline
DDoS attack-LOIC-HTTP   & 5,214                         & 10.428       \\ \hline
Total                    & 50,000                        & 100.000          \\ \hline
\end{tabular}
\end{table}

\subsection{Attack Performance}
We randomly select 50,000 malicious traffic flows from the test set to craft adversarial samples. We summarize the statistics of these samples in Table~\ref{tab:adv_dataset}. We quantify the attack performance of each approach using 4 performance metrics, namely Attack Success Rate (ASR), average benign confidence, MAPE, and average number of queries. The ASR is measured by the ratio between the number successful adversarial samples and the total attack attempts (50,000). This can be an indicator of the effectiveness of an attack approach. An attack attempt is successful if and only if the attack algorithm converges, and the adversarial samples meet the aforementioned constraints. The average benign confidence denotes the probability that the model predicts an adversarial sample $x_{\textup{adv}}$ as benign. Higher confidence implies that the model is more confident about the decision made over a sample. The MAPE is defined in Eq.(\ref{eq:MAPE}) and is only computed over 22 features that allow perturbations. Recall that lower MAPE represent higher similarity between the raw and adversarial samples. The number of queries indicates how many attempts an attacker should perform in order to generate a successful adversarial sample. This can be used to measure the efficiency of an attack approach. Higher number of queries might trigger the NIDS, making the attack easier to be detected. Note that the MAPE, benign confidence and number of queries are averaged over the successful attack attempts for each attack approach and NID model. All attacks are conducted using the original implementations and the Foolbox Python package~\cite{rauber2017foolbox}. In what follows, we show the attack performance of the one-to-all and one-to-one scenarios in this section.

\subsubsection{Attack Performance in One-to-all Scenario}
\begin{figure}[t]
\centering
\includegraphics[width=\columnwidth]{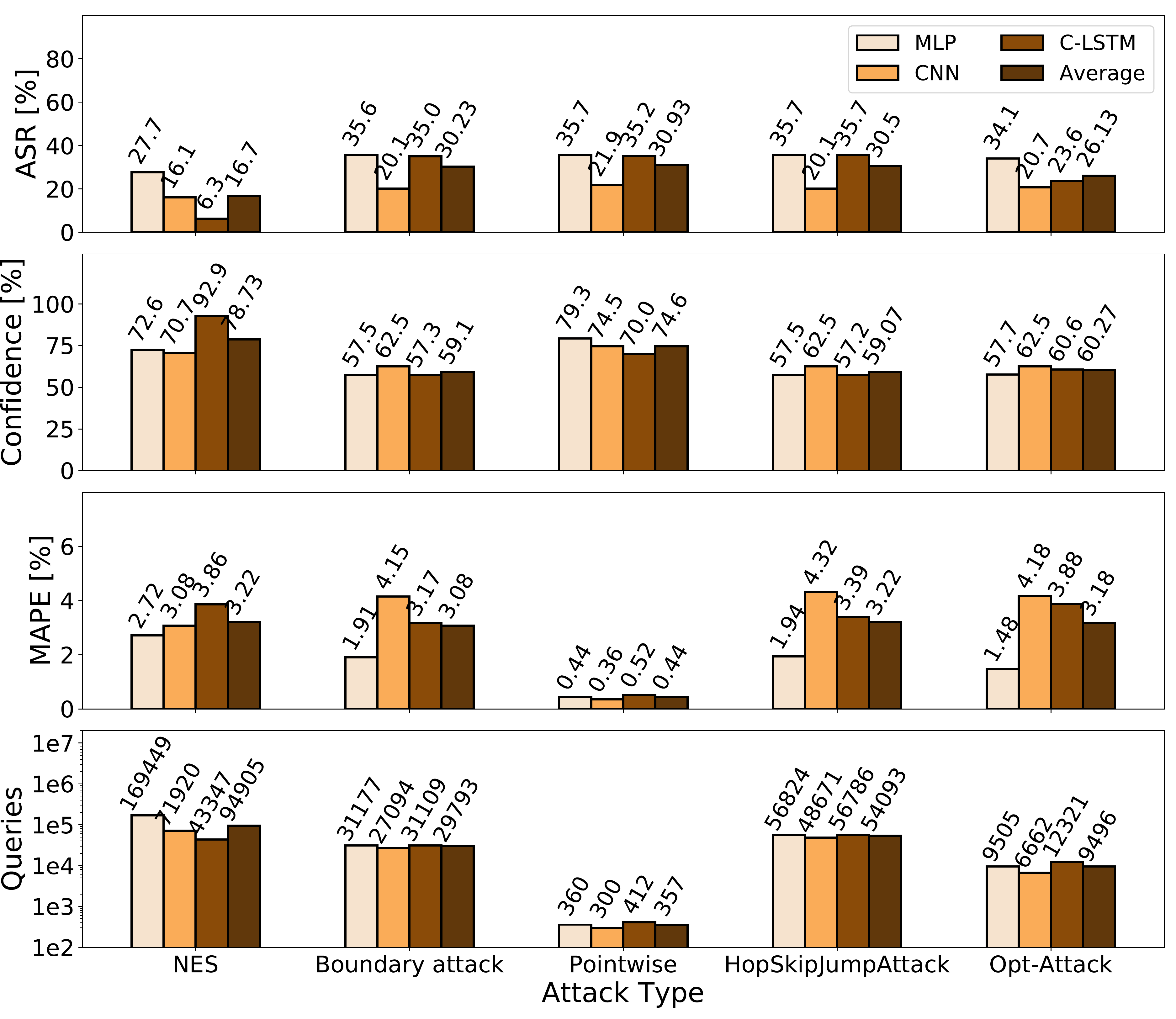}
\caption{ASRs, Confidence, MAPE and number of queries statistics of all attack approaches against 3 NID models considered in this study and their average values in the one-to-all scenario. 
\label{fig:one2all_stats}}
\end{figure}

We show the performance of all attack approaches in one-to-all detection scenarios in Fig.~\ref{fig:one2all_stats}. Observe that the \bd, \pw and \hj obtain similar performance in terms of success rate for all NID models. Worryingly, these approaches can generate adversarial samples with 30\% success rate on average, which make them to be a severe threat to the deep NID system. In particular, the \pw attack achieves highest benign confidence, lowest MAPE and requires fewest number of queries. This implies that, the \pw attack is highly efficient to generate adversarial samples, and more difficult to defend. Though less efficient and effective, the \nes attack generates adversarial samples with high benign confidence, therefore it appears the best cheater against the NID models. On the other hand, the performance of the \opt appears middlebrow, as it does not outstand from any metric. Turning attention to the horizon of NID models, CNN appears to be the most robust model against the black-box attack in the one-to-all scenario, as it gain the lowest ASR (19.78\%) among the three. In addition, attackers are required to made larger changes to the raw samples in order to cheat the CNN, as its average MAPE appears the highest compared to the MLP and C-LSTM. The benign confidence and number of queries are however similar at the model level.

\begin{figure*}[htb]
\centering
\includegraphics[width=\textwidth]{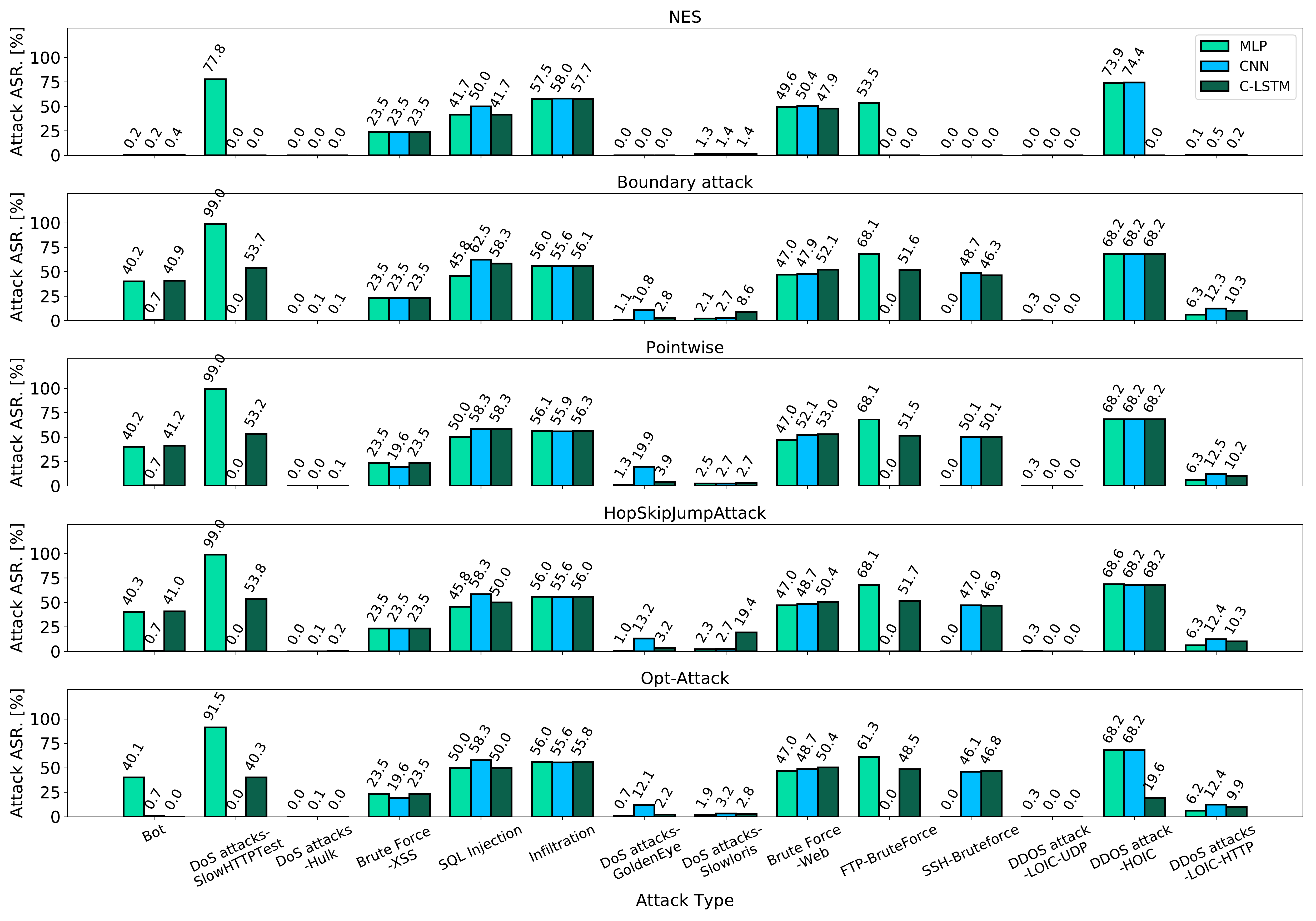}
\caption{ASRs over each type of malicious traffic against all NID models in the one-to-all scenario.
\label{fig:asr}}
\end{figure*}

We delve deeper into the attack performance, by showing the ASR over each type of malicious traffic flow in Fig.~\ref{fig:asr}. Observe that adversarial samples generated for the Dos Attacks with HTTP Unbearable Load King (Dos attacks Hulk) and DDos Low Orbit Ion Cannon UDP (DDos attacks-LOIC-UDP) attack are almost impossible the bypass the NIDs model, as the ASR for them are close to 0\% for all attack methods. On the contrary, adversarial samples for Brute Force-XSS, SQL Injection, Infiltration, Brute Force-Web and DDOS attack-HOIC appears easier to conquer the NID models. Notably, the robustness and vulnerability to a specific type of attack may vary from models. For example, no adversarial samples of the DoS attacks-SlowHTTPTest can bypass CNN, while its ASR for MLP are mostly over 90\%. This information can deliver useful insight for the network service provide to defend specific types of attacks.

\subsubsection{Attack Performance in One-to-one Scenario} 
\begin{figure}[htb]
\centering
\includegraphics[width=\columnwidth]{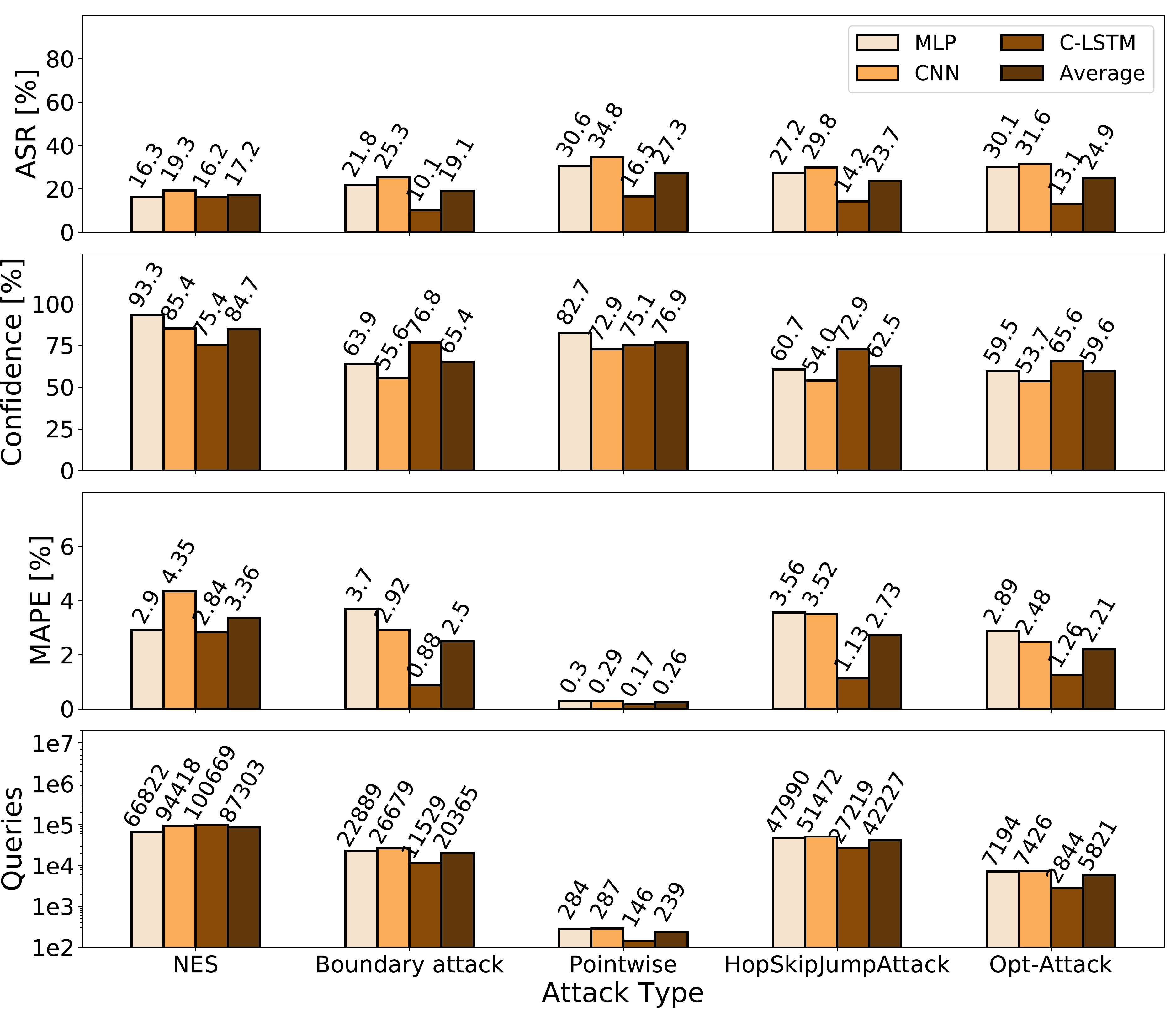}
\caption{ASRs, Confidence, MAPE and number of queries statistics of all attack approaches against 3 NID models considered in this study and their average values in the one-to-one scenario.\label{fig:one2one_stats}}
\end{figure}

We illustrate the statistics of each attack against the different NID models considered for one-to-one scenario in Fig.~\ref{fig:one2one_stats}. Observe that, except for the \nes where the performance is similar among the different NID models, the ASR varies differently among models for all the other attack methods (unlike in the one-to-all scenario discussed in the Appendix). This is because the models work with large number of classes which makes it difficult to craft adversarial samples to match the targeted `benign' label. The \pw method obtains the highest ASR, lowest MAPE, and lower average number of queries. This suggests that this approach is effective and efficient in one-to-one settings. The C-LSTM appears to be the most robust model against adversarial samples, as all attack methods attain the lowest ASR values against this NID model. Although achieving the highest benign confidence with adversarial samples, the \nes obtains the lowest ASR on average. In general, it also requires a large amount of queries to craft an adversarial sample.

\begin{figure*}[htb]
\centering
\includegraphics[width=\textwidth]{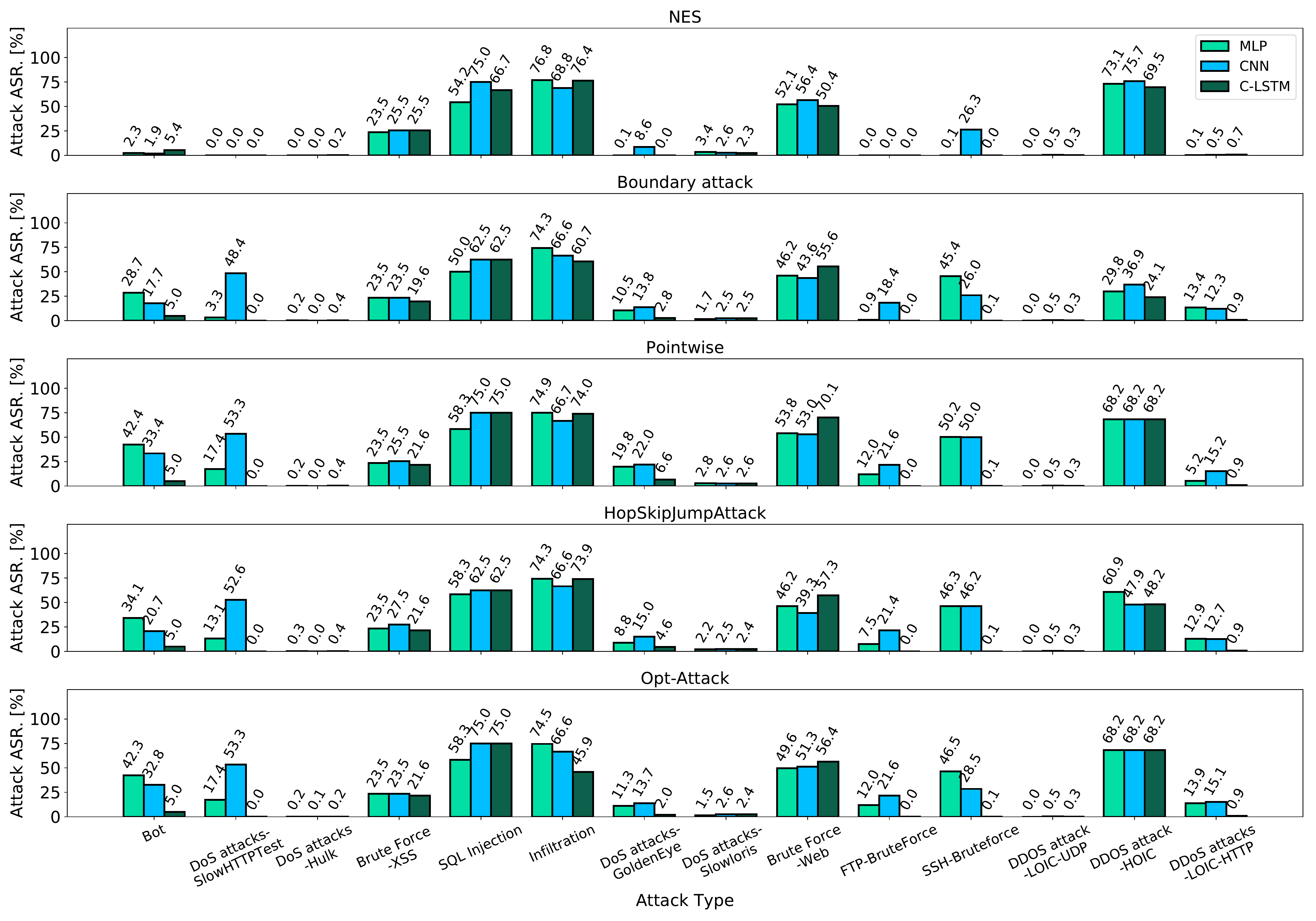}
\caption{ASRs over each type of attack against all NID models in the one-to-one scenario.
\label{fig:asr_mc}}
\end{figure*}

In Fig.~\ref{fig:asr_mc}, we show the ASR for each type of malicious traffic flow considered, in the same one-to-one scenario. Analyzing these results jointly with Fig.~\ref{fig:cm}, observe that anomalies with low detection rate (\ie Brute Force-XSS, SQL Injection, Infiltration, Brute Force-Web) are easier to be disguised by the attackers. This is because the models already have vague decision boundaries for these flow types, thus are easier to be gamed. Attackers obtain the lowest ASR when crafting adversarial samples based on DoS attacks-Hulk, -GoldenEye, -Slowloris, and DDoS attack-LOIC-UDP, as the NID models exhibit high detection rates over these anomalies. Overall, most of attacks achieve similar ASR performance as they obtain in the one-to-all scenario.

\begin{figure}[]
\centering
\includegraphics[width=\columnwidth]{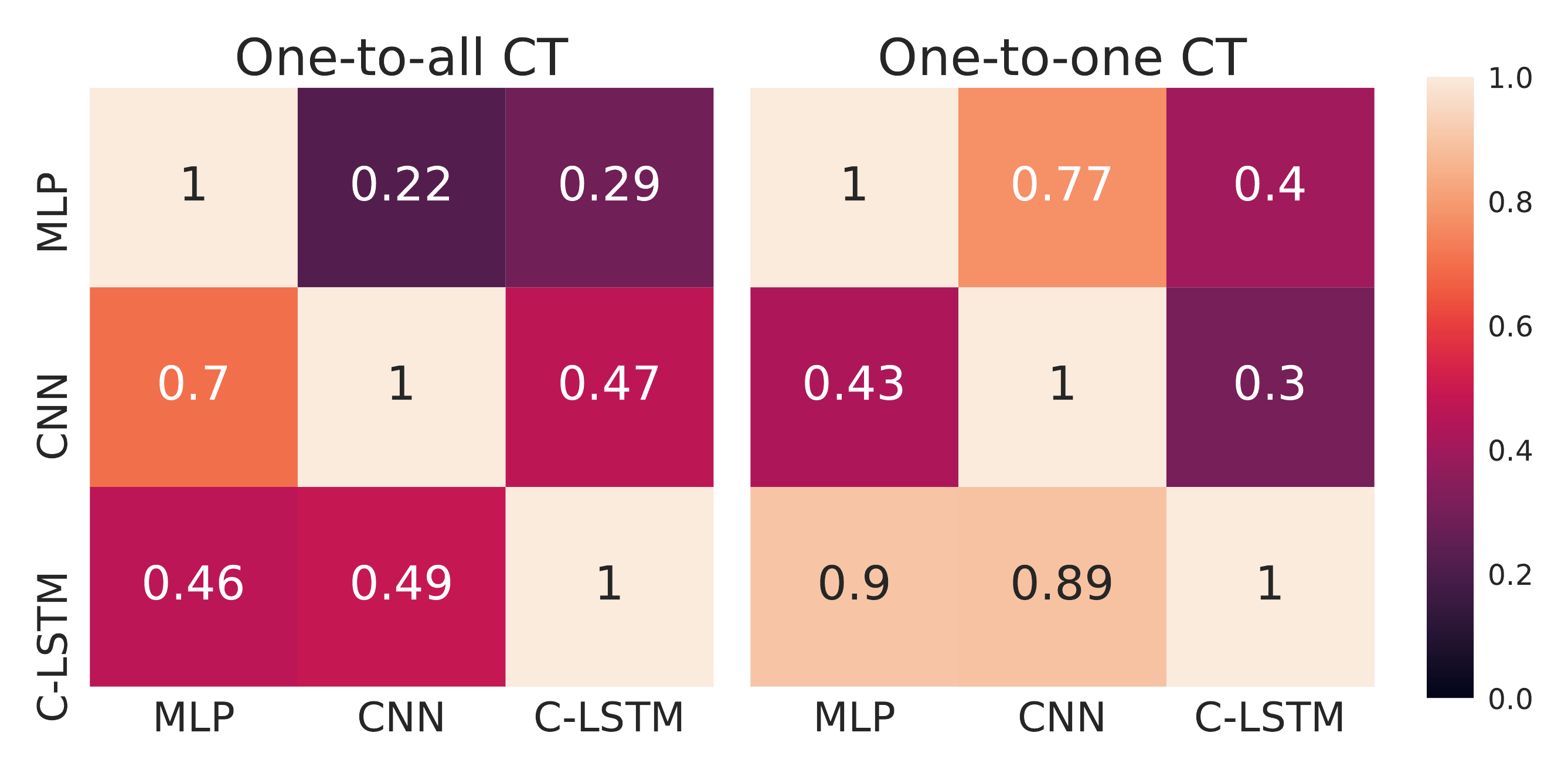}
\caption{The cross-transferability (CT) matrix across all NID models considered.
\label{fig:cv}}
\end{figure}
\subsubsection{Cross-Transferability \label{sec:c-v}} We define the transferred ratio $\varepsilon_{m_1}^{m_2} = \frac{K_{m_1}^{m_2}}{K_{m_1}}$ between models $m_1$ and $m_2$, to evaluate the transferability of adversarial samples across different NID models. Here, $K_{m_1}$ denotes the number of successful adversarial samples crafted from model $m_1$, while $K_{m_1}^{m_2}$ denotes the number of samples among $K_{m_1}$ that can bypass $m_2$ as well. We show the transferred ratios across models as Cross-Transferability (CT) matrices in Fig.~\ref{fig:cv}. The element in the $i$\textsuperscript{th} row and $j$\textsuperscript{th} column represents the value of the $\varepsilon_{m_i}^{m_j}$ (\eg the element at the first row of the second column denotes that 22\% of adversarial samples crafted from MLP can bypass CNN.) Note that $\varepsilon_{m}^{m} = 1$ holds for all $m$.

Observe that large proportion adversarial samples are transferable across NID models for both detection scenarios. This has been also confirmed in the adversarial attack against image classifiers \cite{42503}. This implies that, even with completely different structures, NID models enclose similar blemish, of which attackers can take advantage. For example, we assume that network service provider (NSP) employs the C-LSTM model to perform one-to-one NID. To compromise the system, attackers manage to craft $K$ successful adversarial malicious traffic flows. After being detected, the NSP change the NID model to MLP. Nevertheless, 90\% of the old adversarial traffic can still bypass the new NID system due to the high cross-transferability. This becomes a weakness of deep NID models, which can be exploited by attackers.

Overall, we successfully craft 201,482 adversarial samples for all one-to-all detection models, and respectively 168,454 adversarial samples for all one-to-one models using 5 different attack methods. Among them, 58,928 sample can bypass all one-to-all models, and 66,638 can bypass all three one-to-one models. These account for 29.2\%  and 35.0\% of the total successful adversarial samples for each scenario respectively. 

\subsection{Adversarial Samples Analysis}
\begin{figure*}[htb]
\centering
\includegraphics[width=\textwidth]{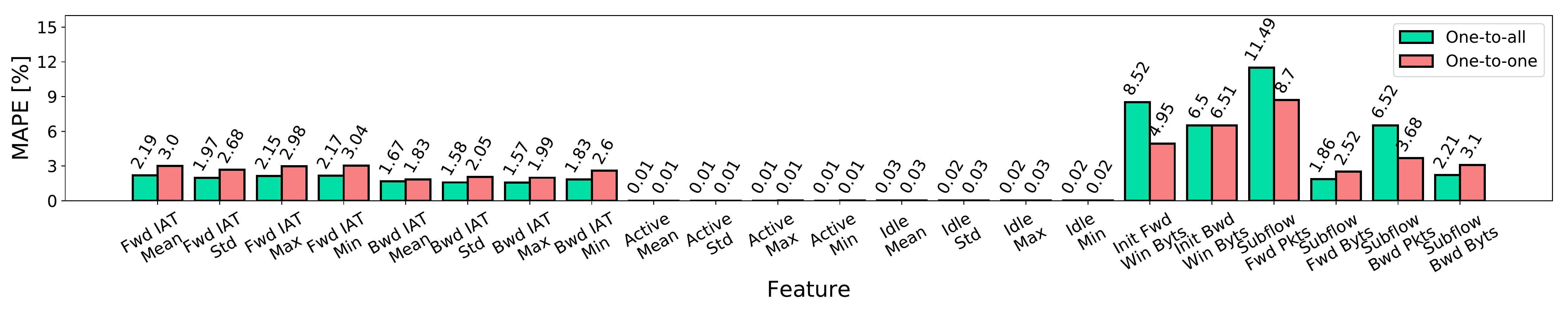}
\caption{The MAPE between original and adversarial samples of each feature that allows perturbations. \label{fig:feature_mape}}
\end{figure*}
\subsubsection{Feature-wise MAPE} We delve deeper into the adversarial samples generated, by showing in Fig.~\ref{fig:feature_mape} the average MAPE of each perturbed feature on all successful attack samples across all NID models and attack approaches. Observe that for both detection scenarios, the Active/Idle Time (\ie the time a flow was idle before becoming active (mean, min, max, std) and amount of time a flow was active before becoming active (mean, min, max, std)) are less affected, as the related features remain almost unchanged in the attack process. In contrast, features that characterize average number of bytes and packets sent in forward and backward direction in the initial window or/and sub flows, are perturbed more significantly. This indicates that these features are the most influential in the decision of NID models, and therefore more likely to be exploited by potential attackers.

\subsubsection{t-SNE Visualization} 
\begin{figure*}[htb]
\centering
\includegraphics[width=\textwidth]{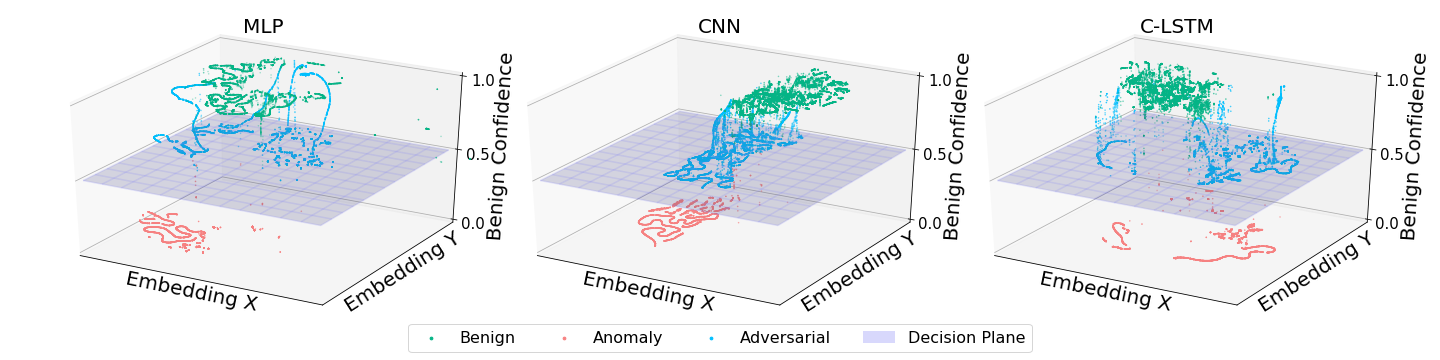}
\caption{Two-dimensional ($x, y$ axis) t-SNE embedding of the representations in the
last hidden layer, along with the benign confidence ($z$ axis) of each NID model. Data generated using 10,000 benign samples (blue), 10,000 adversarial samples (green) produced by all attack methods, and their corresponding original malicious samples (pink). \label{fig:tsne}}
\end{figure*}

We also investigate the inner workings of each NID model, by visualizing the output embedding of their hidden layers, so as to understand better how a neural network ``thinks'' of the benign, malicious, and adversarial samples. To this end, we adopt the t-distributed Stochastic Neighbor Embedding (t-SNE) \cite{maaten2008visualizing} to reduce the dimension of the last hidden layer of each model to 2. In Fig.~\ref{fig:tsne}, we plot the t-SNE embedding ($x, y$ axis) of their hidden representations of 10,000 benign samples (blue), 10,000 adversarial samples (green) generated by each attack method, and their corresponding abnormal samples used to craft them (pink), along with their benign confidence ($z$ axis). Note that the sample will be considered as benign if and only if the the benign confidence is greater than 0.5 (above the decision plane). Typically, the t-SNE approach organizes data points that have higher similarity into nearby embeddings \cite{zhang2018driver}. It can therefore reflect how the model ``thinks'' of the data samples, as similar data representations will be clustered. 

Observe that abnormal samples can be clearly distinguished from benign samples by their t-SNE embeddings for all NID models. The purpose of adversarial attacks is to cause misclassification by bringing abnormal samples across the decision boundary. This is reflected in the Fig.~\ref{fig:tsne}, as the t-SNE embedding of adversarial samples are moved closer to the benign embedding cluster, while they remain anomalies in nature. This successfully confuses the NID models, making the adversarial samples indistinguishable. In addition, adversarial samples with higher benign confidence are in general closer to the benign embedding cluster. This is especially clear in the t-SNE embedding produced by the CNN.

\section{Defending Against Black-box Adversarial Samples\label{sec:def}}
Defense mechanisms against adversarial attacks should improve the robustness of deep learning models to adversarial samples, such that they become less likely to be compromised and the ASR of different attacks is reduced. In general, countermeasures for adversarial examples can be categorized into two types \cite{yuan2019adversarial}: \emph{(i)}~\textbf{Reactive} -- detecting adversarial examples after deep neural networks have been trained; and \emph{(ii)}~\textbf{Proactive} -- improving the robustness of deep neural network models against adversarial examples. In this chapter, we propose three different defense mechanisms, and combine them to counteract the adversarial samples generated by the black-box attack methods discussed in the previous section. These defense mechanisms include:
\begin{enumerate}
    \item \textbf{Model Voting Ensembling (Proactive) \cite{abbasi2017robustness}:} Ensembling pertrained MLP, CNN, and C-LSTM using a voting mechanism to construct stronger models against adversarial samples;
    \item \textbf{Ensemble Adversarial Training (Proactive) \cite{tramer2018ensemble}:} Augmenting the training dataset with adversarial samples, and retraining the NID models, thereby reinforcing their capabilities against adversarial samples;
    \item \textbf{Adversarial Query Detection (Reactive) \cite{chen2019stateful}:} Detecting the query process necessary in the black-box attack process, so as to blacklist the attacker's IP address before success.
\end{enumerate}
In what follows, we detail the proposed defense mechanisms, and demonstrate their effectiveness against black-box attacks.

\subsection{Model Voting Ensembling}

\begin{figure}[t]
\centering
\includegraphics[width=\columnwidth]{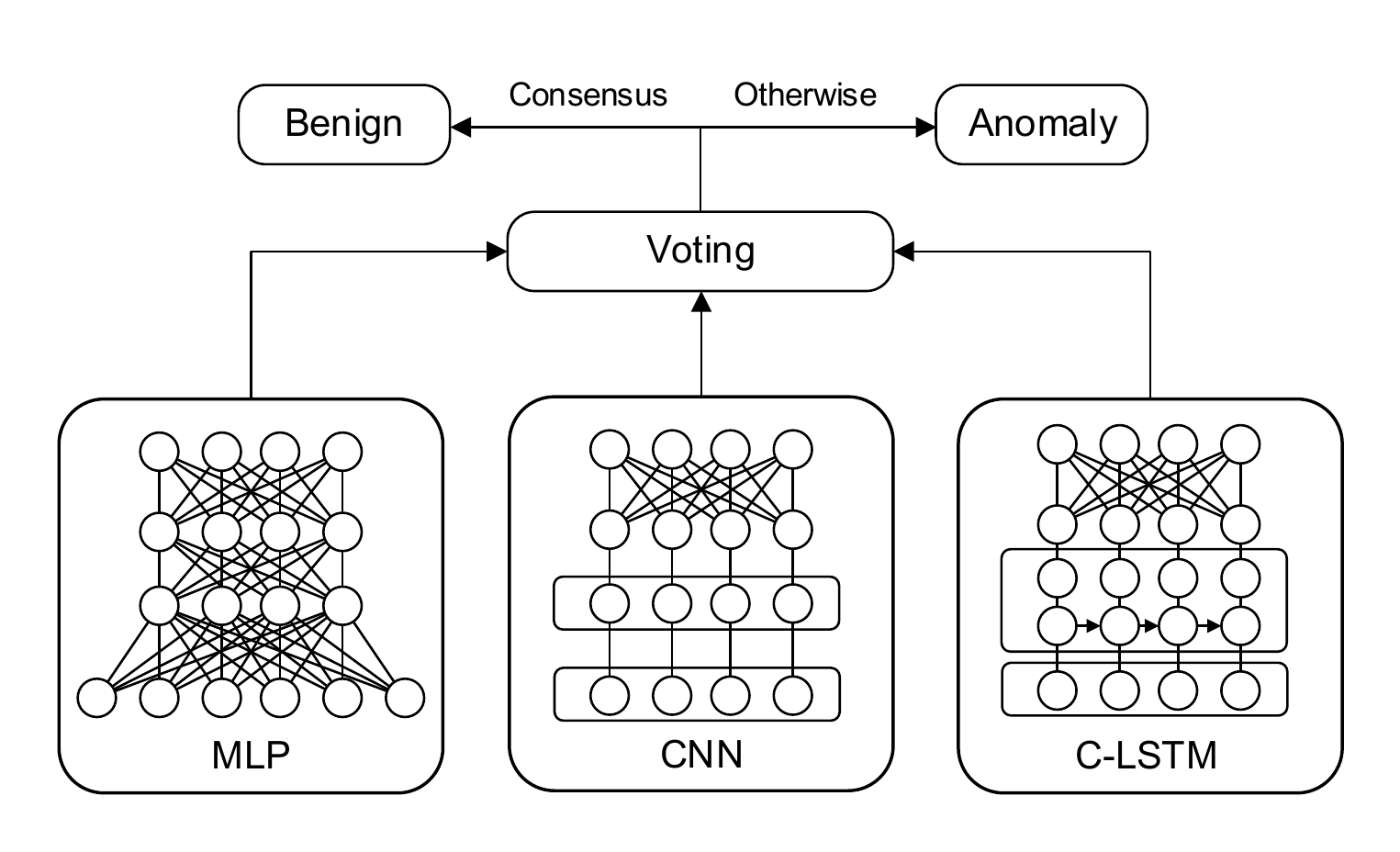}
\caption{An illustration of the voting ensembling defense mechanism.
\label{fig:vote}}
\end{figure}

\begin{table}[t]
\centering
\caption{NID performance of the ensembling model in the one-to-all scenario.\label{tab:e-121} }
\begin{tabular}{|c|c|c|c|c|}
\hline
    \multirow{2}{*}{Model Ensembling} & Accuracy & Precision & Recall & F1 score \\ \cline{2-5}
    & 0.987    & 0.964     & 0.954  & 0.959    \\ \hline
\end{tabular}
\end{table}

\begin{figure}[t]
\centering
\includegraphics[width=1.1\columnwidth]{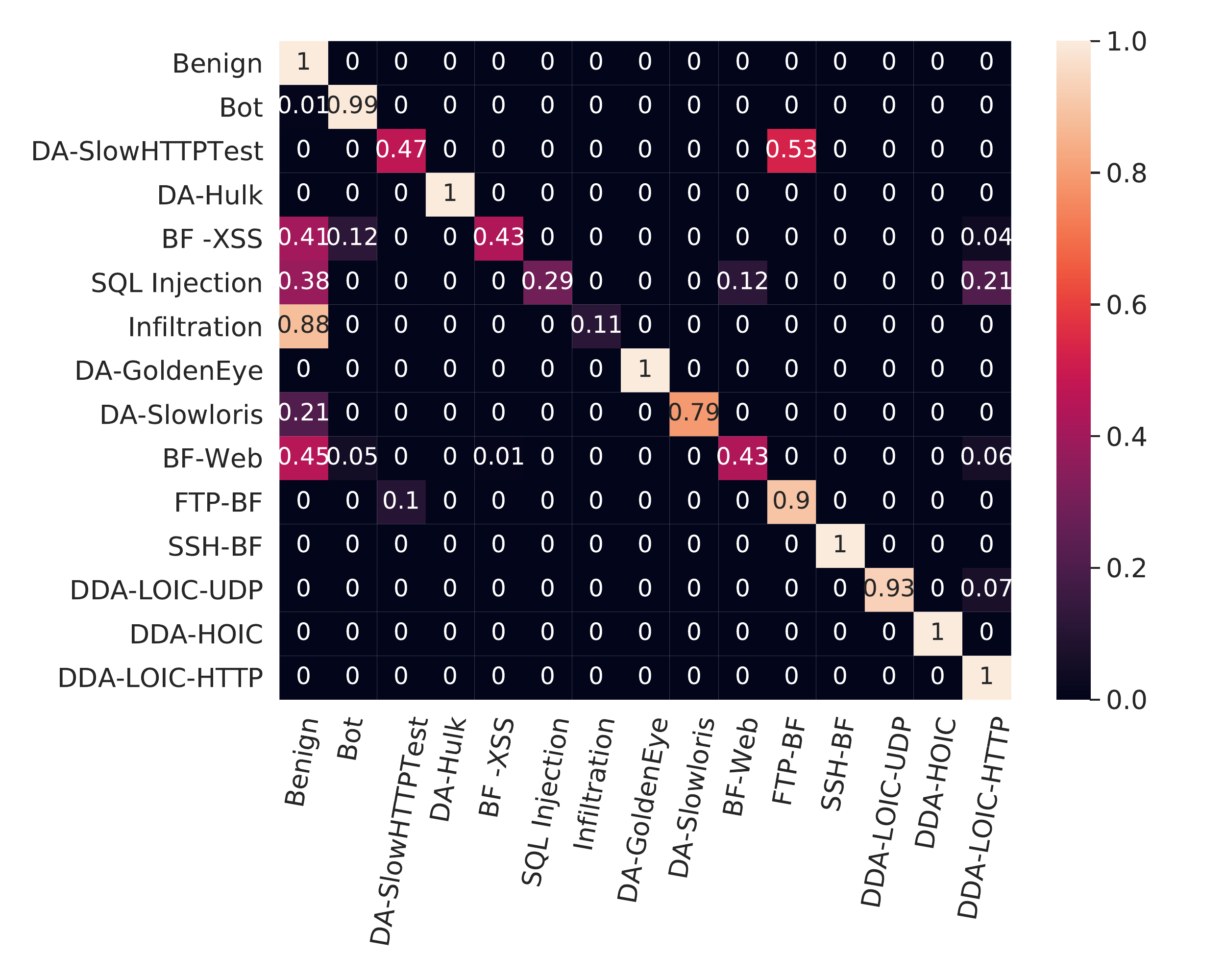}
\caption{The confusion matrix of the ensembling model in the one-to-one detection scenario.
\label{fig:cm_em}}
\end{figure}

\begin{figure}[t]
\centering
\includegraphics[width=\columnwidth]{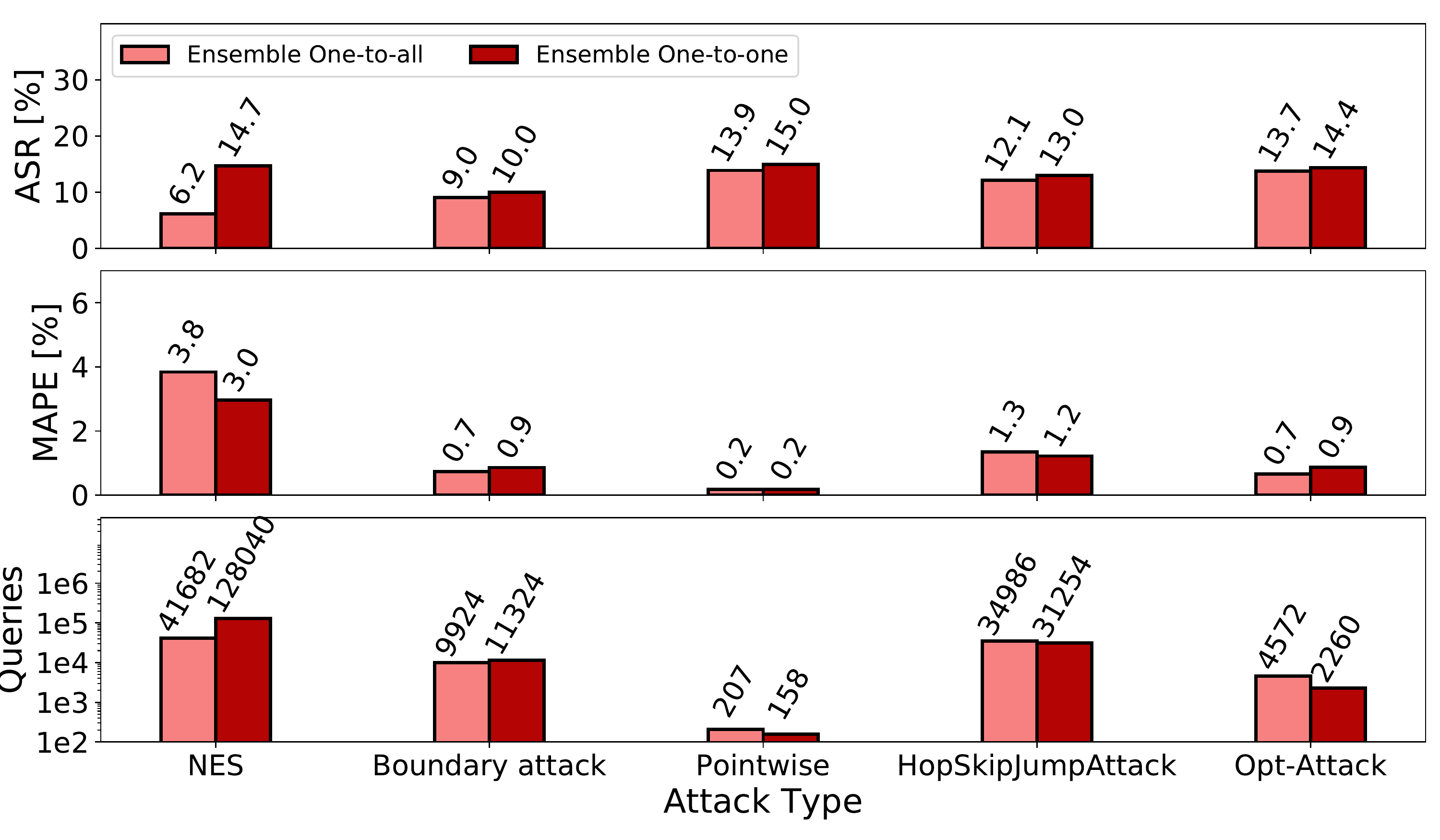}
\caption{ASRs, MAPE, and number of queries statistics of all attack approaches against ensembling models in the one-to-all and one-to-one scenarios.\label{fig:en_stats}}
\end{figure}


\begin{figure*}[]
\centering
\includegraphics[width=\textwidth]{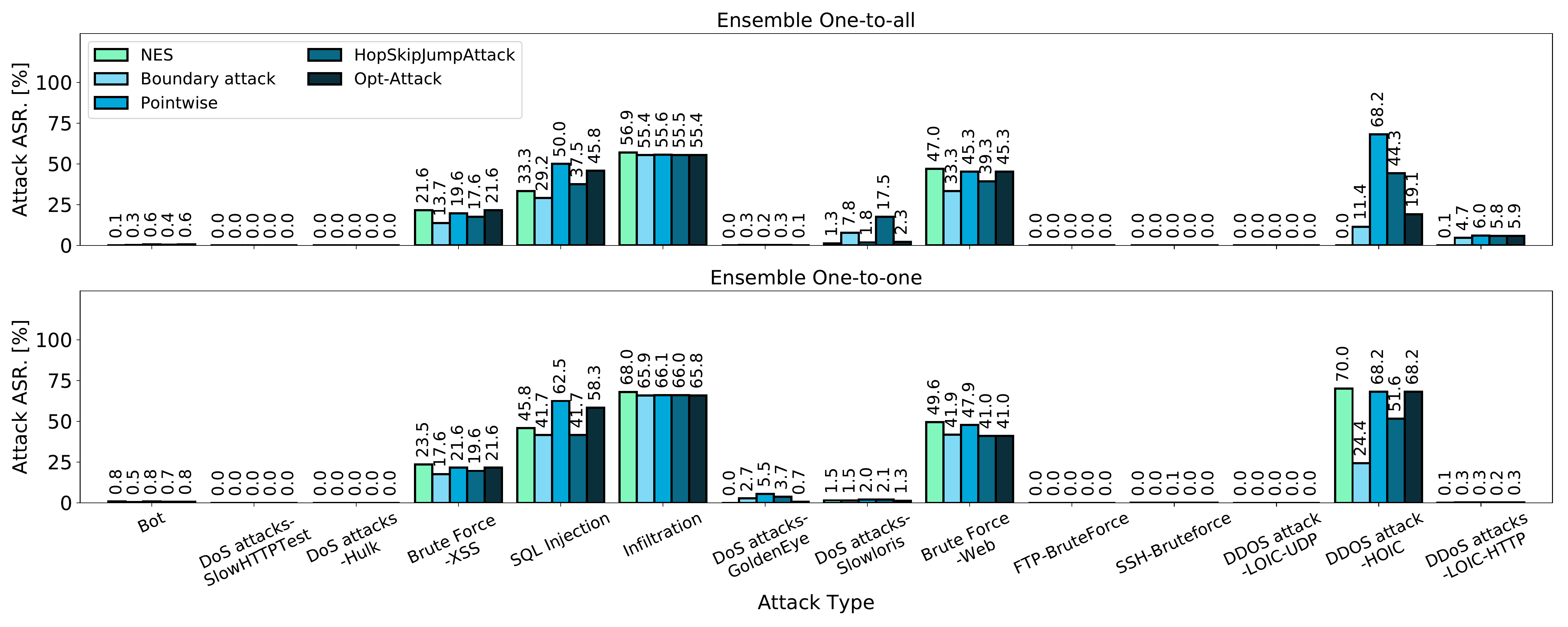}
\caption{ASRs of each type of attack against the model voting ensemble technique in both one-to-one and one-to-all scenarios.
\label{fig:asr_en_both}}
\end{figure*}


The experiments we reported in Sec.~\ref{sec:bba} suggest that an attacker can successfully compromise a NID model with up to 35\% ASR in black-box settings. However, only a small set of adversarial samples can bypass all three NID models simultaneously. This motivates us to construct a new ensembling model \cite{zhang2018long, zhang2017zipnet} by combining all of these structures, to strengthen the barrier against potential attacks. Specifically, for each input traffic flow, we gather the decisions of all NID models individually, and make the classification using a voting process. A traffic flow will be classified as ``benign'' if and only if all models reach consensus, \ie all of them classify it as ``benign''. Otherwise, the traffic flow will be regarded as an ``anomaly''. We show the underlying principle of the model voting ensembling mechanism in Fig.~\ref{fig:vote}. 
We recognize several advantages of using such model voting ensembling for defending agaisnt adversarial attacks. Namely:
\begin{enumerate}
    \item In order to construct a successful adversarial sample, attackers need to defeat all NID models simultaneously, which is much harder than compromising a single one;
    \item The voting mechanism makes the entire model non-differentiable, thus attack approaches that rely on model gradient estimation (\eg \nes) will be obstructed;
    \item The voting mechanism is easy to implement, as it does not require to re-train the original NID models.
\end{enumerate}
The proposed model voting ensembling method is a proactive approach, as it improves the robustness of the pretrained models against adversarial samples. We show the NID performance of the ensembling model for the one-to-all scenario in Table~\ref{tab:e-121} and the confusion matrix for the one-to-one scenario in Fig.~\ref{fig:cm_em}. Revisiting Table~\ref{tab:12a} and Fig.~\ref{fig:cm}, observe that the ensembling model obtains very close performance compared to its individual components in both detection scenarios, while achieving lower false positive rates, since it requires consensus to make the ``benign'' decision. 

We re-run the same 5 black-box attacks considered previously over the same set of 50,000 malicious samples and show their statistics in Fig.~\ref{fig:en_stats}. Note that the benign confidence is abandoned as, outputs of ensembling models are no longer probabilities. Jointly analyzing with Fig.~\ref{fig:one2all_stats} and \ref{fig:one2one_stats}, observe that the ASRs for all attack approach against the ensembling models have dropped compared to attacking each of the its component (\ie MLP, CNN and LSTM). For the best case, the \bd approach obtains 20.1\% ASR in attacking CNN in the one-to-all scenario, while this drops to 8.0\% while attacking the ensembling model. Regarding the one-to-one scenario, the dropping degree of the ASR is also substantial. This indicates that the voting ensembling mechanism operates well as a defense approach. On average, the ensembling models lead to 17.12\% and 9.02\% drop of ASR for one-to-all and one-to-one scenarios respectively. Turning attention to the MAPE, observe that the MAPE of adversarial samples crafted from ensembling models is lower, which suggests that the ensembling mechanism applies hidden and tighter constraints to the adversarial samples, to prevent them deviating excessively from the raw input samples.

We also show in Fig.~\ref{fig:asr_en_both} the ASR on a malicious traffic type bases, when crafting adversarial samples against the ensembling model, for the one-to-one NID scenario. Observe that the voting ensembling mechanism successfully defends 9 type of adversarial samples (\ie Bot, DoS attack-SlowHTTPTest, DoS attack-Hulk, DoS attack-GoldenEye, DoS attack-Slowloris, FTP-BruteForce, SSH-Bruteforce, DDoS attack-LOIC-UDP and DDoS attack-LOIC-HTTP), as the ASR is close to 0\%. For other type of malicious traffic, the ASR also drops by varying degrees, which demonstrates the effectiveness of the proposed mechanism.

\subsection{Ensemble Adversarial Training}
\begin{figure}[]
\centering
\includegraphics[width=\columnwidth]{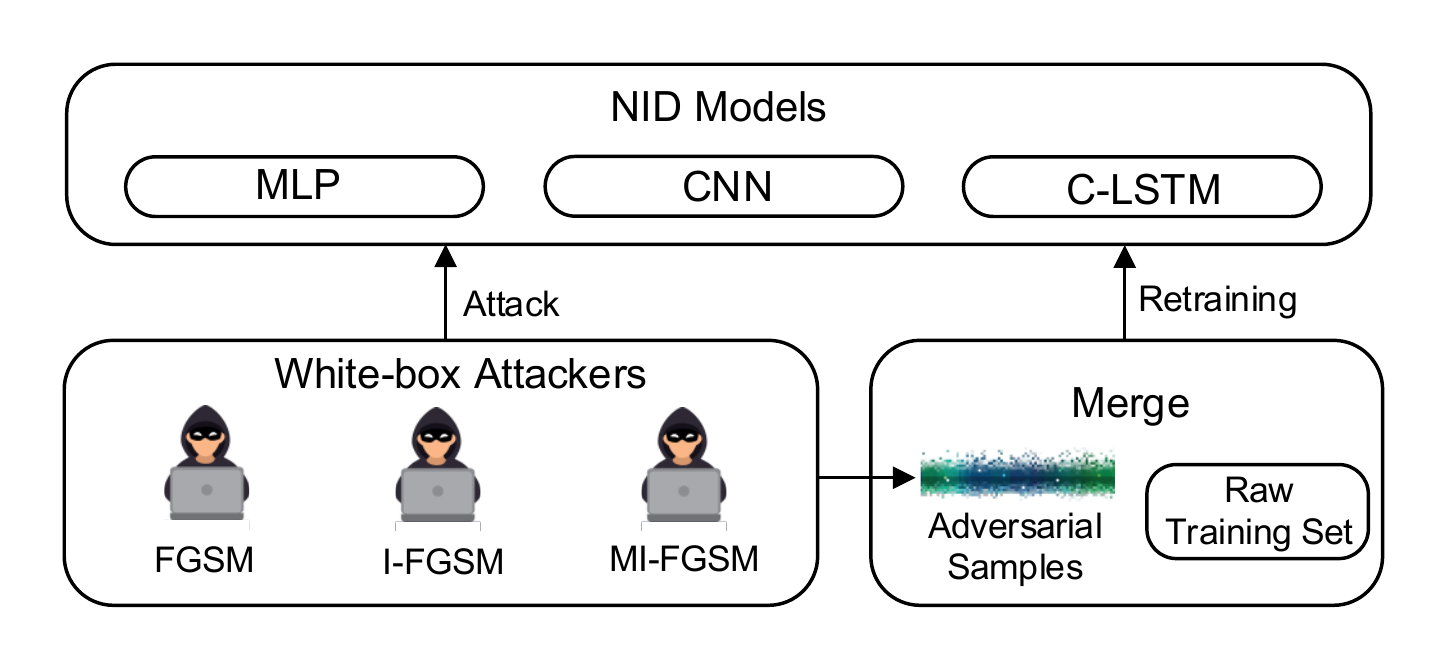}
\caption{An illustration of the EAT defense approach.
\label{fig:eat}}
\end{figure}

As discussed in Sec.~\ref{sec:attack}, white-box attacks are rarely accessible to external adversaries to compromise NIDS, as the training data, model structures and parameters are opaque. However, recent literature confirms that adversarial samples are adaptable across different attack methods and victim models~\cite{tramer2018ensemble, wu2017adversarial}. Therefore, from the defenders' points of view, adversarial samples generated using white-box attacks can be exploited to improve the robustness of NID models, so as to defend against potential black-box adversarial samples. Therefore, we employ the Ensemble Adversarial Training (EAT) as an additional defense approach \cite{tramer2018ensemble}, which augments the training data with adversarial examples generated by white-box attacks crafted on other static pre-trained NID models. Subsequently, the original NID models are reinforced by re-training on the augmented dataset. We show the underlying principle of the EAT mechanism in Fig.~\ref{fig:eat}. We expect that, with the proposed re-training on the augmented dataset, the NID models learn to classify adversarial samples better and thus become more resilient to black-box attacks. 

\subsubsection{Reinforcing NID models with White-box Adversarial Samples}
Specifically, we randomly select 250,000 malicious traffic flow to generate adversarial samples using three state-of-the-art white-box attack approaches. Namely, Fast Gradient Sign Method \cite{43405}, Iterative Attack (I-FGSM) \cite{45818} and Momentum Iterative Fast Gradient Sign Method (MI-FGSM) \cite{dong2018boosting}. The FGSM-based approaches performed one step gradient update along the direction of the sign of gradient at each feature that allows perturbations, and introduce noise following the direction of the gradient. The I-FGSM method extends the FGSM by running a finer optimization for multiple iterations to generate a valid adversarial sample.  The MI-FGSM method introduces the a momentum term into the iterative process of I-FGSM, which help stabilizes the update directions  and escape from poor local maxima. This leads to more transferable adversarial samples.

We show statistics of malicious traffic samples used for white-box attacks, number of successful adversarial samples, and their ratio in Table~\ref{tab:adv_train}. Note that the adversarial sample numbers are summed over all white-box attacks, all models, and both detection scenarios. Due to the \emph{information asymmetry} between attackers and defenders, the defenders do not have knowledge about which features will be perturbed for attack purposes. We therefore relax the features constraints (see Sec.~\ref{sec:constrains}) for perturbations in the white-box setting. However, the constraints over MAPE ($\leq 20\%$) are retained, to restrict the scale of the perturbations. Note that the adversarial samples generated by white-box attacks are not necessarily valid traffic flows, as they are only employed for training purposes. We gather successful adversarial samples generated by all white-box attack methods (\ie FGSM, I-FGSM, and MI-FGSM), crafted with all NID models (\ie MLP, CNN, and C-LSTM) in both detection scenarios (\ie one-to-all and one-to-one) and combine these with the original training data, to build a new augmented dataset for the EAT.

\begin{table}[t]
\centering
\caption{Statistics of the malicious traffic flows used to generate adversarial samples (shared set)/total number of adversarial samples successfully generated by all methods for EAT, and the ratios of each attack with respect to the total, using white-box attacks \label{tab:adv_train}}
\begin{tabular}{|l|r|r|}
\hline
\textbf{Attack Type}     & \textbf{\begin{tabular}[c]{@{}c@{}}Number of \\ Instances\end{tabular}} & \textbf{Ratio [\%]} \\ \hline
Bot                      & 26,601/426,755                         & 10.640/12.356       \\ \hline
DoS attack-SlowHTTPTest & 26,601/317,338                         & 10.640/9.188       \\ \hline
DoS attack-Hulk         & 26,601/311,740                         & 10.640/9.026       \\ \hline
Brute Force-XSS         & 179/2,530                           & 0.072/0.073        \\ \hline
SQL Injection            & 63/1,083                          & 0.025/0.031        \\ \hline
Infiltration            & 26,601/473,116                         & 10.640/13.698       \\ \hline
DoS attack-GoldenEye    & 26,601/349,907                         & 10.640/10.131       \\ \hline
DoS attack-Slowloris    & 8,515/112,142                         & 3.406/3.247        \\ \hline
Brute Force-Web          & 494/8,430                         & 0.198/0.244        \\ \hline
FTP-Brute Force           & 26,601/287,768                         & 10.640/8.332       \\ \hline
SSH-Brute Force           & 26,601/404,402                         & 10.640/11.708       \\ \hline
DDoS attack-LOIC-UDP     & 1,347/2,423                          & 0.539/0.070        \\ \hline
DDoS attack-HOIC         & 26,601/401,923                         & 10.640/11.637       \\ \hline
DDoS attack-LOIC-HTTP   & 26,594/354,402                         & 10.638/10.261       \\ \hline
Total                    & 250,000/3,453,959                       & 100/100          \\ \hline
\end{tabular}
\end{table}

\begin{figure}[tt]
\centering
\includegraphics[width=\columnwidth]{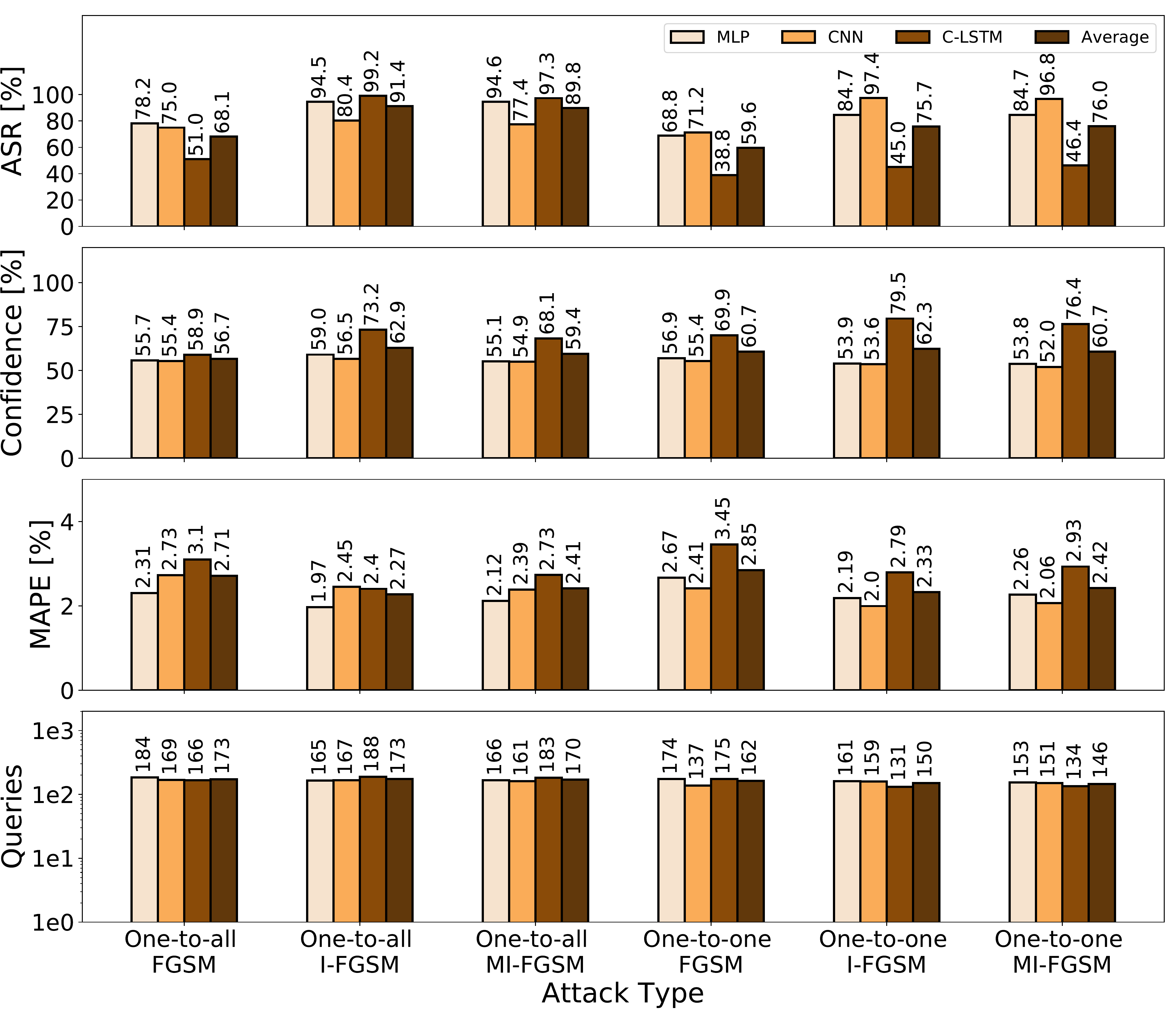}
\caption{ASRs, Confidence, MAPE and number of queries statistics of all white-box attack approaches against 3 NID models considered in this study and their average values for both NID scenarios.\label{fig:whitebox_stats}}
\end{figure}

\begin{figure*}[t]
\centering
\includegraphics[width=\textwidth]{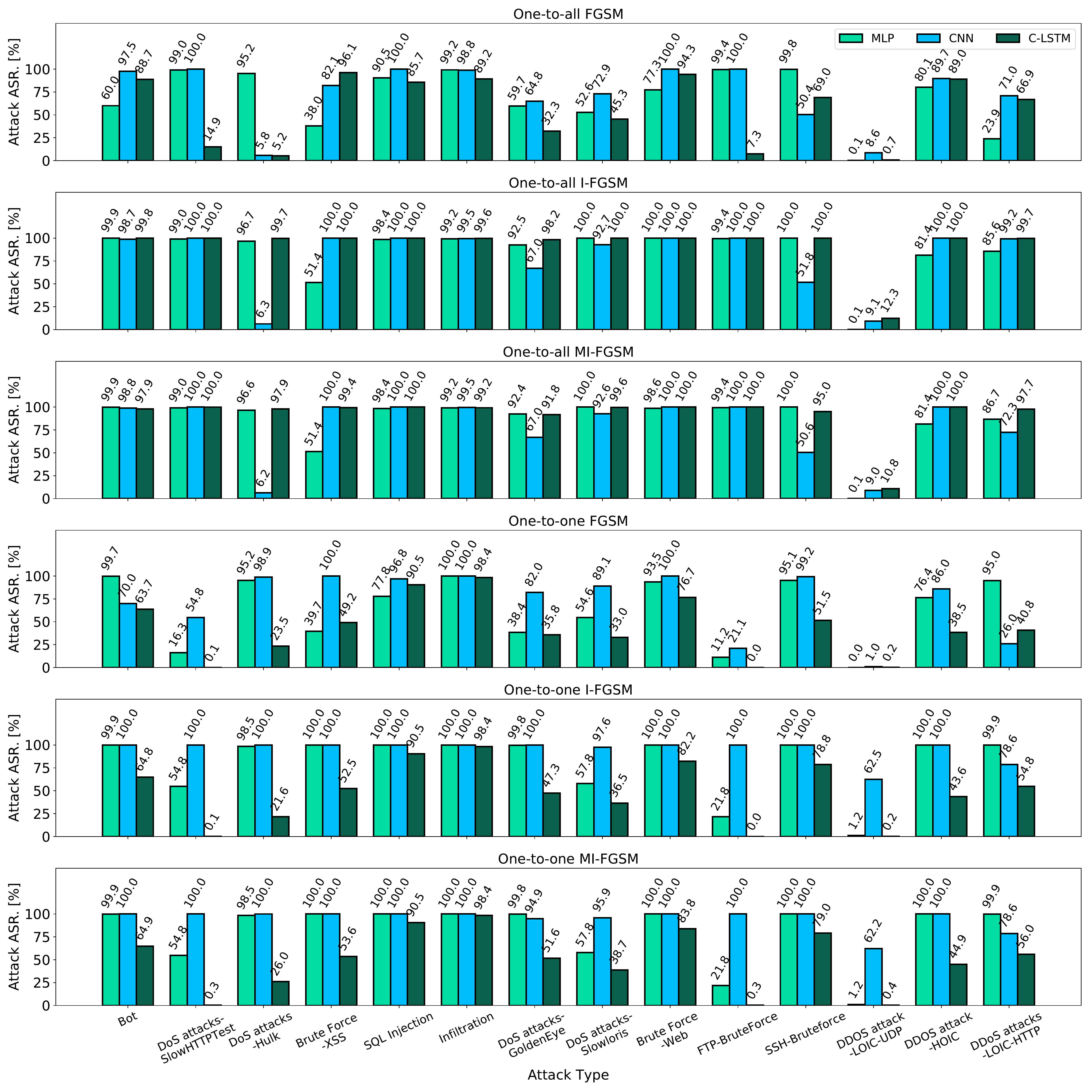}
\caption{ASRs over each type of white-box attack against all models for both NID scenarios.\label{fig:whitebox_asr}}
\end{figure*}

We show the performance of each white-box attack in Fig.~\ref{fig:whitebox_stats}. Observe that since the NID models are transparent, and looser constraints are applied to the adversarial samples, the ASR for all white-box attacks is significantly higher than their black-box counterparts. White-box attacks also require fewer queries to generate adversarial samples. Fortunately, attackers normally do not have access to the NID models.  The ASRs for crafting each type of anomaly is shown in \ref{fig:whitebox_asr}. Observe that ASR for many types of anomalies are close to 100\%. 

\subsubsection{NID Performance of Post-EAT Models}
\begin{figure*}[t]
\centering
\includegraphics[width=1.1\columnwidth]{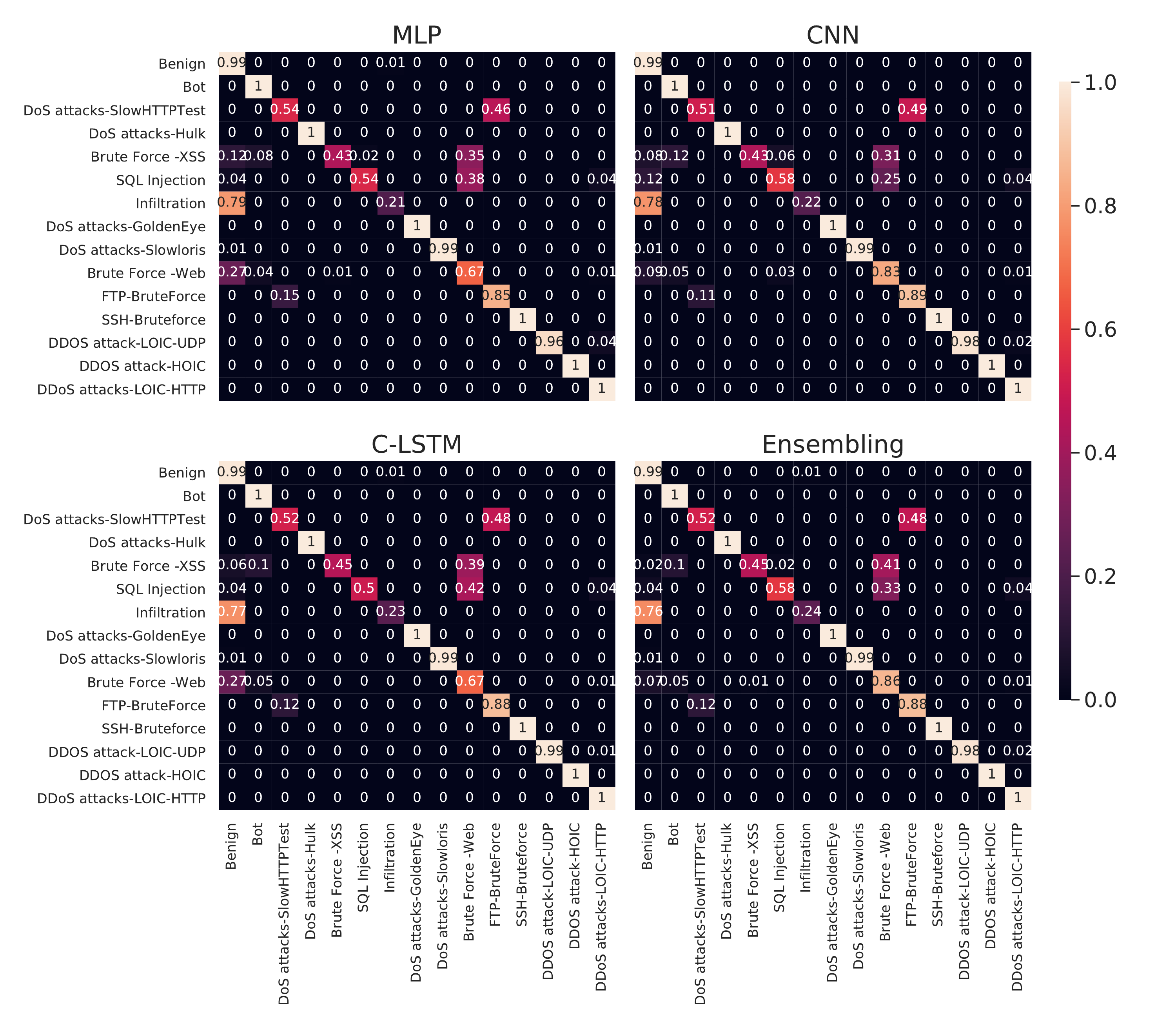}
\caption{The confusion matrix of the MLP, CNN and C-LSTM on one-to-one NID after the EAT. 
\label{fig:cm_aug}}
\end{figure*}

We show the detection performance on the same test set after EAT, for the one-to-all scenario, in Table~\ref{tab:stat_eat}. Compared to NID models prior to the EAT (See Tables~\ref{tab:e-121}, \ref{tab:12a}), the detection performance of the newly trained models has dropped slightly in terms of accuracy, precision, and F1 score. However, the recall rate of each model has improved. This indicates that the models are prone to classify some ambiguous samples as anomalies, which results in higher false positive rate and lower false negative rate. Similar phenomena are also observed in the one-to-one scenario. The accuracy for the MLP, CNN, C-LSTM, and the ensembling model are 98.13\%, 98.12\%, 98.10\% and 97.96\% respectively, which appears worse than what was achieve prior to EAT. However, by taking a closer look at their confusion matrices in Fig.~\ref{fig:cm_aug}, post-EAT models achieve high detection rate on most of the anomalies that fail to be detected previously (\ie Brute Force-XSS, SQL Injection, and Brute Force-Web). This suggests that the EAT has improved the robustness of each NID model, making them more sensitive to abnormal traffic flows that are difficult to classify.

\subsubsection{Robustness to Old Adversarial Samples}
In Table~\ref{tab:ratio_bypass}, we further show the ratio of adversarial samples crafted from the models before EAT, which can compromise the corresponding post-EAT models. Observe that EAT also makes each model more resilient to old adversarial samples, as those ratios are significantly below 100\%. In particular, only 38.06\% of adversarial samples crafted from the old C-LSTM can bypass the EAT C-LSTM. This means that the EAT enables each model to learn to characterize adversarial samples generated using white-box attacks, and therefore fix some ``bugs'' in the old setting. 

\begin{figure*}[t]
\centering
\includegraphics[width=\textwidth]{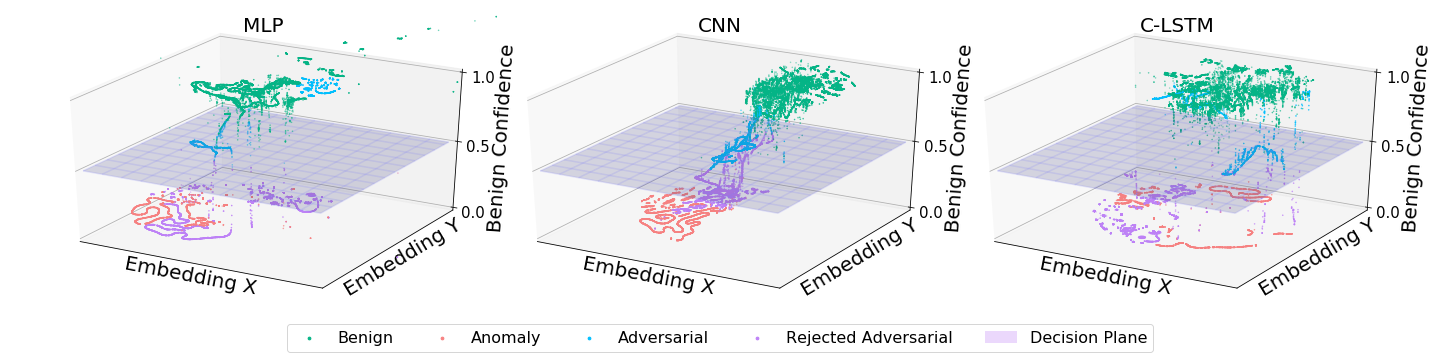}
\caption{Two-dimensional ($x, y$ axis) t-SNE embedding of the representations in the last hidden layer along with the benign confidence ($z$ axis) of each NID models after the EAT. Data generated using 10,000 benign samples (blue), 10,000 adversarial samples produced by all attack methods that successfully bypass the model (green) and are rejected by the model (purple),  with their corresponding malicious samples they craft from (pink). \label{fig:tsne_aug}}
\end{figure*}

This effect is confirmed by their t-SNE embedding. In Fig.~\ref{fig:tsne_aug}, we show the two-dimensional ($x, y$ axis) t-SNE embedding of the representations in the last hidden layer along with the benign confidence ($z$ axis) of each NID models after the EAT, as similar to Fig.~\ref{fig:tsne}. Note that we employ the same set of samples in Fig.~\ref{fig:tsne} to generate the new Fig.~\ref{fig:tsne_aug}. Observe that after the EAT, some old adversarial samples are rejected by the new models (purple), as they are below the decision boundary. It appears that the EAT pushes certain of the adversarial samples away from the benign cluster, such that they become more separable. Even though some adversarial samples still escape, their benign confidences become lower compared to Fig.~\ref{fig:tsne}. These means that the new models are more suspicious on these data after the EAT. One can raise the decision boundary to filter those samples, though it might result in higher false negative rate.

\begin{table}[t]
\centering
\caption{The performance of MLP, CNN, C-LSTM and the ensembling model after the ensemble adversarial training under the one-to-all scenario.\label{tab:stat_eat}}
\begin{tabular}{|c|c|c|c|c|}
\hline
\textbf{Model}& \textbf{Accuracy} & \textbf{Precision} & \textbf{Recall} & \textbf{F1 score}  \\ \hline
MLP        & 0.987    & 0.968     & 0.953  & 0.960    \\ \hline
CNN        & 0.986    & 0.959     & 0.954  & 0.956    \\ \hline
C-LSTM     & 0.985    & 0.953     & 0.955  & 0.954    \\ \hline
Ensembling & 0.983    & 0.943     & 0.956  & 0.949    \\ \hline
\end{tabular}
\end{table}

\begin{table}[t]
\centering
\caption{Ratio of adversarial samples that can bypass each NID model after the EAT. \label{tab:ratio_bypass}}
\begin{tabular}{|c|c|c|c|c|}
\hline
 \textbf{Scenario}  & \textbf{MLP}     & \textbf{CNN}     & \textbf{C-LSTM} & \textbf{Ensembling}  \\ \hline
One-to-all & 40.04\% & 53.15\% & 48.43\% & 81.54\%\\ \hline
One-to-one & 42.45\% & 38.06\% & 38.26\% & 43.31\%\\ \hline
\end{tabular}
\end{table}

\begin{figure}[t]
\centering
\includegraphics[width=\columnwidth]{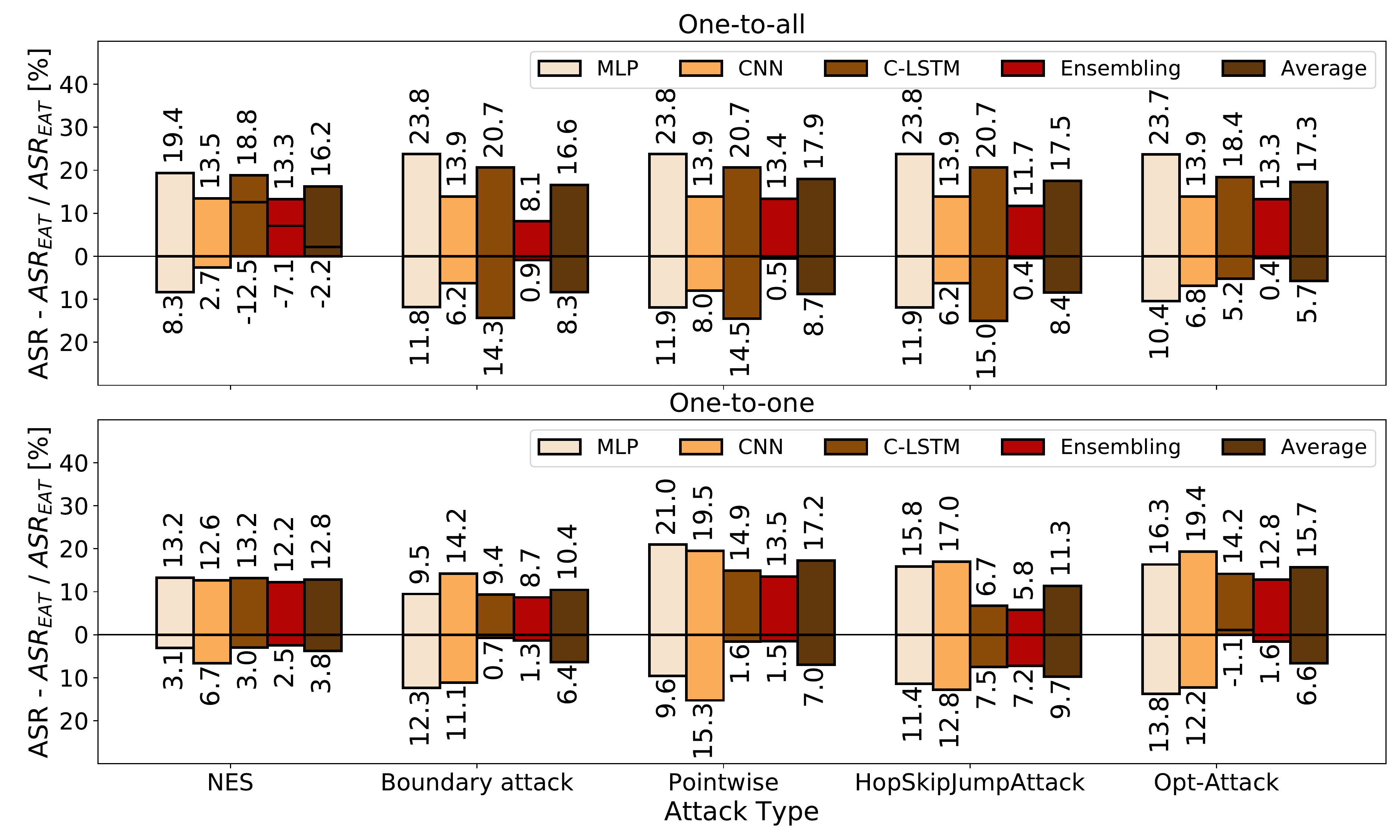}
\caption{The ASR of each attack after the EAT (bars on the top), and the ASR reduction compared to whn no further defense is applied (bars on the bottom) for both NID scenarios.\label{fig:asr_12both}}
\end{figure}


\subsubsection{The Effect of EAT}
In Fig.~\ref{fig:asr_12both}, we show the ASR for each attack after EAT (ASR\textsubscript{EAT}, bars in the upper part of the plot), and the ASR reduction compared to the case before EAT was applied (ASR - ASR\textsubscript{EAT}, bars in the lower part) for the both NID scenarios. In the figure, positive numbers below the x-axis indicate that the ASR\textsubscript{EAT} has dropped after EAT was employed. We observe that ASR of each attack drops for most of the models. This means that EAT successfully improves the robustness of each model, making them more difficult to be compromised. On average, the ASR drops to 6.70\% and 5.78\% for one-to-one and one-to-all NID scenarios, respectively. 

On the other hand, we also observe that the EAT is not a silver bullet for all the cases. For example, the ASR\textsubscript{EAT} of \nes increases by 12.5\% when crafting from LSTM in the one-to-all scenario. This also weakens the ensembling model, as its ASR\textsubscript{EAT} increases accordingly. Nevertheless, the EAT remains a effective defense approach, as it reinforces each NID model and block attempts from black-box attack for most of the case. In what follow, we introduce the query detection defense, which helps resolve the adversarial samples that can not be defended by the model voting ensembling and EAT.

\subsection{Adversarial Query Detection}
Recall that all black-box attack methods rely on continuous queries to the victim model and feedback received. Based on the feedback, the attackers learn to adjust the perturbations added to the input, so as to compromise detection. The scale of perturbations is usually small, so that they to do not change the essence of the original input. Therefore, the queries in the same attack round are typically with high similarity. This inherent similarity between queries can be harnessed to detect an attack. Therefore, we explore query detection \cite{chen2019stateful} as the final defense mechanism. Once queries have been discovered, the NIDS can blacklist the attackers' IP addresses, to prevent potential threats.

\begin{figure}[]
\centering
\includegraphics[width=\columnwidth]{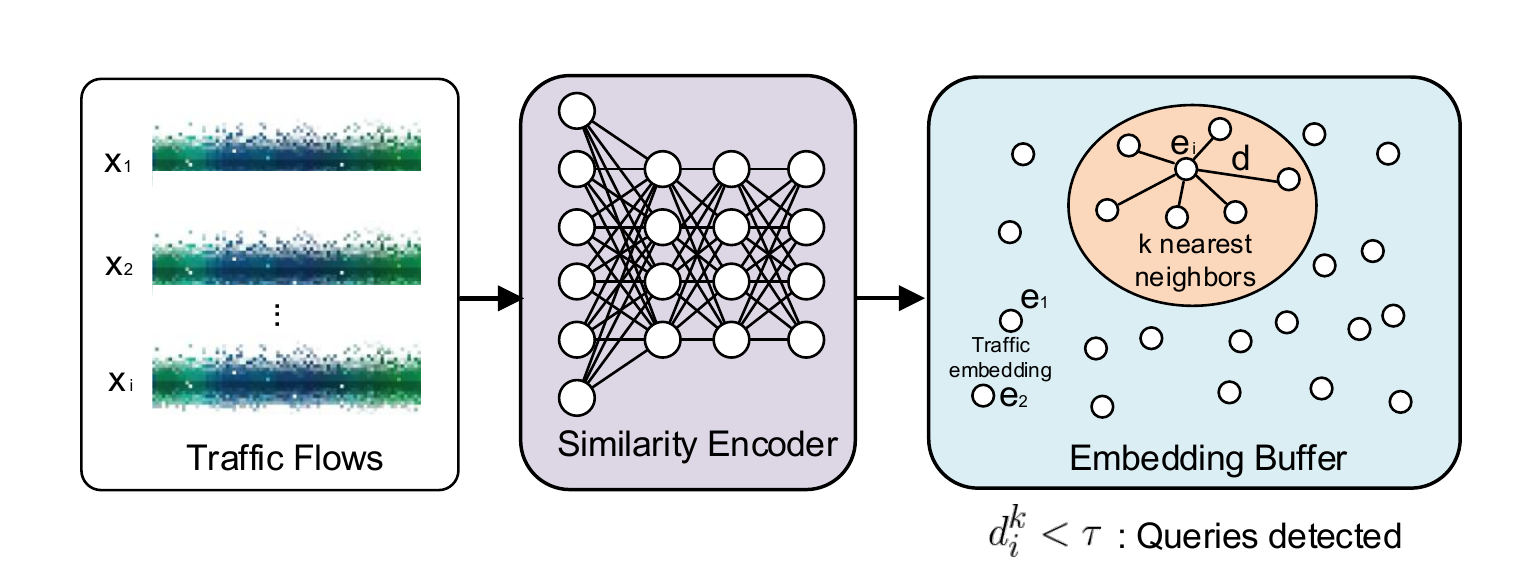}
\caption{An illustration of the query detection defense mechanism using a deep similarity encoder.
\label{fig:detection}}
\end{figure}

Specifically, for each IP address, we construct a buffer with size $B$ to store the features of the traffic flows originating from that address in a pre-defined period. To reduce the dimension of the features saved and model the similarity degree between flows, we employ a deep similarity encoder (DSE) \cite{bell2015learning}, encoding similar traffic flow in a lower-dimensional space with shorter $l_2$ distance. More precisely, for each new flow $x$ sent from that IP address, we compute the pairwise distance between the embedding of this flow and others in the buffer, calculating the $k$ nearest neighbor average distance $d_x^k$. If $d_x^k$ is lower than a threshold $\tau$, \ie $d_x^k < \tau$, this suggests that particular IP address has sent an excessive number of similar traffic flows, which can be considered as queries in an ongoing attack. When this happens, the IP address can be blacklisted and thus the potential threat eliminated. We show the underlying principle of the query detection mechanism in Fig.~\ref{fig:detection}.

After an attack is detected, the buffer associated to the specific IP address can be cleared. In addition, when query detection suggests a potentially malicious actor, their IP address can be banned either immediately, or after subsequent queries, as suggested in \cite{bell2015learning}. This can minimize an attacker's knowledge of the time when their attack was detected, therefore reducing the probability of compromising the query detection mechanism.

\subsubsection{Deep Similarity Encoder}
The core component of the query detection based defense mechanism is the deep similarity encoder (DSE) \cite{bell2015learning}, which is a neural network that reduces the dimension of the input data. After embedding by a DSE, dissimilar flows will be far away from each other in the encoded space, while similar queries will be close to each others. Therefore, queries and traffic flows become more distinguishable.

For the DSE, we employ a CNN similar to that in Fig.~\ref{fig:model_nid}, only replacing the last layer with 3 units. This means that the embedding of each traffic flow is a 3-dimensional vector. We denote $e_i = \mathrm{DSE}(x_i)$, as the embedding of the input sample $x_i$. The DSE can be trained via minimizing the following contrastive loss function:
\begin{equation}\label{eq:loss}
    \mathit{L(x_i, \Tilde{x}_i, x_m, x_n ;\theta)} = ||e_i-\Tilde{e}_i||^2_2 + \max (0, \varpi^2 - ||e_m-e_n||^2_2)
\end{equation}
Here, $x_i, \Tilde{x}_i$ are a pair of similar traffic flows, while $x_m, x_n$ are traffic flows which are dissimilar. $\theta$ is the trainable parameter set of the DSE, and $\varpi$ is a constant penalty, which regularizes the scale of $||e_m-e_n||^2_2$. We choose $\varpi=0.5$ in our experiments. The first term of the function ensures that the $l_2$ distances of the similar traffic flows are minimized, while the second term guarantees that distances of dissimilar traffic pairs are maximized but limited to the $\varpi$.

\begin{figure}[]
\centering
\includegraphics[width=\columnwidth]{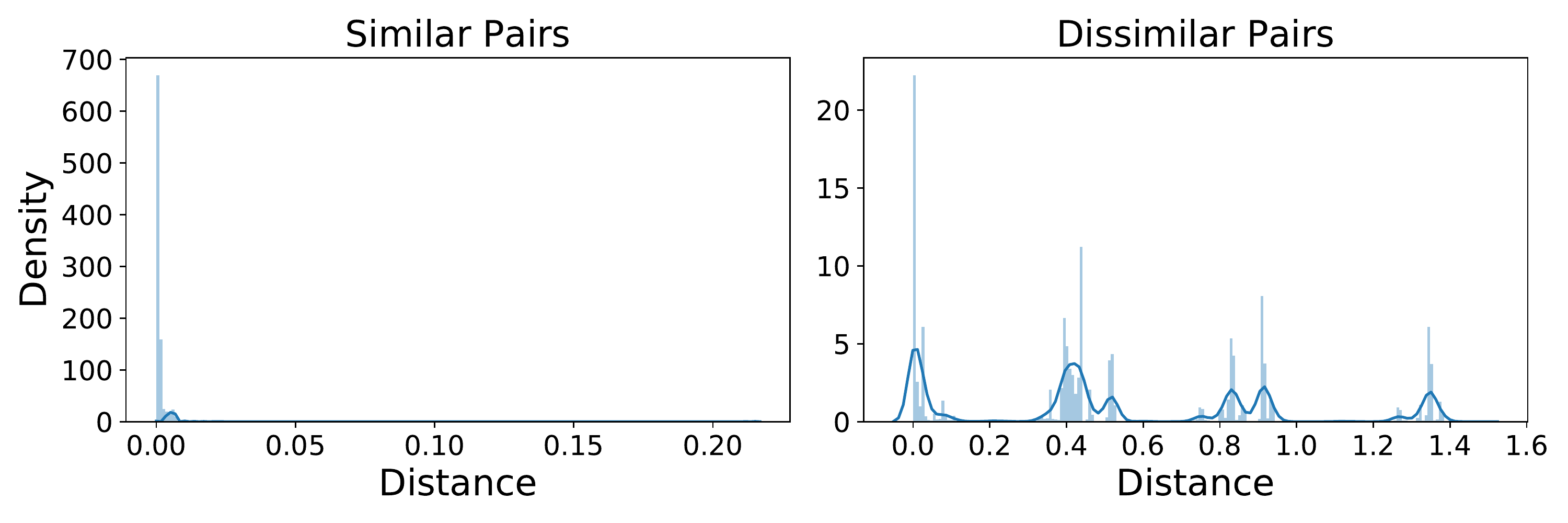}
\caption{Histograms of $l_2$ distances of DSE embeddings between similar flow pairs (left) and dissimilar flow pairs (right) generated using the training set. \label{fig:hist_dist}}
\end{figure}

\begin{figure}[]
\centering
\includegraphics[width=\columnwidth]{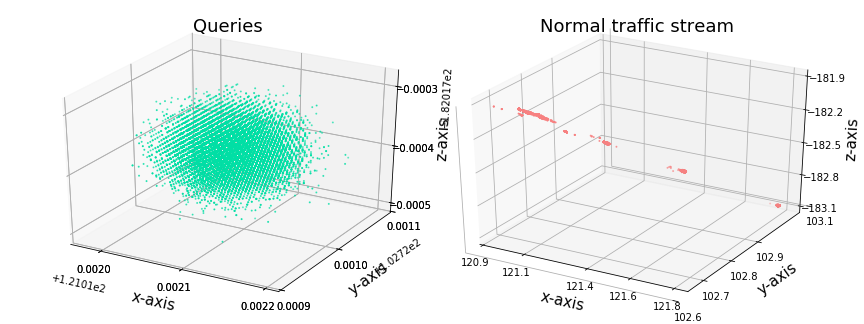}
\caption{Two set of traffic sample embedding generated by the deep similarity encoder. Left: sample embedding of the queries process of the \hj crafting from the MLP NID model. Right:  samples of routine traffic flows emulating with the training set. \label{fig:dse_embedding}}
\end{figure}

We train the DSE using the same training set sampled from the CSE-CIC-IDS2018 dataset as used by other NID models. For the training purpose, we construct the $\Tilde{x}_i$ by adding Gaussian noise $\sigma_i \sim N(0, \alpha |{x}_i|)$ to each sample ${x}_i$, \ie $\Tilde{x}_i = {x}_i + \sigma_i$. Here, $\alpha$ controls the standard deviation of the Gaussian noise and we choose $\alpha = 0.15$. $x_m, x_n$ are sampled from the training set distinct from $x_i$. After training, we use the full training set to randomly generate 13,153,902 pairs of similar and dissimilar flows and visualize the distributions of the $l_2$ distances of their embedding by histograms, as shown in Fig.~\ref{fig:hist_dist}. Observe that most of of the the $l_2$ distances between similar flows pairs are close to 0, while they have multiple peaks away from 0 for dissimilar flows pairs which are farther from the origin. This indicates that the DSE successfully learn the similarity knowledge between traffic flows, and therefore can operate effectively for the query detection purpose.

To evaluate the effectiveness of our DSE, in Fig.~\ref{fig:dse_embedding} we show the DSE-embedding of the 35,215 queries of a shot of \hj crafted from the MLP model, and the DSE-embedding of 35,215 benign samples. These benign samples can be viewed as a stream of routine traffic in real life. Observe that the embeddings of the query set congregate in a fairly small region, and are close to each other. In contrast, the embeddings of the normal traffic appear more dispersed and separable. This further proves that our DSE can effectively learn the similarity between each traffic sample.

\subsubsection{Hyper-parameters Selection}
There are three important hyper-parameters to be configured for query detection, namely \emph{(i)} the detection threshold $\tau$; \emph{(ii)} the number of neighbors $k$ used for the detection; and \emph{(iii)} the size of the buffer $B$, which stores the traffic flows sent from the same IP address. These parameters will significantly affect the performance of the query detection, \rv{and we optimize such hyper-parameters via a separate validation set.}. First, we select $\tau=0.00157$, since 10\% of dissimilar pairs and 86.4\% of similar pairs in the training set are below this threshold. This provides an appropriate decision boundary to discriminate normal traffic flows and attack queries. \rv{Note that a high threshold will block benign traffic and a low threshold might lead to leakage of adversarial queries. The values of $k$ and $B$ affect the robustness of detection and also the computational and storage cost of the NIDS. We select $B=500$ and $k=200$, as these numbers allow efficient detection and yield 0 false positive rates when operating with traffic streams simulated with the entire training set.}

\begin{figure}[t]
\centering
\includegraphics[width=\columnwidth]{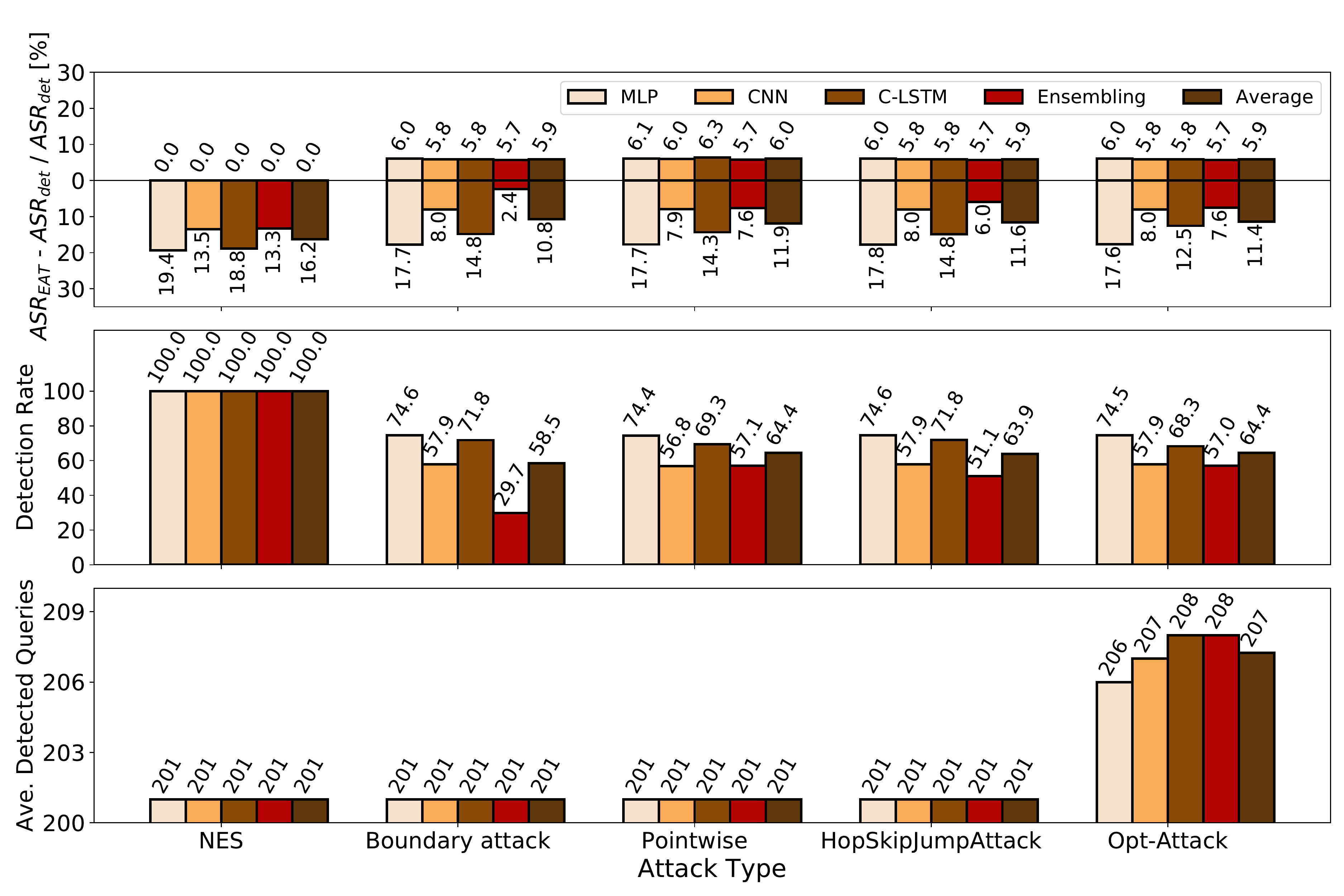}
\caption{Performance statistics of the query detection defense for the one-to-all scenario. Upper: The ASR of each attack after the query detection (higher bar), and the ASR reduction compared to removing the query detection (lower bar). Middle: the query detection rate of each adversarial attack. Lower: the average number of query when the query detector is alarmed.\label{fig:qd_12a}}
\end{figure}

\begin{figure}[t]
\centering
\includegraphics[width=\columnwidth]{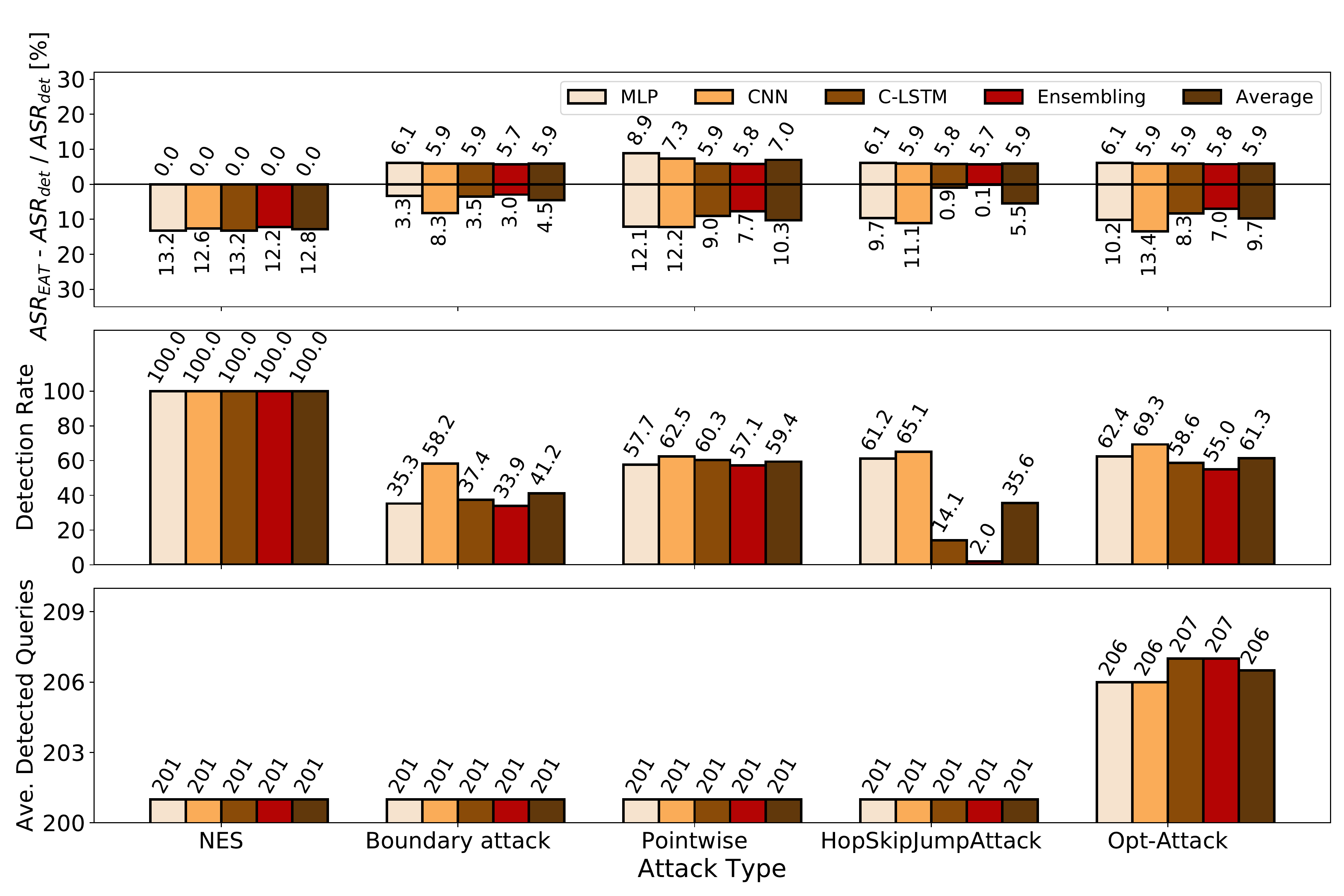}
\caption{Performance statistics of the query detection defense for the one-to-one scenario. Upper: The ASR of each attack after the query detection (higher bar), and the ASR reduction compared to removing the query detection (lower bar). Middle: the query detection rate of each adversarial attack. Lower: the average number of query when the query detector is alarmed.\label{fig:qd_121}}
\end{figure}

\subsubsection{Query Detection Defense Performance}
We show the ASR of each attack after the query detection, the ASR reduction compared to when query detection is not employed (bars bellow the x-axis in the top sub-plot), the query detection rate (middle), and the average number of queries (bottom) when the attack is detected, for each attack method for both NID scenarios in Fig.~\ref{fig:qd_12a} and~\ref{fig:qd_121} respectively. Observe that the ASR has dropped significantly after the query detection is employed. In particular, the ASR of NES reaches 0 for all models for both NID scenarios, and the detection rates therefore become 100\%. On average, the query detection defense reduces the ASR of attacks by 12.38\% and 8.56\% for each NID scenario respectively, and their average detection rate are 70.24\% and 59.50\%. This means that majority of the adversarial attack are detected during their query process.

Taking a closer look at the average detected query, we observe that \nes, \bd, \pw, and \hj attack attempts are detected at their 201\textsuperscript{st} query. Recall that the $k$ neighbor size selected for the query detection is 200, hence the detection alarm will only be triggered when the buffer has more than 200 samples. This means that the attack is detected immediately after the buffer has $k$ neighbor samples. Regarding the \opt attack, this is detected always 208 queries. This is due to the initial phrase of the attack, when it injects a few benign traffic flows to learn the direction of perturbation to be added to the adversarial samples. These samples are normally dissimilar, which slightly increases the detection time. Note that, despite the efficiency of the query detection mechanism, a larger buffer size ($B=500$) is still needed for tolerance, as the attacks may fill the buffer with queries (similar samples) and garbled traffic (dissimilar samples) alternately, to compromise the defense.

\begin{figure*}[t]
\centering
\includegraphics[width=\textwidth]{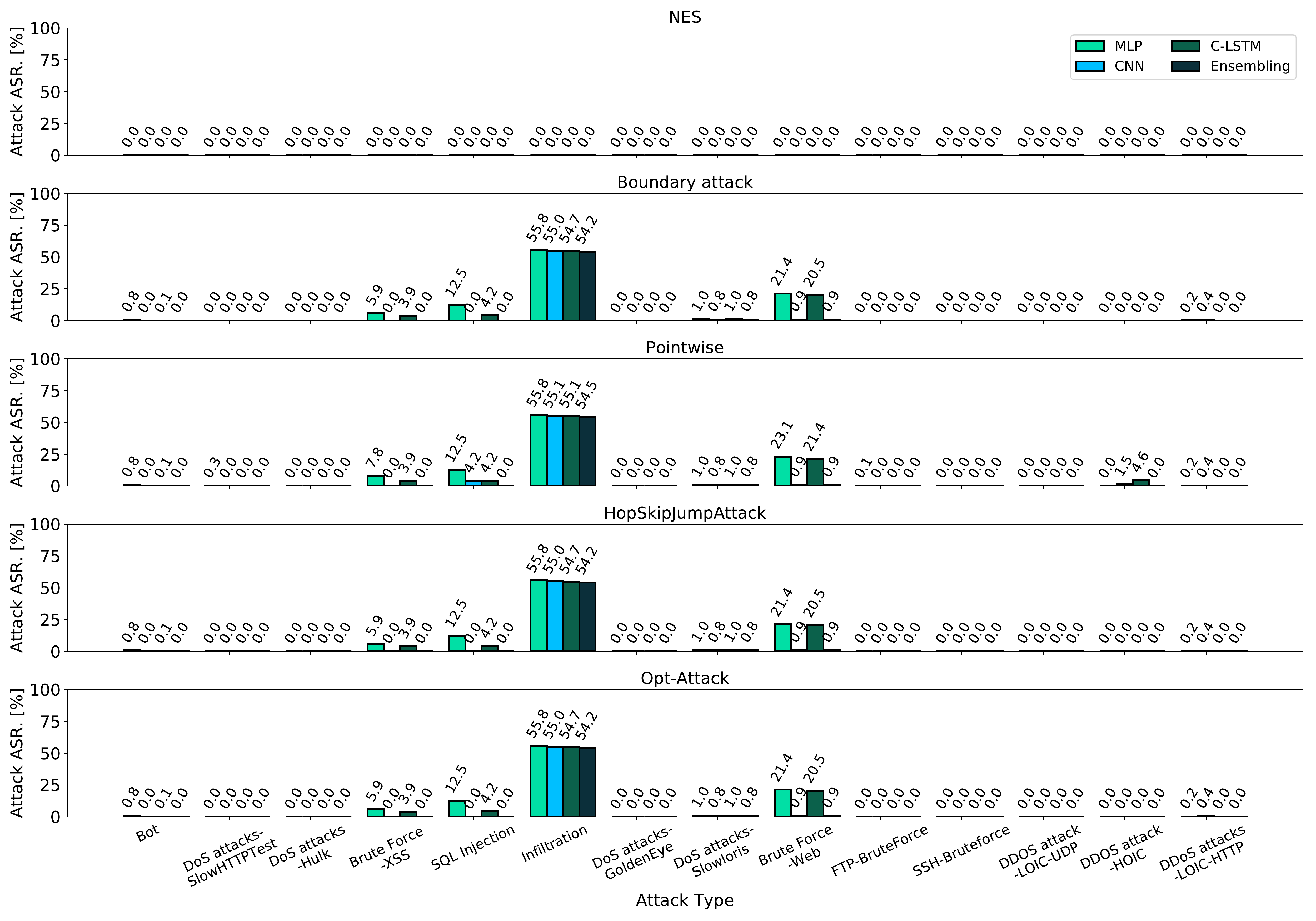}
\caption{ASRs over each type of attack after the EAT and query detection against all models in the one-to-all scenario. \label{fig:detection_asr}}
\end{figure*}

\begin{figure*}[t]
\centering
\includegraphics[width=\textwidth]{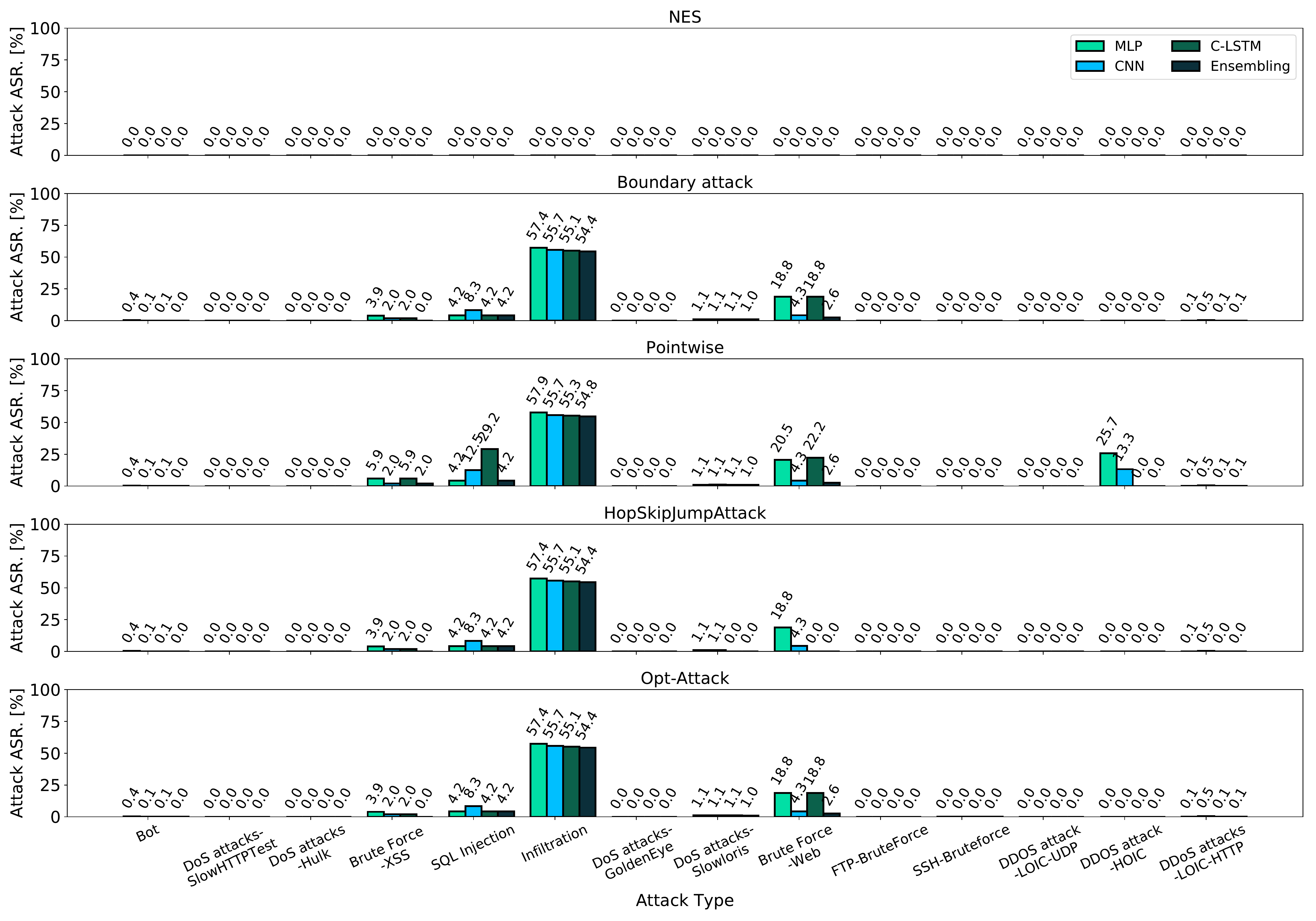}
\caption{ASRs over each type of attack after the EAT and query detection against all models in the one-to-one scenario. \label{fig:detection_asr_mc}}
\end{figure*}

In Fig.~\ref{fig:detection_asr} and~\ref{fig:detection_asr_mc} we show the ASR for each type of attack for both NID scenarios after the query detection is employed. We observe that the ASR drops significantly. In particular, except for the Infiltration traffic, the ASR for all type of malicious traffic are close to 0. This means that our defense mechanisms work fairly well in preventing all black-box attacks for each type of malicious traffic. Overall, by combining the model voting ensembling, EAT and query detection mechanisms, our proposal can successfully prevent 5 types of mainstream black-box attacks from compromising neural-network based NID systems. 

\begin{figure*}[t]
\centering
\includegraphics[width=\textwidth]{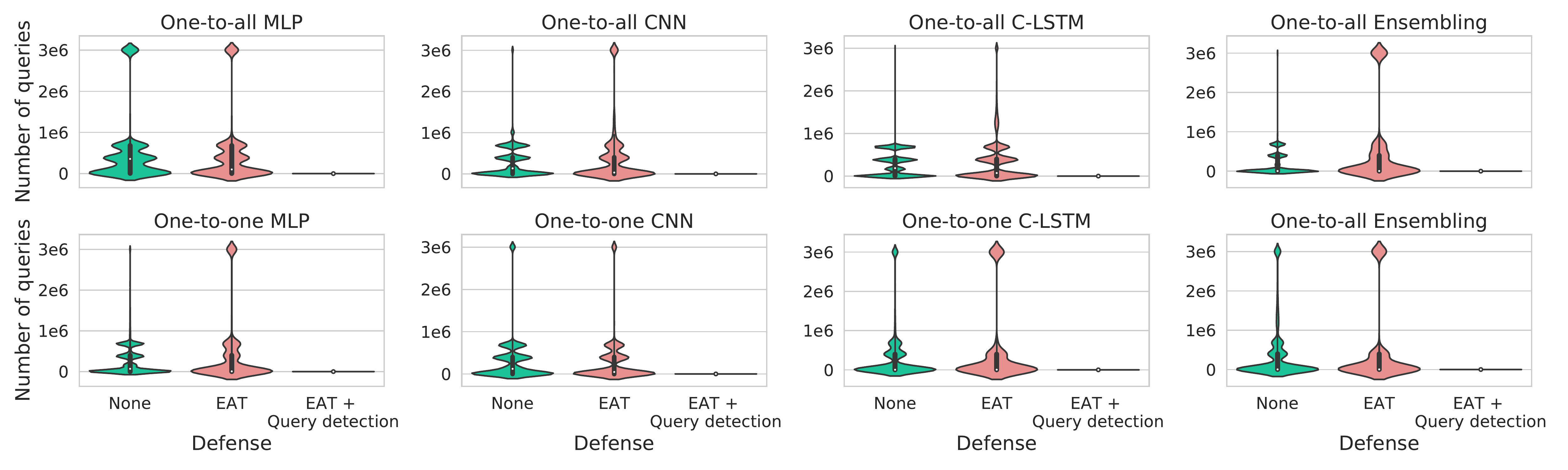}
\caption{The violin plots of the query distribution of all attacks at both NID scenarios. \label{fig:violin}}
\end{figure*}

\begin{table}[t]
\centering
\caption{The performance of MLP, CNN, C-LSTM and the ensembling model pre-EAT/post-EAT on IDS 2017 dataset in the one-to-all scenario.\label{tab:stat_2017}}
\begin{tabular}{|c|c|c|c|c|}
\hline
\textbf{Model}& \textbf{Accuracy} & \textbf{Precision} & \textbf{Recall} & \textbf{F1 score} \\ \hline
MLP        & 0.993/0.996    & 0.966/0.987     & 0.997/0.994  & 0.981/0.991    \\ \hline
CNN        & 0.997/0.998    & 0.988/0.993     & 0.997/0.996  & 0.992/0.994    \\ \hline
C-LSTM     & 0.996/0.998    & 0.984/0.990     & 0.998/0.999  & 0.991/0.994    \\ \hline
Ensembling & 0.992/0.997    & 0.963/0.984     & 0.999/0.999  & 0.980/0.992    \\ \hline
\end{tabular}
\end{table}

\begin{figure*}[t]
\centering
\includegraphics[width=\textwidth]{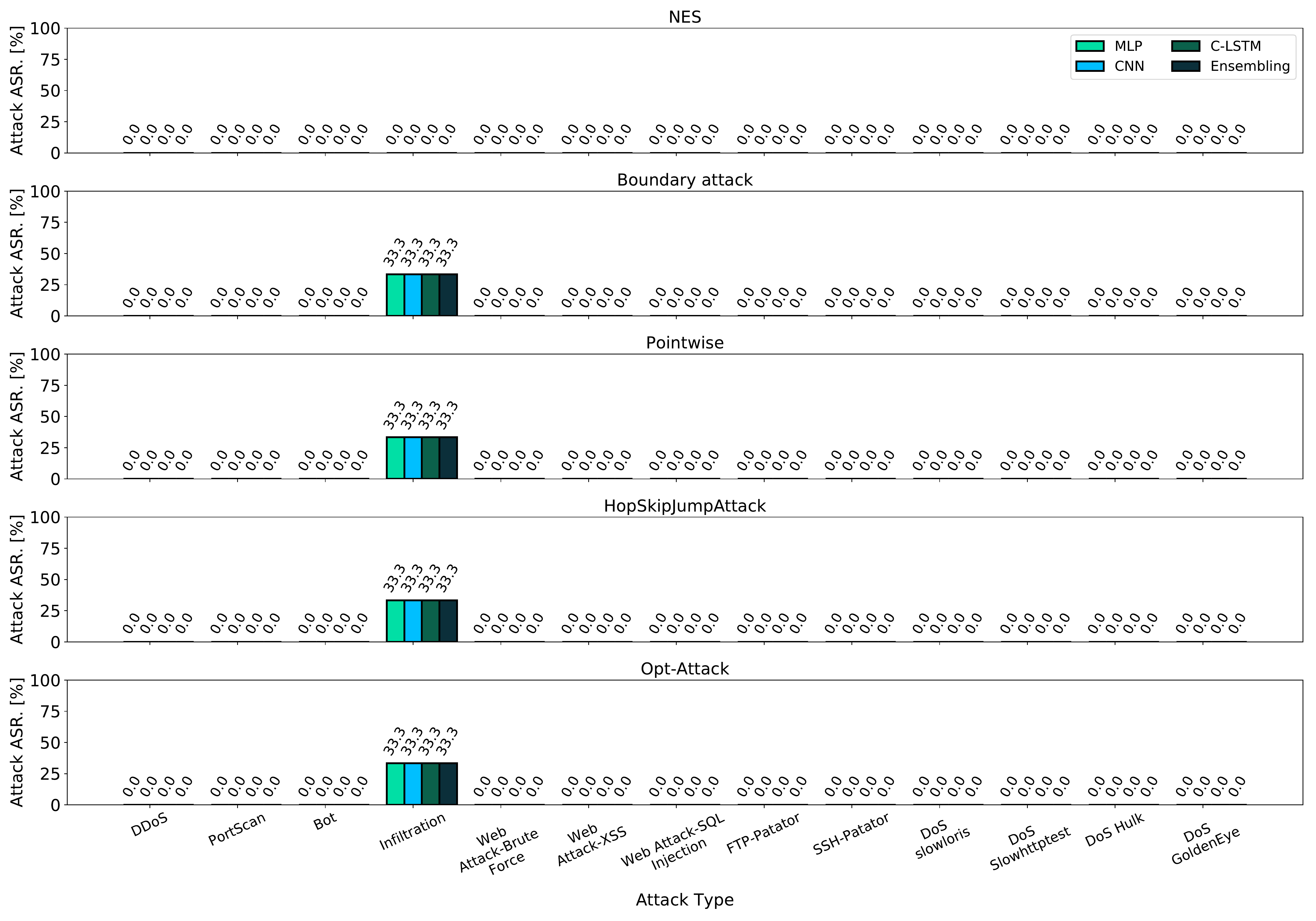}
\caption{ASRs over each type of attack after the EAT and query detection against all models in the one-to-all scenario on the IDS 2017 dataset. \label{fig:detection_asr_2017}}
\end{figure*}

\subsection{Zooming in on Infiltration Traffic}
To understand the reason why infiltration traffic samples escape detection, in Fig.~\ref{fig:violin} we examine the distribution of the number of queries of all successful attack attempts at each defense stage. We observe that while EAT does not change the query distribution significantly, most of the adversarial attacks bypassing the query detection only require 1 query. This means that the original traffic is already misclassified. The confusion matrices in Fig.~\ref{fig:cm_aug} further confirm this, as the Infiltration traffic yields both high misclassification rate and ASR. 


Recall that we removed three features (\emph{i.e.} flow duration, total time between two packets sent in the forward and backward direction) that may be affected by perturbations during training and evaluation. The total time between two packets sent in the backward direction is however essential for identifying the infiltration traffic, according to our experiments. Once this feature is added back, the detection rate increases to over 97\% and our defense mechanisms operate well in this case. We leave this as future work, which requires further in-depth study.

\section{Effectiveness in Different Landscapes}
To further demonstrate the effectiveness of our defense mechanism, we re-stage the \name flow (\ie NID models training, 5 black-box attacks, and 3 defense mechanism) on a different dataset, namely CICIDS2017 \cite{sharafaldin2018toward}. The CICIDS2017 dataset is the predecessor of the CSE-CIC-IDS2018 dataset, where similar benign/abnormal traffic flows were collected and similar features extracted. 

We show the pre-EAT and post-EAT NID performance of all models considered in Table.~\ref{tab:stat_2017}. Observe that all NID models obtain excellent performance, as they all achieve over 98\% in terms of F1 score. After EAT, their performance are further improved which demonstrate that the EAT can effectively improve the robustness of NID models. We employ the aforementioned black-box attacks to craft adversarial samples on 10,000 malicious traffic flows. After applying all defense mechanism proposed, we obtain the ASR for each types of attacks in CICIDS2017 dataset in Fig.~\ref{fig:detection_asr_2017}. Observe that except for Infiltration, the ASR for all attacks are 0\% for all models. This complies with the performance we obtain on the CSE-CIC-IDS2018 dataset, which demonstrates that our defense methods can generalize well, thus being a reliable mechanism for the defense of deep learning driven NID systems. On the other hand, we observe that the defense performance obtained in this dataset appears better than the CSE-CIC-IDS2018. The reason is that the traffic patterns are less diverse in the CICIDS2017 dataset, thus it is easier for NID models to learn. This is reflected by the detection accuracy, where the F1 score of each NID models in the CICIDS2017 is higher. As a result, this naturally hinders a proportion of adversarial samples, which fall in similar traffic patterns.

\section{Summary}
In this chapter, we introduced \tita, a framework for defending against adversarial attacks on deep learning based NIDS. We trained 3 state-of-the-art deep learning models (MLP, CNN and C-LSTM) on a publicly available dataset, then employed 5 classes of black-box, decision-based adversarial attacks to compromise the neural models. Experiments show that despite having high detection rates, deep learning based NIDS are vulnerable to adversarial samples. To strengthen NIDS against such threats, we proposed three defense methods, namely model voting ensembling, ensembling adversarial training, and query detection. The combination of these can reduce the success rate of all attacks considered, bringing detection rates close to 100\% on most malicious traffic. Future work will focus on handling infiltration traffic, which appears more resilient to NID models and defense approaches. 

%% file: chap8.tex
\chapter{Future Work and Conclusion\label{chap:conclude}}

\section{Future Research Perspectives}
As deep learning is achieving increasingly promising results in the mobile networking domain, several important research issues remain to be addressed in the future. Some of which we discuss next, before we conclude this thesis. 

\subsection{Serving Deep Learning with Massive High-Quality Data}
Deep neural networks rely on massive and high-quality data to achieve good performance. When training a large and complex architecture, data volume and quality are very important, as deeper models usually have a huge set of parameters to be learned and configured. This issue remains true in mobile network applications. Unfortunately, unlike in other research areas such as computer vision and NLP, high-quality and large-scale labeled datasets still lack for mobile network applications, because service provides and operators keep the data collected confidential and are reluctant to release datasets. While this makes sense from a user privacy standpoint, to some extent it restricts the development of deep learning mechanisms for problems in the mobile networking domain. Moreover, mobile data collected by sensors and network equipment are frequently subject to loss, redundancy, mislabeling and class imbalance, and thus cannot be directly employed for training purpose.

To build intelligent 5G mobile network architecture, efficient and mature streamlining platforms for mobile data processing are in demand. This requires considerable amount of research efforts for data collection, transmission, cleaning, clustering, transformation, and annonymization. Deep learning applications in the mobile network area can only advance if researchers and industry stakeholder release more datasets, with a view to benefiting a wide range of communities.

\subsection{Deep Learning for Spatio-Temporal Mobile Data Mining}\label{sec:st-traffic}
\begin{figure}[t]
\begin{center}
\includegraphics[width=\columnwidth]{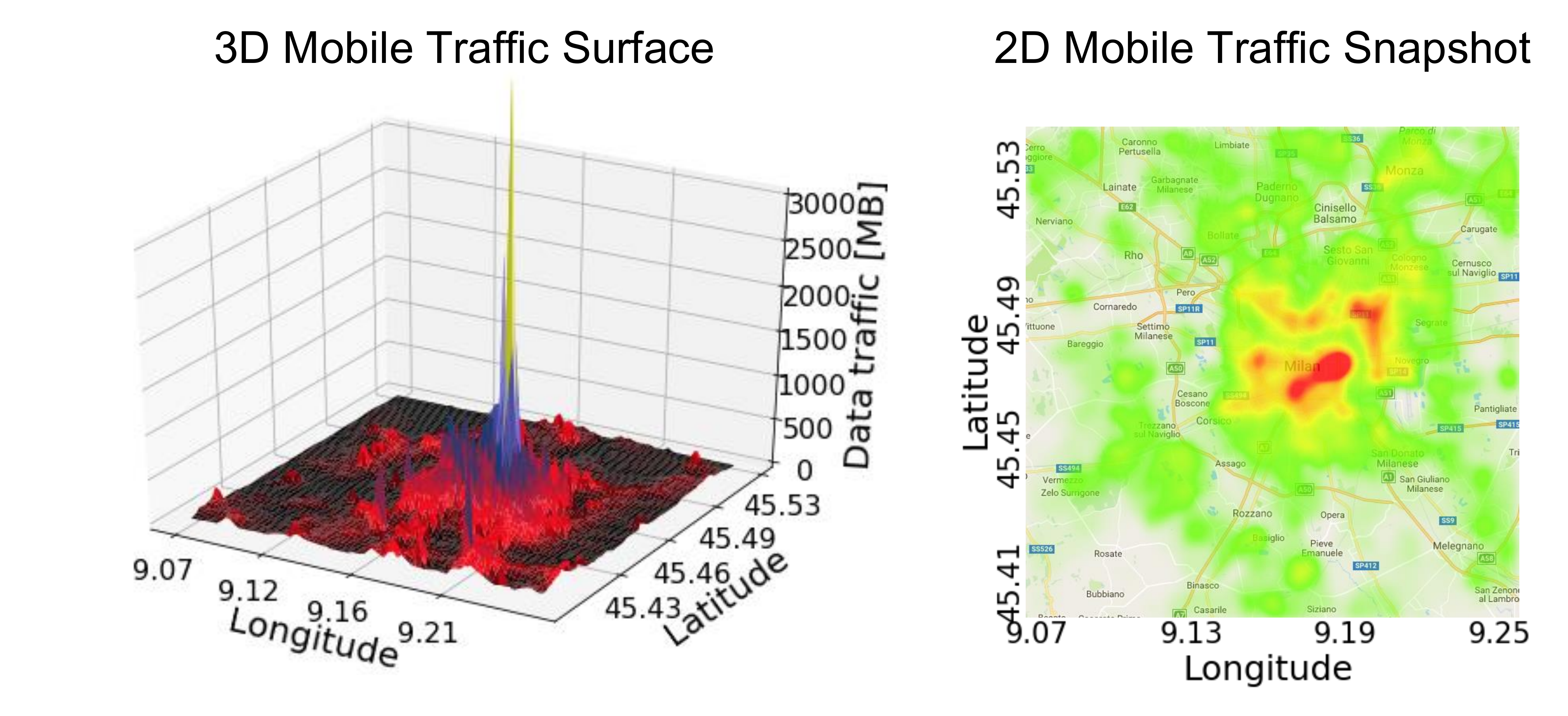}
\end{center}
\caption{\label{fig:mtraffic} Example of a 3D mobile traffic surface (left) and 2D projection (right) in  Milan, Italy. Figures adapted from \cite{zhang2017zipnet} using data from \cite{barlacchi2015multi}.}
\end{figure}

Accurate analysis of mobile traffic data over a geographical region is becoming increasingly essential for event localization, network resource allocation, context-based advertising and urban planning \cite{furno2017joint}. However, due to the mobility of smartphone users \cite{li2019learning, ouyang2016deepspace}, the spatio-temporal distribution of mobile traffic \cite{wang2015understanding} and application popularity \cite{marquez2017apps} are difficult to understand (see the example city-scale traffic snapshot in Fig.~\ref{fig:mtraffic}).
Recent research suggests that data collected by mobile sensors (e.g. mobile traffic) over a city can be regarded as pictures taken by panoramic cameras, which provide a city-scale sensing system for urban surveillance \cite{liu2015urban}. These traffic sensing images enclose information associated with the movements of individuals \cite{naboulsi2016large}. 

\begin{figure}[htb]
\begin{center}
\includegraphics[width=\columnwidth]{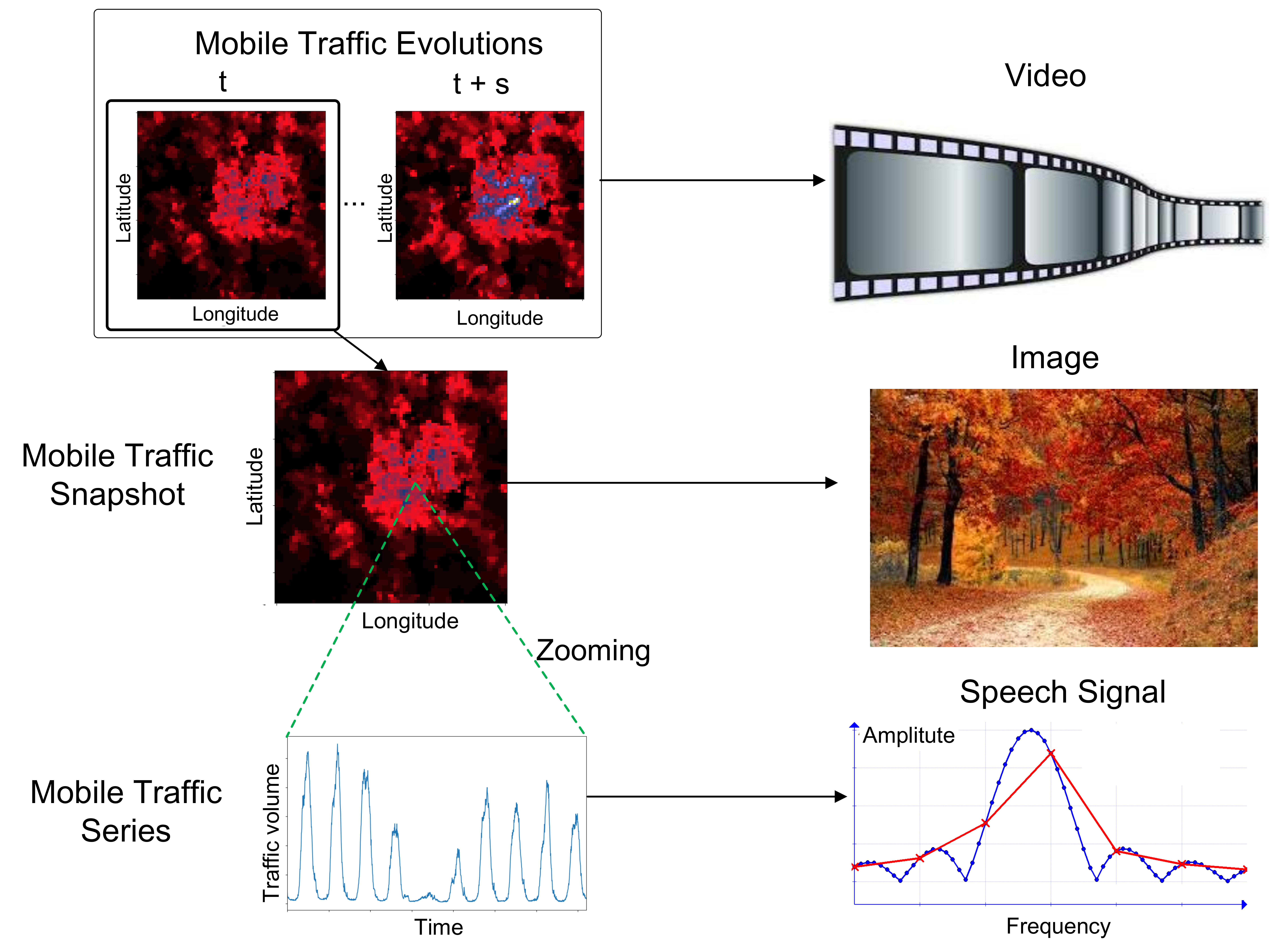}
\end{center}
\caption{\label{fig:compare} Analogies between mobile traffic data consumption in a city (left) and other types of data (right). }
\end{figure}

From both spatial and temporal dimensions perspective, we recognize that mobile traffic data have important similarity with videos or speech, which is an analogy made recently also in \cite{zhang2017zipnet} and exemplified in Fig.~\ref{fig:compare}.
Specifically, both videos and the large-scale evolution of mobile traffic are composed of sequences of ``frames''. Moreover, if we zoom into a small coverage area to measure long-term traffic consumption, we can observe that a single traffic consumption series looks similar to a natural language sequence. These observations suggest that, to some extent, well-established tools for computer vision (e.g. CNN) or NLP (e.g. RNN, LSTM) are promising candidate for mobile traffic analysis.

Beyond these similarity, we observe several properties of mobile traffic that makes it unique in comparison with images or language sequences. Namely,

\begin{enumerate}
\item The values of neighboring `pixels' in fine-grained traffic snapshots are not significantly different in general, while this happens quite often at the edges of natural images.
\item Single mobile traffic series usually exhibit some periodicity (both daily and weekly), yet this is not a feature seen among video pixels.
\item Due to user mobility, traffic consumption is more likely to stay or shift to neighboring cells in the near future, which is less likely to be seen in videos.
\end{enumerate}
Such spatio-temporal correlations in mobile traffic can be exploited as prior knowledge for model design. We recognize several unique advantages of employing deep learning for mobile traffic data mining:

\begin{enumerate}
\item CNN structures work well in imaging applications, thus can also serve mobile traffic analysis tasks, given the analogies mentioned before.
\item LSTMs capture well temporal correlations in time series data such as natural language; hence this structure can also be adapted to traffic forecasting problems. 
\item GPU computing enables fast training of NNs and together with parallelization techniques can support low-latency mobile traffic analysis via deep learning tools.
\end{enumerate}
In essence, we expect deep learning tools tailored to mobile networking, will overcome the limitation of traditional regression and interpolation tools such as Exponential Smoothing~\cite{tikunov2007traffic}, Autoregressive Integrated Moving Average model~\cite{Kim2011}, or unifrom interpolation, which are commonly used in operational networks.

\subsection{\rev{Deep learning for Geometric Mobile Data Mining}}
\begin{figure*}[t]
\begin{center}
\includegraphics[width=\columnwidth]{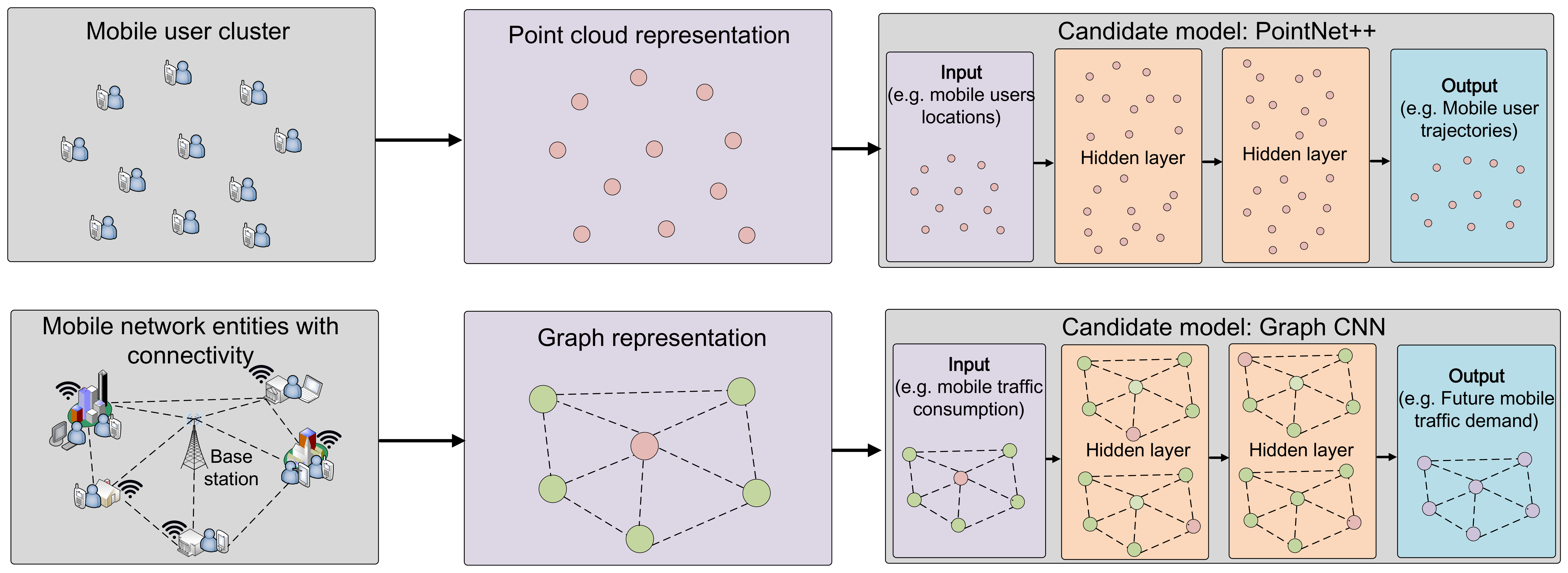}
\end{center}
\caption{\label{fig:geometric} Examples of mobile data with geometric properties (left), their geometric representations (middle) and their candidate models for analysis (right). PointNet++ could be used to infer user trajectories when fed with point cloud representations of user locations (above); A GraphCNN may be employed to forecast future mobile traffic demand at base station level (below).}
\end{figure*}

\rev{As discussed in Sec.~\ref{sec:adv}, certain mobile data has important geometric properties. For instance, the location of mobile users or base stations along with the data carried can be viewed as point clouds in a 2D plane. If the temporal dimension is also added, this leads to a 3D point cloud representation, with either fixed or changing locations. In addition, the connectivity of mobile devices, routers, base stations, gateways, and so on can naturally construct a directed graph, where entities are represented as vertices, the links between them can be seen as edges, and data flows  may give direction to these edges. We show examples of geometric mobile data and their potential representations in Fig.~\ref{fig:geometric}. At the top of the figure a group of mobile users is represented as a point cloud. Likewise, mobile network entities (e.g. base station, gateway, users) are regarded as graphs below, following the rationale explained below. Due to the inherent complexity of such representations, traditional ML tools usually struggle to interpret geometric data and make reliable inferences.}

\rev{In contrast, a variety of deep learning toolboxes for modeling geometric data exist, albeit not having been widely employed in mobile networking yet. For instance, PointNet \cite{charles2017pointnet} and the follow on PointNet++ \cite{qi2017pointnet} are the first solutions that employ deep learning for 3D point cloud applications, including classification and segmentation \cite{ioannidou2017deep}. We recognize that similar ideas can be applied to geometric mobile data analysis, such as clustering of mobile users or base stations, or user trajectory predictions. Further, deep learning for graphical data analysis is also evolving rapidly \cite{scarselli2009graph}. This is triggered by  research on Graph CNNs \cite{kipf2016semi}, which brings convolution concepts to graph-structured data. The applicability of Graph CNNs can be further extend to the temporal domain \cite{yuan2017temporal}. One possible application is the prediction of future traffic demand at individual base station level. We expect that such novel architectures will play an increasingly important role in network graph analysis and applications such as anomaly detection over a mobile network graph.}

\subsection{Deep Unsupervised Learning in Mobile Networks}
We observe that current deep learning practices in mobile networks largely employ supervised learning and reinforcement learning. However, as mobile networks generate considerable amounts of unlabeled data every day, data labeling is costly and requires domain-specific knowledge. To facilitate the analysis of raw mobile network data, unsupervised learning becomes essential in extracting insights from unlabeled data~\cite{usama2017unsupervised}, so as to optimize the mobile network functionality to improve QoE. 

The potential of a range of unsupervised deep learning tools including AE, RBM and GAN remains to be further explored. In general, these models require light feature engineering and are thus promising for learning from heterogeneous and unstructured mobile data. For instance, deep AEs work well for unsupervised anomaly detection~\cite{zhou2017anomaly}. Though less popular, RBMs can perform layer-wise unsupervised pre-training, which can accelerate the overall model training process. GANs are good at imitating data distributions, thus could be employed to mimic real mobile network environments. Recent research reveals that GANs can even protect communications by crafting custom cryptography to avoid eavesdropping~\cite{abadi2016learning}. All these tools require further research to fulfill their full potentials in the mobile networking domain.

\subsection{\rv{Deep Reinforcement Learning for Mobile Network Control}}
Many mobile network control problems have been solved by constrained optimization, dynamic programming and game theory approaches. Unfortunately, these methods either make strong assumptions about the objective functions (e.g. function convexity) or data distribution (e.g. Gaussian or Poisson distributed), or suffer from high time and space complexity. As mobile networks become increasingly complex, such assumptions sometimes turn unrealistic. The objective functions are further affected by their increasingly large sets of variables, that pose severe computational and memory challenges to existing mathematical approaches.   

In contrast, deep reinforcement learning (DRL) does not make strong assumptions about the target system. It employs function approximation, which explicitly addresses the problem of large state-action spaces, enabling reinforcement learning to scale to network control problems that were previously considered hard. Inspired by remarkable achievements in Atari~\cite{mnih2015human} and Go \cite{Silver1140} games, a number of researchers begin to explore DRL to solve complex network control problems. However, these works only scratch the surface and the potential of DRL to tackle mobile network control problems remains largely unexplored. We recognize several potential of employing DRL to address problems in the mobile networking domain. \rv{Specifically, DRL allows network entities to learn knowledge over complex and high-dimensional mobile network environments, which enables network controllers to deliver superior performance without complete and accurate network information. In addition, DRL can make instant decisions with modern computing machines, which is essential for reducing network latency.} 

\rv{There have already been many encouraging DRL applications in the networking domain.} 
For instance, as DeepMind trains a DRL agent to reduce Google's data centers cooling bill,\footnote{DeepMind AI Reduces Google Data Center Cooling Bill by 40\% \url{https://deepmind.com/blog/deepmind-ai-reduces-google-data-centre-cooling-bill-40/}} DRL could be exploited to extract rich features from cellular networks and enable intelligent on/off base stations switching, to reduce the infrastructure's energy footprint. In addition, such exciting applications make us believe that advances in DRL that are yet to appear can revolutionize the autonomous control of future mobile networks, \rv{and we will further explore this area in the future.}

\section{Conclusions}
This thesis address several problems in the mobile networking domain using deep learning. We surveyed research at the intersection between deep learning and mobile networking domains. We then presented two deep learning frameworks, \ie STN and CloudLSTM tailored to the precise city-scale mobile traffic forecasting with different data structures (\ie city grid and geospatial point cloud). We introduced the MTSR technique, which enables to reduce mobile traffic data post-processing overhead by inferring fine-grained mobile traffic distribution from its coarse-grained counterparts using a novel GAN based architecture. Subsequently, a MTD technique was proposed to infer the city-scale service-wise mobile traffic consumption from aggregates, thus reducing the cost and privacy concerns raised by traditional deep packet inspection. We then studied the vulnerability of deep anomaly detectors against black-box adversarial attacks, and propose three defense mechanisms to minimize the risks of this type of threats.

Overall, we believe this thesis not only represents a pioneering step towards building deep neural intelligence to automate the analytics and management of mobile networks, but also can be used as an valuable handbook for those who intend to further explore this territory.